\documentclass[11pt]{madoc}
\usepackage{german}
\usepackage{graphicx}
\usepackage{color}
\usepackage{amsmath}
\usepackage{amssymb}
\usepackage{graphpap}
\newtheorem{algorithm}{Algorithm}[chapter]
\begin{document}

\newcommand{\bra}[1]{\left< #1 \right|}
\newcommand{\ket}[1]{\left| #1 \right>}
\newcommand{\sprod}[2]{\left< #1 | #2 \right>}
\newcommand{\figpanel}[2]{
\begin{tabular}{@{\hspace{0em}}c@{\hspace{0em}}}
#1 \\
#2 
\end{tabular}
}
\newcommand{\AB}{{\em AB}}
\newcommand{\A}{{\em A}}
\newcommand{\B}{{\em B}}
\newcommand{\cdtxt}{$c(2\times2)_{\mathrm{Cd}}$}
\newcommand{\cdtxo}{$(2\times1)_{\mathrm{Cd}}$}
\newcommand{\tetxo}{$(2\times1)_{\mathrm{Te}}$}
\newcommand{\zntxt}{$c(2\times2)_{\mathrm{Zn}}$}
\newcommand{\setxo}{$(2\times1)_{\mathrm{Se}}$}
\newlength{\ptlng}
\newsavebox{\ptbox}
\settowidth{\ptlng}{$+\hspace{\ptlng}\mbox{\scalebox{1.2}{$\times$}}$}
\addtolength{\ptlng}{-1.5\ptlng}
\savebox{\ptbox}{$+\hspace{\ptlng}\raisebox{-0.5pt}{\mbox{\scalebox{1.2}{$\times$}}}$}
\newcommand{\plustimes}{\usebox{\ptbox}}
\newcommand{\botbasebox}[2]{
\raisebox{\depth}{\parbox{#1}{#2}}   
}
\newcommand{\botbase}[1]{
\raisebox{\depth}{#1}
}	
\newlength{\leinh}
\setlength{\leinh}{11pt}
\newlength{\zabst}
\newcommand{\rettezeilen}{ %
\setlength{\zabst}{\baselineskip} %
\addtolength{\zabst}{-12pt} %
\lineskip 2\zabst}
%
%
%
\newcommand{\linksbuend}{\rightskip=0pt plus2em \spaceskip=.3333em%
\xspaceskip=.5em\relax}
\newlength{\temp}
\newlength{\lbreit}
\newlength{\llbreit}
\settowidth{\temp}{September 2002\,--\,September 2002\hspace{5mm}}
\settowidth{\llbreit}{September 2002\,--\,September 2002}
\setlength{\lbreit}{\textwidth}
\addtolength{\lbreit}{-\temp}
\newcommand{\ltitel}[1]{\linebreak \rule{0pt}{3.5\leinh}{\bf #1}\ \leaders\hrule\hfill\hspace{0.001pt}\vspace{1.25\leinh}\linebreak}
\newcommand{\leintrag}[2]{{\parbox[t]{\llbreit}{%
\linksbuend #1}\hfill \parbox[t]{\lbreit}{\linksbuend #2}}}
\newcommand{\ltrenn}{\vspace{\leinh}\linebreak}

\raggedbottom
\setlength{\unitlength}{0.01\textwidth}
\originalTeX

\begin{titlepage}
\begin{center}
\newfont{\riesig}{cmssbx10 at 40pt}
{\riesig Surface properties\\
\rule{0pt}{45pt}of epitaxially grown\\
\rule{0pt}{45pt}crystals}\\ 
\normalfont\sffamily\huge  \rule{0pt}{72pt}Dissertation zur Erlangung des\\
naturwissenschaftlichen Doktorgrades\\
der Bayerischen Julius-Maximilians-Universit\"{a}t\\
W\"{u}rzburg\\
\rule{0pt}{72pt}vorgelegt von\\
\rule{0pt}{40pt}\normalfont\sffamily\Huge\bfseries Martin Ahr\\
\rule{0pt}{40pt}\normalfont\sffamily\huge aus Griesbach i. Rottal\\
\rule{0pt}{72pt}W\"{u}rzburg 2002
\end{center}
\end{titlepage}

\thispagestyle{empty}
\germanTeX
\vspace*{\fill}
\begin{flushleft}
Eingereicht am: \leaders\hbox{\,.}\hfill \ 18.\ April 2002\rule{5cm}{0pt}\\
\rule{0pt}{11pt}\\
bei der Fakult"at f"ur Physik und Astronomie\\
\rule{0pt}{11pt}\\
1. Gutachter: \leaders\hbox{\,.}\hfill \ Privatdozent Dr.\ Michael Biehl\rule{5cm}{0pt}\\
\rule{0pt}{11pt}\\
2. Gutachter: \leaders\hbox{\,.}\hfill \ Professor Dr.\ Wolfgang Kinzel\rule{5cm}{0pt}\\
der Dissertation\\
\rule{0pt}{11pt}\\
1. Pr"ufer: \leaders\hbox{\,.}\hfill \ Privatdozent Dr. Michael Biehl\rule{5cm}{0pt}\\
\rule{0pt}{11pt}\\
2. Pr"ufer: \leaders\hbox{\,.}\hfill \ Professor Dr. Eberhard Umbach\rule{5cm}{0pt}\\
der m"undlichen Pr"ufung\\
\rule{0pt}{11pt}\\
Tag der m"undlichen Pr"ufung: \leaders\hbox{\,.}\hfill \ 27.\ Juni 2002\rule{5cm}{0pt}\\
\rule{0pt}{11pt}\\
Doktorurkunde ausgeh"andigt am: \leaders\hbox{\,.}\hfill\rule{5cm}{0pt}\\
\vspace*{-\footskip}
\end{flushleft}
\pagebreak

\originalTeX
\chapter*{Abstract}

In this PhD thesis, we develop models of surfaces of epitaxially
grown crystals. In the introductory chapter \ref{einleitungskapitel},
we introduce the most important physical processes which occur on
crystal surfaces in a molecular beam epitaxy (MBE) environment. In the
first part of this work (chapters
\ref{dreikapitel}--\ref{zinkblendenmodell}) we model the (001) surface
of the II-VI semiconductors CdTe and ZnSe, our main focus being on
CdTe. In the second part (chapters \ref{waveletkapitel} and
\ref{coarsekapitel}), we study generic features of epitaxial growth
which are not specific to one particular material. These are kinetic
roughening and the formation of mounds. Both effects can be
investigated only after the deposition of a thick film on the
substrate. Therefore, we must restrict ourselves to simple models
which can be simulated with moderate computational effort.

In chapter \ref{dreikapitel}, we introduce a lattice gas model of flat
(001) surfaces of CdTe and ZnSe in thermal equilibrium. Both the
arrangement of metal atoms in vacancy structures and the dimerization
of nonmetal atoms are considered. The surface is represented by a
two-dimensional square lattice. We introduce anisotropic interactions between
nearest and next nearest neighbour sites which reflect the known
properties of the surface. The phase diagram of this model is
determined by means of transfer matrix calculations and Monte Carlo
simulations. The phases of the model are identified with different
reconstructions which have been observed in experiments. In
particular, the transition between a Cd terminated $c(2\times 2)$
reconstruction at low temperature and a Cd terminated $(2\times 1)$
reconstruction at high temperature which occurs on the CdTe surface
can be explained as an encompanying effect of an order-disorder phase
transition. Here, the small energy difference between both
reconstructions plays an important role.

Crystal surfaces in an MBE environment are not in thermal
equilibrium. Instead, non-equilibrium processes like growth and
sublimation occur. Are the essential features of our equilibrium model
preserved under these conditions or are there completely different,
nonequilibrium effects? To answer this question, we need to extend our
two-dimensional lattice gas to a model of a three-dimensional crystal
which can be used to simulate growth and sublimation. The basic idea
is that there is an isotropic attraction between particles in the bulk
of the crystal while particles on the surface interact with the
anisotropic interactions of the lattice gas.

Such a model which, however, contains several simplifications compared
to a realistic model of CdTe, is investigated in chapter
\ref{rekkapitel}. We neglect the dimerization of surface Te atoms and
simulate a simple cubic lattice instead of the zinc-blende lattice of
CdTe. In spite of these simplifications, many effects which have been
observed on sublimating CdTe(001) surfaces in vacuum and under an
external flux of pure Cd or pure Te can be reproduced. Additionally,
the behaviour of this model agrees qualitatively with that of a
simplified version of our lattice gas where the dimerization of Te
atoms is neglected. Thus, we have shown that the investigation of
two-dimensional lattice gas models in thermal equilibrium is an
appropriate tool to obtain insight into the behaviour of reconstructed
surfaces in an MBE environment. On the other hand, the non-equilibrium
conditions of sublimation induce characteristic deviations from the
behaviour of the equilibrium model which should be considered in
precise investigations.

Starting from this result, we develop a model of the CdTe(001)
surface which is closer to reality. It is introduced in section
\ref{zinkblendenmodell}. We simulate a zinc-blende lattice instead of
a simple cubic lattice. However, we still neglect the dimerization of
Te atoms. An appropriate choice of the parameter set yields a rough
quantitative agreement with experimental results. This indicates, that
an extension towards a truly realistic model of CdTe(001) should be
possible as soon as experimental data are available which allow for an
adaptation of the model. We perform simulations of atomic layer
epitaxy (ALE). Although this process is widely used in technological
applications and basic research, to our knowledge it has never been
investigated by means of Monte Carlo simulations before. We find a
growth rate of one half monolayer per cycle and an alternation between
a rough and a flat surface in subsequent cycles. Both effects have
also been observed in experiments.

In chapter \ref{waveletkapitel}, we investigate the fractal properties
of kinetically roughened surfaces. We consider a model of an
unreconstructed surface of a simple cubic crystal during MBE growth at
low temperature. This model yields self-similar, fractal surfaces. We
apply the WTMM method to obtain the first complete singularity spectra
of such surfaces. The statistical properties of growing surfaces are
invariant under dynamic scaling. We identify the dynamic exponent
$\alpha$ with the H\"{o}lder exponent which maximizes the singularity
spectrum. This scaling behaviour is a generalization of the well-known
Family-Vicsek scaling. However, we measure a strong dependency of the
scaling exponents on the desorption rate which indicates that they are
non-universal.

Finally, in chapter \ref{coarsekapitel} we consider the coarsening of
mounded surfaces which are created during growth if the diffusion of
particles across steps is suppressed by a strong Schwoebel barrier. A
continuum model of this process which was studied first by Siegert
et.\ al.\ predicts a dependency of the dynamic exponents on the
symmetry of the surface. To test this prediction, we simulate growth
in the presence of an infinite Schwoebel barrier on different lattice
structures. We use an efficient simulation algorithm which traces the
motion of a single particle from deposition until it has reached a
state where it is strongly bound. We find, contrary to the results of
the continuum theory, that the scaling exponents are independent of
the surface symmetry. Instead, the exponents depend on the speed of
the diffusion of particles along step edges. The exponents which are
measured in the limit of strong step edge diffusion can be explained
by means of a simple consideration.

In the two appendices we explain the techniques which we have used to
investigate the models which are considered in this thesis. These are
the transfer matrix method and continuous time Monte Carlo simulations. 

\germanTeX
\selectlanguage{ngerman}

\chapter*{Zusammenfassung}

In dieser Doktorarbeit werden Modelle f\"{u}r die Oberfl\"{a}chen
epitaktisch gewachsener Kristalle entwickelt. Im Einleitungskapitel
\ref{einleitungskapitel} werden die wichtigsten Prozesse auf
Kristalloberfl\"{a}chen in einer f\"{u}r die Molekularstrahlepitaxie
(MBE) typischen Umgebung vorgestellt. Im ersten Teil der Arbeit
(Kapitel \ref{dreikapitel} -- \ref{zinkblendenmodell}) pr\"{a}sentieren
wir Modelle f\"{u}r die (001)-Oberfl\"{a}chen der II-VI Halbleiter
CdTe und ZnSe. Haupts\"{a}chlich besch\"{a}ftigen wir uns dabei mit
CdTe. Im zweiten Teil (Kapitel \ref{waveletkapitel} und
\ref{coarsekapitel}) werden 
allgemeine Eigenschaften des Wachstumsprozesses untersucht, die nicht
an ein spezielles Material gebunden sind.
Dies
sind das kinetische Aufrauen wachsender Oberfl\"{a}chen und das
H\"{u}gelwachstum. Da beide Effekte nur nach dem Aufdampfen dicker
Schichten beobachtet werden k\"{o}nnen, m\"{u}ssen wir uns bei ihrer
Untersuchung auf einfache Modelle beschr\"{a}nken, die mit relativ
wenig Rechenaufwand simuliert werden k\"{o}nnen.

In Kapitel \ref{dreikapitel} stellen wir ein Gittergasmodell f\"{u}r
flache CdTe(001)- und ZnSe(001)-Ober\-fl\"{a}\-chen im thermischen
Gleichgewicht vor. Sowohl die Anordnung der Metallatome in
Leerstellenstrukturen als auch die Dimerisierung der Nichtmetallatome
werden ber\"{u}cksichtigt. Die Oberfl\"{a}che wird durch ein
zweidimensionales Quadratgitter dargestellt. Wir f\"{u}hren anisotrope
Wechselwirkungen zwischen n\"{a}chsten und \"{u}bern\"{a}chsten
Nachbarpl\"{a}tzen ein, in denen die bekannten Eigenschaften der
Oberfl\"{a}che ber\"{u}cksichtigt werden. Das Phasendiagramm des
Modells wird mit Transfermatrixrechnungen und Monte Carlo Simulationen
bestimmt. Die einzelnen Phasen werden mit verschiedenen experimentell
untersuchten Rekonstruktionen identifiziert. Insbesondere kann der
\"{U}bergang von einer Cd-terminierten $c(2\times 2)$-Rekonstruktion
bei tiefen Temperaturen zu einer ebenfalls Cd-terminierten $(2\times
1)$-Rekonstruktion bei hohen Temperaturen, der auf
CdTe-Oberfl\"{a}chen beobachtet wurde, als Folge eines
Ordnungs-Unordnungs-Phasen\"{u}bergangs erkl\"{a}rt werden. Dabei
spielt der kleine Ener\-gie\-un\-ter\-schied zwischen den beiden
Rekonstruktionen eine wesentliche Rolle.

Kristalloberfl\"{a}chen in einer MBE-Anlage befinden sich nicht im
thermischen Gleichgewicht. Stattdessen finden
Nichtgleichgewichtsprozesse wie Wachstum und Sublimation
statt. Bleiben die wesentlichen Eigenschaften unseres
Gleichgewichtsmodells unter diesen Bedingungen erhalten oder treten
vollkommen neue Effekte auf? Um diese Frage zu beantworten, m\"{u}ssen
wir unser zweidimensionales Gittergas zu einem Modell eines
dreidimensionalen Kristalls erweitern, mit dem Wachstum und Sublimation
simuliert werden k\"{o}nnen. Die grundlegende Idee dabei ist, dass
sich Atome im Inneren des Kristalls isotrop anziehen, w\"{a}hrend
zwischen Teilchen an der Oberfl\"{a}che die anisotropen
Wechselwirkungen des Gittergases wirken.

Ein solches Modell, das im Vergleich zu einer realistischen Simulation
des CdTe allerdings stark vereinfacht ist, betrachten wir in Kapitel
\ref{rekkapitel}. Wir vernachl\"{a}ssigen die Dimerisierung von
Te-Atomen an der Oberfl\"{a}che und simulieren ein einfach kubisches
Gitter anstelle des Zinkblendengitters des CdTe. Trotz dieser
Vereinfachungen k\"{o}nnen mit diesem Modell viele Effekte
reproduziert werden, die auf sublimierenden CdTe(001)-Oberfl\"{a}chen
im Vakuum und unter einem externen Fluss aus reinem Cd oder reinem
Te beobachtet wurden. Weiterhin stimmt das Verhalten des Modells
qualitativ mit dem einer vereinfachten Version unseres Gittergases
\"{u}berein, in dem die Dimerisierung von Te-Atomen nicht
ber\"{u}cksicht wird. Damit ist gezeigt, dass die Untersuchung
zweidimensionaler Gittergasmodelle im thermischen Gleichgewicht
prinzipiell ein geeignetes Werkzeug zum Verst\"{a}ndnis
rekonstruierter Oberfl\"{a}chen unter MBE-Bedingungen
ist. Andererseits verursachen die Nichgleichgewichtsbedingungen der
Sublimation aber auch charakteristische Abweichungen vom Verhalten des
Gleichgewichtsmodells, die in genauen Untersuchungen
ber\"{u}cksichtigt werden sollten.

Ausgehend von diesem Ergebnis entwickeln wir ein
realit\"{a}tsn\"{a}heres Modell der CdTe(001)-Oberfl\"{a}che, das in
Kapitel \ref{zinkblendenmodell} vorgestellt wird. Wir simulieren das
Zinkblendengitter anstelle eines einfach kubischen Gitters. Die
Dimerisierung von Te-Atomen wird aber weiterhin vernachl\"{a}ssigt.
Durch Wahl eines geeigneten Parametersatzes k\"{o}nnen experimentelle
Ergebnisse nicht nur rein qualitativ, sondern auch grob quantitativ
reproduziert werden. Das deutet darauf hin, dass eine Erweiterung
zu einem wirklich realistischen Modell m\"{o}glich sein sollte, sobald
geeignete experimentelle Daten zur Anpassung vorliegen. Mit diesem
Modell werden Simulationen der Atomlagenepitaxie (ALE)
durchgef\"{u}hrt. Obwohl dieser Prozess in Technik und
Grundlagenforschung h\"{a}ufig angewendet wird, wurde er unseres
Wissens nach noch nie in Monte Carlo Simulationen untersucht. Wir
beobachten eine Wachstumsrate von einer halben Monolage pro Zyklus und
einen Wechsel zwischen rauen und glatten Oberfl\"{a}chen in
aufeinanderfolgenden Zyklen. Beide Effekte wurden auch in Experimenten
beobachtet.

In Kapitel \ref{waveletkapitel} werden die fraktalen Eigenschaften
kinetisch aufgerauter Oberfl\"{a}chen untersucht. Wir betrachten ein
Modell einer unrekonstruierten Oberfl\"{a}che eines einfach kubischen
Kristalls w\"{a}hrend des MBE-Wachstums bei niedriger
Temperatur. Dabei entstehen selbst\"{a}hnliche Oberfl\"{a}chen, deren
multifraktale Spektren mit Hilfe der WTMM-Methode erstmals
vollst\"{a}ndig bestimmt werden konnten. Die statistischen
Eigenschaften wachsender Oberfl\"{a}chen sind invariant unter der
dynamischen Skalentransformation. Wir identifizieren den dynamischen
Exponenten $\alpha$ mit demjenigen H\"{o}lder-Exponenten, der das
multifraktale Spektrum maximiert. Dieses Skalenverhalten stellt eine
Verallgemeinerung der aus der Literatur bekannten
Family-Vicsek-Skaleninvarianz dar. Allerdings messen wir eine starke
Abh\"{a}ngigkeit der Skalenexponenten von der Desorptionsrate, was
darauf hindeutet dass diese nicht universell sind.

Abschlie{\ss}end untersuchen wir in Kapitel \ref{coarsekapitel} die
Vergr\"{o}berungsdynamik h\"{u}geliger Ober\-fl\"{a}\-chen, wie sie
w\"{a}hrend des Wachstums entstehen, wenn die Teilchendiffusion
\"{u}ber Stufen hinweg durch eine starke Schw\"{o}belbarriere
unterdr\"{u}ckt wird. Ein Kontinuumsmodell dieses Vorgangs, das zuerst
von Siegert et.\ al.\ untersucht wurde, sagt eine Abh\"{a}ngigkeit der
dynamischen Skalenexponenten von der Symmetrie der Oberfl\"{a}che
voraus. Um diese Vorhersage zu \"{u}berpr\"{u}fen, simulieren wir
das Wachstum mit unendlich hoher Schw\"{o}belbarriere auf verschiedenen
Gitterstrukturen. Dabei wird ein effizienter Simulationsalgorithmus
verwendet, in dem jeweils die Bewegungen eines einzelnen Teilchens von
der Deposition bis zum Erreichen eines fest gebundenen Endzustands
verfolgt werden. Wir stellen fest, dass die Skalenexponenten,
anders als von der Kontinuumstheorie vorhergesagt, nicht von der
Oberfl\"{a}chensymmetrie abh\"{a}ngen. Stattdessen beobachten wir
eine Abh\"{a}ngigkeit der Exponenten von der Geschwindigkeit der
Diffusion entlang von Stufenkanten. Die bei starker Kantendiffusion
gemessenen Skalenexponenten k\"{o}nnen mit einer einfachen
Absch\"{a}tzung erkl\"{a}rt werden.

In den beiden Anh\"{a}ngen werden die Techniken, mit denen die in
dieser Arbeit betrachteten Modelle untersucht wurden, also die
Transfermatrixmethode und die Monte Carlo Simulation mit
kontinuierlicher Zeit, vorgestellt. 

\originalTeX

\tableofcontents

\chapter{The physics of crystal surfaces in an MBE environment
 \label{einleitungskapitel}}

\begin{figure}[hbt]
\botbase{\includegraphics{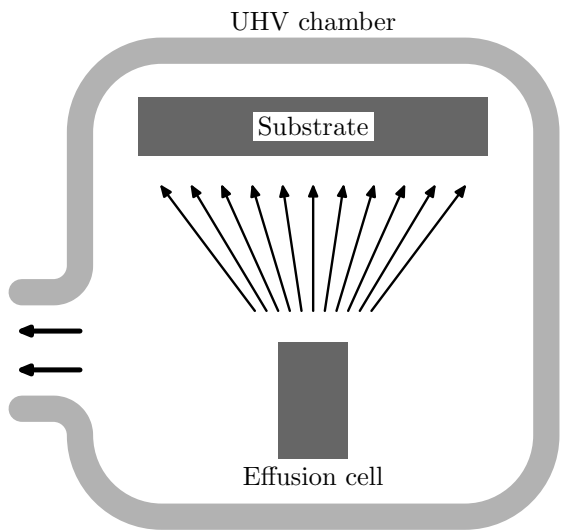}} \hfill
\botbasebox{0.5\textwidth}{
\caption{Schematical sketch of an MBE growth system. In an UHV chamber,
material which is evaporated in an effusion cell creates an atomic or
molecular beam. Particles from the beam are attached to the substrate
and create a film.\label{skizzembe}}}
\end{figure}

In the fabrication of modern semiconductor devices like computer chips
or laser diodes one frequently is interested in depositing thin films
on a monocrystalline substrate. Important techniques to achieve this
goal are molecular beam epitaxy (MBE) and related methods \cite{hs96}.
A sketch of an MBE chamber is shown in figure \ref{skizzembe}. The
whole process is performed in ultra high vacuum (UHV) to avoid the
intrinsic instabilities of crystal growth in a liquid or gassy medium
which result from diffusive and convective processes
\cite{pv97}. Material is evaporated in an effusion cell and creates an
atomic or molecular beam which is directed towards the substrate. The
mean free path of the particles in the beam is large compared to the
distance between the effusion cell and the substrate such that the
particles move on straight lines. Thus, the whole surface is exposed
to an even material flux. The success of MBE depends on the ability to
control the physical processes which occur on the surface of the
substrate itself. This might be done e.g.\ by a regulation of
temperature and particle flux.

MBE growth leads to the formation of a rich variety of surface
morphologies - to the regret of many practicians who are interested in
smooth films. From the theoretical point of view, these phenomena pose
a challenge to our understanding.  First, the physics of crystal
surfaces is determined by a complex interplay of various effects like
surface reconstruction and the adsorption, diffusion and desorption of
adatoms. Second, a crystal under MBE conditions is in a state far from
thermal equilibrium. In this case, no general theoretical framework
like equilibrium statistical mechanics is available. Instead, the
behaviour of the system might be determined by the details of
microscopic processes which occur on atomic lengthscales and short
timescales of $10^{-12} \,\mathrm{s}$. However, large surface formations
like mounds have extensions of up to $1 \mu\mathrm{m}$ and their
formation takes several minutes to hours. This wide range of relevant
length- and timescales has inspired a variety of approaches which
range from first principles methods based on quantum mechanics to
macroscopic differential equations. In this thesis, we study lattice
gas models which build a bridge between a purely microscopic and a
purely macroscopic point of view. In contrast to differential
equations, they consider the atomistic nature of matter. On the other
hand, their conceptual simplicity permits the investigation of large
systems on long timescales, which is not possible by means of
molecular dynamics or density functional methods. A large fraction of
our work is devoted to II-VI semiconductors, in particular CdTe and
ZnSe.  However, we also investigate more abstract models which aim at
an understanding of general, material-independent features.

In the description of a non-equilibrium process like growth or
sublimation of a crystal, two different aspects have to be
considered. On the one hand, systems which are not in thermal
equilibrium tend to approach to it. Therefore, an understanding of the
evolution of such a system requires an understanding of its
equilibrium configuration. In this thesis, we present the results of
models of pure surfaces, i.e.\ surfaces without adsorbates. Here, we
have to consider surface reconstructions and the roughening
transition. On the other hand, the way the system evolves depends on
the kinetics of processes which occur on the surface like deposition,
diffusion and desorption of atoms and molecules.

\section{Surface reconstruction \label{reconstintro}}

In the groundstate, i.e.\ at temperature $T = 0$, atoms in a
crystal arrange in such a way that the energy of the system is minimized. This
principle determines the material-specific lattice structure. Atoms at
the surface of the crystal have a smaller number of binding partners
than atoms in the bulk. Therefore, it is frequently energetically
favourable if the surface atoms arrange in a structure which has a
lower degree of symmetry than a planar section of the bulk. This
phenomenon is called {\em surface reconstruction}.

Reconstructions have been observed in various materials.  In II-VI
semiconductors, a variety of reconstructions consisting of vacancy
structures, dimers and trimers has been found \cite{ct97}. These will
be discussed in detail in chapers
\ref{dreikapitel}--\ref{zinkblendenmodell} where we present models of
reconstructed (001) surfaces of CdTe and ZnSe.

In general, a theoretical prediction whether a surface of a particular
material reconstructs and which reconstructions can be found is a hard
problem even at $T = 0$. It requires an investigation of the
properties of the chemical bonds in the crystal. For this purpose,
molecular statics using empirical potentials \cite{s99} and density
functional theory (DFT) \cite{gn94,gffh99,pc94} have been
employed. The computational burden of these methods is comparatively
high, which restricts their practical applicability to systems
consisting of only a few atoms. Due to the periodicity of crystal
surfaces this is not a severe restriction if one is interested in
ground state properties.

\begin{figure}[htb]
\botbase{\includegraphics{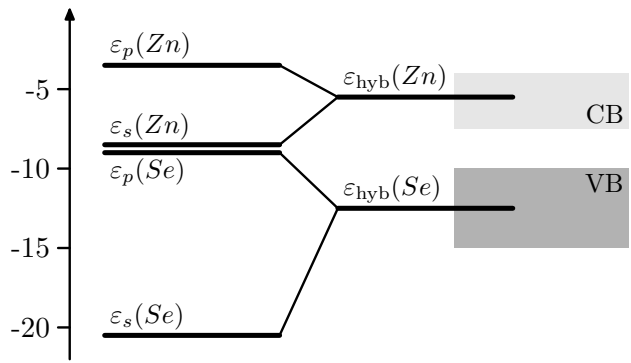}} \hfill
\botbasebox{0.42\textwidth}{\caption{Electronic states in ZnSe
according to \cite{p89}. All energy levels are given in
eV. $\varepsilon_s$ and $\varepsilon_p$ are the energies of the $s$- and
$p$-orbitals, respectively. These hybridize to form $sp^3$ orbitals
which have an energy $\varepsilon_{\mathrm{hyb}}$. The dangling bond
energy of anions is inside the valence band, that of cations is in the
conduction band.\label{energielevelintro}}}
\end{figure}

In the case of polar semiconductors like III-V and II-VI compounds,
there is a simple consideration which allows to {\em exclude}
energetically unfavourable surface configurations \cite{h79,p89}.
Consider the electronic structure of a compound semiconductor. A
schematical sketch of the energy levels in ZnSe is shown in figure
\ref{energielevelintro}. Other compound semiconductors like GaAs and
CdTe show a similar behaviour. From the $s$- and $p$-orbitals of the
valence electrons $sp^3$ hybrids are created which have an energy
$\varepsilon_{\mathrm{hyb}}$. The energy bands of the crystal are
obtained from linear combinations of these atomic orbitals. The
bonding combinations yield the valence band, the antibonding ones the
conduction band. At $T = 0$, the valence band is fully occupied and
the conduction band is empty.

In the bulk of the crystal, the four $sp^3$ orbitals of each atom
overlap with orbitals of neighbours and create chemical bonds. Each
bond is occupied by two electrons. Due to the different
electronegativities of the elements the metal atoms carry positive
charge (cations) and the atoms of the nonmetallic elements are
negatively charged (anions). Atoms at the surface have a smaller
number of nearest neighbours such that some of their orbitals are
directed out of the crystal. These are denoted as {\em dangling
bonds}. Cation dangling bonds are empty since their energy level is in
the conduction band. On the contrary, the binding energy of electrons
in the anion dangling bonds is below the valence band maximum such
that each of these orbitals is occupied by two electrons. The total
charge of each atom is the sum of its valence nuclear charge, the
charge of the bonding electrons and the charge of the electrons in the
dangling bonds. In general, it is nonzero.

In the majority of all conceivable arrangements of atoms in the
terminating layer of the crystal there is a net charge of the surface
which creates an electric field in the bulk. Energetically, this is
extremely unfavourable \cite{h79}. Consequently, in nature only surface
reconstructions without a net charge are realized. This condition is
fulfilled if the number of electrons is just sufficient to fill the
chemical bonds and the anion dangling bonds. In the literature, this
law is known as {\em electron counting rule} \cite{p89}.

At higher temperature, the properties of the surface are influenced by
thermodynamic effects. Frequently, phase transitions between different
reconstructions are observed.  Strictly speaking, phase transitions
exist only in the thermodynamic limit of an infinite number of degrees
of freedom. Therefore, their theoretical investigation requires the
study of systems with a large number of atoms which is beyond the
scope of first principles methods or molecular dynamics simulations
using realistic empirical potentials. In this case, effective models
which neglect several details but preserve the relevant features of
the atomic interactions are required. The models of CdTe and ZnSe
which we present in chapters
\ref{dreikapitel}--\ref{zinkblendenmodell} belong to this class.

\section{The roughening transition}

In thermal equilibrium at low temperature, surfaces of homogeneous
crystals which are oriented in a high symmetry direction are 
essentially flat apart from the pattern of a possible reconstruction. 
Since
atoms on the surface are bound more weakly than atoms in the bulk, the
system minimizes its energy by reducing the area of the surface. The
most frequent thermal excitations are single atoms or small groups of
atoms diffusing around on the surface and the vacancies they
have left when jumping out of the terrace.

The creation of a step of atomic height and length $l$ costs an energy
$\varepsilon_{\mathrm{step}} l$. In good approximation, the step energy
per unit length $\varepsilon_{\mathrm{step}}$ is independent of
temperature. Since atoms at the step edge have a smaller number of
binding partners than atoms which are incorporated in a terrace,
$\varepsilon_{\mathrm{step}}$ is positive. However, at the creation of
the step an entropy $s_{\mathrm{step}} l$ is gained since there are
many step configurations with equal energy. Consequently, the free
energy per unit length, $\gamma = \varepsilon_{\mathrm{step}} - T
s_{\mathrm{step}}$ decreases with increasing temperature. $\gamma$ is
denoted as {\em line tension of steps}.  At a critical temperature
$T_R$, there is a Kosterlitz-Thouless phase transition
\cite{n90,pv97}. At higher temperature, we have $\gamma = 0$. Then,
steps are created spontaneously by thermal fluctuations and the
surface becomes rough. This phase transition is called {\em roughening
transition}.

A quantitative measure of the roughness of a surface is the
height-height correlation function $G(\vec{l})$. Consider a system of
cartesian coordinates $(x, y, z)$ the $z$-axis of which is
perpendicular to the surface. In the remainder of this thesis we will
use this orientation of the coordinate system. We define the {\em
height} of the surface $h(\vec{x})$ as function of the vectors
$\vec{x}$ in the two-dimensional $(x, y)$ space which returns the
$z$-coordinate of the surface at $\vec{x}$.  Then, $G(\vec{l})$ is
defined as
\begin{equation}
G(\vec{l}) := \left< \left( h(\vec{x}) - h(\vec{x} + \vec{l})
\right)^2 \right>_{\vec{x}}.
\label{hoehendifferenzkorrelation}
\end{equation}
Below $T_R$, one finds that $G(\vec{l})$ remains finite in the limit
of an infinite distance $|\vec{l}|$. This expresses the fact that the
surface is essentially flat. On the contrary, at $T > T_R$ we have 
$G(\vec{l}) \sim \ln(|\vec{l}|)$. This logarithmic divergence
corresponds to a surface with strong fluctuations. 

Usually, molecular beam epitaxy is performed at temperatures far below
the roughening transition. This is also the case in the growth models
which are investigated in this thesis. Consequently, surfaces of
homogeneous crystals in thermal equilibrium should be flat at typical
MBE temperatures.

\section{Strained overlayers}

In {\em heteroepitaxy}, the deposited film consists of a different
material than the substrate. If substrate and adsorbate have different
lattice constants $a_s$ and $a_a$, equilibrium configurations where
the surface is not flat can be found even at low temperature. At the
adsorbate-substrate interface, the interaction of the adsorbate
particles with the substrate forces the adatoms to adapt to the
substrate lattice constant $a_s$. However, the adatom-adatom
interaction favours a distance $a_a$ between adsorbate particles. This
induces strain in the adsorbate. The equilibrium configuration depends
on the relative strength of the competing forces and the relative
lattice mismatch $(a_a - a_s)/a_s$. 

Three different growth modes have been observed
\cite{pv97,pgmpv00}. If the attraction between adsorbate particles is
much stronger than the adsorbate-substrate interaction and/or the
lattice mismatch is large, one obtains small droplets of the
adsorbate. This is called the {\em Volmer-Weber} growth mode. Conversely,
if the binding of the adsorbate atoms to the substrate is stronger
than the adsorbate-adsorbate bond, small amounts of the adsorbate form
a flat layer. The behaviour of a thicker adsorbate layer depends on
whether the adsorbate-adsorbate interaction is strong enough to
outweigh the elastic energy of the strain. If this is the case, thick
layers are flat. This is the {\em Frank-van der Merwe}
mode. Otherwise, in the {\em Stranski-Krastanov} mode, adsorbate
islands are formed on top of a thin wetting layer.  In very thick
films, dislocations are created. Their creation costs energy, however,
they permit the adsorbate to adapt to its own lattice constant
$a_a$. Recently, specialized simulation techniques have been developed
which allow for the simulation of strained heteroepitaxial growth
\cite{m00,mabk01,s99}. They are computationally expensive, which restricts
their practical applicability to $1+1$-dimensional models.

In the remainder of this thesis we restrict ourselves to
{\em homoepitaxial} growth where substrate and adsorbate particles are
identical.

\section{Motion of atoms on crystal surfaces \label{mobilityintro}}

Simulations of surfaces in an MBE environment are based on rules which
describe the motion of atoms. In growth models, the atomic nuclei are
treated as classical particles the motion of which is, in principle,
described by Newton's laws. Of course, the forces which act between the
particles are determined by the electronic structure which follows
quantum-mechanical laws. In {\em molecular dynamics} (MD) simulations,
the equation of motion of the particles is solved numerically. The
forces are calculated by means of empirical potentials or DFT.

In the energy landscape of a large number of atoms in phase space,
there are many local minima. These correspond to configurations where
the atoms arrange e.g.\ in a regular crystal lattice. If the heights of
the barriers between the minima are noticeably higher than $k T$, most
of the time the actual state of the system will fluctuate around one
of them. These are the thermally excited lattice vibrations of the
crystal. Very rarely, the system leaves a minimum and jumps into a
neighbouring one. These are processes where the arrangement of atoms
changes like e.g.\ a diffusion jump of an adatom on the surface. In
a MD simulation, the lattice vibrations are considered
explicitly. This is the reason why this method is computationally
extremely expensive.  On modern computers, the motion of several
thousand atoms can be traced for at most some nanoseconds, which is by
far too short to investigate MBE.

In Monte Carlo simulations, we investigate the system on a coarse
grained scale. Only changes of the arrangement of atoms are considered
explicitly, while the chaotic motion of atoms due to lattice
vibrations is considered implicitly in a statistical manner. It
determines the rate $\rho$ of events where the atomic arrangement
changes. One can show quite generally \cite{pon90} that for thermally
excited events $\rho$ is given in good approximation by an Arrhenius
law \cite{a1889}
\begin{equation}
\rho = \nu \exp \left( - \frac{E}{k T} \right). 
\label{arrheniusgesetz}
\end{equation}
$E$ is the height of the energy barrier which must be overcome to
change from the initial to the final minimum.  $\nu$ is an attempt
frequency which is on the order of magnitude of the Debye frequency of
the crystal. In general, it is reasonable to assume that it is
identical for all processes which occur on the surface. Compared to
MD, the Monte Carlo technique reduces the computational effort by many
orders of magnitude and makes it feasible to consider the typical
timescales of MBE.

Frequently, the position of the atoms in the relevant local energy
minima can be assigned to sites in a (potentially distorted) crystal
lattice. In this case, {\em lattice gas} models are appropriate.  We
consider a fixed crystal lattice.  Each lattice site is either
occupied by an atom or empty. At the deposition of an atom, the state
of one site changes from empty to occupied. If an atom is desorbed,
its site is switched from occupied to empty. At the diffusion of an
atom, the states of two sites are exchanged. In the presence of
surface reconstructions, the topology of the lattice at the surface
might be different from that in the bulk. It is possible to extend a
lattice gas model to include this effect by adding additional discrete
degrees of freedom.  Examples are the formation of Si dimers on
Si(001) in the simulations of Rockett et.\ al.\ \cite{br88,r90} and the
dimerization of Te atoms in our model of the CdTe(001) surface which
will be presented in section \ref{dreikapitel}.

In addition to the application in Monte Carlo simulations, lattice
gases and Arrhenius activated processes are the basis of most theories
which consider the dynamics of processes which occur at crystal
surfaces. Examples include nucleation theory \cite{vsh84,w95} and BCF
theory \cite{pv97,pv94}.

\subsection{Deposition \label{depositionintro}}

Atoms impinge on the substrate with a kinetic energy $\sim k
T_{\mathrm{eff}}$, where $T_{\mathrm{eff}}$ is the temperature of the
effusion cell. Typically, $T_{\mathrm{eff}}$ is higher than the
substrate temperature. Additional energy is gained from the formation
of chemical bonds to the surface. Consequently, the energy of the
newly arriving particles is considerably higher than that of
particles on the surface which are already thermalized at substrate
temperature. This leads to an increased mobility of adatoms until the
excess energy is dissipated.  The particles get trapped at
energetically favourable sites with high coordination number. Clearly,
such sites will be found preferentially in layers close to the
substrate. Therefore, the motion of the adatoms is biased in the
``downhill'' direction. The momentum of the particles in the beam
which is directed towards the substrate enhances this bias. In the
literature \cite{gs91,yhd98} results of molecular dynamics simulations
have been reported which confirm these qualitative considerations.

In our simulations, we consider this effect by means of effective
rules termed {\em incorporation} and {\em downhill funneling}. In
incorporation, a particle moves to the lattice site at the lowest
height within an {\em incorporation radius} $r_{i}$ around the
position where it hits the surface \cite{skr00}. The simulations
presented in \cite{gs91,yhd98} suggest that reasonable values of $r_i$
should be on the order of magnitude of a few lattice constants. In
downhill funneling \cite{e91,estp90}, adatoms perform a biased random
walk to nearest neighbour sites lying lower than the current site. The
particle stops when the height of all neighbour sites is equal to or
greater than the current height.

At the deposition of particles, {\em precursor states} can play an
important role. Particles which land on the surface first enter a
state where they are only weakly bound to the surface. Later, a strong
chemical bond is formed. Meanwhile, they may diffuse over a
considerable distance. In chapter \ref{zinkblendenmodell}, we
investigate a model where such weakly bound states play an important
role.

\subsection{Diffusion}

To understand the properties of diffusion, the concept of the {\em
potential energy surface} (PES) or {\em surface energy map} is
employed. It is assumed that the energy barriers at zero temperature
are a reasonable approximation for the energy barriers at growth
temperature. Consider an atom close to a surface. We fix its position
in the $x$- and $y$-direction at $ \vec{x} = (x, y)$. There is no
constraint in the $z$-coordinate of this test atom and the positions
of the atoms in the crystal. We determine the equilibrium
configuration at $T = 0$ by minimizing the total energy with respect to these
remaining degrees of freedom. Measuring the total energy of the system
as a function of $\vec{x}$, we obtain the PES of an adatom.  Local
minima of the PES correspond to sites where adatoms are stable with
respect to small perturbations e.g.\ from lattice vibrations.

In the investigation of the hopping of adatoms by means of Monte Carlo
simulations, we consider processes where atoms move between these
sites. The energy barrier that a single atom must overcome to hop to a
neighbouring site equals the difference between the energy of the
initial configuration and the saddle point of the PES which lies
between the initial and the final site. This information is sufficient
to calculate the ratios between the rates of various diffusion
processes which occur on the surface from the Arrhenius law (equation
\ref{arrheniusgesetz}). Interactions between atoms are
short-ranged. Consequently, the rates of diffusion processes depend
only on the local environment of the diffusing atom within a range of
a few lattice constants such that it is sufficient to consider a
limited number of typical configurations.

\begin{figure}[htb]
\botbase{\resizebox{0.65\textwidth}{!}{\includegraphics{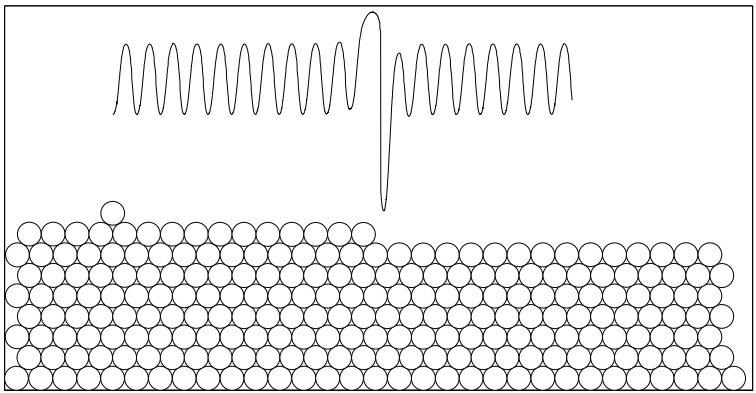}}}
\hfill \botbasebox{0.33\textwidth}{\caption{Potential energy of a test
atom on the surface of a two-dimensional Lennard-Jones crystal at zero
temperature as a function of its position.  By courtesy of
F. Much.\label{ljonespesintro}}}
\end{figure}
\vspace{-4mm}
At typical MBE growth temperatures, bulk diffusion can be
neglected. The dominant mechanism of material transport is {\em
surface diffusion}.   
We discuss some important effects in the simplified case of the
surface of a two-dimensional crystal of Lennard-Jones particles
\cite{m00,mabk01}. This is an example of a system without surface
reconstruction. Its PES is shown in figure \ref{ljonespesintro}. As
expected, the potential energy of the test particle has local minima
at lattice sites. Therefore, at the deposition of atoms the crystal
grows in its own lattice structure such that a lattice gas modelling
of this system would be appropriate. At a step edge,
where it is surrounded by a large number of neighbours an atom is
bound stronger than on a flat surface.  Consequently, there is a high
energy barrier for the detachment of an atom from the step edge.
Similarly, adatoms are bound to other adatoms. 

An adatom wich crosses a step edge from above must overcome a greater
energy barrier than that for diffusion on a flat terrace. This
so-called {\em Schwoebel effect} \cite{s69,ss66} originates from the
fact that a particle at the rim of the step has only a small number of
binding partners. Therefore, hopping down a step is suppressed
significantly compared to terrace diffusion.

The same effects are found in the physically relevant case of the
surface of a three-dimensional crystal. Additionally, we have to
consider the diffusion of atoms along step edges. In many materials,
{\em step edge diffusion} at straight steps is as fast as diffusion on
a flat terrace or even faster. Atoms in kinks of an edge have a high
coordination number. Consequently, particles get trapped at these {\em
kink sites}. Similar to the Schwoebel effect, energy barriers which
suppress step edge diffusion around corners are possible.

In addition to the hopping of atoms, diffusion can take place via
exchange processes. An adatom pushes another atom out of the surface
and occupies its site. The atom pushed out becomes an adatom at a
neighbouring site. From the viewpoint of Monte Carlo simulations of
homoepitaxy both processes are equivalent, since initial and final
state of hopping and exchange diffusion are identical. However, the
energy barriers and therefore the rates of both processes are
different. Whether diffusion occurs preferentially via hopping or
exchange is material dependent.

The PES of a reconstructed surface has the symmetry of the
reconstruction. This may lead to strong anisotropies in the diffusion
rates.

In the literature, PES and diffusion barriers for various materials
have been published which have been calculated by means of empirical
potentials \cite{jjs94,saan99,s99,wr91} and density functional theory
\cite{f98,kms99,kps01,ss94}. However, currently no such data are
available for the materials which we investigate in this
thesis. Therefore, in our models we will use parameterizations of
energy barriers which consider the effects which have been discussed
in this section.

\subsection{Desorption}

Similar to diffusion, the desorption of atoms from the surface is a
thermally activated process the rate of which is given by an Arrhenius
law. In general, the higher the coordination number of an atom, the
greater is the energy barrier for desorption. Consequently, the
desorption of single adatoms which diffuse on a flat terrace occurs at
the highest rate, while atoms which are attached to step edges or
incorporated in a flat surface are much more stable. In most cases it
is reasonable to assume that the energy barrier which must be overcome
to remove an atom from the surface equals the difference between the
energy of the surface without the adatom and a single atom in vacuum
and the energy of the crystal with the adatom at the surface.

\section{Epitaxial crystal growth \label{epitaxintro}}

In the following, we measure lengthscales in appropriate units such
that the rate at which particles are deposited at one site equals the
particle flux $F$, while diffusion constants of adatoms are identical
to hopping rates. This can be done by setting the unit length to a
small multiple of the lattice constant of the substrate which depends
on the geometry of the crystal lattice. The average time an atom stays
on the surface before it is incorporated into the crystal is
$1/F$. Diffusion and desorption processes which occur on much longer
timescales can be neglected. Typically, this is the case for the
detachment of particles from step edges and for the decay of islands
which consist of more than $i^{*}$ atoms. The {\em critical nucleus
size} $i^{*}$ is small at typical MBE temperatures.  Frequently, even
pairs of adatoms are stable such that $i^{*} = 1$.  Adatoms which are
deposited on the surface diffuse around until they stick to a step
edge or collide with more than $i^{*}$ other adatoms and form a stable
island. The latter process is called {\em nucleation}. Once an island
is formed, it captures more and more adatoms and grows. Thereby, the
adatom density in the vicinity of the island is reduced which
suppresses further nucleation events. Thus, islands keep a typical
distance $l_i$ to their neighbours.

Nucleation theory \cite{vsh84,w95} states that the minimal value of
$l_i$ which is obtained during the deposition of the first monolayer
on a perfectly flat surface has a powerlaw dependence on the ratio of
the adatom diffusion constant $D$ and $F$:
\begin{equation}
l_i \propto \left( \frac{D}{F} \right)^\kappa 
\end{equation} 
The value of $\kappa$ depends on various parameters like $i^{*}$, the
fractal dimension of islands, the mobility of small clusters of
adatoms and on whether a significant fraction of adatoms is
desorbed. In the simplest case of $i^{*} = 1$, compact islands and
vanishing desorption, we have $\kappa = 1/6$. We refer the reader to
refs. \cite{vsh84,w95} for a discussion of this subject. In the late
phase of this {\em submonolayer regime}, neighbouring islands coalesce
and $l_i$ increases again.

A perfectly flat surface is never realized in experiments although it
is the equilibrium configuration at growth temperature. On the one
hand, this is due to the fact that preparation techniques yield an
intrinsic surface morphology and thermal smoothing is extremely
slow. On the other hand, steps are created during growth by kinetic
roughening and the Schwoebel instability which will be discussed in
sections \ref{kineticroughenintro} and \ref{intromound}. 
As long as the average distance $l_s$
between steps is much greater than $l_i$, the influence of the steps
is weak and growth proceeds essentially the same way as on a flat
surface. However, if $l_s \ll l_i$ all adatoms are captured by the
steps and no islands nucleate. In this case, growth proceeds in {\em
step flow mode}. The influence of steps on growth can be investigated
particularly well on a {\em vicinal surface} which consists of flat
terraces separated by regularly spaced straight steps which are
parallel to each other. Experimentally, vicinal surfaces can be
created by cleaving a crystal in a small {\em miscut angle} to a high
symmetry surface.

\subsection{Kinetic roughening \label{kineticroughenintro}}

Clearly, the deposition of material on a substrate in an MBE chamber
is a stochastic process. Adatoms impinge on random positions. There
are random time intervals between successive deposition events at one
site. This randomness tends to create rough surfaces with a strongly
fluctuating height function $h(\vec{x})$ \cite{sb85,sm83}. On the
other hand, diffusion tends to smooth the surface. In the following,
we assume that growth starts at time $t = 0$. Since particles diffuse
at a finite speed, diffusion can level out fluctuations only at
lengthscales smaller than the {\em correlation length} $\xi(t)$ which
increases with time.  Properties of regions which are separated by a
greater distance are uncorrelated. In particular, there is a random
height difference between points at great distance. Consequently,
growing surfaces are always rough at large lengthscales. This
phenomenon is denoted as {\em kinetic roughening}, since it is induced
by the kinetics of diffusing adatoms. In contrast to the roughening of
surfaces in thermal equilibrium which occurs at high temperature,
kinetic roughening is most effective at low temperature where
diffusion is slow.

The appearance of growing surfaces on small lengthscales depends on
the properties of the diffusion process. In the {\em absence} of a
Schwoebel barrier, surfaces with fractal properties can be found. In
the modelling of these surfaces, frequently the concept of
statistically {\em self-affine} surfaces is employed
\cite{bs95}. Mathematically, a self-affine surface is defined as a
single-valued height function $h(\vec{x})$ the properties of which
are invariant under a simultaneous rescaling $\vec{x} \rightarrow b
\vec{x}$ and $h \rightarrow h_b(h)$ \cite{s90}. Here, $b$ is an
arbitrary real number. Clearly, an iterated transformation of the
rescaled system with a scaling factor $b'$ should yield the same
result as a rescaling of the originial system with a factor $b
b'$. This {\em group property} implies that $h_{b'}(h_b(h)) = h_{b
b'}(h)$, which is fulfilled for $h_b(h) = b^H h$.  Thus, the general
form of the scale transformation is
\begin{equation}
\vec{x} \rightarrow b \vec{x}; \; \; h \rightarrow b^H h.
\label{selbstaffinintro}
\end{equation}
The exponent $H$ is denoted as {\em Hurst exponent}. In the special
case $H = 1$, the surface is invariant under an isotropic rescaling of
lateral extension and height. This property is denoted as {\em
self-similarity}. 

In {\em statistically self-affine} surfaces, invariance under
rescaling is not fulfilled for $h(\vec{x})$
itself. Instead, we assume that there is a stochastic process which
creates surfaces such that scale invariance is fulfilled for 
measurends averaged over an ensemble of independent realizations. 

Scale invariance implies conditions for the functional form of various
quantities which can be used to measure the Hurst exponent. If
$\vec{x}$ is scaled by a factor $b$, the height-height correlation
$G(\vec{l})$ which was defined in equation
\ref{hoehendifferenzkorrelation} behaves like the square of the
surface height such that $G(b \vec{l}) = b^{2 H} G(\vec{l})$. This is
fulfilled if and only if
\begin{equation}
G(\vec{l}) =  g\left(\frac{\vec{l}}{l}\right) l^{2 H} 
\label{selbstaffing}
\end{equation}
where we have introduced $l = |\vec{l}|$.  Additionally, we consider
the standard deviation of the surface height in a square of size $L
\times L$,
\begin{equation}
w(L) = \left[ \frac{1}{L^2} \int_0^L dx \int_0^L dy \left( h(\vec{x})
- \left< h(\vec{x}) \right> \right)^2 \right]^{\frac{1}{2}}.
\label{ofbreite}
\end{equation}
Under the scale transformation \ref{selbstaffinintro},
a square of size $L$ is transformed into a square  of size $b
L$. $w(L)$ scales like the height function such that $w(b L) = b^H
w(L)$, which implies
\begin{equation}
w(L) = w_0 L^H. 
\label{selbstaffinw}
\end{equation}
Physical crystal surfaces are never statistically self-affine in the
strict mathematical sense. First, there cannot be any structures which
are smaller than the lattice constant $a$. This {\em lower cutoff}
implies that invariance under the transformation
\ref{selbstaffinintro} breaks down for too large $b$. Additionally,
there is an {\em upper cutoff} due to the finite correlation length
$\xi(t)$. At greater lengthscales, surface heights are uncorrelated
which implies that $w(L) = w_\infty = \mbox{const.}$ for $L \gg
\xi(t)$ and $G(\vec{l}) = 2 w_\infty^2$ for $l \gg \xi(t)$. However,
at late time where $\xi(t)$ is large, in several growth models there
is a wide range of lengthscales $a \ll l, L \ll \xi(t)$ where
equations \ref{selbstaffing} and \ref{selbstaffinw} are fulfilled in
good approximation. Surfaces with these properties are denoted as {\em
physical fractals}. However, to fulfil symmetry properties like
equation \ref{selbstaffinintro}, the mean surface height must be
subtracted. Therefore, we consider the {\em reduced surface height}
$f(\vec{x}) := h(\vec{x}) - \left< h(\vec{x}) \right>$ instead of
$h(\vec{x})$ itself in the investigation of symmetry properties of
growing surfaces.

Careful investigations have shown, that some growth models yield
surfaces with more complicated fractal properties which are denoted as
{\em multifractal} \cite{bbjkvz92,dslkg96,dsp97,k94}. In chapter
\ref{waveletkapitel}, we will investigate such a model.

\subsection{Unstable growth \label{intromound}}

On materials with a Schwoebel barrier, epitaxially
grown surfaces have a mounded morphology. In the submonolayer regime,
islands are formed at a typical distance $l_i$. Adatoms which land on
top of one of these islands diffuse around. Due to the Schwoebel
barrier there is only a small probability that they cross the island
edge and are incorporated in the layer below. Instead, they stay on
top of the island until they either desorb or meet other adatoms and
nucleate.  This process creates islands in the second layer before the
first layer is completed. Then, islands in the third layer nucleate
on top of these islands and so on. Thus, in the course of time mounds
build up. Their initial perimeter equals the distance between islands
in the submonolayer regime.

At the flank of a mound, there is a train of steps similar to a
vicinal surface. Adatoms which approach a step edge from below are
captured by the step. The attachment of atoms to steps from above is
hindered by the Schwoebel effect. Nucleation on terraces can be
neglected since the terrace width is necessarily smaller than $l_i$.
Thus, the majority of the adatoms on a terrace is attached to the
upper step.  This yields a {\em Schwoebel current} in the uphill
direction which tends to further increase the slope. However,
processes like incorporation and downhill funneling keep the mounds
from steepening infinitely. In these nonthermal processes, adatoms
which impinge close to a step edge move in the downhill
direction. This downhill current compensates the Schwoebel current at
a {\em magic slope} $m_0$ at which the steepening of mounds
stops. Frequently, neighbouring mounds coalesce such that the average
perimeter of the mounds increases with time. This process is denoted
as {\em coarsening}.

We consider the initial phase of mound formation where the slope $m$ is
much smaller than $m_0$ in the case $l_i \gg
1$. Then, there are large terraces such that the influence of
incorporation or downhill funneling can be neglected. In a simplifying
manner we assume that the Schwoebel barrier is infinite such that
there is no interlayer diffusion. In the first layer, $F \sigma$
particles per unit time and unit area are deposited, where $\sigma$ is the
fraction of the substrate which is not covered by the adsorbate. These
particles cover the substrate such that
\begin{equation}
\frac{d \sigma}{d t} = - F \sigma \; \; \Longrightarrow \; \; 
\sigma(t) = \exp (- F t). 
\end{equation} 
Consequently, the first layer is not completed as long as $m \ll
m_0$. This is the so-called {\em Zeno effect} \cite{ev94}. The mounds
are separated by deep grooves and there is no coarsening. In case of a
finite Schwoebel barrier, there is an initial coarsening
regime. However, coarsening stops at a critical mound size and deep
grooves between the mounds are formed. Such a behaviour was predicted
from continuous differential equations \cite{p97,pv96} and Monte Carlo
simulations \cite{sp00,kscm99}.  Experimentally, it has been observed
in homoepitaxial growth on Pt(111) \cite{kscm99}.

When $m$ approaches $m_0$, incorporation processes become
important. Then, the grooves are filled up and the flanks of the
mounds assume their stable slope. Simultaneously, coarsening
starts. In chapter \ref{coarsekapitel}, we will investigate this process. 

\subsection{Dynamic scaling \label{introdynamicscale}}

In the investigation of crystal growth, one frequently finds that
surface snapshots are similar to rescaled surface snapshots taken at
different time. In the growth of fractal surfaces, the correlation
length $\xi(t)$ increases with time while the fractal properties do
not change (figure \ref{wavesnapshot}). In the coarsening process of
mounded surfaces, the size of the mounds grows but their shape remains
the same (figure \ref{coarsefigb}). Therefore, it is plausible to
assume that the statistical properties of the surface are invariant
under a simultaneous transformation of spatial extension $\vec{x}$,
time $t$ and the reduced surface height $f(\vec{x}, t)$ \cite{bs95}:
\begin{equation}
\vec{x} \rightarrow b \vec{x}; \; \; \; t \rightarrow b^z t; \; \; \;
f \rightarrow b^\alpha f.
\label{dynamicscalingintro}
\end{equation}  
The powerlaw dependence of the scaling factors for $t$ and $f$ on $b$
is due to the group property of the transformation. This symmetry
property is denoted as {\em dynamic scaling}. It is {\em independent}
of the scale invariance \ref{selbstaffinintro} of self-affine
surfaces. In many models of growth on an initially flat surface, it is
valid after a transient phase like e.g.\ the initial formation of
mounds in unstable growth. It is widely believed that the dynamic
exponents $\alpha$ and $z$ are {\em universal}. That means, that
$\alpha$ and $z$ are independent of ``details'' of the model, in
particular the numerical values of parameters. Then, the values of
$\alpha$ and $z$ determine the {\em universality class} of a model.

Dynamic scaling implies conditions for the time dependence of the {\em
surface width} $W(t) := \sqrt{\left< f(\vec{x}, t)^2 \right>}$ and the
correlation length $\xi(t)$. Under the transformation
\ref{dynamicscalingintro}, the correlation length scales like
$\vec{x}$ such that $\xi \rightarrow b \xi$.  The surface width scales
like $f$ such that $W \rightarrow b^z W$. These conditions are
fulfilled if $W$ and $\xi$ have a powerlaw dependence on time:
\begin{eqnarray}
\xi(t) &=& \xi_0 \; t^{\frac{1}{z}} \label{xidynamicscaling} \\ 
W(t) &=& W_0 \; t^\beta \; \; \mbox{where} \; \; \beta =
\frac{\alpha}{z}.
\label{Wdynamicscaling}
\end{eqnarray}
Similarly, on mounded surfaces the average perimeter of the mounds
increases $\propto t^{1/z}$, while their height is $\propto t^\beta$.
If the growth process selects one particular slope $m_0$, the ratio of
height and perimeter of the mounds is constant, which implies $1/z =
\beta = \alpha/z \Rightarrow \alpha = 1$.

Additionally, there are conditions for the functional form of the
height-height correlation. Since dynamic scaling links properties of
the surface at different times, we will use the notation $G(\vec{l},
t)$ for the height-height correlation at time $t$ in the
following. Since $G(\vec{l}, t)$ scales like the square of $f$, we
have $G(b \vec{l}, b^z t) = b^{2 \alpha} G(\vec{l}, t)$. This is
fulfilled if
\begin{equation}
G(\vec{l}, t) = l^{2 \alpha} u \left( \frac{\vec{l}}{t^{1/z}} \right).
\label{gdynamikscaling}
\end{equation}
Since surface heights at distances greater than $\xi(t)$ are
uncorrelated, $G(\vec{l}, t)$ saturates at large distance
$l$. Consequently, $u(\vec{x}) \sim |\vec{x}|^{-2 \alpha}$ for large
$|\vec{x}|$. 

Similarly, the standard deviation $w(L, t)$ of $f(\vec{x}, t)$ in a
square of size $L \times L$ must fulfil
\begin{equation}
w(L, t) = L^\alpha v \left(\frac{t}{L^{z}} \right).
\label{wdynamicscaling}
\end{equation}
For small $t$ and large $L$ where $\xi(t) \ll L$, the finiteness of
$L$ is irrelevant such that $w(L, t) = W(t)$. This implies that $v(x)
\sim x^{\beta}$ for small $x$. On the other hand, for $\xi(t) \gg L$,
we expect $w(L, t)$ to saturate such that $v(x) \sim \mbox{const.}$
for large $x$.  Then, $w(L, t)$ assumes an asymptotic value
\begin{equation} 
w_{\mathrm{late}} (L) \propto L^\alpha.
\label{wasympdynamic}
\end{equation}

In Monte Carlo simulations, only systems of finite size can be
considered. In a square system of size $N \times N$, the deviations
from the physically relevant case of an infinite system are small as
long as $\xi(t) \ll N$. As soon as the correlation length is
comparable to $N$, finite size effects become relevant. In general, the
properties of a finite system are similar to those of a section of an
infinite system. Thus, the surface width of a finite system resembles
the properties of $w(N, t)$.

So far, no assumptions on the morphology of the surface have been
made. If we assume that the surface has self-affine properties on
lengthscales smaller than $\xi(t)$, additional conclusions on the
scaling behaviour can be drawn. Comparing equations \ref{selbstaffinw}
and \ref{wasympdynamic}, we find that the exponent $\alpha$ and the
Hurst exponent $H$ must be identical. Additionally, from equation
\ref{selbstaffing} we expect the height-height correlation $G(\vec{l},
t)$ to increase $\sim l^{2\alpha}$ for small $l$. Consequently, we
have $u(\vec{x}) \sim \mbox{const.}$ for small $|\vec{x}|$.  In the
literature, the dynamic scaling properties of self-affine surfaces are
kown as {\em Family-Vicsek scaling} \cite{fv85}. For historical
reasons, deviations from this behaviour are denoted as {\em anomalous
scaling} \cite{l99,lr96,lrc97,rlr00}.

\section{Atomic layer epitaxy \label{aleintro}}

Kinetic roughening and unstable growth make it difficult to grow
perfectly flat layers by means of MBE. There is an alternative called
{\em atomic layer epitaxy} (ALE) which, however, requires specific
material properties \cite{hs96}. ALE works for a binary compound \AB{}
which consists of alternating layers of \A{} and \B{} atoms which are
parallel to the surface. The chemical bond between \A{} and \B{} atoms
must be considerably stronger than \A{}--\A{} and \B{}--\B{} bonds. These
conditions are fulfilled e.g.\ for (001) surfaces of II-VI
semiconductors.

In contrast to MBE of a compound material where the constituents are
deposited simultaneously, in ALE the surface is exposed to alternating
pulses of pure \A{} and \B{}.  Between the \A{} and \B{} phases, there
is a {\em dead time} $t_d$ without any particle flux.  We start with
the deposition of \A{} on a flat \B{}-terminated substrate. Particle
flux $F$ and pulse time $t_p$ are chosen such that the amount of
deposited material is sufficient for more than one layer of
$A$. \A{}-atoms in the first layer are bound via the strong \A{}--\B{}
bond. Atoms in the second or a higher layer have only weak \A{}--\A{}
bonds. Temperature and dead time are chosen such that these surplus
particles desorb, while atoms bound to the substrate remain on the
surface. Since the rate at which atoms with \A{}--\B{} bonds desorb is
much smaller than the desorption rate of atoms with \A{}--\A{} bonds
only, no fine-tuning of the parameters $F$, $t_p$ and $t_d$ is
required for that purpose. Instead, we obtain the deposition of
exactly one layer of \A{} in a self-regulated manner.  In the
following \B{}-phase and the subsequent dead time, the same process
repeats with the roles of \A{} and \B{} interchanged.  At the end of
the cycle, one monolayer of \AB{} has been deposited. The whole
procedure may be repeated an arbitrary number of times to obtain a
film of the desired thickness.

This behaviour might be complicated by the presence of surface
reconstructions. In many compound semiconductors a surface
terminated with a complete layer of one particle species is
unstable. In particular, cation terminated surfaces of II-VI
semiconductors like CdTe are characterized by vacancy structures with
submonolayer coverage. Such effects may lead to ALE growth at rates
$\neq 1 \,\mathrm{ML}/\mathrm{cycle}$. This feature has been used to
investigate the stoichiometry of CdTe(001) and ZnTe(001) surfaces in
experiments \cite{dbt96,fs90,vadt96}. In chapter
\ref{zinkblendenmodell}, we will simulate ALE of reconstructed
surfaces within the framework of our model of CdTe.

\section{Sublimation \label{sublimintro}}

Occasionally, atoms at the surface of a crystal desorb.  At a solid in
thermal equilibrium with its vapour, this loss of material is
compensated by the adsorption of an equal number of particles. In
ultra high vacuum, this is not the case such that the crystal shrinks
in the course of time. This non-equilibrium process is denoted as {\em
sublimation}.

The {\em sublimation rate} $r_{\mathrm{sub}}$ is defined as the number
of monolayers which desorb per unit time. If $r_{\mathrm{sub}}$ is
measured as a function of $T$, one finds that in a wide temperature
range $r_{\mathrm{sub}}(T)$ is described well by an Arrhenius law
(equation \ref{arrheniusgesetz}) with attempt frequency
$\nu_{\mathrm{sub}}$ and activation energy $E_{\mathrm{sub}}$. Monte
Carlo simulations \cite{schi99,sk98} have shown, that the {\em
macroscopic} parameters $\nu_{\mathrm{sub}}$ and $E_{\mathrm{sub}}$
are not equal to the corresponding parameters of some rate limiting
{\em microscopic} process like e.g.\ the desorption of a single
adatom. Instead, they are functions of all model
parameters. Consequently, to understand sublimation we have to
consider the interplay of various physical processes on the surface
similar to the investigation of crystal growth.

At moderate temperature, the desorption rate of atoms bound to steps
and islands can be neglected compared to that of single adatoms.
Adatoms diffuse over a typical distance $l_{\mathrm{des}}$, the {\em
desorption length} before they are desorbed. An adatom is created when
an atom detaches from a step edge or when an atom jumps out of a
closed layer. In general, the activation energy of the first process
is smaller. At the latter process, a vacancy is created
simultaneously. An atom at a neighbouring site in the same layer which
diffuses into the vacancy leaves a new vacancy behind at the site where
it came from. Conversely, one might say that the original vacancy has
moved. Therefore, this process is called {\em vacancy diffusion}.

Vacancies diffuse around until they are either filled by an adatom,
are captured by a step edge or meet other vacancies and nucleate a
{\em vacancy island}. This is a hole in the terminating layer of the
crystal which exceeds some critical size such that it is unlikely to
be closed again. Similar to the typical island distance $l_i$ in
submonolayer growth, in the sublimation of a flat surface there is a
typical distance $l_v$ between neighbouring vacancy islands. Vacancy
islands grow as they capture vacancies diffusing around and adatoms
detach from their edges and desorb. Finally, they coalesce with their
neighbours and one layer of the crystal is gone. This process is
denoted as {\em layer-by-layer} sublimation.

Similar to crystal growth, sublimation is influenced strongly by the
presence of steps on a vicinal surface. If the terrace width
$l_s$ is smaller than $l_v$, the creation of vacancy islands is
unlikely since vacancies are captured by the steps. Instead,
sublimation proceeds in {\em step-flow} mode where the steps move back
at a speed $v_{\mathrm{step}}$. At the desorption of one monolayer,
the steps move over a distance $l_s$ such that $r_{\mathrm{sub}} =
v_{\mathrm{step}}/l_s$. Pimpinelli and Villain \cite{pv94} have
investigated step-flow sublimation by means of BCF theory.

If the spacing between steps is large compared to the desorption
length, it is unlikely that an adatom which detaches from a step
reaches a neighbouring step. Consequently, the interaction between
steps is weak and $v_{\mathrm{step}}$ is independent of $l_s$. This
case is denoted as {\em free step-flow}. In {\em hindered step-flow},
we have $l_s < l_{\mathrm{des}}$. Then, there is a high probability
that adatoms which detach from a step are captured by another step
instead of being desorbed. Consequently, at small interstep spacing,
$v_{\mathrm{step}}$ decreases if $l_s$ is reduced.

Vicinal surfaces where the width of all terraces is equal sublimate
either in layer-by-layer mode or in step-flow mode. However,
techniques of surface preparation frequently yield a wide distribution
of terrace widths. In this case, there might be a temperature range
where $l_v$ is greater than the width of small terraces, but there are
also large terraces which are wider than $l_v$. Then, small terraces
sublimate in step-flow mode. Simultaneously, there is layer-by-layer
sublimation on large terraces. This effect has been observed
experimentally on (001) surfaces of CdTe \cite{nskts00}.

In general, $E_{\mathrm{sub}}$ depends on the mode of
sublimation. Additionally, different surface reconstructions yield
different values of $E_{\mathrm{sub}}$. This problem will be addressed
in chapters \ref{rekkapitel} and \ref{zinkblendenmodell}.

\chapter{Flat (001) surfaces of II-VI semiconductors: A lattice gas
model \label{dreikapitel}}

Within the last years, potential technological applications of
electronic devices based on II-VI semiconductors \cite{si00} have
inspired basic research concerning surfaces of these materials. In
this context, various studies have addressed the properties of surface
reconstructions. Experimental studies have investigated which
reconstructions are present \cite{n98,ntss00,tdbev94} and how the
reconstruction of the surface is influenced by parameters like
temperature and particle flux in an MBE environment
\cite{dbt96,vadt96,wewr00}. The majority of this work has been devoted
to CdTe and ZnSe, where a fairly complete qualitative overview over
the phase diagram has been gained. An overview over the properties of
CdTe can be found in \cite{ct97}. On the other hand, there have been
theoretical investigations of the reconstructions of CdTe
\cite{gffh99} and ZnSe \cite{gn94,g97,pc94} using density functional
theory. In these studies, knowledge about the chemical bonding of
surface atoms and ground state energies of various reconstructions has
been gained.

In this and the following two chapters, we present models which aim at
a theoretical understanding of the reconstructed (001) surface of CdTe
and ZnSe under typical MBE conditions. As explained in section
\ref{reconstintro}, to study thermodynamic effects like phase
transitions we need to consider systems with a large number of
atoms. On the other hand, the numerical effort should not exceed the
capabilities of today's computers. Therefore, efficient representations
of atomic interactions are required. In many cases, two-dimensional
lattice gases have been used successfully to model atoms adsorbed on a
singular crystal surface or the terminating layer of such a crystal
\cite{baksv01,ksb82,lbadebt00,s80}. In spite of the conceptual
simplicity of such models the interplay of attractive and repulsive
short range interactions can result in highly nontrivial critical
behaviour and complex phase diagrams. In this chapter, we will follow
this approach to model flat $(001)$ surfaces of CdTe
and ZnSe in thermal equilibrium, our main focus being on CdTe.
In the following two chapters, we will extend this model to consider
three-dimensional crystals in nonequilibrium situations like growth
and sublimation.  

The outline of this chapter is as follows: In section
\ref{dreioverview}, we review the known facts about $(001)$
surfaces of CdTe and ZnSe. In section \ref{dreilgasbasics}, we
introduce a lattice gas model which considers the occupation of Cd
sites and the dimerization of Te atoms and discuss its phase
diagram. We conclude with a comparison of the features of our model
with experimental results in section \ref{dreiexpcmp}.

\section{Surface reconstructions of CdTe and ZnSe \label{dreioverview}}

\begin{figure}[htb]
\botbase{\includegraphics{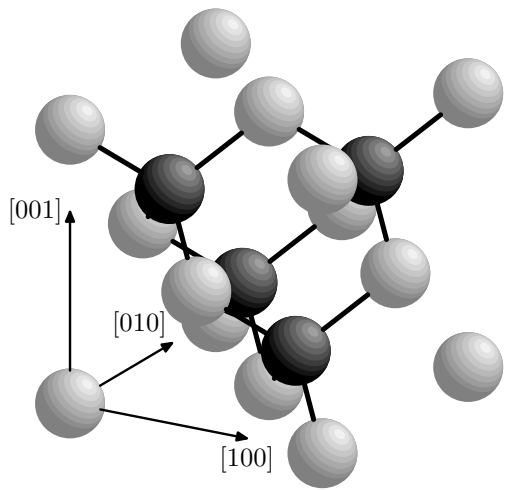}} \hfill
\botbasebox{0.5\textwidth}{\caption{Unit cell of the zinc-blende
lattice. Anions are shown in light grey, cations in dark
grey.\label{zbeinheitszell}}}
\end{figure}

Both CdTe and ZnSe crystallize in the zinc-blende lattice. A sketch of
the unit cell is shown in figure \ref{zbeinheitszell}. This lattice is
equivalent to a diamond lattice apart from the fact that the lattice
sites are occupied alternately with the two elements. Each atom is
surrounded by four nearest neighbours of the other species the nuclei
of which lie on a tetrahedron. The zinc-blende lattice can be
decomposed into two intersecting fcc lattices the sites of which are
occupied with particles of one species. Thus, it is composed of
alternating half-layers of metal atoms (cations) and nonmetal atoms
(anions) which are parallel to the $(001)$ surface. Consequently, an
ideal $(001)$ surface would be terminated by a complete layer of one
particle species. The positions of the atoms in one layer lie on a
regular square lattice the axes of which are oriented in the $[110]$
direction and the $[1\overline{1}0]$ direction. The lattice constant
of this square lattice is $a/\sqrt{2}$, where $a$ is the lattice
constant of the zinc-blende lattice itself.

The orientation of the bonds of the atoms in one layer breaks the
four-fold symmetry of the surface. Each surface atom is bound to two
atoms of the opposite species in the half-layer below. In cations,
these bonds point in the $[1\overline{1}0]$ direction. Conversely, surface
anions have bonds in the $[110]$ direction. Therefore, the
symmetry of the $(001)$ surface of the zinc-blende lattice is
two-fold. 

\begin{figure}
\begin{picture}(100, 30)(0, 0)
\put(0, 0){\resizebox{!}{0.3\textwidth}{\includegraphics{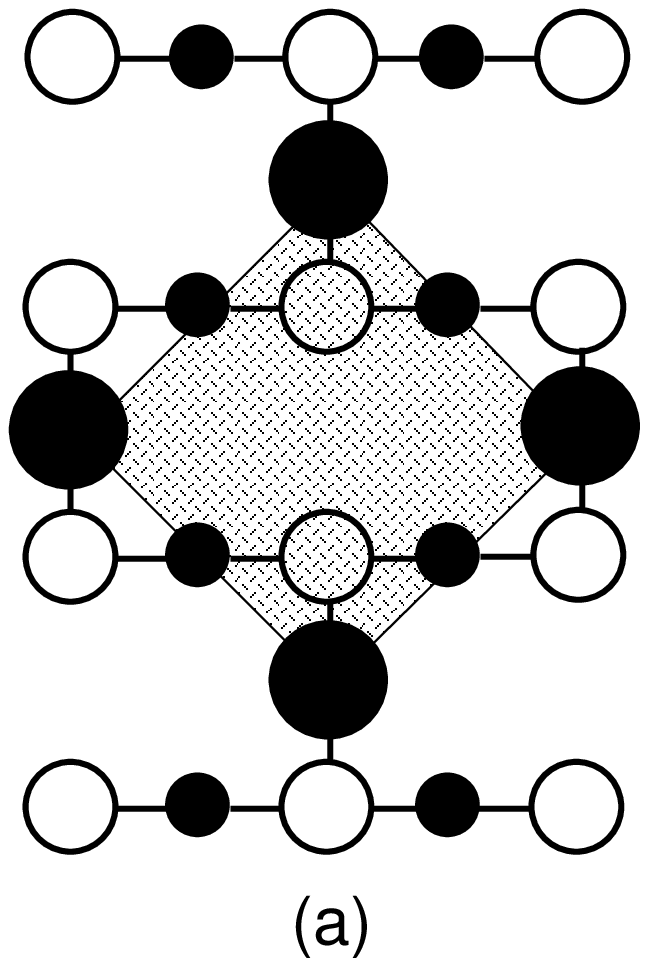}}}
\put(26, 0){\resizebox{!}{0.3\textwidth}{\includegraphics{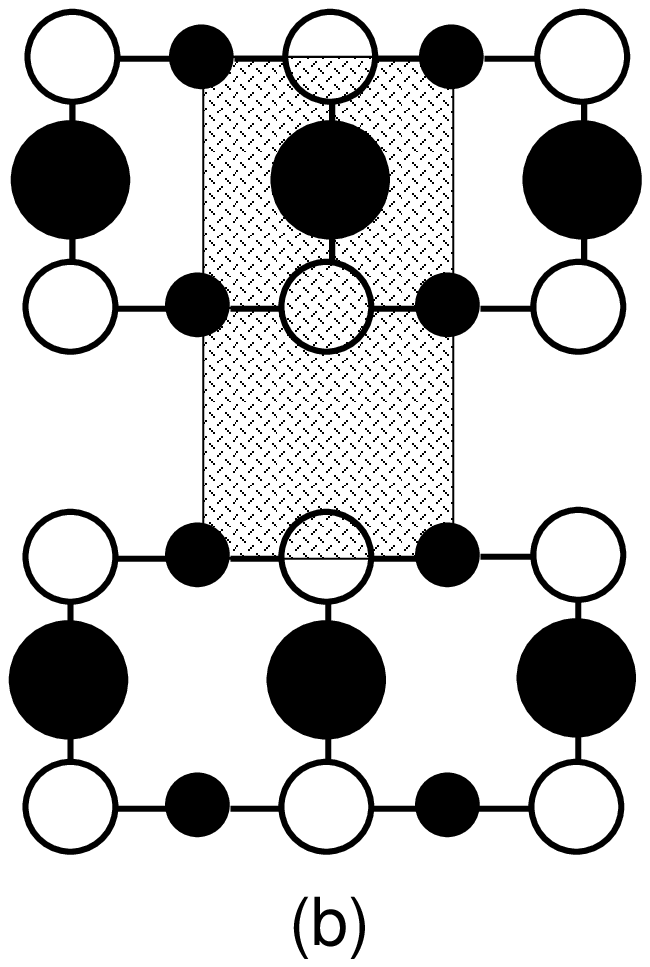}}}
\put(54,0){\resizebox{!}{0.3\textwidth}{\includegraphics{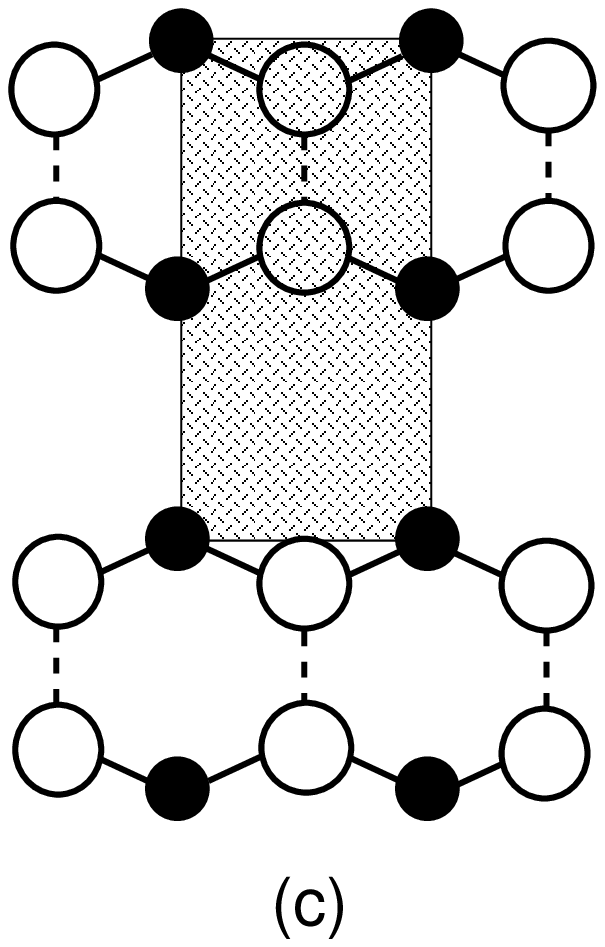}}}
\put(80,0){\resizebox{!}{0.3\textwidth}{\includegraphics{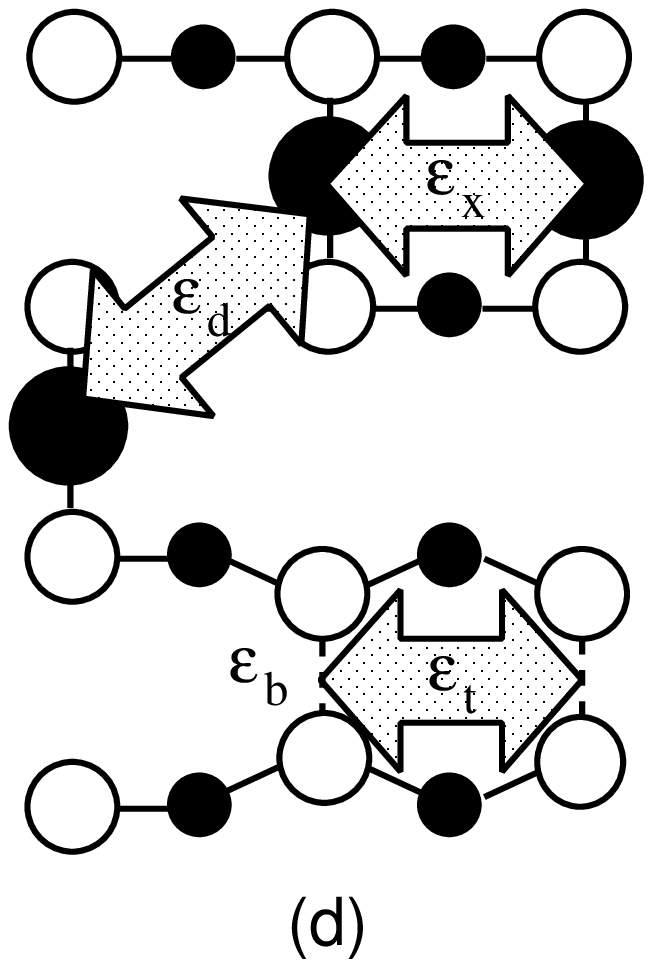}}}
\end{picture}
\caption{Panels (a),(b),(c): Sketches of the reconstructions of CdTe(001)
which are discussed in this work. Larger and smaller filled circles denote
Cd atoms in the first and third layer, open circles represent 
Te atoms in the second layer. The grey rectangles show the surface
unit cells. The $[110]$ axis is aligned horizontally.  (a), (b):
Cd terminated reconstructions. Panel (a) shows the \cdtxt{}
reconstruction, panel (b) the \cdtxo{} reconstruction. Panel (c) shows
the \tetxo{} reconstrution of surfaces terminated with a complete
monolayer of Te. Panel (d) shows the attractive couplings in our
model. \label{dreifigrek}}
\end{figure} 
The stability of the surface reconstructions can be understood from
the electron counting rule \cite{h79,p89} which was introduced in
section \ref{reconstintro}. Each metal atom provides two electrons. In
each bond to a bulk atom, $1/2$ electron is engaged. A nonmetal
atom brings six electrons, $3/2$ of which are engaged in each
bond to an atom in the bulk. The electron counting rule states that
the remaining electrons must be sufficient to fill each anion dangling
bond and every bond of a surface dimer with two electrons, whereas the
cation bonds must remain empty. In particular, a surface terminated
with a complete layer of undimerized atoms is unstable.

Under vacuum, the $(001)$ surface of CdTe is Cd terminated. The
surface is characterized by vacancy structures where less than one
half of the potential Cd sites in the top layer are occupied
\cite{dbt96,vadt96}. At low temperature, one finds a \cdtxt{}
reconstruction \cite{ct97,tdbev94}, where Cd atoms and vacancies
arrange in a checkerboard pattern (figure
\ref{dreifigrek}a). Frequently, a contribution of a \cdtxo{}
arrangement can be found. In the \cdtxo{} structure, the Cd atoms
arrange in rows along the $[110]$ direction which alternate with rows
of vacancies (figure \ref{dreifigrek}b). Density functional
calculations \cite{gffh99} have shown that the surface energies of the
two competing reconstructions are nearly degenerate and differ only by
a small amount $\Delta E \approx 0.008 \,\mathrm{eV}$ per $(1 \times
1)$ surface unit cell\footnote{In \cite{gffh99} a greater value of
$\Delta E = 0.016 \,\mathrm{eV}$ is given. However, $\Delta E$ is
rather sensitive to some approximations. We use the unpublished result
of a more precise calculation by S. Gundel.}. At a temperature $T
\approx 570 \,\mathrm{K}$, there is a phase transition \cite{tdbev94}
above which the \cdtxo{} arrangement of the Cd atoms dominates the
surface. An analysis of high resolution low energy electron
diffraction (HRLEED) peaks \cite{n98,ntss00} has shown, that there is
a high degree of disorder in this high temperature phase. One finds
elongated domains with a large correlation length of $\approx
375\,\mbox{\AA}$ in the $[1\overline{1}0]$ direction and a domain
width as small as $22\,\mbox{\AA}$ in the $[110]$ direction.

\begin{figure}
\botbase{\includegraphics{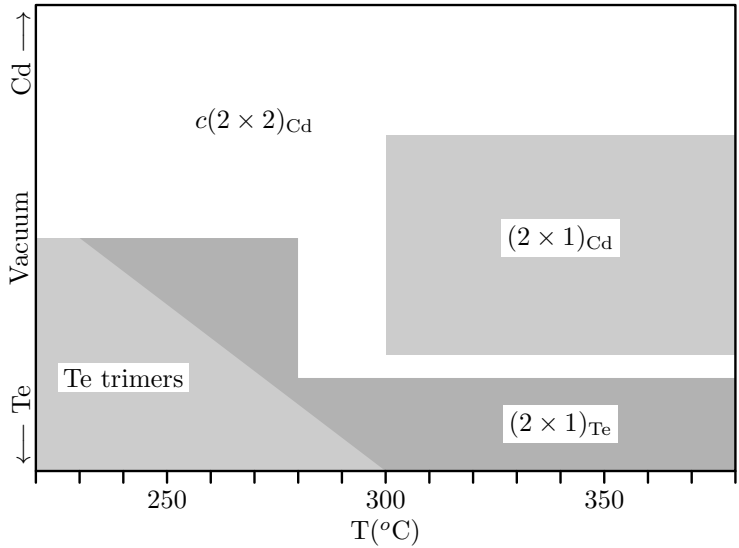}} \hfill
\botbasebox{0.33\textwidth}{\caption{Schematical surface phase diagram
of CdTe in an MBE environment according to \cite{ct97}. Shown is the
predominant reconstruction as a function of temperature and particle
flux. The properties of the surface in vacuum are shown in the
middle. The upper part displays the reconstruction under a Cd flux,
the lower part shows the features of the surface under a Te
flux.\label{cdtephasdschematisch}}}
\end{figure}

If CdTe is exposed to an external Cd flux in an MBE chamber, the
\cdtxt{} reconstruction is stabilized at temperatures above the
transition in vacuum.  Under a Te flux, the surface is Te terminated
with a \tetxo{} reconstruction. At small Te fluxes, the surface is
terminated by a complete monolayer of Te. The Te atoms on the surface
form dimers which arrange in rows (figure \ref{dreifigrek}c). At high
Te flux and low temperature, one additional Te atom is incorporated
into each dimer, such that the Te coverage of the surface is 1.5. The
symmetry of this reconstruction is still $(2 \times 1)$, since the Te
trimers tend to arrange in rows.  A schematical phase diagram of the
surface following \cite{ct97} is shown in figure
\ref{cdtephasdschematisch}.
 
Qualitatively, the properties of the reconstructions of ZnSe
\cite{ccd88,wewr00} are quite similar to those of CdTe, where the Zn
atoms are the counterparts of the Cd atoms and the Se atoms those of
Te, respectively. There is one important exception: to date, no $(2
\times 1)_{\mathrm{Zn}}$ reconstructed surface has been found at high
temperature. Density functional calculations \cite{gn94,g97,pc94}
yield a higher energy difference $\Delta E \approx 0.03 \,\mathrm{eV}$
per $(1 \times 1)$ surface unit cell between ideal $c(2\times 2)$ and
$(2\times1)$ reconstructed surfaces, which is approximately twice the
value calculated for CdTe. As we will show, this greater energy
difference might explain the different behaviour of CdTe and ZnSe.

\section{The lattice gas model \label{dreilgasbasics}}

The basic structure of our model of (001) surfaces is the same for
different II-VI semiconductors.  The differences between the materials
are represented by the numerical values of parameters. In the
following, for simplicity we will loosely speak of Cd and Te atoms
instead of ``cations'' and ``anions'' without restricting ourselves to
a modelling of CdTe only.

We present a lattice gas model with effective interactions which include
effects of surface strain and, at elevated temperature, lattice
vibrations. Therefore, there is no simple mapping between ground state
energies of the lattice gas model and surface energies determined from
density functional theory. 

We model a flat $(001)$ surface of CdTe, i.e.\ we neglect the influence
of steps on the reconstruction. This is a reasonable approximation if
the typical distance between steps is much greater than the size of
the unit cell of the reconstruction. In thermal equilibrium, this is
fulfilled at temperatures well below the roughening transition. In
this case, in the absence of bulk vacancies or other defects the
crystal is uniquely described by the state of the topmost monolayer
which consists of one Cd half-layer and a Te half-layer. In our model,
we consider such a monolayer with the Cd atoms on top. For simplicity,
we assume the Te layer to be fully occupied. This is not a severe
restriction, since the removal of Te atoms will uncover Cd atoms in
the layer below. Within the limit of a model of a flat surface, we
simply do not distinguish whether a Cd terminated surface has been
created by removing a Te layer or by depositing additional Cd atoms
onto an intact Te layer.

We use cartesian coordinates where the $x$-axis points in the $[110]$
direction, the $y$-axis points in the $[1\overline{1}0]$ direction and
the $z$-axis points in the $[001]$ direction. The origin of the
coordinate system is at the centre of a Cd atom. Since we consider a
flat surface, the $z$-coordinate of all atoms of one species is
identical and will be omitted in the following.  Measuring the lattice
constant in appropriate units, the Cd atoms are at positions $(x, y)$
with integer $x, y$. The Te atoms are at positions $(x, y+ 1/2)$,
where $x, y \in \mathbb{Z}$. Tellurium dimers are created by the
formation of a chemical bond between neighbouring Te atoms in
$y$-direction ($[1\overline{1}0]$). Since this is also the direction
of the bonds of a surface Cd atom, a Te dimer is formed by a pair of
atoms which might also be the binding partners of a Cd atom above
them. This suggests a simple lattice gas representation both of the
occupation of Cd sites and Te dimerization: We consider a rectangular
array $x$ of integers $\{x_{i,j}\}_{i, j = 1}^{L, N}$. $x_{i,j} = 1$
represents a Cd atom at site $(i, j)$ which is bound to the Te atoms
at sites $(i, j-1/2)$ and $(i, j+1/2)$. $x_{i,j} = 2$ corresponds to a
dimerization of Te atoms $(i, j-1/2)$ and $(i, j+1/2)$. Otherwise,
$x_{i, j} = 0$.

In principle, one might also consider the formation of Te trimers by
introducing a fourth state $x_{i,j} = 3$. This corresponds to a trimer
which consists of the Te atoms at $(i, j-1/2)$ and $(i, j+1/2)$ and an
additional Te atom at $(i, j)$.  However, the formation of Te trimers
plays a role only if the surface is exposed to a strong flux of pure
Te at comparatively low temperature. Therefore, we have neglected this
effect to reduce the number of parameters of our model and the
numerical effort of its investigation.

\subsection{Interactions of atoms and dimers}

The detailed representation of the surface energies of a II-VI
compound certainly would require long-range interactions and terms
which depend on the simultaneous occupation of three or more
sites. However, it is plausible to assume that the dominant
contribution to the surface energy stems from pairwise interactions of
atoms at short distances. In this section, we introduce a Hamiltonian
which considers pairwise interactions between $x_{i,j}$ on nearest
neighbour sites and diagonal neighbour sites. These reflect the known
properties of the reconstructions of CdTe and ZnSe.

Due to the electron counting rule, surfaces with Cd coverages greater
than $1/2$ are unstable while both a \cdtxt{} reconstruction and a
\cdtxo{} reconstruction are permitted. This feature can be captured by
introducing a hardcore repulsion between Cd atoms on neighbouring
sites in the $y$-direction. In the $x$-direction, an attractive
interaction favours the occupation of nearest neighbour pairs the
strength of which is denoted by $\varepsilon_x < 0$. An attractive
interaction $\varepsilon_d < 0 $ between diagonal neighbours
stabilizes the \cdtxt{} reconstruction. The parameters $\varepsilon_x$
and $\varepsilon_d$ are chosen such that the energy difference $\Delta
E = |2 \varepsilon_d - \varepsilon_x|/2$ between these two
reconstructions is small compared to the total surface energy per
site.

The electron counting rule favours Te dimerization, but forbids the
formation of additional bonds of dimerized Te atoms. Therefore, we
forbid the simultaneous occupation of neighbour sites in the
$y$-direction with dimers (formation of chains of Te-Te-bonds) and Cd
atoms next to a dimer. These rules permit both a \tetxo{}
reconstruction, where the dimers arrange in rows and a
$c(2\times2)_{\mathrm{Te}}$ reconstruction, where they arrange in a
checkerboard pattern. Density functional calculations \cite{gffh99}
have shown that the surface energy of \tetxo{} is significantly lower
than that of a $c(2\times2)_{\mathrm{Te}}$ reconstruction. This may be
understood from the more efficient relaxation of surface strain in
\tetxo{}: Since the lattice is deformed in the same direction by
dimers on neighbouring sites in $x$-direction, such an arrangement
will be energetically favourable. We consider this fact by an
attractive nearest neigbour interaction $\varepsilon_t$ between dimers
in $x$-direction. Additionally, a dimer contributes a binding energy
$\varepsilon_b$ to the surface energy. Apart from the hardcore
repulsion, we consider no interaction between Cd atoms and dimers. An
overview over the couplings in our model can be found in figure
\ref{dreifigrek}d.

This yields the Hamiltonian
\begin{equation}
H = \sum_{i,j = 1}^{L, N} \varepsilon_x c_{i,j} c_{i+1,j} +
\varepsilon_d c_{i,j} \left( c_{i+1,j+1} + c_{i+1,j-1} \right) +
\varepsilon_t d_{i,j} d_{i+1,j} + \varepsilon_b d_{i,j} - \mu c_{i,j},
\label{dreienergie}
\end{equation}
where we have introduced $c_{i,j} = \delta_{x_{i, j}, 1}$ and $d_{i,j}
= \delta_{x_{i,j}, 2}$. The boundary conditions are assumed to be
periodic. $c_{i, j}$ represents the occupation of lattice sites with
Cd atoms: $c_{i,j} = 1$ at the positions of Cd atoms and zero
otherwise. Similarly, the positions of dimers are given by $d_{i,j} =
1$. $\mu$ is the effective Cadmium chemical potential, which includes
the binding energy of surface Cd atoms to the substrate. The
groundstate of the system at $T = 0$ is determined by the chemical
potential $\mu$. For $\mu > \mu_0 = 2 \varepsilon_d - \varepsilon_t -
\varepsilon_b$, the surface configuration with minimal energy is
\cdtxt{}. For $\mu < \mu_0$ the groundstate is \tetxo{}.

In the following, we measure energy in dimensionless units which have
been adjusted such that $\varepsilon_d = -1$. Additionally, we set
Boltzmann's constant $k = 1$ such that temperature is measured in
units of $|\varepsilon_d|$.

\subsection{Characterization of the surface configuration 
\label{dreicharakterisier}}

To characterize the surface reconstruction quantitatively, we evaluate
the mean Cd coverage $\rho_{Cd} = \left< c_{i, j} \right>_{i,j}$ and
the correlations
\begin{eqnarray}
C_{Cd}^{d} &=& \frac{1}{2} \left< c_{i, j} \left(c_{i+1,j+1} +
c_{i+1,j-1} \right) \right>_{i,j} \label{defccdd} \\ C_{Cd}^{x} &=&
\left<c_{i,j} c_{i+1,j} \right>_{i,j}.
\label{defccdx}   
\end{eqnarray} 
$C_{Cd}^{d}$ measures the probability to find two diagonal neighbour
sites which are simultaneously occupied by Cd atoms. This is a measure
for the fraction of the surface which is covered by regions with a
{\em local} \cdtxt{} order.  Similarly, $C_{Cd}^{x}$ measures the
contribution of locally \cdtxo{} reconstructed regions in the system.

The {\em long range order} of the Cd atoms is measured by the order
parameters
\begin{eqnarray}
M_{Cd}^{(2\times 2)} &=& \frac{1}{L N} \sum_{i,j = 1}^{L,N} c_{i,j} \cos 
\left( \pi \left(i + j \right) \right) \label{defmcdtxt} \\
M_{Cd}^{(2\times 1)} &=& \frac{1}{L N} \sum_{i,j = 1}^{L,N} c_{i,j} \cos 
\left( \pi j \right) \label{defmcdtxo}.
\end{eqnarray} 
$M_{Cd}^{(2\times 2)}$ is the staggered magnetization of a system of
Ising variables $\{s_{i,j}\}_{i,j=1}^{L,N}$, where $s_{i,j} = 1$ if
$c_{i,j} = 1$ and $-1$ otherwise. Large absolute values indicate a
long range order of the \cdtxt{} reconstruction. Its counterpart
$M_{Cd}^{(2 \times 1)}$ measures the long range order of \cdtxo{}.
  
Further, we introduce the mean Tellurium dimerization $\rho_{D} :=
\left< d_{i,j} \right>_{i,j}$ and the dimer correlation $C_{D}^{x} :=
\left< d_{i,j} d_{i+1,j} \right>_{i,j}$, which characterizes the
number of dimers which are incorporated in {\em locally} $(2\times1)$
ordered areas. Their long range order is characterized by
\begin{equation}
M_{D}^{(2\times1)} = \frac{1}{L N} 
\sum_{i, j = 1}^{L, N} d_{i, j} \cos \left( \pi j \right) .
\label{dreimddef}
\end{equation}

\subsection{Methods of investigation}

We have investigated this model by means of the transfer matrix method
and Monte Carlo simulations. An introduction to the transfer matrix
technique is presented in appendix \ref{tmappendix}. This method
allows for a numerical calculation of the free energy of lattice
systems with short-range interactions. In general, the system size
must be finite in all directions but one. In our investigations, we
choose the $y$-direction as the infinite direction.  Since the
computational effort increases exponentially fast with the system size
$L$ in the $x$-direction, only comparatively small $L$ are feasible.
However, this is sufficient to investigate the phase transitions of
the system in the limit $L \rightarrow \infty$ via finite size scaling
techniques.  To this end, we have applied the method discussed in
section \ref{finitesizefirst} to determine the loci of first order
phase transitions and to compute coverage discontinuities.  Continuous
transitions have been investigated by means of phenomenological
renormalization group theory, which will be introduced in section
\ref{finitesizekontinuierlich}. Coverage and correlations can be
obtained from proper derivatives of the free energy or directly from
the relevant eigenvector of the transfer matrix. We have refrained
from calculating order parameters by means of the transfer
matrix. This would require the introduction of staggered fields in the
Hamiltonian \ref{dreienergie}. A system with this modified energy
function is numerically intractable for reasonable values of $L$.

Additionally, we have performed Monte Carlo simulations of our
model. To obtain a reasonably fast equilibration, we have applied
continuous time algorithms which will be discussed in more detail in
appendix \ref{mcappendix}. We have simulated both the grand-canonical
ensemble where $\rho_{Cd}$ is controlled by $\mu$ and the canonical
ensemble where the number of Cd atoms in the system is fixed. In
general, the results of the transfer matrix calculations and the Monte
Carlo simulations are in good agreement (figures \ref{dreismuconst} 
and \ref{dreifig1}).
 
\subsection{Simplified model without Te dimerization \label{dreizwei}}

In this section we present a simplified version of our model where we
do not consider the dimerization of Te atoms, i.e.\ the possible
values of the $x_{i,j}$ are constrained to the values $0$ and
$1$. Then, the state of the system is uniquely described by
$\{c_{i,j}\}_{i,j=1}^{L,N}$. The Hamiltonian of the simplified model
is
\begin{equation} 
H_s = \sum_{i,j = 1}^{L, N} \varepsilon_x c_{i,j} c_{i+1,j} + \varepsilon_d
c_{i,j} \left( c_{i+1,j+1} + c_{i+1,j-1} \right) - \mu c_{i,j}.
\label{zweihamilton}
\end{equation} 
\begin{figure}
\botbase{
\begin{picture}(48, 33)(0, 0)
\put(0,33){\resizebox{0.48\textwidth}{!}{\rotatebox{270}{\includegraphics{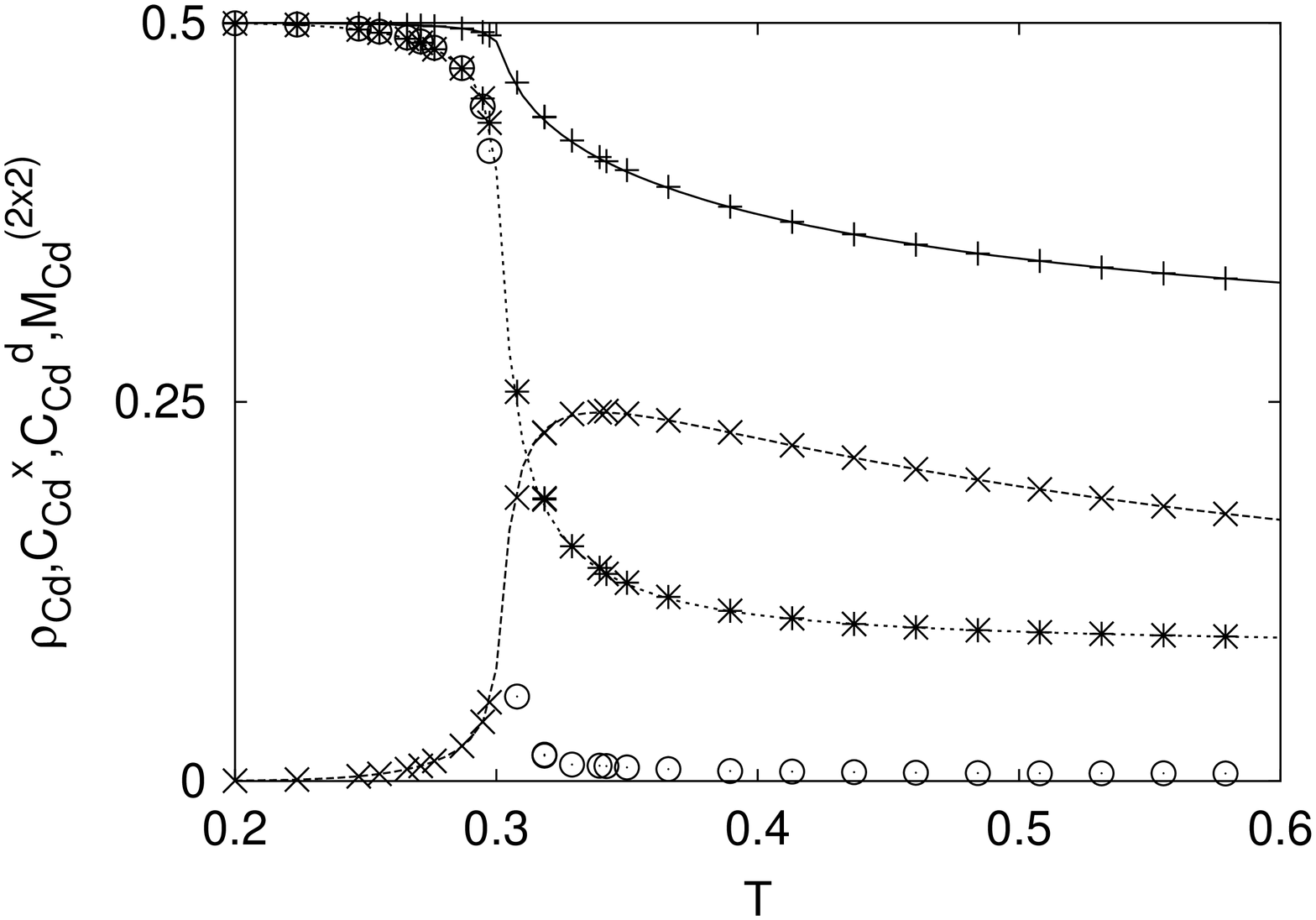}}}}
\put(0,30){(a)}
\end{picture}
\hspace{-2mm}
}
\hfill
\botbase{
\hspace{-6mm}
\figpanel{\resizebox{0.23\textwidth}{!}{\includegraphics{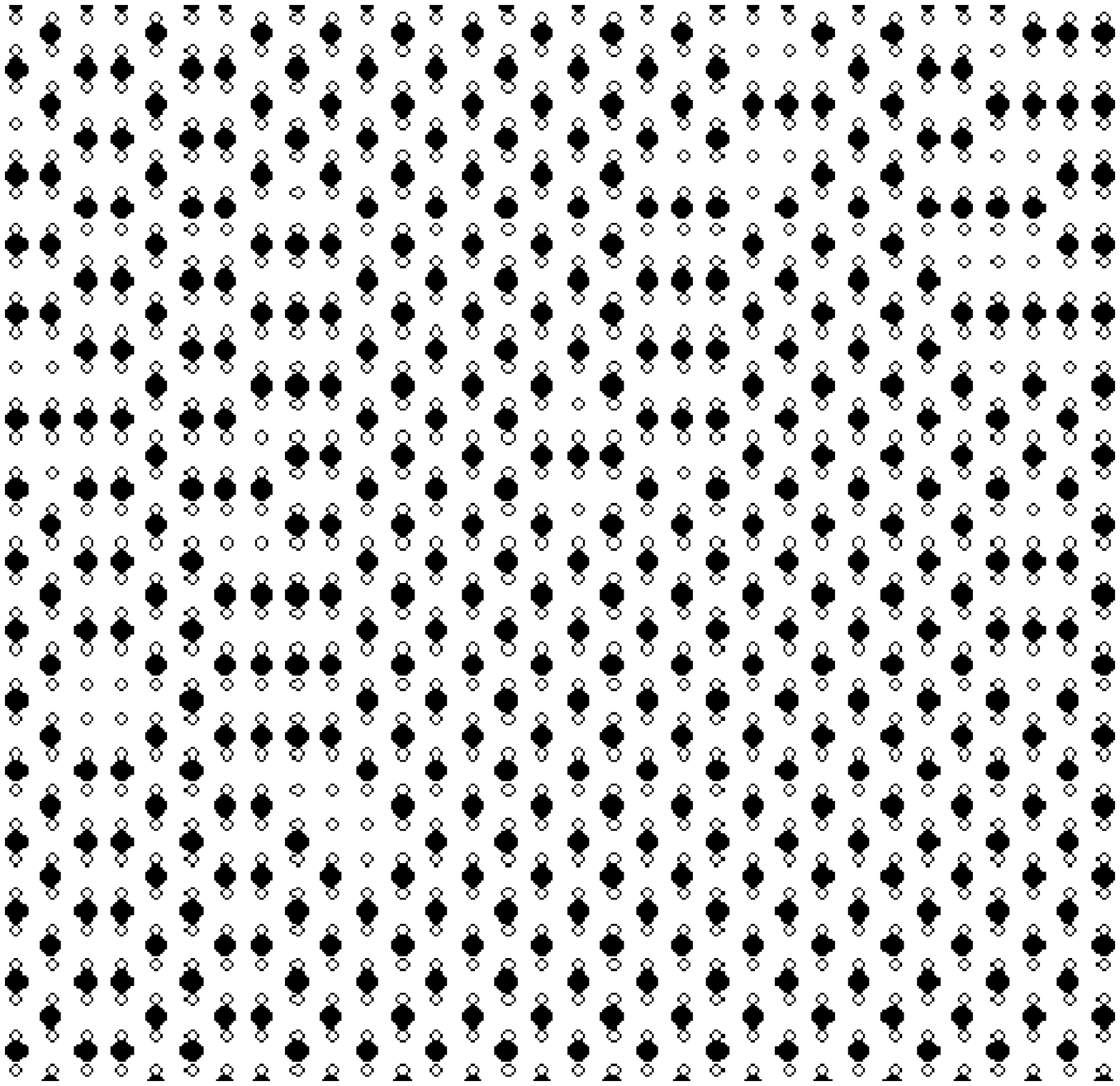}}}{(b)}
\hspace{-6mm}
}
\hfill
\botbase{
\hspace{-7mm}
\figpanel{\resizebox{0.23\textwidth}{!}{\includegraphics{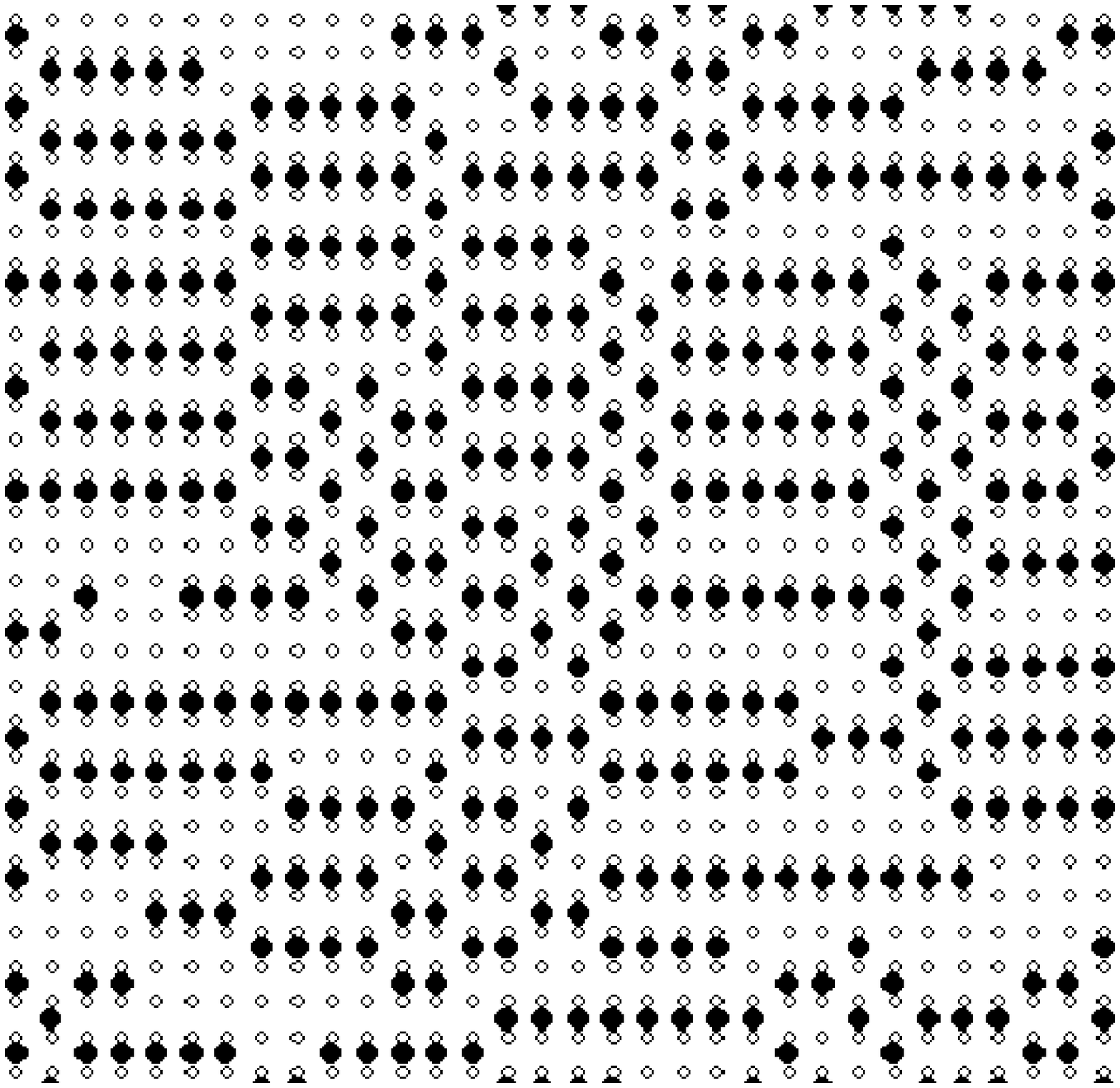}}}{(c)}
\hspace{-6mm}
}
\caption{Temperature dependent behaviour of the simplified model at
$\varepsilon_d = -1$, $\varepsilon_x = \mu = -1.96$. Panel (a): Cd
coverage $\rho_{Cd}$ ($+$), correlations $C_{Cd}^{x}$ ($\times$),
$C_{Cd}^{d}$ ($\plustimes$) and mean absolute of the order parameter
$M_{Cd}^{(2\times 2)}$ ($\odot$). The symbols show results of Monte
Carlo simulations of a $128 \times 128$ system. $10^4 \cdot L N$
events have been performed both for equilibration and for
measurement. The lines show results of a transfer matrix calculation
at a strip width $L = 18$.  Panels (b), (c) show sections of $16
\sqrt{2} \times 16 \sqrt{2}$ zinc-blende lattice constants of the
surface at the end of the simulation run at $T = 0.29$ (b) and $T =
0.33$ (c).
\label{dreismuconst}}
\end{figure}
For $\mu > -2$, the groundstate is a \cdtxt{} reconstruction, whereas
for $\mu < -2$, the system is empty at zero temperature. Figure
\ref{dreismuconst} shows the temperature dependent behaviour of the
correlations and order parameters of this model with the parameter set
$\varepsilon_d = -1$, $\varepsilon_x = -1.96$ in the grand canonical
ensemble at a chemical potential $\mu = -1.96$. In the Monte Carlo
simulations, the system was initialized with a perfect \cdtxt{}
reconstruction at all investigated temperatures. At low temperature,
we obtain values of $M_{Cd}^{(2\times 2)}$ and $C_{Cd}^{d}$ close to
0.5.  $C_{Cd}^{x}$ is small and $M_{Cd}^{(2\times 1)}$ is zero apart
from small fluctuations which are finite size effects. This is the
signature of a long-range ordered \cdtxt{} reconstructed phase. At $T
\approx 0.3$, the system loses its long-range order such that at high
temperature we have $M_{Cd}^{(2\times 2)} \approx M_{Cd}^{(2\times 1)}
\approx 0$. Simultaneously, the Cd coverage jumps to a smaller value.
These features indicate a first order transition from the \cdtxt{}
phase at low $T$ to a disordered regime at high temperature.

At the phase transition, $C_{Cd}^{d}$ decreases rapidly. Conversely,
$C_{Cd}^{x}$ displays a sudden increase such that $C_{Cd}^{x} >
C_{Cd}^{d}$ in the high temperature regime. Consequently, in 
the vicinity of the phase
transition not only the long range order, but also the {\em local
arrangement} of the Cd atoms changes. In the disordered phase they
prefer to arrange in rows which are characteristic of the \cdtxo{}
reconstruction.

\begin{figure}
\begin{picture}(100, 33)(0, 0)
\put(0,33){\resizebox{0.48\textwidth}{!}{\rotatebox{270}{\includegraphics{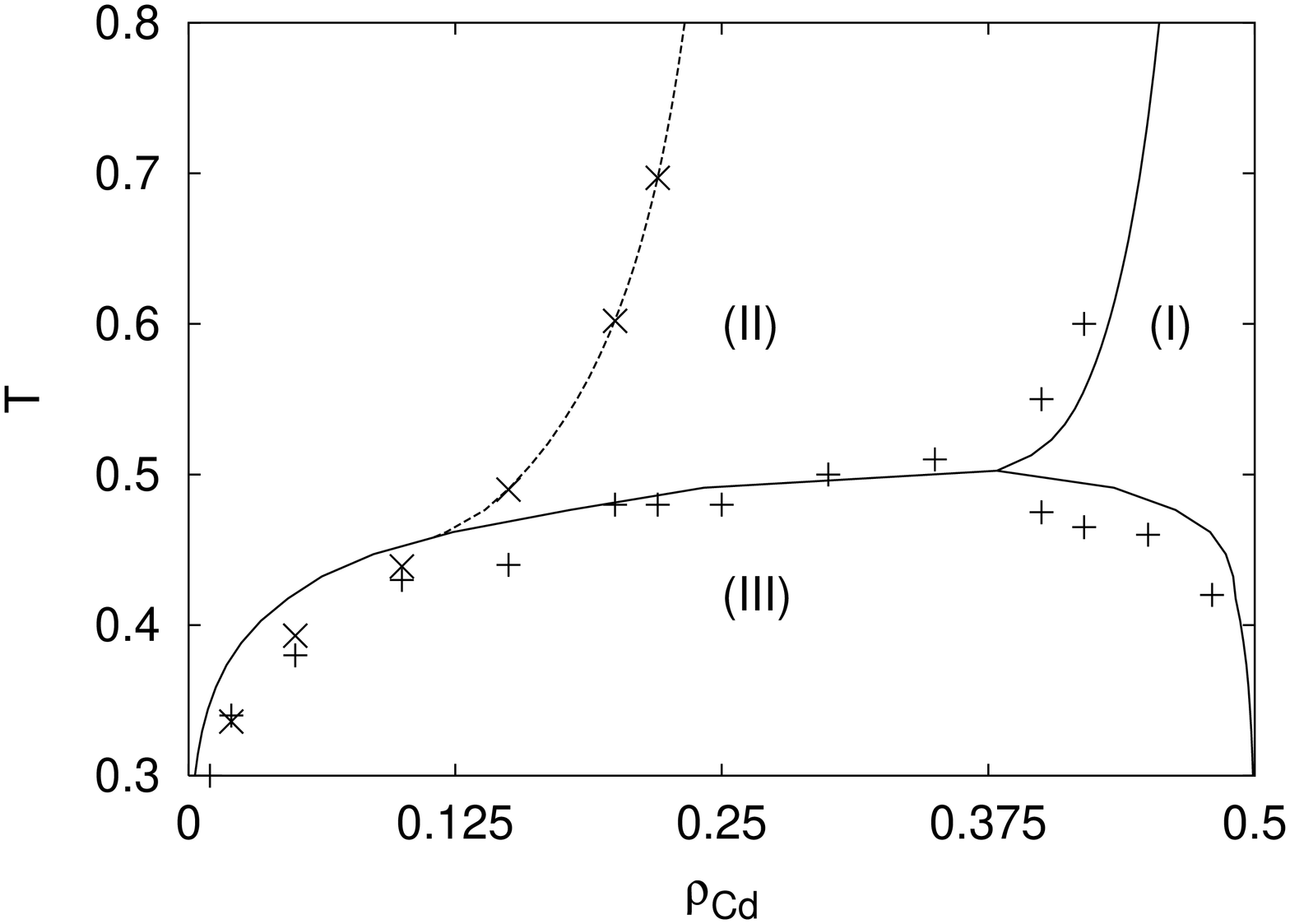}}}}
\put(50,33){\resizebox{0.48\textwidth}{!}{\rotatebox{270}{\includegraphics{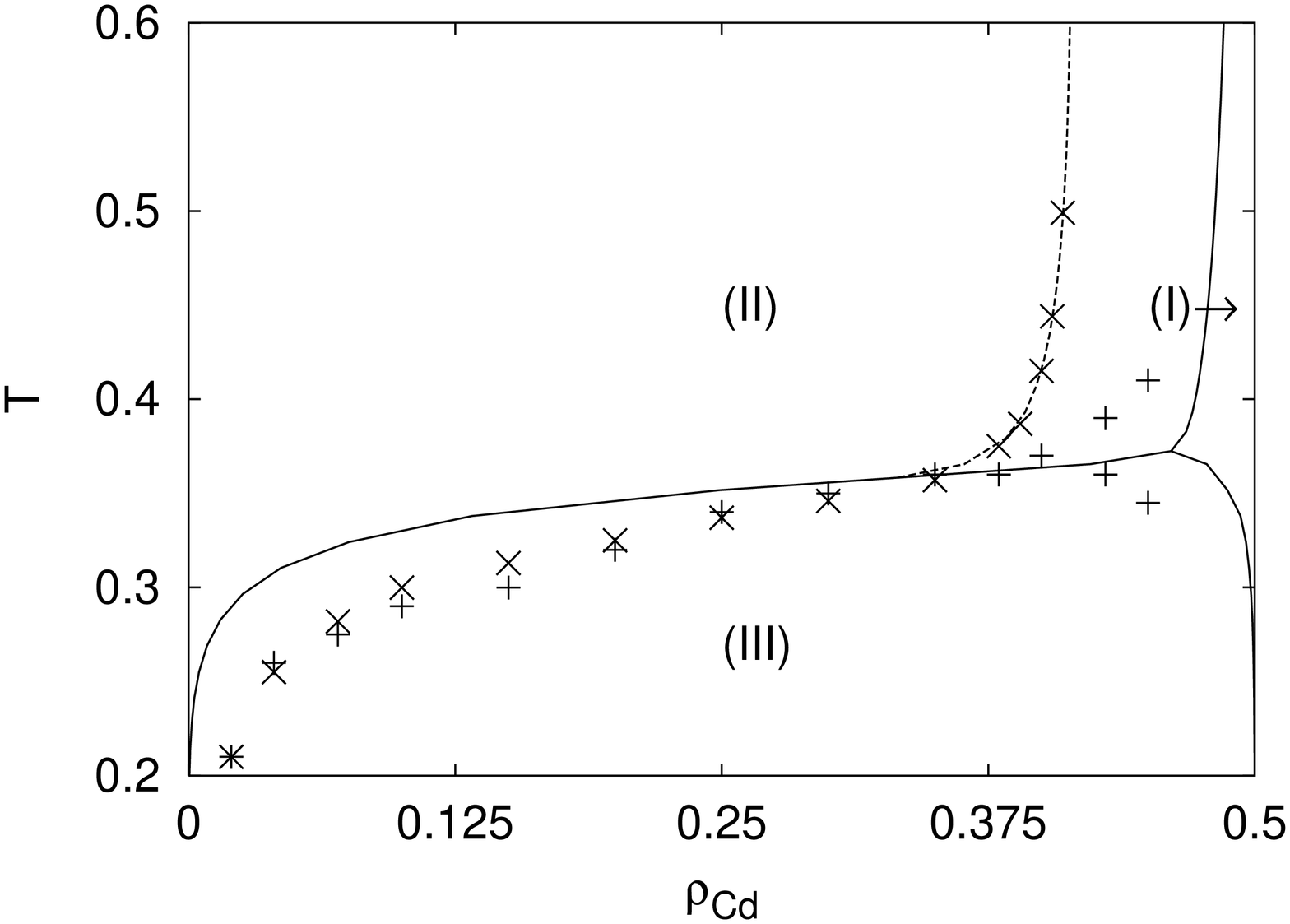}}}}
\put(0,30){(a)}
\put(50,30){(b)}
\end{picture}
\caption{Phase diagrams of the simplified model without Te
dimerization in the $\rho_{Cd}$-$T$ plane at $\varepsilon_x = -1.6$
(panel (a)) and $\varepsilon_x = -1.9$ (panel (b)). In both figures we
have $\varepsilon_d = -1$. The solid lines show the phase boundaries
which have been extrapolated from transfer matrix calculations with
strip widths $L = 14$, $16$ and $18$. The crosses show results of
Monte Carlo simulations of a $128 \times 128$ system at fixed
$\rho_{Cd}$. Phase (I) is homogeneously ordered with a \cdtxt{}
reconstruction whereas phase (II) is disorderd. In region (III),
phases (I) and (II) coexist. At the dashed line, we have $C_{Cd}^{x} =
C_{Cd}^{d}$. The symbols ($\times$) show the respective results of the
Monte Carlo simulations. Statistical errors of the simulation data are
on the order of 0.05. \label{dreisphas}}
\end{figure}
In figure \ref{dreisphas}, we show phase diagrams of the simplified
model at $\varepsilon_x = -1.6$ (figure \ref{dreisphas}a) and
$\varepsilon_x = -1.9$ (figure \ref{dreisphas}b) in the
$\rho_{Cd}$-$T$ plane. In addition to the transfer matrix
extrapolation, we obtain a rough estimate of the phase diagram from
canonical Monte Carlo simulations at constant Cd
coverage\footnote{These simulations have been performed by
T. Volkmann. See \cite{v00} for additional information.}. For this
purpose, a non-local algorithm which exchanges empty and occupied
sites according to Kawasaki rates \cite{nb99,v00} is applied. A rapid
decrease of $M_{Cd}^{(2\times 2)}$ with $T$ indicates the transition
into the disordered phase. A search for a pronounced maximum in the
fluctuation of the order parameters yields results which are identical
within errorbars. The symbols in figure \ref{dreisphas} show the
results for $N = L = 128$ which are in reasonable agreement with the
transfer matrix prediction.

At low temperature and high $\rho_{Cd}$, the system is in the
homogeneously ordered phase (I). The disordered phase (II) is found at
extremely low Cd coverage. In the regime (III) at intermediate Cd
coverages, the phases (I) and (II) coexist. $\rho_{Cd}$ in both phases
is given by the right and the left boundary of region (III),
respectively. The coexistence regime (III) disappears at a tricritical
point. At high temperature the system is either homogeneously ordered
or disordered, depending on coverage. The transition between these
regimes is continuous in terms of $M_{Cd}^{(2\times 2)}$ and should
belong to the Ising universality class. Our transfer matrix
calculations yield a temperature exponent\footnote{The exponents $y_T$
and $y_h$ are defined in appendix \ref{finitesizekontinuierlich}.}
$y_T = 1$ which is consistent with this picture. However, a
calculation of the exponent $y_h$ is not possible without the
introduction of staggered fields. In the limit $T \rightarrow \infty$,
the line of the transition between the phases (I) and (II) approaches
the line $\rho_{Cd} = 1/2$. This can be understood from the fact that
at very high temperature the infinite repulsion between nearest
neighbours in $y$-direction is the only relevant interaction. Then,
colums of lattice sites aligned in $y$-direction decouple such that
the system is disordered at arbitrary $\rho_{Cd}$. This is in contrast
to isotropic hard square lattice gases \cite{ber86,ks81} where an
extended regime (I) persists for arbitrary temperature.

To characterize the short range order of the Cd atoms, we have
determined the line where $C_{Cd}^{x} = C_{Cd}^{d}$. Since this is not
a line of phase transition, it is not possible to apply finite size
scaling theory to extrapolate its position to the limit of an infinite
system. However, the results of transfer matrix calculations for
various $L$ and the canonical Monte Carlo simulations at $N = L = 128$
agree well such that finite size effects are probably weak. At the
right of the dashed line in figure \ref{dreisphas}, the \cdtxt{}
ordering is prevalent. At smaller $\rho_{Cd}$, the Cd atoms arrange
preferentially in a \cdtxo{} order.

Figure \ref{dreismuconst} demonstrates the crucial role that the
difference between the surface energies of perfect \cdtxt{} and \cdtxo{}
reconstructions plays for the phase diagram. With increasing $\Delta
E$, the tricritical point shifts to smaller coverage and higher
temperature. Even more so does the line which separates \cdtxt{} from
\cdtxo{} prevalence.

\subsection{Results of the model with Te dimerization \label{dreiresults}}

To get insight into the typical behaviour of the model, we will first
present a detailed investigation using the parameter set $\varepsilon_d =
\varepsilon_b = -1$, $\varepsilon_x = \varepsilon_t = -1.9$. The choice
$\varepsilon_x = -1.9$ reflects the fact that the energy difference
between \cdtxt{} and \cdtxo{} is small. The choice $\varepsilon_b = -1$
yields low concentrations of Te atoms which are neither dimerized nor
bound to Cd atoms. This is consistent with experimental results.  For
simplicity, we have chosen $\varepsilon_t = \varepsilon_x$ as a typical
example of the case where both couplings are of the same order of
magnitude.  However, since there is no next-nearest neighbour
interaction between dimers, there is a significant difference in the
surface energies of \tetxo{} and $c(2\times2)_{\mathrm{Te} }$, which is
consistent with the results of \cite{gffh99}. 

Below, we discuss the influence of the
numerical values of the parameters on the phase diagram. We will show,
that a smaller value of $\varepsilon_t$ affects the phase diagram
qualitatively. In particular, this concerns the high temperature
phases. 

\subsubsection{Grand-canonical ensemble}
 
With the above parameter set, the Cd and the Te terminated
groundstates are separated by $\mu_0 = 0.9$.
\begin{figure}
\begin{center}
\begin{picture}(100, 66)(0, 0)
\put(0, 66){\resizebox{0.48\textwidth}{!}{\rotatebox{270}{\includegraphics{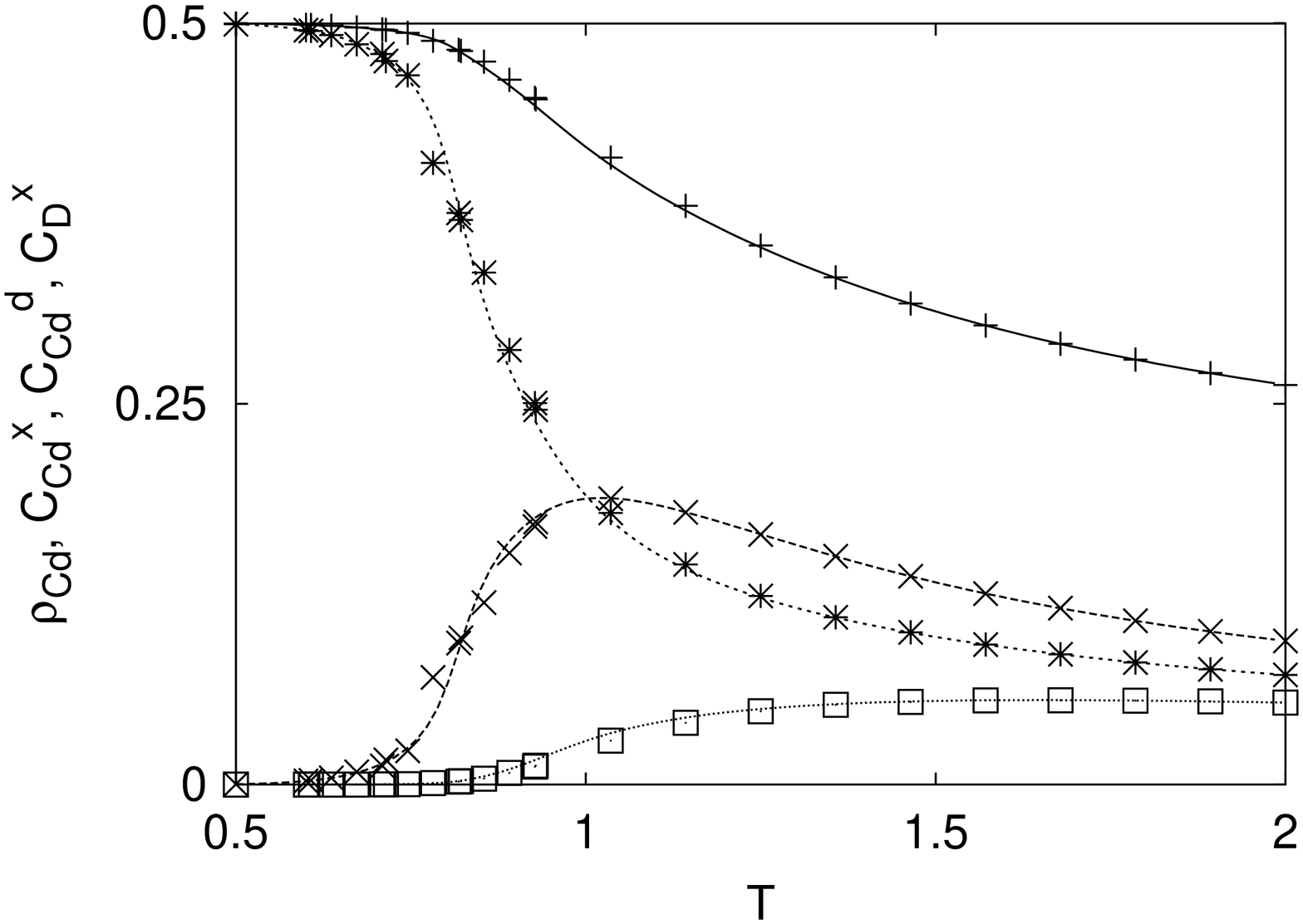}}}}
\put(50, 66){\resizebox{0.48\textwidth}{!}{\rotatebox{270}{\includegraphics{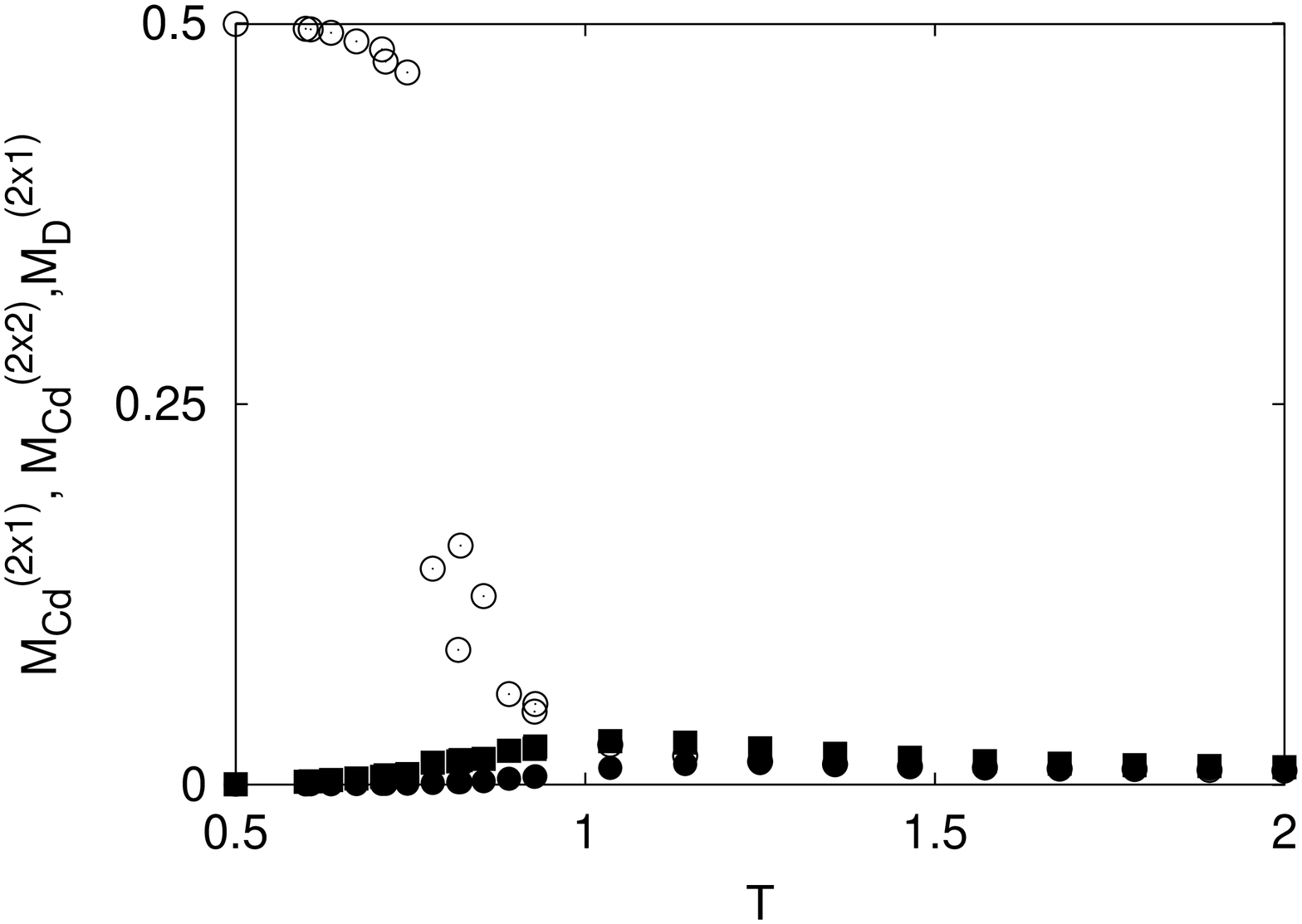}}}}
\put(0, 63){(a)}
\put(50, 63){(b)}
\put(0, 33){\resizebox{0.48\textwidth}{!}{\rotatebox{270}{\includegraphics{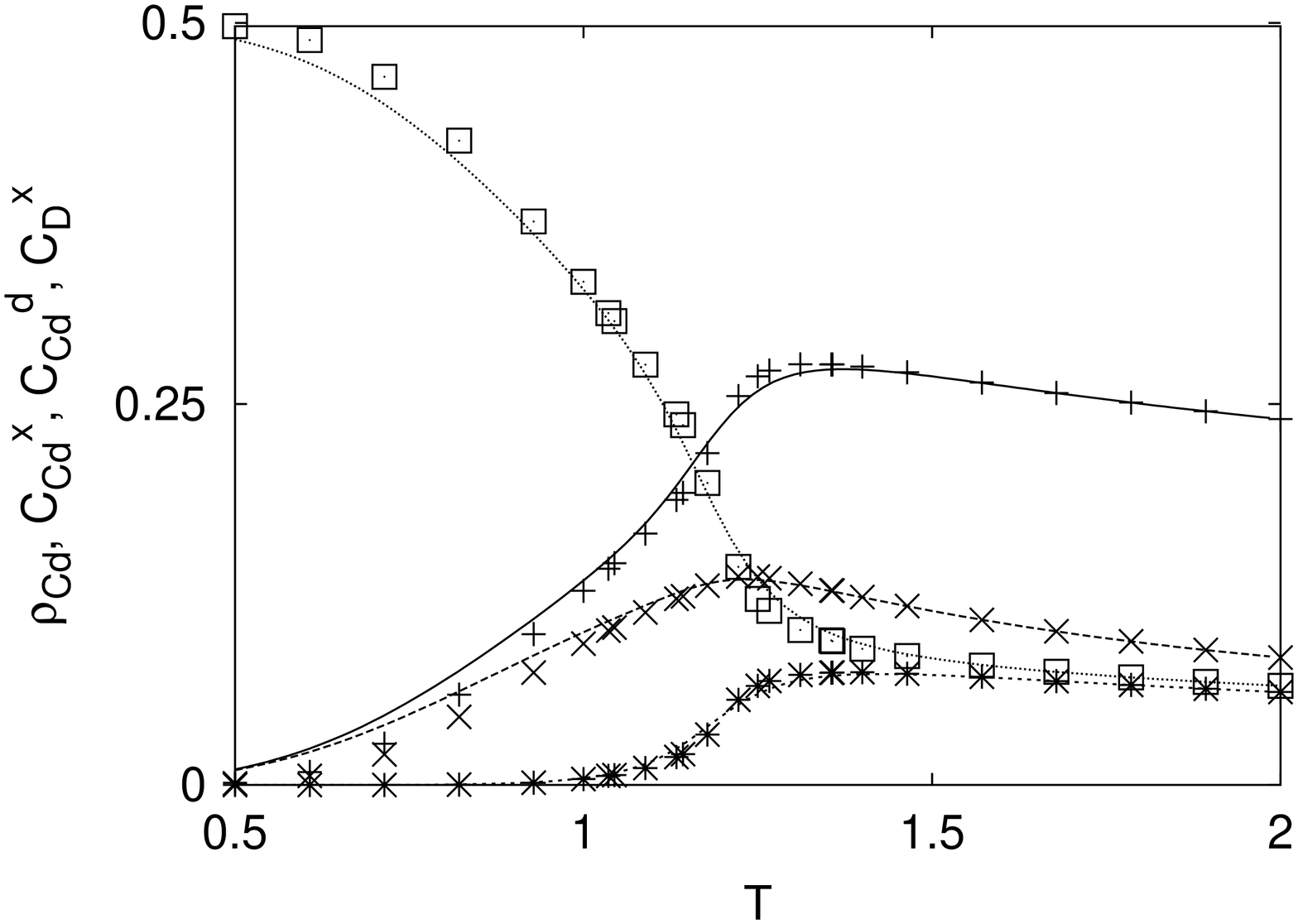}}}}
\put(50, 33){\resizebox{0.48\textwidth}{!}{\rotatebox{270}{\includegraphics{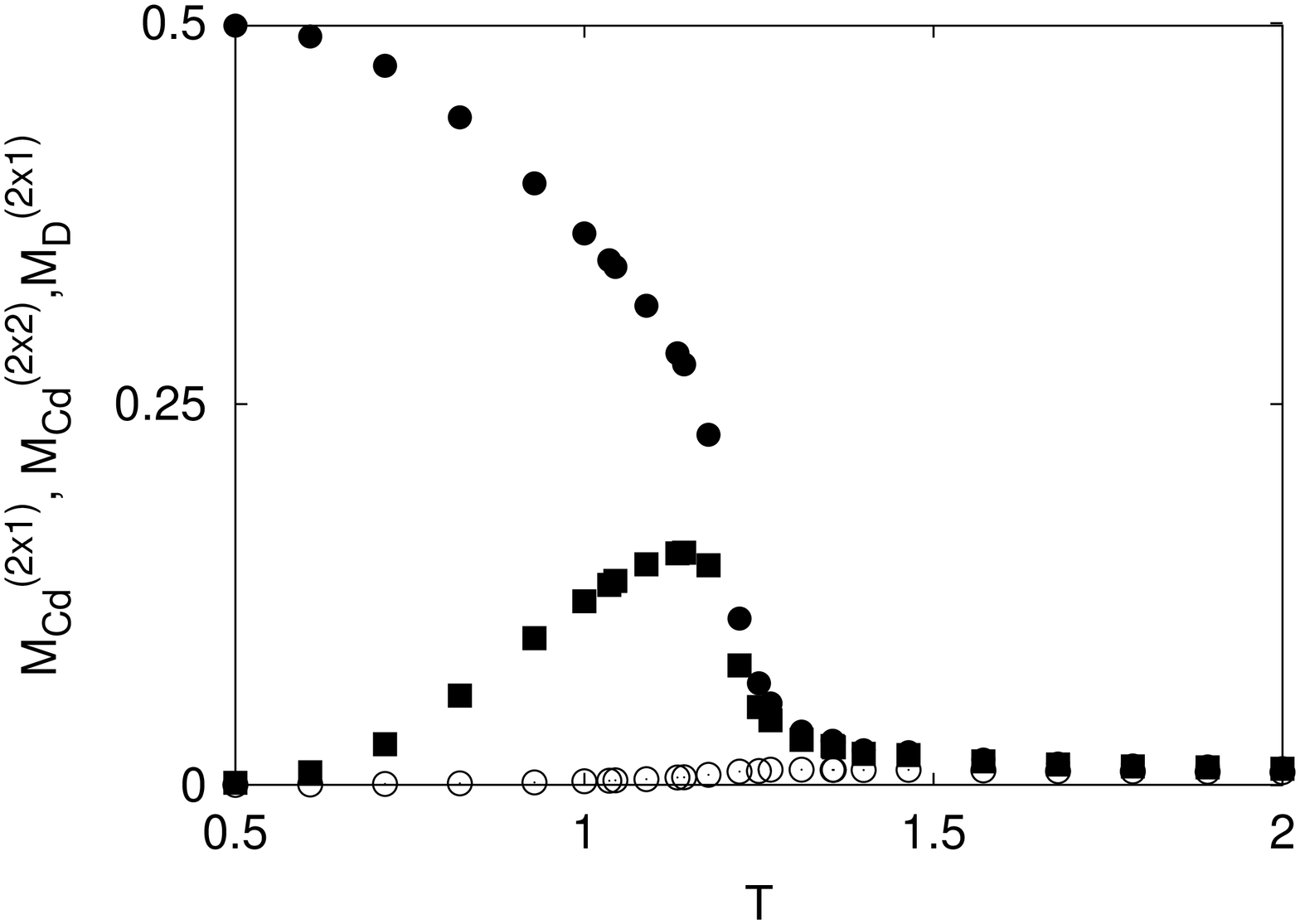}}}}
\put(0, 30){(c)}
\put(50, 30){(d)}
\end{picture}
\end{center}
\caption{Simulations at constant chemical potential. Panels (a), (b)
show the behaviour of the system with the couplings $\varepsilon_d =
\varepsilon_b = -1$, $\varepsilon_x = \varepsilon_t = -1.9$ and a cadmium
chemical potential $\mu = 1$. With these parameters, the ground state
of the system is \cdtxt{}. Panel (a) shows the Cd coverage $\rho_{Cd}\;
(+)$ and the correlations $C_{Cd}^{x} \; (\times)$, $C_{Cd}^{d}\;
(\plustimes)$ and $C_{d}^{x}\; (\boxdot)$. Panel (b) shows the mean absolute
of the order parameters $M_{Cd}^{(2\times2)}\; (\odot)$,
$M_{Cd}^{(2\times1)} \; (\blacksquare)$ and $M_{D}^{(2\times1)}\;
(\bullet)$.  The symbols show data from a simulation run at a system
size $L = N = 64$.  $4 \cdot 10^4 \cdot L N$ events have been
performed both for equilibration and for measurement. The lines have
been obtained by means of a transfer matrix calculation at a strip
width $L = 10$. The data shown in panels (c), (d) have been obtained
at $\mu = 0.8$, where the ground state is \tetxo{}. All other parameters
are identical to those used in (a), (b). The meaning of the symbols is
the same as in panels (a) and (b).\label{dreifig1}}
\end{figure}
In figure \ref{dreifig1} the temperature dependent behaviour of the
system at constant chemical potential is shown for $\mu = 1$ (figure
\ref{dreifig1}a,b) and $\mu = 0.8$ (figure \ref{dreifig1}c,d), which
are examples for both cases. At $\mu = 1$, the ordered \cdtxt{} phase
at low $T$ manifests itself in a high Cd coverage $\rho_{Cd}$ and
values of the correlation $C_{Cd}^{d}$ and the order parameter
$M_{Cd}^{(2\times2)}$ close to $0.5$. In figure \ref{dreifig2}d a
surface snapshot in this phase at $T = 0.71$ is shown. As expected,
the Cd atoms arrange preferentially in a checkerboard configuration.
At a temperature $T = 0.83$, there is a first order phase transition
to a disordered phase. This is indicated by a decrease of the order
parameter $M_{Cd}^{(2\times2)}$ and the correlation
$C_{Cd}^{d}$. Simultaneously, $C_{Cd}^x$ starts to increase. At $T =
1$, both lines cross such that the high temperature behaviour of the
system is dominated by a local $(2\times1)$ ordering of the Cd
atoms. The simulation data shown in figure \ref{dreifig1}a,b have been
taken from two independent simulation runs. The system was initialized
with a perfect \cdtxt{} configuration at the lowest temperature
investigated. Then, as successively higher temperatures were imposed,
the surface configuration was kept as initial state of the next
simulation. This reflects in a small hysteresis effect in
$M_{Cd}^{(2\times2)}$ due to the first order nature of the phase
transition. Qualitatively, the behaviour of the Cd atoms is quite
similar to that found in the simplified model (figure
\ref{dreismuconst}).

At $\mu = 0.8$, we obtain a completely different behaviour. Since the
groundstate is a \tetxo{} reconstruction, at small $T$ we measure low Cd
coverages and values of $C_{D}^{x}$ and $M_{D}^{(2\times1)}$ close to
$0.5$. The most frequent thermal excitations are Cd adatoms, the
density $\rho_{Cd}$ of which increases with $T$. Figure \ref{dreifig2}c
shows a surface snapshot at $T = 0.93$. The Cd atoms preferentially
arrange in rows such that $C_{Cd}^{x} \gg C_{Cd}^{d}$.  These rows
adapt to the structure of the \tetxo{} reconstruction. This yields
nonzero values of the order parameter $M_{Cd}^{(2\times1)}$, which
indicate a global ordering of Cd atoms in a $(2\times1)$ arrangement.
However, the interactions between the Cd atoms themselves are
insufficient to stabilize this global order. Instead, it is purely
induced by the environment of the Te dimers.

At $T = 1.23$, there is a first order phase transition above which the
system is in a disordered phase similar to that found at $\mu =
1$. Remarkably, we observe high $\rho_{Cd} \approx 0.2$ at
temperatures slightly below the phase transition. At the phase
transition, $\rho_{Cd}$ jumps to an even higher value.  Above the
transition it decreases slightly with $T$.

\subsubsection{Phase diagram}

\begin{figure}
\begin{picture}(100, 63)(0, 0)
\put(0,63){\resizebox{0.48\textwidth}{!}{\rotatebox{270}{\includegraphics{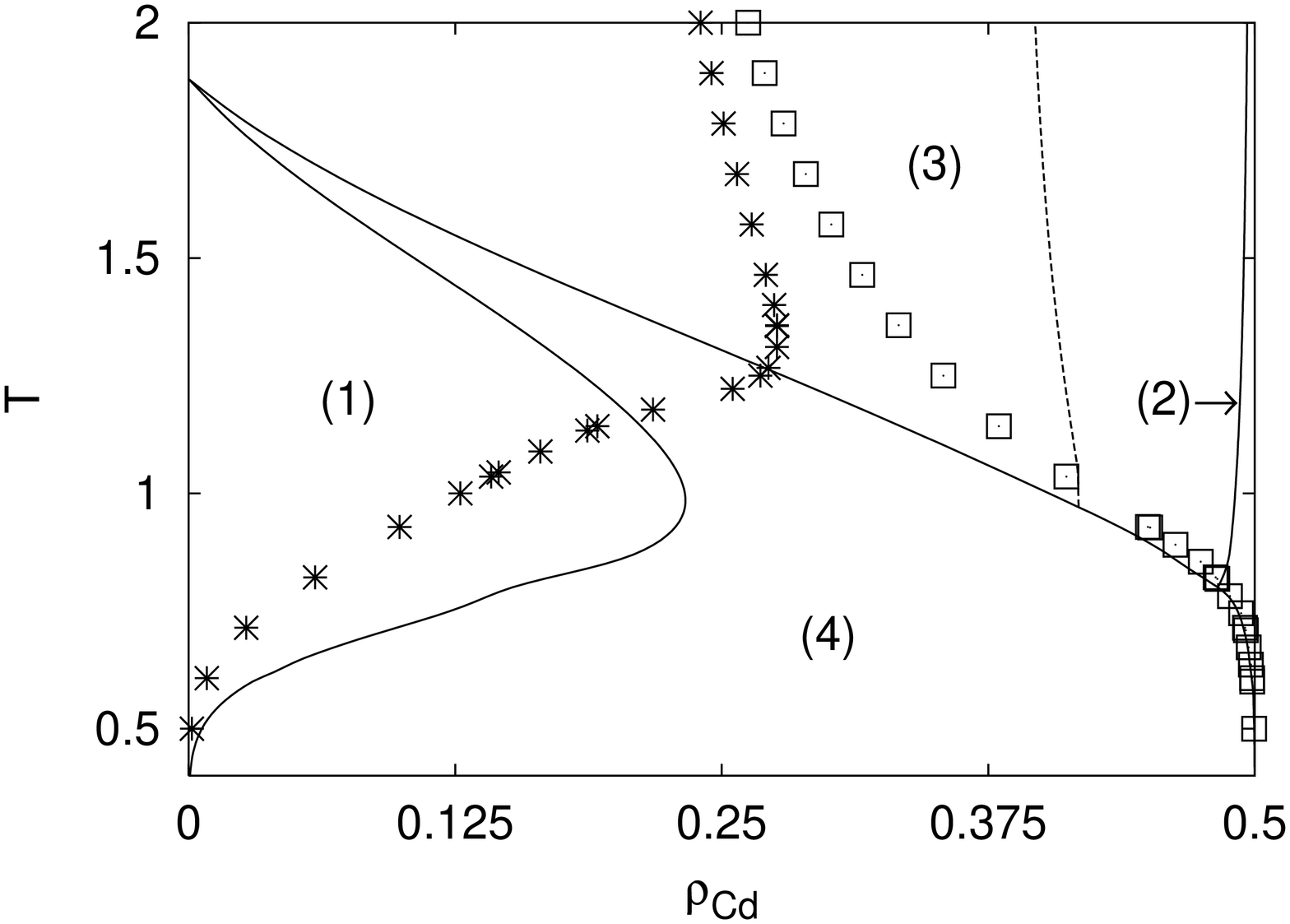}}}}
\put(50,63){\resizebox{0.48\textwidth}{!}{\rotatebox{270}{\includegraphics{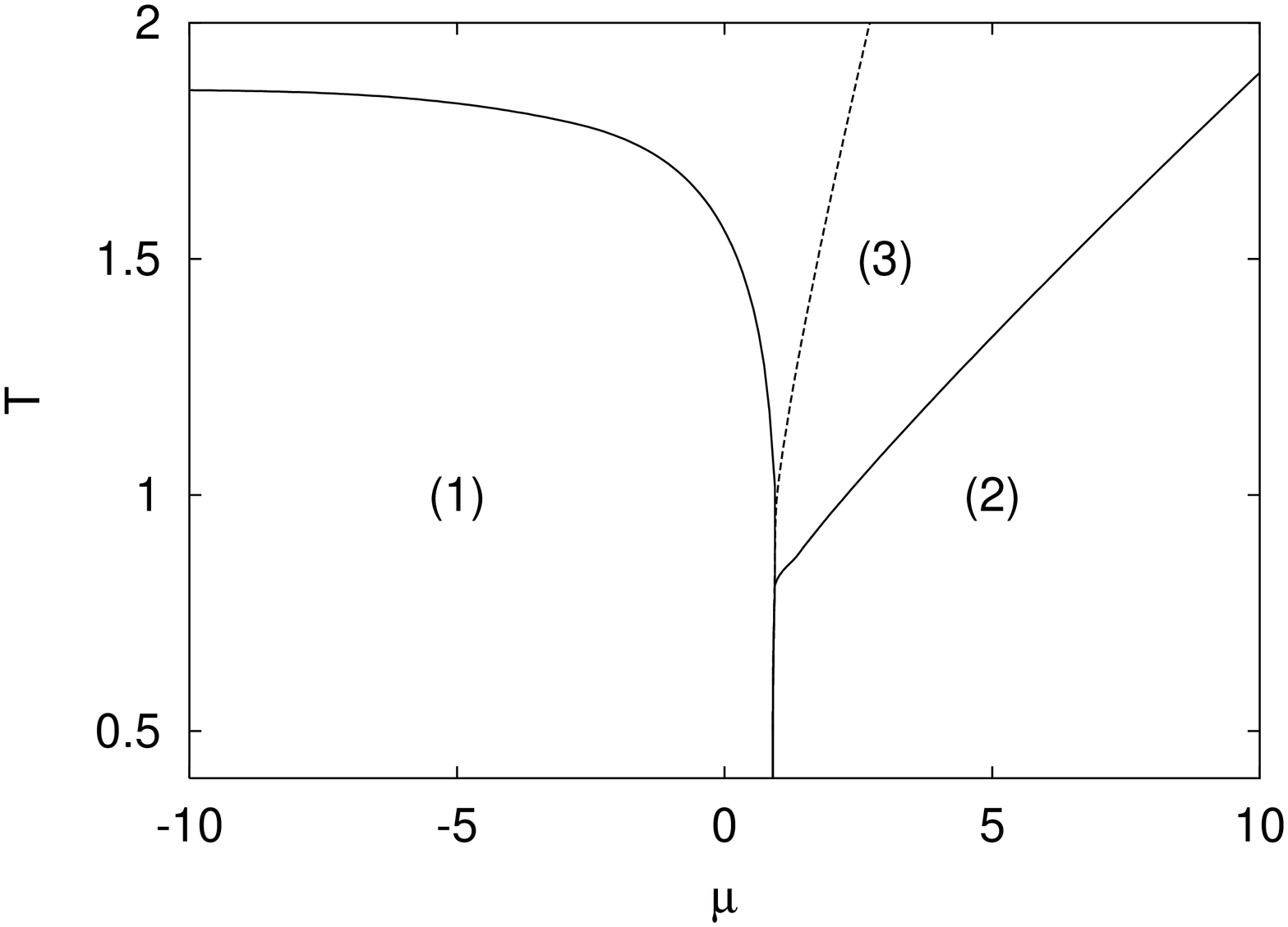}}}}
\put(0,60){(a)}
\put(50,60){(b)}
\put(0, 3){\resizebox{0.23\textwidth}{!}{\includegraphics{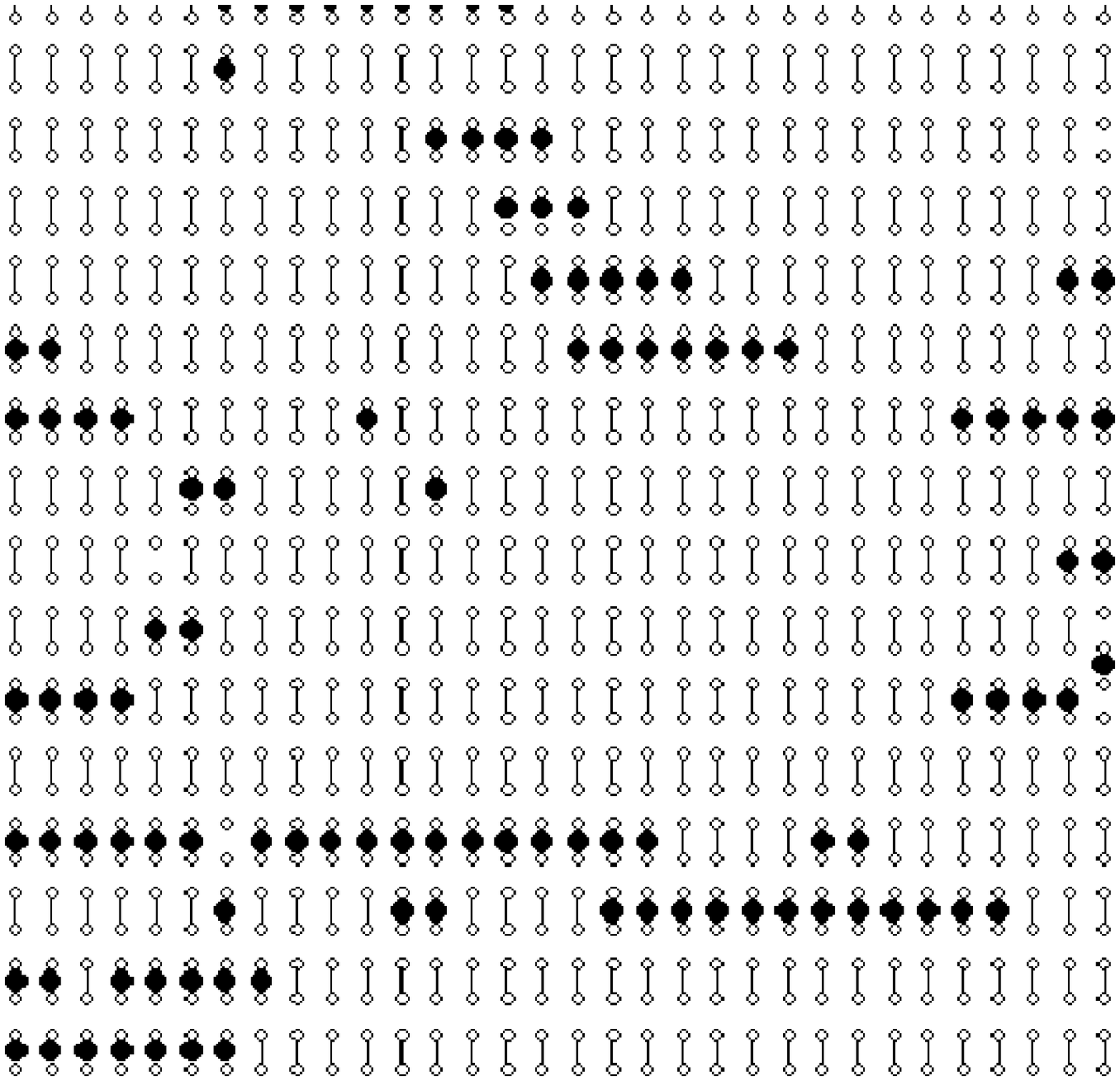}}}
\put(25, 3){\resizebox{0.23\textwidth}{!}{\includegraphics{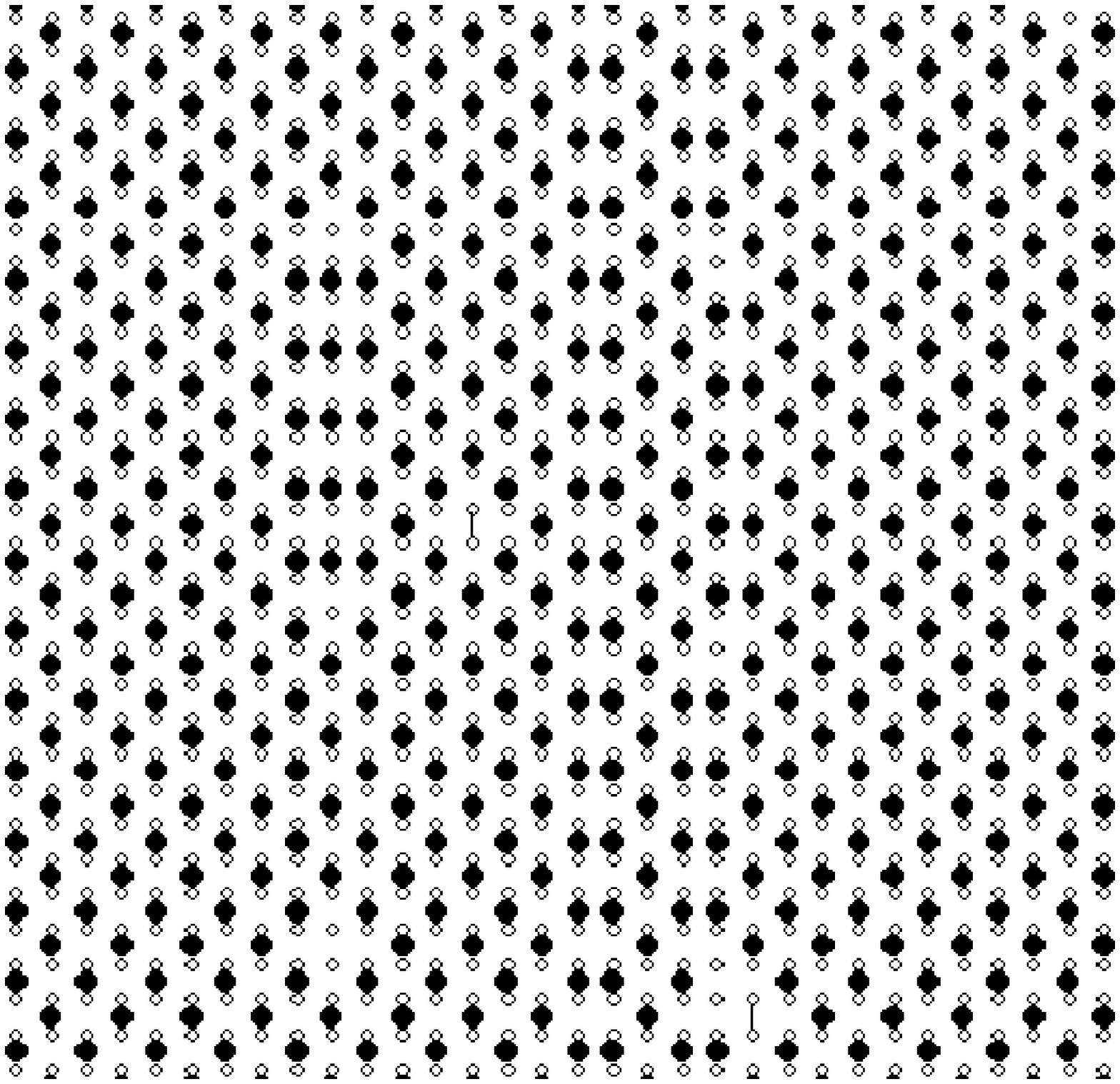}}}
\put(50, 3){\resizebox{0.23\textwidth}{!}{\includegraphics{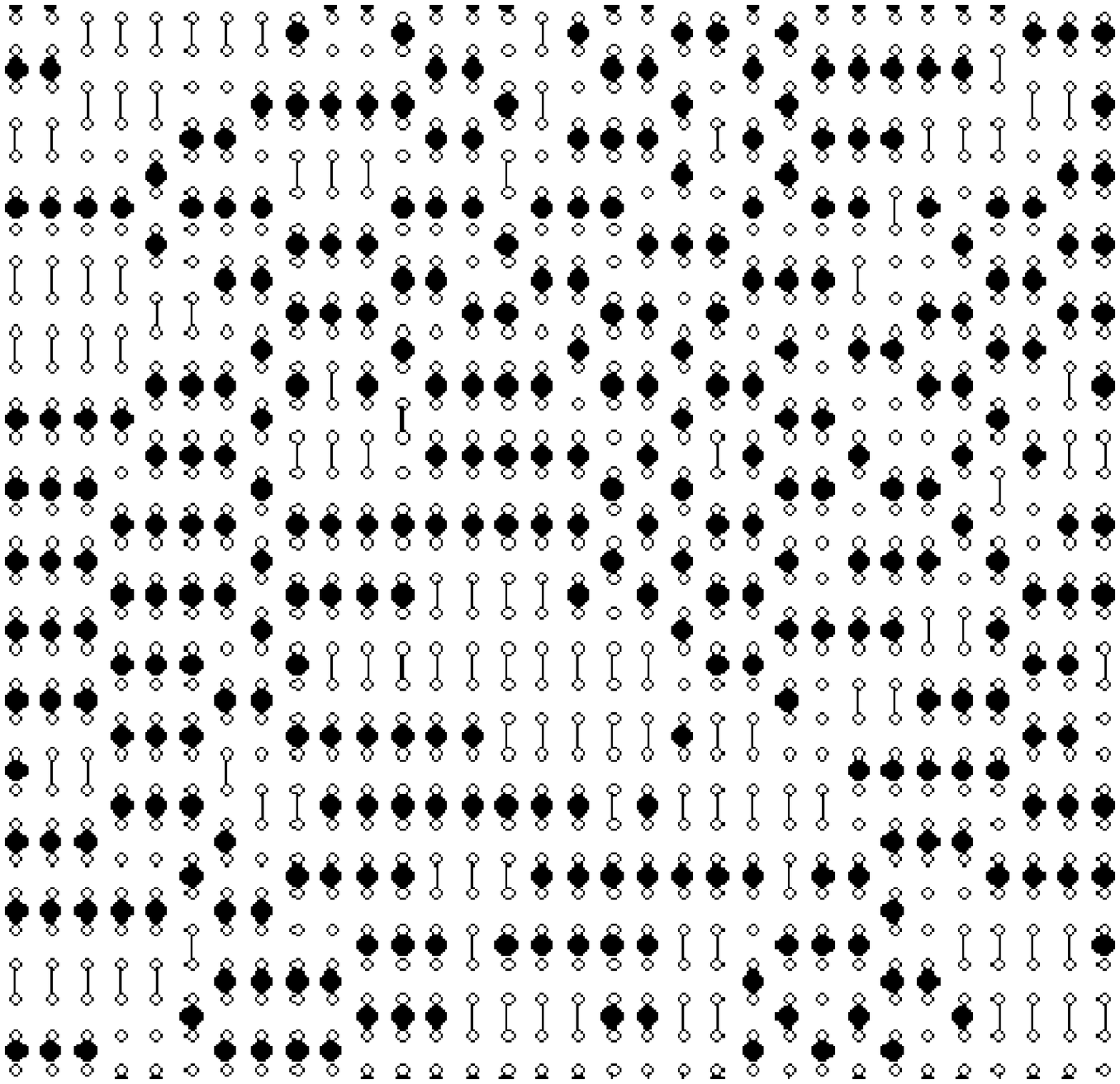}}}
\put(75, 3){\resizebox{0.23\textwidth}{!}{\includegraphics{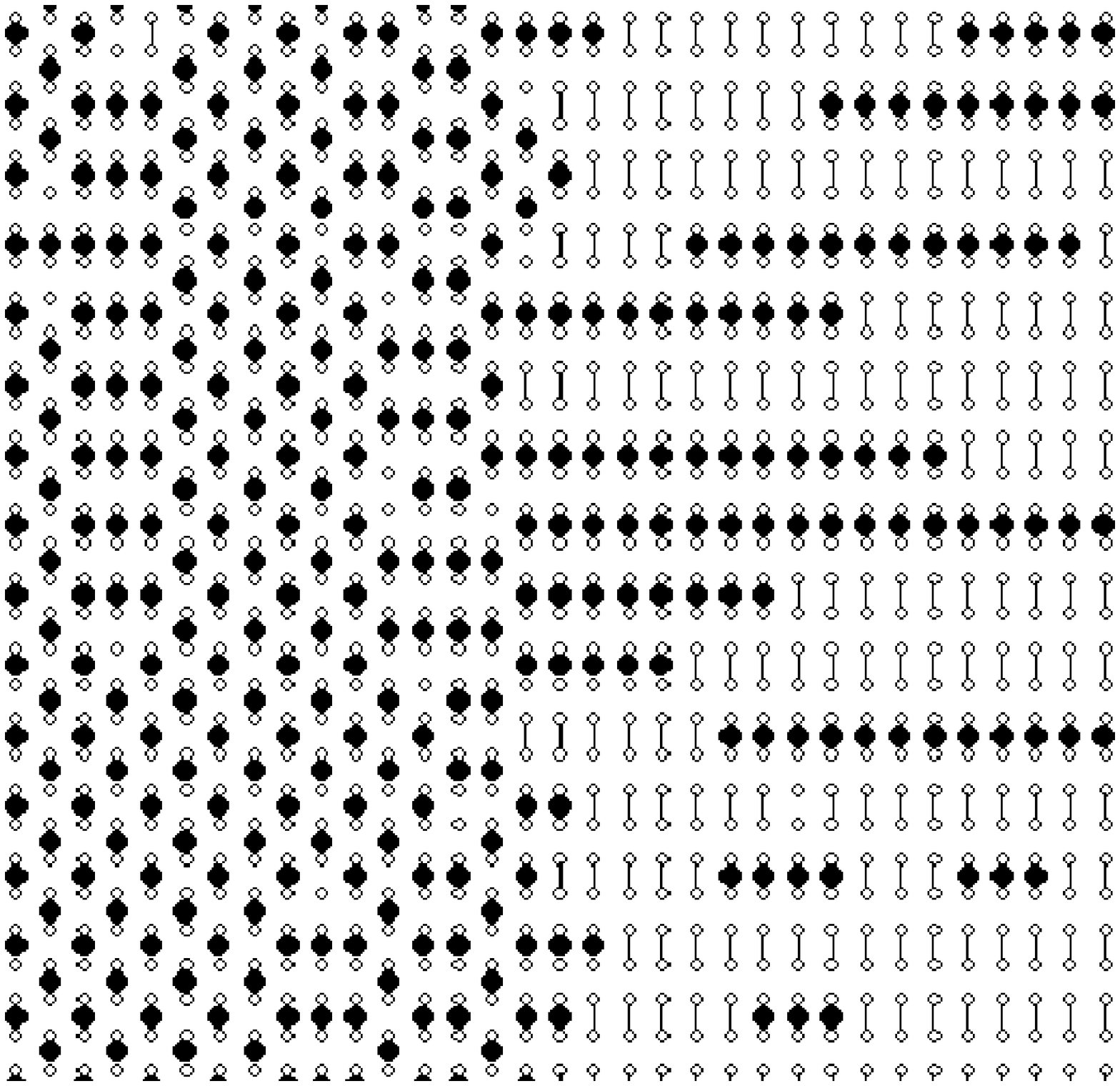}}}
\put(10, 0){(c)}
\put(35, 0){(d)}
\put(60, 0){(e)}
\put(85, 0){(f)}
\end{picture}
\caption{Phase diagram with parameters $\varepsilon_d = \varepsilon_b
= -1$, $\varepsilon_x = \varepsilon_t = -1.9$. The solid lines in
panel (a) show the lines of phase transitions as functions of
$\rho_{Cd}$ and $T$, in panel (b) they are shown in the $\mu$-$T$
plane. Note the offset in the temperature axes. In region (1), the
system is in a homogeneously ordered phase with a \tetxo{}
reconstruction, in region (2) it is homogeneously ordered and \cdtxt{}
reconstructed. Region (3) is the disordered phase. On the left side of
the dashed line, the Cd atoms show preferentially a local $(2\times1)$
ordering ($C_{Cd}^{x} > C_{Cd}^{d}$), while on its right side a
$c(2\times2)$ ordering dominates. Due to the coverage discontinuity
between (1) and (2),(3), there is a coexistence regime where regions
with high and low $\rho_{Cd}$ coexist (4). The symbols show lines of
constant $\mu=0.8$ ($\plustimes$) and $\mu = 1$ ($\boxdot$) (same data
as in figure \ref{dreifig1}a,c). Panels (c)-(f) show typical surface
snapshots. (c)-(e) correspond to grand-canonical simulations. The
snapshots show surface configurations after $8 \cdot 10^4 \cdot L N$
events. (c): $T = 0.93$, $\mu = 0.8$ (phase 1), (d): $T = 0.71$, $\mu
= 1$ (phase 2), (e): $T = 1.14$, $\mu = 1$ (phase 3). (f) displays a
surface configuration after $2 \cdot 10^4 \cdot L N $ events in a
canonical simulation at $T = 0.73$, $\rho_{Cd} = 0.35$ (Region 4,
coexistence of phases 1 and 2).  All snapshots show sections of
$16\sqrt{2} \times 16\sqrt{2}$ zinc-blende lattice constants of
systems of size $L = N = 64$. \label{dreifig2}}
\end{figure}
Figure \ref{dreifig2}a,b shows the phase diagram of the model, which
has been extrapolated from transfer matrix calculations with strip
widths $L$ of 6, 8 and 10. In figure \ref{dreifig2}b, the lines of
phase transitions in the $\mu$-$T$ plane have been plotted. At low
temperature, the system is either in an ordered \tetxo{} (1) or an
ordered \cdtxt{} phase (2). The line of phase transition between these
phases starts at zero temperature and $\mu = \mu_0$ where the energies
of both reconstructions are degenerate. For $0 < T < T_t = 0.84$, it
remains at the same chemical potential $\mu_0$, apart from small
numerical uncertainties of extrapolation. In consequence, a phase
transition between a \cdtxt{} and a \tetxo{} reconstruction cannot be
observed at constant chemical potential. The disordered phase (3)
exists for $T \geq T_t$.  At the point $(T = T_t, \mu = \mu_0)$ five
phases coexist: the disordered phase, and \cdtxt{} and \tetxo{}
reconstructed phases in two sublattices, corresponding to positive and
negative values of the order parameters $M_{Cd}^{(2\times2)}$ and
$M_{D}^{(2\times1)}$, respectively. The \tetxo{} reconstructed phase
(1) exists only at temperatures below a critical temperature
$T_{c}^{1} \approx |\varepsilon_t|$. At $T_{c}^{1}$, the line of the
phase transition to the disordered phase (3) diverges to $\mu = -
\infty$. On the contrary, the \cdtxt{} reconstructed phase (2) may
exist at arbitrary temperature, if the Cd chemical potential is large
enough.  The dashed line in figure \ref{dreifig2}b shows the chemical
potential at which the correlations $C_{Cd}^{d}$ and $C_{Cd}^{x}$ in
the disordered phase are equal. For smaller $\mu$, the local ordering
of Cd atoms is dominated by a $(2\times1)$ arrangement, while for
larger $\mu$ they prefer a local $c(2\times2)$ ordering.

The phase transition from phase (2) to phase (1) occurs at a
temperature independent chemical potential $\mu_0$. At the transition
between the phases (2) and (3), $\rho_{Cd}$ varies continuously. 
Therefore, there is no coverage discontinuity if temperature
is varied at a constant chemical potential $\mu > \mu_0$ where the
groundstate is a \cdtxt{} reconstruction.

Figure \ref{dreifig2}a shows the phase diagram in the $\rho_{Cd}$-$T$
plane. There is a coverage discontinuity at the phase transition
between the ordered phases (1) and (2) and at the transition between
phases (1) and (3). These discontinuities yield a coexistence regime
(4). Here, for $T < T_t$, \cdtxt{} and \tetxo{} reconstructed phases
coexist, while for $T > T_t$ the \tetxo{} reconstructed phase coexists
with the disordered phase. Figure \ref{dreifig2}f shows a surface
snapshot from a simulation which was performed at a constant Cd
density $\rho_{Cd} = 0.35$ and a temperature $T = 0.73$. This is a
typical surface configuration in the regime where the ordered phases
(1) and (2) coexist. The system is separated in two phases, one with a
high $\rho_{Cd}$ and a \cdtxt{} reconstruction, and another one with a
\tetxo{} reconstruction and a low concentration of Cd adatoms. The
local values of $\rho_{Cd}$ in both regions are given by the left and
the right boundary of the coexistence regime. At low temperature one
obtains $\rho_{Cd} \sim 0.5$ in the \cdtxt{} phase and $\rho_{Cd} \sim
0$ in the \tetxo{} phase. At temperatures $T \gtrapprox 0.6$,
$\rho_{Cd}$ in the \tetxo{} phase increases strongly with $T$ and
obtains its maximal value $0.23$ at $T = 0.98$. At even higher
temperature it decreases with $T$ and becomes zero at $T_{c}^{1}$. On
the contrary, $\rho_{Cd}$ in the \cdtxt{} phase remains high for $T <
T_t$. At $T > T_t$, the Cd-rich phase is disordered. Then, $\rho_{Cd}$
at the right boundary of the coexistence regime decreases with $T$.
At $T_{c}^{1}$, it becomes zero and the coexistence regime
disappears. The dashed line in figure \ref{dreifig2}a marks the values
of $\rho_{Cd}$ at which $C_{Cd}^{x} = C_{Cd}^{d}$. For smaller
coverages, $C_{Cd}^{x} > C_{Cd}^{d}$ such that the local ordering of
the Cd atoms is dominated by a $(2\times1)$ arrangement, while for
greater coverages $C_{Cd}^{x} < C_{Cd}^{d}$. The values of $\rho_{Cd}$
which have been measured in the simulations shown in figure
\ref{dreifig1} are plotted in the phase diagram as examples of lines
of constant chemical potential. The loci of the phase transitions and
the point where $C_{Cd}^{x} = C_{Cd}^{d}$ (at $\mu = 1$) are in good
agreement with the results of the transfer matrix extrapolation.
 
\subsubsection{Canonical ensemble \label{dreicanonicalresults}}

\begin{figure}
\begin{picture}(100, 33)(0, 0)
\put(0,33){\resizebox{0.48\textwidth}{!}{\rotatebox{270}{\includegraphics{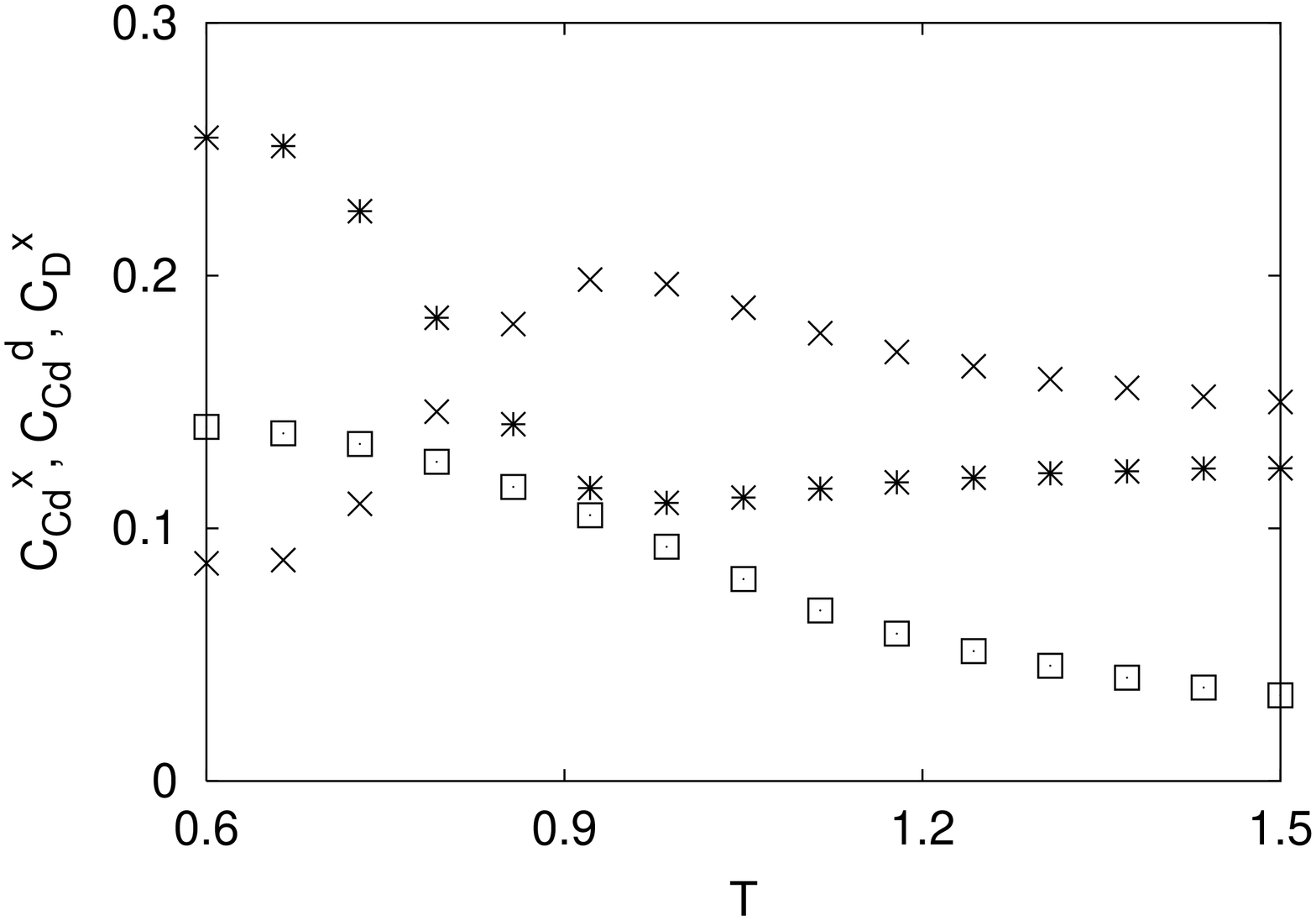}}}}
\put(50,33){\resizebox{0.48\textwidth}{!}{\rotatebox{270}{\includegraphics{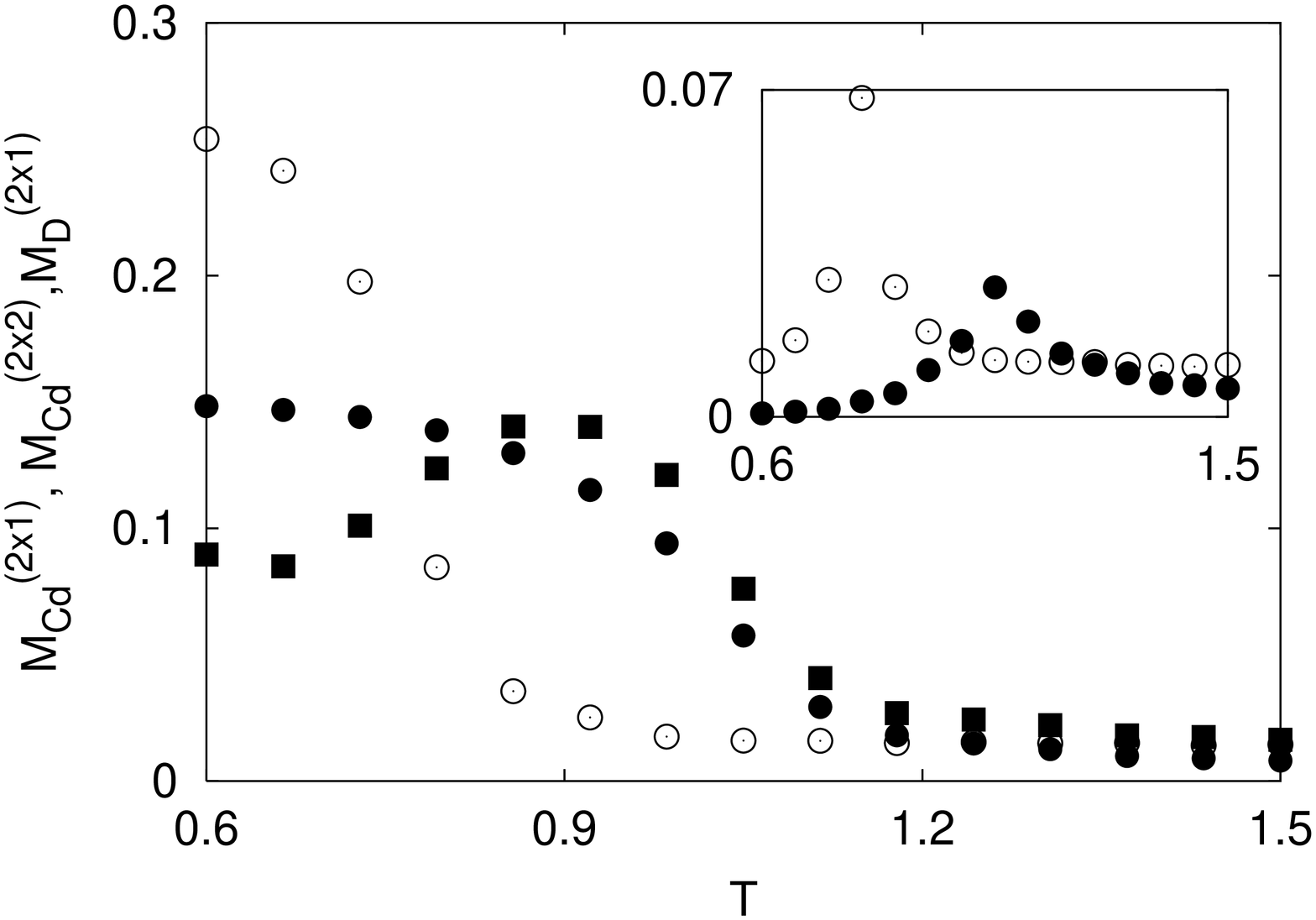}}}}
\put(0,30){(a)}
\put(50,30){(b)}
\end{picture}
\caption{Results of a simulation of a $64\times64$ system at conserved
$\rho_{Cd} = 0.35$ and couplings $\varepsilon_d = \varepsilon_b = -1$,
$\varepsilon_x = \varepsilon_t = -1.9$. $10^4 \cdot L N $ events have
been performed both for equilibration and for data sampling. Panel (a)
shows the correlations $C_{Cd}^{x} \; (\times)$, $C_{Cd}^{d}\;
(\plustimes)$ and $C_{d}^{x}\; (\boxdot)$.  Panel (b) shows the mean
absolute of the order parameters $M_{Cd}^{(2\times2)}\; (\odot)$,
$M_{Cd}^{(2\times1)} \; (\blacksquare)$ and $M_{D}^{(2\times1)}\;
(\bullet)$. The inset shows standard deviations of order
parameters. The meaning of the symbols is the same as in the large
picture. \label{dreifig3}}
\end{figure}

The temperature dependent behaviour of the model at a constant Cd
adatom density $\rho_{Cd} = 0.35$ is shown in figure
\ref{dreifig3}. This is an example which shows the typical behaviour
of the model under the conditions of phase separation. At low
temperature, most of the Cd atoms are concentrated in a Cd rich phase
with a \cdtxt{} reconstruction. The remaining area of the system is
covered with a \tetxo{} reconstructed phase. Since both phases are
long-range ordered, we obtain large values of the order parameters
$M_{Cd}^{(2\times2)}$, $M_{D}^{(2\times1)}$ and the corresponding
correlations. The fraction of Cd atoms which occupy sites in the Te
rich phase yields nonzero values of the order parameter
$M_{Cd}^{(2\times1)}$. As temperature increases, more and more Cd
atoms pass into the Te rich phase. This yields an increase in
$C_{Cd}^{x}$ and $M_{Cd}^{(2\times1)}$ and a decrease of $C_{Cd}^{d}$
and $M_{Cd}^{(2\times2)}$ due to the $(2\times1)$ ordering of the Cd
atoms {\em in the Te rich phase}. At the temperature $T_t$, the Cd
rich phase undergoes the order-disorder transition where
$M_{Cd}^{(2\times2)}$ drops to zero. At this phase transition, the
fraction of Cd atoms which are incorporated in locally $(2\times1)$
ordered configurations {\em in the Cd rich phase} increases. Thus,
there are two independent effects which lead to a dominance of
$C_{Cd}^{x}$ over $C_{Cd}^{d}$ at higher temperature: The specific
shape of the left boundary of the coexistence regime and the
order-disorder transition of the Cd rich phase. At a temperature
$T_c^4$, the system is leaving the coexistence region into phase
(3). At this phase transition, the separation between a Te rich and a
Cd rich phase disappears, such that at high temperature the system is
in a homogeneous, disordered state. The order parameters
$M_{D}^{(2\times1)}$ and $M_{Cd}^{(2\times1)}$ become zero and the
local correlation between dimers, $C_{D}^{(2\times1)}$, decreases. In
the simulations we have determined the critical temperatures of the
phase transitions from the standard deviations of the order parameters
$M_{Cd}^{(2\times2)}$ and $M_{D}^{(2\times1)}$. $M_{Cd}^{(2\times2)}$
is peaked at $T_t$, where the long range order of the Cd atoms is
lost, while $M_{D}^{(2\times1)}$ is peaked at $T_{c}^4$, where the
system leaves the coexistence regime and the dimers lose their long
range order. We obtain $T_t = 0.79 \pm 0.2$ and $T_c^4 = 1.0 \pm 0.2$,
where the uncertainty is due to the temperature spacing between the
single simulations. These values are systematically lower than the
theoretical results of the transfer matrix extrapolation ($T_t =
0.84$, $T_c^4 = 1.11$). However, the theoretical results are valid in
the limit of an infinite system size, where the free energy of the
phase boundary can be neglected compared to that of the bulk of the
phases. We have verified that the systematic deviation between theory
and simulations decreases with the system size.

\subsubsection{The influence of the parameter set on the phase diagram
\label{dreiparaminfl}}

\begin{figure}
\begin{picture}(100, 33)(0, 0)
\put(0,33){\resizebox{0.48\textwidth}{!}{\rotatebox{270}{\includegraphics{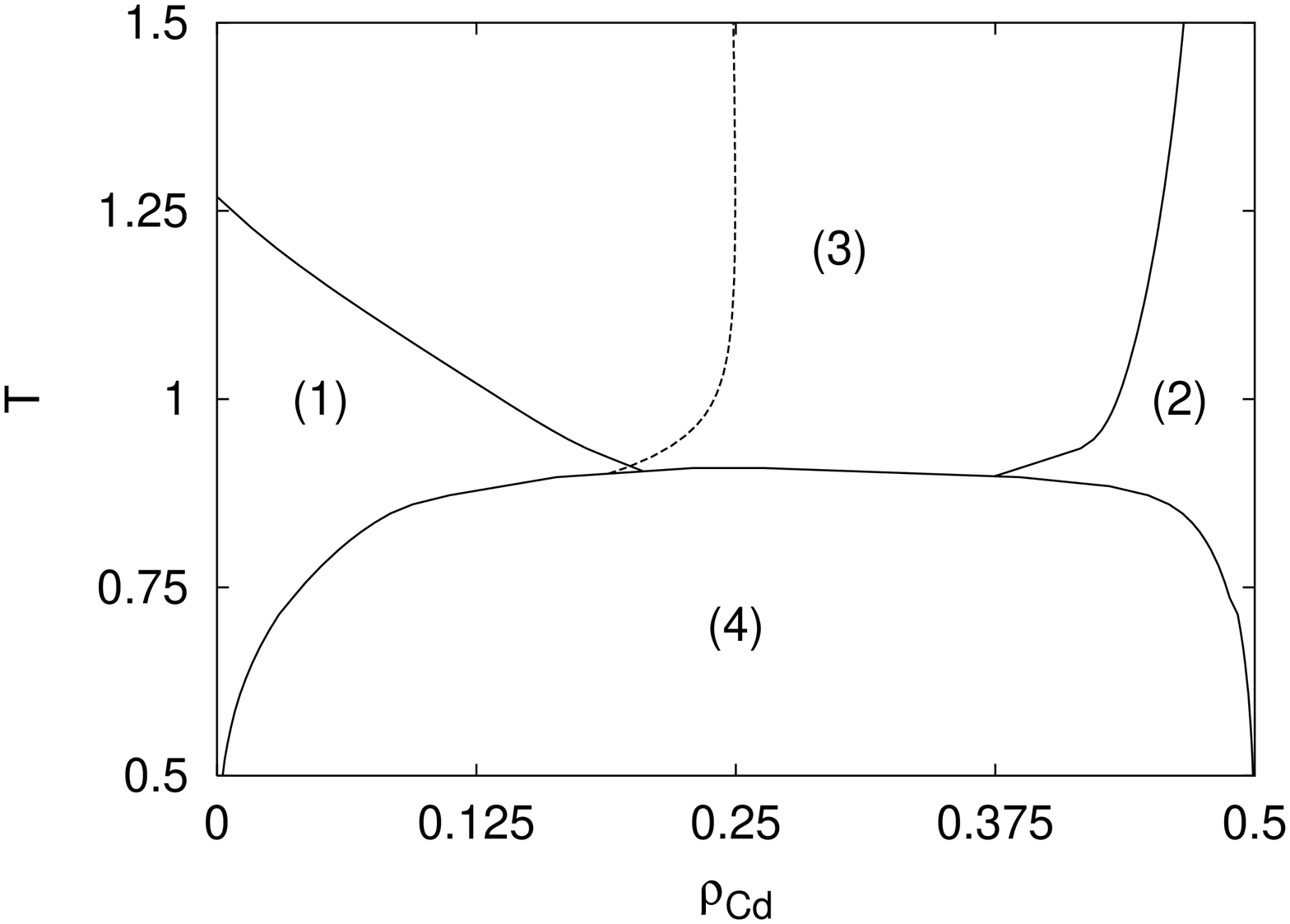}}}}
\put(50,33){\resizebox{0.48\textwidth}{!}{\rotatebox{270}{\includegraphics{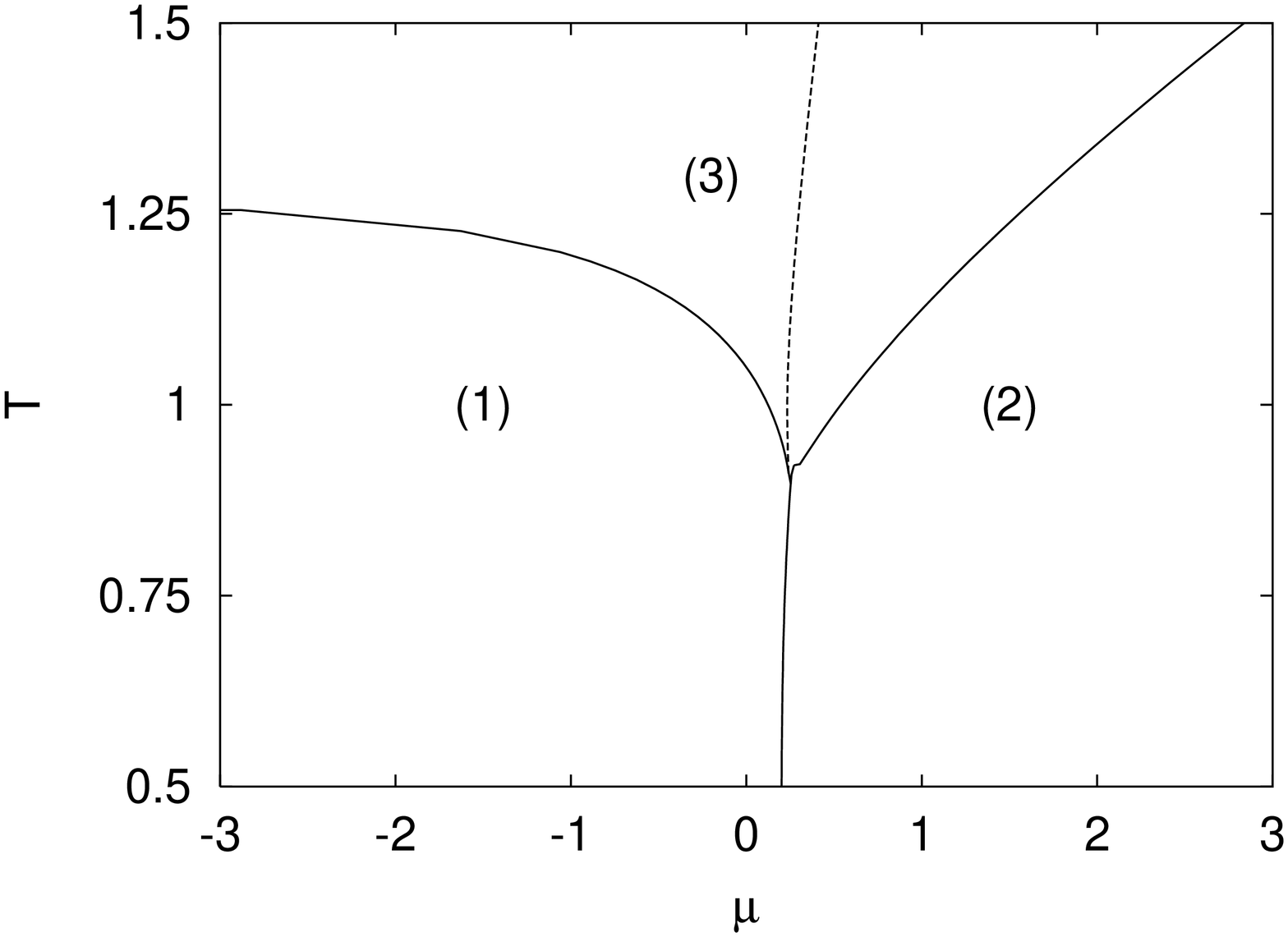}}}}
\put(0,30){(a)}
\put(50,30){(b)}
\end{picture}
\caption{Phase diagram with the parameter set $\varepsilon_d =
\varepsilon_b = -1$, $\varepsilon_x = -1.6$, $\varepsilon_t =
-1.2$. Compared to the phase diagram shown in figure \ref{dreifig2},
the energy difference between the \cdtxt{} and the \cdtxo{}
reconstruction is greater and the interaction between dimers is
weaker. With these parameters, the transition between phases (1) and
(3) is continuous. Panel (a) shows the phase diagram in the
$\rho_{Cd}$-$T$ plane, in panel (b) it is shown in the $\mu$-$T$
plane.
\label{dreifig4}}
\end{figure}

Finally, we discuss the influence of the parameter set on the phase
diagram. The main effect of a variation of the binding energy of the
Te dimers $\varepsilon_b$ is to shift the chemical potential at which
the phase transitions occur. The influence on the shape of the phase
diagram is small as long as $\varepsilon_b$ is sufficiently large such
that the density of Te atoms which are neither dimerized nor bound to
Cd atoms is small. Due to the electron counting rule, a state where a
large fraction of the Te atoms remains unbound should be irrelevant
for CdTe.

A smaller energy difference $\Delta E$ between a perfect \cdtxt{} and
a perfect \cdtxo{} reconstruction increases the tendency of Cd atoms
to arrange in a $(2\times1)$ order. If $\varepsilon_x = -1.95$ and all
other parameters are identical to our standard parameter set, the
dashed line where $C_{Cd}^{x} = C_{Cd}^{d}$ is shifted to a higher
$\rho_{Cd} \approx 0.45$. Additionally, we obtain a higher value of
0.28 for the maximal Cd coverage in the Te rich phase, which is
achieved at a lower temperature $T = 0.9$. The temperature $T_t = 0.7$
above which the disordered phase is stable is also slightly lower.  In
general, a greater value of $|\varepsilon_x|$ increases both the
tendency of the Cd atoms to arrange in a $(2\times1)$ order in the
disordered phase and the concentration of Cd atoms in the \tetxo{}
reconstructed phase. The temperature of the order-disorder transition
in the Cd rich phase is lowered. Conversely, a {\em greater} energy
difference between perfect \cdtxt{} and \cdtxo{} reconstructed phases
alters the temperature of this transition and leads to a preferential
$c(2\times2)$ arrangement of the Cd atoms in the disordered phase.
 
The main effect of a variation of the interaction $\varepsilon_t$
between Te dimers is to change the properties of the transition
between the phases (1) and (3). As long as $|\varepsilon_t|$ is
sufficently high, the phase diagram is qualitatively similar to that
shown in figure \ref{dreifig2}. In this case, the main effect of a
variation of $\varepsilon_t$ is to shift the temperature $T_c^1
\approx |\varepsilon_t|$ where the \tetxo{} phase vanishes. However,
at low $|\varepsilon_t|$ this transition becomes a {\em continuous}
phase transition without any coverage discontinuity. As an example, in
figure \ref{dreifig2} a phase diagram with $\varepsilon_t = -1.2$,
$\varepsilon_b = \varepsilon_d = -1$ and $\varepsilon_x = -1.6$ is
shown. The coexistence regime (4) vanishes at $T_t = 0.9$ such that
there is no phase separation between the long-range ordered \tetxo{}
phase and the disordered phase. This parameter set yields a
comparatively high $\Delta E$. Therefore, the line where $C_{Cd}^{x} =
C_{Cd}^{d}$ is at relatively low Cd coverages.  In contrast to the
situation at large $|\varepsilon_t|$, the chemical potential at which
the transition from phase (1) to phase (2) occurs is not independent
of temperature. Instead, the transition is at slightly greater $\mu$
at higher temperature. In consequence, there is a small range $0.2 <
\mu < 0.27$ where there is a phase transition from phase (2) to phase
(1) if $T$ is increased at constant chemical potential. Qualitatively,
the phase diagram at small $|\varepsilon_t|$ is reminiscent of that of
the simplified model as far as the arrangement of Cd atoms is
concerned.

\section{Comparison with experimental results \label{dreiexpcmp}}
 
The results presented in section \ref{dreiresults} suggest an
interpretation of the experimentally observed crossover from a
\cdtxt{} reconstruction to a \cdtxo{} reconstruction as an
accompanying effect of an order-disorder phase transition. At low
temperature, there is a long-range ordered Cd-rich phase with a
\cdtxt{} reconstruction. At a critical temperature, the Cd atoms lose
their long-range order and arrange preferentially in a $(2 \times 1)$
pattern. This picture is consistent with the experimental observation
of small domains in the \cdtxo{} reconstruction \cite{n98,ntss00}
which indicate a high degree of disorder.

Strictly speaking, a CdTe surface under vacuum is not in thermal
equilibrium.  At the temperature of the \cdtxt{} - \cdtxo{}
transition, sublimation plays an important role. As we will show in
chapter \ref{rekkapitel}, important features of the simplified model
which neglects Te dimerization are preserved under the conditions of
step-flow sublimation. This is the dominant sublimation mechanism for
CdTe (001) \cite{nskts00}. However, the Cd coverage $\rho_{Cd}$ is
determined by the sublimation process. Therefore, within the limit of
an equilibrium model it is not possible to calculate the path in the
phase diagram that CdTe follows.

Auger measurements \cite{n98,ntss00} yield $\rho_{Cd} \approx 0.35$ at
the transition temperature in vacuum. This suggests that the system is
in the coexistence regime for a wide range of temperatures. Then, the
behaviour of CdTe in vacuum should be similar to our results at
constant $\rho_{Cd} = 0.35$ (figure \ref{dreifig3}). Electron
diffraction techniques investigate large regions on the surface and
hence yield averages over all coexisting phases. Consequently, it
should be reasonable to compare results of these experiments with
quantities which are averaged over the whole system in our canonical
simulations. As discussed in section \ref{dreicanonicalresults}, there
are two independent effects which lead to a clear dominance of the
arrangement of Cd atoms in rows: the order-disorder transition in the
Cd-rich phase and the fact that an increasing fraction of the Cd atoms
is dissolved in the Te-rich phase. Electrons are diffracted both from
the Cd atoms and the Te dimers which have a $(2 \times 1)$ order.  The
superposition of both effects should yield the pronounced $(2 \times
1)$ diffraction peaks which have been observed in
\cite{n98,ntss00,tdbev94}.

In the \cdtxt{} reconstructed phase, at temperatures well below the
phase transition one frequently finds collective thermal excitations
of adatoms, where a row of several Cd atoms is shifted by one lattice
constant in the $y$-direction.  Due to the repulsion between Cd atoms
in this direction, this is possible only if one Cd atom per excitation
is missing. An example is shown in figure \ref{dreifig2}d. This effect
has been observed experimentally by Seehover et.\ al.\
\cite{sfjetbd95} at room temperature using STM microscopy.  There is a
striking similarity between figure 3 in \cite{sfjetbd95} and figure
\ref{dreifig2}d.

The phase diagram of our model explains the properties of the
reconstructions of the CdTe (001) surface under an external particle
flux. If the Cd coverage of the surface is increased by deposition of
Cd, the state of the system moves into regions of the phase diagram
where the Cd atoms arrange preferentially in a $c(2 \times 2)$
pattern. Depending on temperature, this is either the ordered \cdtxt{}
phase (2) or the region of the disorded phase (3) on the right side of
the dashed line where $C_{Cd}^{d} > C_{Cd}^{x}$. Indeed, experiments
\cite{ct97,tdbev94} have shown that an external Cd flux restores the
\cdtxt{} reconstruction at high temperatures where a $(2 \times 1)$
order is found under vacuum conditions.  Clearly, a strong Te flux
induces the formation of a long-range ordered \tetxo{} phase at
temperatures below $T_{c}^{1}$.

On ZnSe, a Zn terminated $(2 \times 1)$ reconstruction has not been
observed yet. Our model offers an explanation of this fact which is
consistent with the results of density functional theory
\cite{gn94,g97,gffh99,pc94}. These calculations show, that for ZnSe
the difference between the surface energies of perfect cation
terminated $c(2\times 2)$ and $(2\times 1)$ reconstructions is
significantly greater than for CdTe ($0.03\,\mathrm{eV}$ per $(1\times
1)$ surface unit cell versus $0.008\,\mathrm{eV}$). In our model,
this greater energy difference corresponds to a smaller value of
$|\varepsilon_x|$ which shifts the line where $C_{Cd}^{x} =
C_{Cd}^{d}$ to smaller coverages. Consequently, a \zntxt{} arrangement
dominates in a much wider range of coverages.

Wolfframm et.\ al.\ \cite{wewr00} have measured the locus of the
transition beween the \zntxt{} reconstruction and the \setxo{}
reconstruction as a function of temperature and the composition of an
external particle flux by means of reflection high energy electron
diffraction (RHEED). They find that at higher temperature a greater Se
flux is required to obtain a $(2 \times 1)$ diffraction pattern.  At
temperatures above 450$^{o}$C no \setxo{} reconstruction could be
observed even under extremely Se-rich conditions. This is reminiscent
of our observation that the anion-rich phase (1) vanishes at a
temperature $T_{c}^{1}$.

Unfortunately, the available experimental data are insufficient for a
systematical fit of the model parameters. However, some rough
estimates show that at least the orders of magnitude are
reasonable. As discussed above, the \cdtxt{} -\cdtxo{} transition of
CdTe in vacuum should be at a temperature close to $T_t$. Identifying
this with the experimental value of \mbox{570 K}, we obtain the value
of our energy unit $|\varepsilon_d| \approx 0.06 \,\mathrm{eV}$ in
physical units. This yields a value $\Delta E \approx 0.003
\,\mathrm{eV}$ for the difference in the surface energies of \cdtxt{}
and \cdtxo{}, which is a bit less than $1/2$ of the value of $0.008
\,\mathrm{eV}$ which has been obtained by means of DFT
calculations. This shows at least that the qualitative agreement
between experiments and our model has been obtained in a physically
reasonable region of the parameter space.

In our model, phase (1) vanishes at a temperature $T_c^1 \approx
|\varepsilon_t|$. Identifying this with the value of $450 ^{o}C$ which
has been measured in experiments on ZnSe, we obtain that
$|\varepsilon_t| \approx 0.06 \,\mathrm{eV}$. This is the same order
of magnitude as our estimate of $|\varepsilon_d|$ in CdTe.

These considerations suggest that it should be possible to obtain
quantitative agreement between experiments and our model both for CdTe
and ZnSe with values of the model parameters on the order of magnitude
of a few ten meV.

\chapter{Reconstructions on non-equilibrium crystal surfaces 
\label{rekkapitel}}

In section \ref{dreikapitel}, we have presented a lattice gas model
which displays an order-disorder phase transition the properties of
which are reminiscent of the \cdtxt{}-\cdtxo{} reordering of the Cd
atoms at the CdTe(001) surface.  However, a crystal surface in an MBE
chamber is not in thermal equilibrium. In the presence of a particle
flux, the crystal grows. Under vacuum, there is sublimation (section
\ref{sublimintro}).

In the literature, a few models of crystal growth have been presented
which take into account the effects of surface reconstructions. On the
one hand, there have been models which aim at a detailed understanding
of the features of specific materials, in particular Si
\cite{br88,r90} and GaAs \cite{ibajjv98,kms99,ks02}. On the other
hand, Chin and den Nijs \cite{cn01} have investigated a simple model
to address the fundamental question whether there might be a
reconstruction order on growing surfaces. They find, that a long range
order in the strict sense which is present in thermal equilibrium
\cite{n90} is absent during growth. However, the order of the
reconstruction is preserved on small and intermediate lengthscales.

To the best of our knowledge, phase transitions between different
reconstructions have not been considered in growth models, so far.
Consequently, a theoretical understanding of the interplay of this
phenomenon with the dynamics of growing or sublimating surfaces has
not yet been achieved.  We address this important problem in the
context of a model of a compound material that displays a competition
of different vacancy structures in the terminating layer similar to
the lattice gas model which was introduced in section \ref{dreizwei}.
In particular, we address the following questions: (1) It has been
claimed \cite{n98,ntss00} that the surface layer is in effective
thermal equilibrium under the non-equilibrium conditions of
sublimation. Is this justifed? If not, which features of an
equilibrium phase transition are preserved? \mbox{(2) What} is the effect of
an external flux of particles of one species on the surface
reconstruction? These simulations model situations which are given in
experiments where phase diagrams like that shown in figure
\ref{cdtephasdschematisch} are determined.

\section{The model \label{rekmodel}}

Our model is an extension of the simplified planar lattice gas model
introduced in section \ref{dreizwei} to a model of a three-dimensional
crystal. In this chapter, we aim at an understanding of fundamental
properties of nonequilibrium crystal surfaces rather than a realistic
modelling of CdTe. To this end, we consider a lattice gas model of a
compound ``\AB{}\/'' with a cubic lattice and a comparatively simple
potential energy surface. The \A{} (\B{}) atoms correspond to Cd
(Te). We impose the constraint that there are no overhangs and
dislocations. This {\em solid-on-solid} condition makes is possible to
describe the configuration of the crystal uniquely by a
two-dimensional array $h$ of integers $\{h_{i, j}\}_{i,j=1}^{N}$ which
denote the height of the column of atoms at $x$-coordinate $i$ and
$y$-coordinate $j$. Thus, we achieve an effectively two-dimensional
description of a three-dimensional crystal which reduces the
complexity of our model considerably.

\begin{figure}[htb]
\botbase{\includegraphics{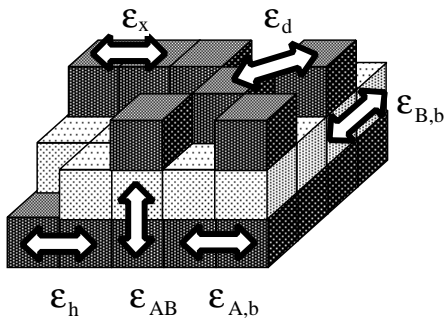}} \hfill
\botbasebox{0.57\textwidth}{\caption{Sketch of a surface showing the
interactions between the particles. \A{} particles are shown as dark
cubes, \B{} particles in light grey. The $x$-axis points from left to
right, the $y$-axis from front to back.\label{rekfig3}}}
\end{figure}
To model the layered structure of the CdTe crystal, we assign the odd
heights to \A{} particles, and the even heights to \B{} particles. In
the following, the term monolayer (ML) will denote one \AB{} layer,
i.e.\ the height difference between adjacent monolayers is $2$. Inside
the bulk of the crystal, there is an attractive interaction between
the particles and their nearest neighbours which is isotropic in
directions parallel to the surface. While there is no difference in
the interaction of \B{} particles between the bulk and the surface,
\A{} particles on the surface interact with the anisotropic
interactions of the simplified version of our lattice gas model of the
CdTe(001) surface introduced in section \ref{dreizwei}. The other
parameters of the model are defined in Figure \ref{rekfig3}. Thus, the
Hamiltonian of the crystal becomes
\begin{equation}
H = \varepsilon_{AB} n_{AB} + \varepsilon_{B,b} n_{B,b} +
\varepsilon_{A,b} n_{A,b} + \varepsilon_{h} n_{A,h}+ \varepsilon_{d}
n_{A,d} + \varepsilon_{x} n_{A,x}
\end{equation}
Here, $n_{AB}$ is the number of \A{}--\B{} bonds between the layers,
$n_{B,b}$ is the number of nearest neighbour pairs of \B{} atoms,
$n_{A,b}$ the number of \A{} nearest neighbour pairs inside the bulk,
$n_{A,h}$ the number of surface \A{} atoms next to a higher column and
$n_{A,d}$ ($n_{A,x}$) the number of bonds of surface \A{} atoms to
diagonal neighbours (nearest neighbours in $x$-direction) at the same
height. The occupation of neighbouring sites in the $y$-direction with
\A{}-atoms is forbidden, independent of the height difference between
both atoms. Additionally, atoms must not stay in the wrong sublattice.

\begin{figure}
\includegraphics{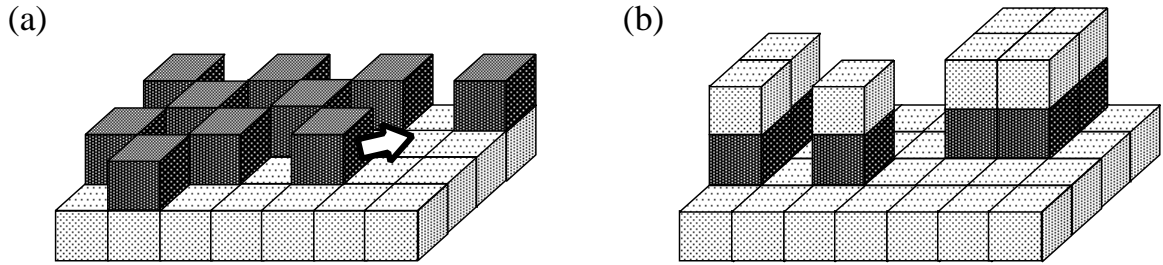}
\caption{Motivation for diffusion to diagonal neighbour sites and
diffusion of \AB{} pairs. \A{} atoms are shown as dark cubes, \B{}
particles in light grey. The $x$-axis points from left to right, the
$y$-axis from front to back. Panel (a): Diffusion of \A{} atoms on a
\B{} terminated surface. The diffusion of an \A{} atom (arrow) along
the edge of a cluster of other \A{} atoms is severely restricted
without diffusion to diagonal neighbour sites. Panel (b): Coalescence
of islands is impossible without \AB{} diffusion. \label{rekmotivier}}
\end{figure}

In addition to diffusion to nearest neighbour sites we permit
diffusion to diagonal neighbour sites. This process is important to
obtain a reasonable mobility of \A{} atoms on a \B{} terminated
surface. An example is shown in figure \ref{rekmotivier}a. In the
absence of diagonal neighbour diffusion, due to the repulsive
interaction in $y$-direction an \A{} atom at the edge of an
\A{}-cluster cannot diffuse along the cluster edge. Consequently, an
equilibration of the cluster configuration is possible only via the
detachment of atoms from the cluster. However, this process has a
large activation energy since all bonds in the horizontal direction
must be broken.

Additionally, we permit the diffusion of an \AB{} pair with the \B{}
atom on top of the \A{} atom, if the diffusion of the \B{} atom alone
would end in the wrong sublattice. This process is required to
preserve the ergodicity of the system. Consider the surface
configuration shown in figure \ref{rekmotivier}b. Since atoms must not
stay in the wrong sublattice the \B{} atoms on top of the islands
cannot diffuse across the gap between the islands. Consequently, the
energetically favourable coalescence of the islands is suppressed
unless \AB{} pairs are allowed to diffuse together. However, it is not
necessary to permit also the diffusion of \AB{} pairs with the \A{}
atom on top since the vacancy structure of \A{} terminated islands is
more flexible than a flat \B{} terminated layer. We have performed
some trial simulations where the diffusion of both arrangements of
\AB{} pairs was permitted. The results are identical to those of the
standard model.

Diffusion is an Arrhenius activated process (equation
\ref{arrheniusgesetz}). For simplicity, we assume an equal activation
energy $B_0$ for all diffusion processes where the energy of the
system does not increase. Then, due to the detailed balance condition
\ref{balance} which will be discussed in appendix \ref{mcappendix},
the barrier height for the diffusion of a particle to a site which is
energetically unfavourable compared to the initial site is necessarily
$B_0 + \Delta H$. Here, $\Delta H$ is the energy difference between
the initial and the final state of the system. In the literature, 
a dynamics with these rates is
denoted as ``Kawasaki dynamics'' \cite{nb99}.  The desorption of a
particle requires an additional activation energy $\varepsilon_{v}$.
This dynamics does not depend on the parameter $\varepsilon_{AB}$, so
it needs not to be specified, and $\varepsilon_{B,b}$ and
$\varepsilon_{A,b}$ enter only via the sum $\varepsilon' :=
\varepsilon_{B,b} + \varepsilon_{A,b}$. The constant prefactor
$1/t_{0} := \nu \exp(-B_{0}/kT)$ sets the timescale of the model and
needs not to be specified, if we measure time in units of
$t_{0}$. Particles from an external source arrive at each lattice site
with equal probability. If the adsorption of the particle at this site
would lead to a forbidden state, it is reflected.

The simulations are performed using continuous time Monte Carlo
techniques. A detailed discussion of the algorithm can be found in
appendix \ref{mcappendix}.

For simplicity, we measure energy in units of $|\varepsilon_d|$ and
set Boltzmann's constant to unity. In the following, we use the
parameter set $\varepsilon' = \varepsilon_{d} = -1$, $\varepsilon_{h}
= 0$, $\varepsilon_{x} = -1.97$, $\varepsilon_{v} = 1.5$.  With these
parameters, the difference in the surface energies between perfectly
$c(2\times2)$ and $(2\times1)$ reconstructed surfaces is small
compared to the surface energies themselves. This is consistent with
our finding in chapter \ref{dreikapitel} that a small $\Delta E$ is an
essential feature of the \cdtxt{}-\cdtxo{} reordering. We have
verified that the behaviour of our model remains qualitatively the
same for a wide range of parameters, as long as the basic property of
a small energy difference between the \A{} terminated reconstructions
is preserved.

\section{Characterization of the surface \label{rekcharakterisier}}

The surface width $W := \left< (h_{i,j} - \left< h_{i,j} \right>)^2
\right>$ is a measure for the roughness of the surface. To
characterize the arrangement of \A{} atoms quantitatively, we
investigate correlations and an order parameter similar to the
quantities which were defined in section \ref{dreicharakterisier}. To
this end, we consider an array $c$ of integers $c_{i,j} := h_{i,j} \,
\mbox{mod}\, 2$. $c_{i,j} = 1$ if the column of atoms at $(i, j)$ is
terminated by an \A{} atom and zero otherwise. These variables can be
used to calculate the \A{} coverage $\rho_A := N^{-2}\sum_{i,j =
1}^{N} c_{i,j}$ and the normalized correlations
\begin{eqnarray}
K_A^d & := & \frac{1}{2 \rho_A N^2} \sum_{i,j=1}^{N} c_{i,j} \left(
c_{i+1,j+1} + c_{i+1,j-1} \right) \\ K_A^x & := & \frac{1}{\rho_A N^2}
\sum_{i,j=1}^{N} c_{i,j} c_{i+1,j}.
\end{eqnarray}  
These definitions are similar to those of $C_{Cd}^{d}$ and
$C_{Cd}^{x}$ in equations \ref{defccdd} and \ref{defccdx}. We have
normalized the correlations with the \A{} coverage such that $K_A^d$
and $K_A^{x}$ are equal to the fraction of \A{} atoms which are
incorporated in locally $c(2\times 2)$ and $(2\times 1)$ reconstructed
regions of the surface, respectively. The long range order of the \A{}
atoms is measured by the order parameter $M_A^{(2\times 2)}$ the
definition of which is identical to that of $M_{Cd}^{(2 \times 2)}$ in
equation \ref{defmcdtxt}.

As we will show, there is no long range order in the strict sense in
our model such that $M_A^{(2\times 2)}$ is zero even at low
temperature. However, there might be correlations between \A{} atoms
at intermediate distance. These are characterized by means of the
autocorrelation function
\begin{equation}
C(\vec{x}) := \left< c_{i,j} c_{i+x,j+y} \right> - \rho_A^2
\label{rekautokorr}
\end{equation}
of $c$. If the positions of \A{} atoms at a distance $\vec{x} = (x, y)$
are uncorrelated, we have $C(\vec{x}) = 0$. Ornstein-Zernike theory
\cite{cf72} states that in thermal equilibrium the asymptotical
behaviour of $|C(\vec{x})|$ for large $|\vec{x}|$ is
\begin{equation}
\left| C(\vec{x}) \right| \sim C_{\infty} + \lambda
|\vec{x}|^{-(d-1)/2} \exp(-|\vec{x}|/\xi)
\label{ornstein}
\end{equation}
where $\xi$ is the correlation length and $d$ is the dimension of the
system. A nonzero value of $C_{\infty}$ indicates long range
order. This equation is not valid at the locus of a phase transition
where the correlation length diverges. Note, that in a system with
anisotropic interactions the constants $C_{\infty}$, $\lambda$ and
$\xi$ may have a directional dependence. If neighbouring sites are
correlated, $C(\vec{x})$ is always positive. If the values of
$c_{i,j}$ on neighbouring sites are anticorrelated, $C(\vec{x})$
oscillates between positive and negative values.  In our model, this
is necessarily the case if $\vec{x}$ is parallel to the $y$-direction
since there is a hard repulsion between \A{} atoms on neighbouring sites
in this direction. However, if $\vec{x}$ points in the $x$-direction,
the configuration of neighbouring sites is anticorrelated only if the
\A{} atoms arrange preferentially in a $c(2\times 2)$ pattern. On a
$(2\times 1)$ reconstructed surface neighbouring sites in the
$x$-direction are correlated. If $K_A^x \approx K_A^d$, these effects
compensate each other. Consequently, on such a surface we expect a
very small correlation length in the $x$-direction.

\section{Layer-by-layer sublimation \label{reksub}}

\begin{figure}
\botbase{\resizebox{0.48\textwidth}{!}{\rotatebox{270}{\includegraphics{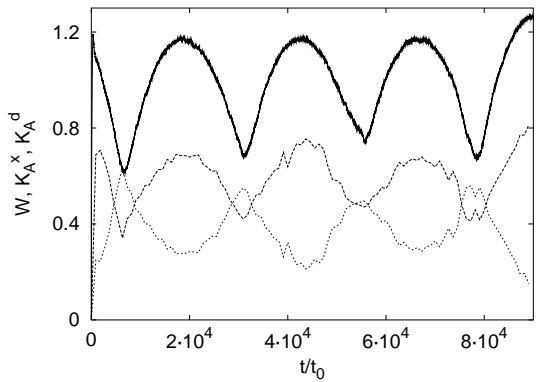}}}}
\hfill \botbasebox{0.47\textwidth}{\caption{Sublimation starting from
a flat \B{} terminated surface at $T = 0.5$. The system size is $N =
128$. Solid line:\ $W$, dashed:\ $K_{A}^{x}$, dotted:\
$K_{A}^{d}$. For clarity of plotting, the last two curves have been
smoothed; $W$ shows the natural fluctuations. Each oscillation period
corresponds to the desorption of one monolayer.\label{rekfig1}}}
\end{figure}

Figure \ref{rekfig1} shows the evolution of an initially
\B{} terminated, flat surface under vacuum at a temperature $T =
0.5$. At the beginning of the simulation, $\rho_{A}$ increases 
from zero to an asymptotic value
close to 0.5.  Thus, the sublimating surface is \A{} terminated.

We find, that this increase can be fitted with an exponential
relaxation, $\rho_{A} = \rho_{A}^{\infty} ( 1 -
\exp(-t/\tau))$. Here, $\tau$ is the time constant of the decay
of the \B{} terminated surface. The temperature dependence
of the corresponding desorption probability $p \propto 1/\tau$ follows
an Arrhenius law, $p \propto \exp(- E_{\mathrm{sub}}/T)$, with an activation
energy $E_{\mathrm{sub}} = B_{0} + 3.0 \pm 0.1$ 

After this onset, \A{} and \B{} evaporate stoichiometrically, and
sublimation proceeds in layer-by-layer mode (see section
\ref{sublimintro}). This is reflected by the oscillations of the
surface width $W$ (solid line in figure \ref{rekfig1}). Whenever the
surface is atomically flat, $W$ is minimal. The maxima of $W$
correspond to a configuration where 50\% of the surface are covered
with islands of one monolayer thickness. The dependence of the
sublimation rate on $T$ follows an Arrhenius law with an activation
energy $B_{0} + 6.0 \pm 0.2$ which is greater than that found for the
decay of the \B{} terminated surface. However, in our model the {\em
microscopic} energy barriers for desorption of \A{} and \B{} are
equal. Consequently, the difference in the {\em macroscopic}
activation energies is solely a consequence of the stabilizing effect
of the reconstruction of the \A{} terminated surface.

\begin{figure}
\begin{center}
\begin{picture}(100,45)(0,0)
\put(5,0){\resizebox{0.45\textwidth}{!}{\includegraphics{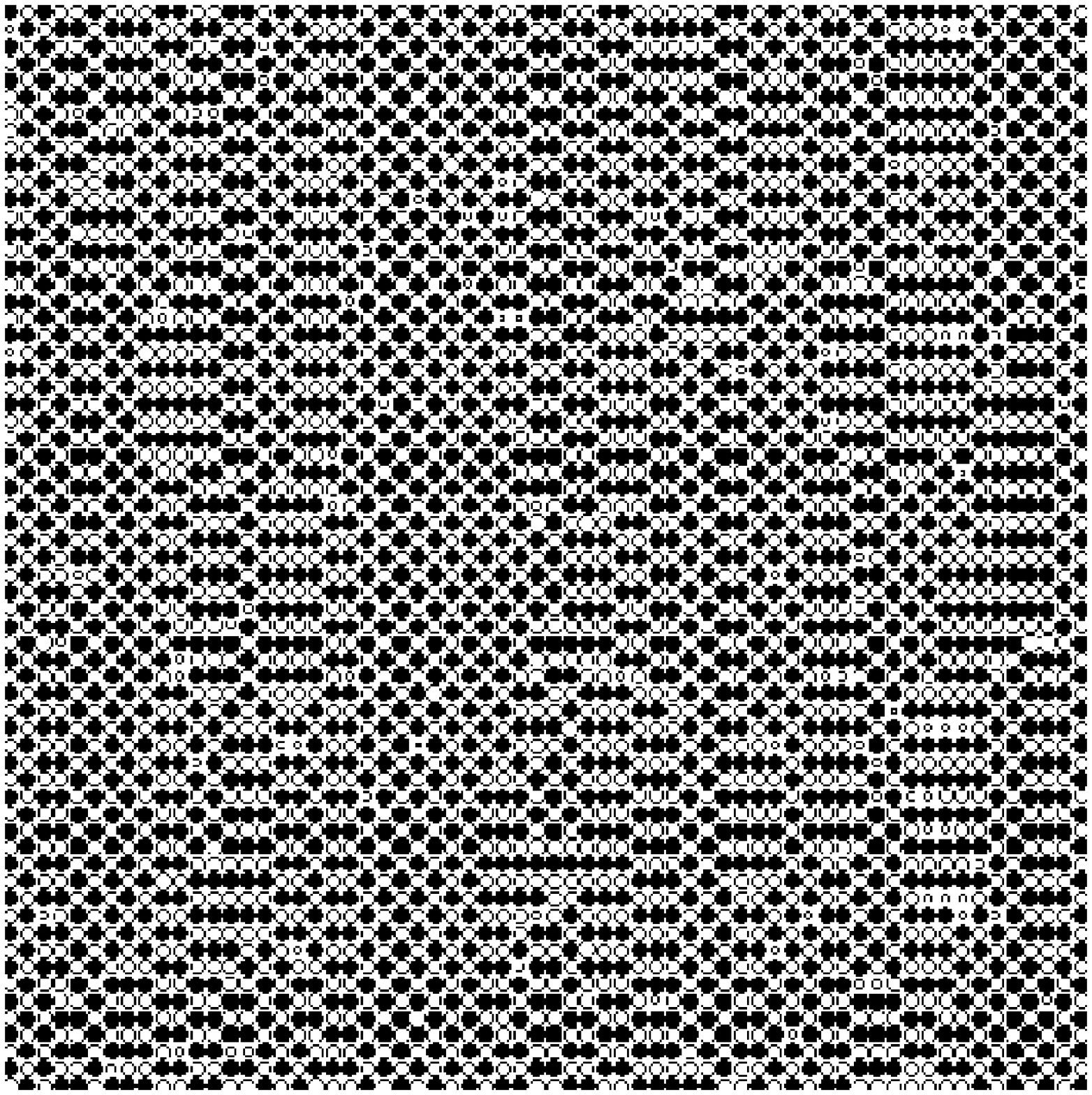}}}
\put(1,43){(a)}
\put(55,0){\resizebox{0.45\textwidth}{!}{\includegraphics{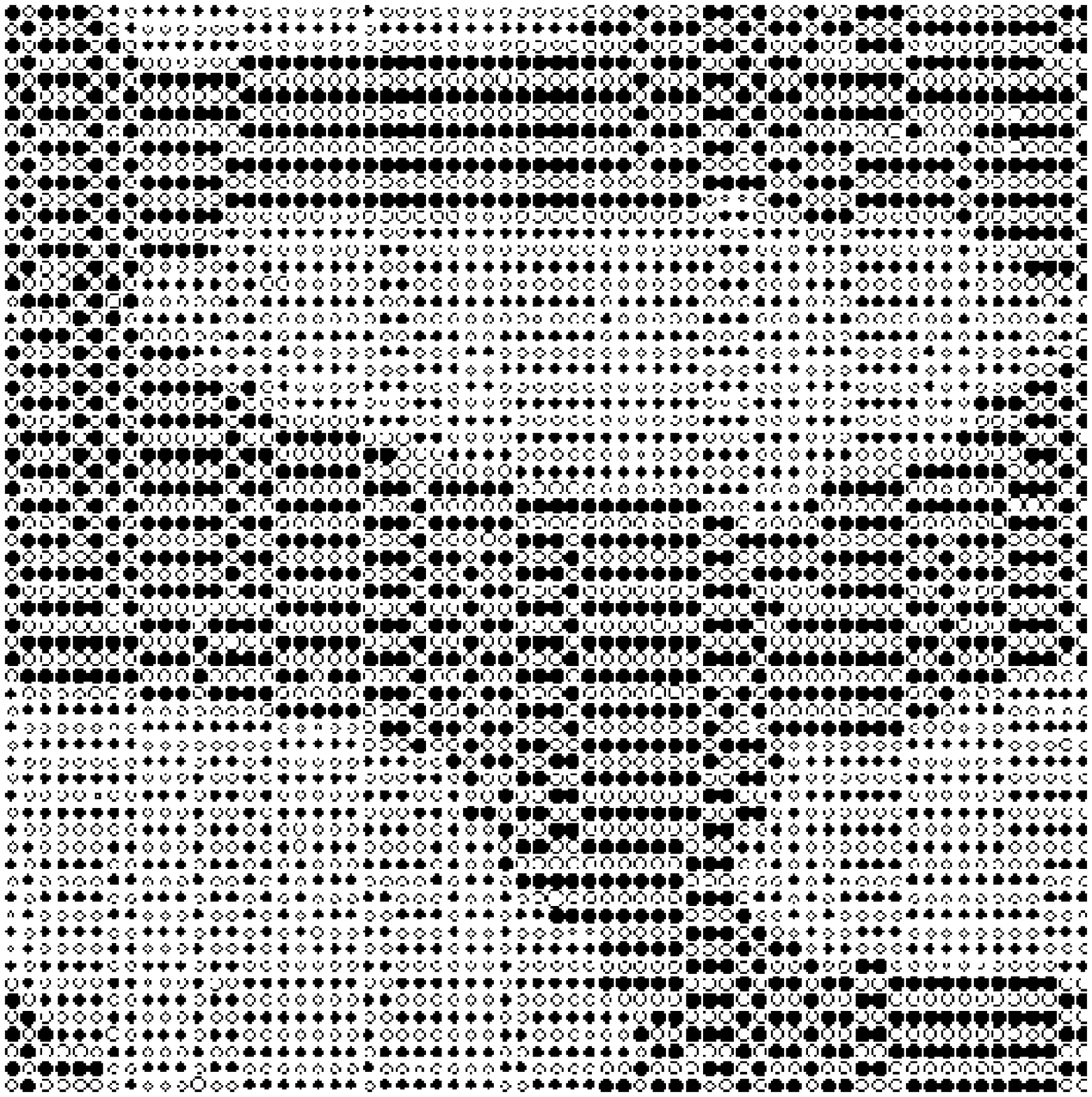}}}
\put(51,43){(b)}
\end{picture}
\end{center}
\caption{Evolution of the surface of a $64 \times 64$ system in
layer-by-layer sublimation. \A{} atoms are shown as filled circles, \B{}
atoms as open circles. Greater symbols denote greater surface
heights. Panel (a) shows a surface snapshot at $t = 7000 \cdot t_0$
where the surface roughness is minimal, panel (b) shows the surface at
$t = 19000 \cdot t_0$ where the roughness is
maximal. \label{reklayer}}
\end{figure}
Surprisingly, $K_{A}^{d}$ and $K_{A}^{x}$ oscillate during
layer-by-layer sublimation. Each time a complete \AB{} layer has
desorbed and the surface is atomically flat (minima of $W$) the
$c(2\times2)$ reconstructed fraction of the surface is maximal. On the
contrary, a rough surface with a large number of islands (maxima of
$W$) seems to prefer the $(2\times1)$ reconstruction. This can be
understood from the fact that the attractive lattice gas interactions
are present only between particles in the same layer.  Thus, the
island edges impose open boundary conditions to the lattice gas of \A{}
atoms on the island. In contrast to a $c(2\times2)$ reconstructed
domain, a $(2\times1)$ terminated island can {\em reduce} the energy
of its boundary by elongating in $x$-direction. Since the ground state
energies of both structures are nearly degenerate, the formation of
$(2\times1)$ may reduce the surface free energy. This picture is
confirmed by the fact that islands on sublimating surfaces are indeed
elongated (see figure \ref{reklayer}b).

\section{Step-flow sublimation}

Clearly, an oscillatory behaviour cannot be understood within the
framework of equilibrium thermodynamics. However, pure layer-by-layer
sublimation is rarely ever observed in experiments. In particular,
investigations of CdTe \cite{nskts00} have shown, that step-flow
sublimation is the dominant mechanism. 

We determine the energy of steps in our model. This is done
by calculating the energy which is needed to shift one half of a large
crystal by the thickness of one monolayer in the $z$-direction. In
this process, two steps are created: one at the upper side of the
crystal and a second one at the lower side. Consequently, the step
energy is one half of the total energy costs. Normalizing this
with the length of the step, we obtain the step energy per unit
length. On perfectly $c(2 \times 2)$ reconstructed surfaces, we find 
\begin{equation}
\gamma_{\mathrm{diag}} = \frac{-1}{\sqrt{2}} \left(
\varepsilon_x + \varepsilon' \right)
\end{equation}
for diagonal steps and 
\begin{equation}
\gamma_{\mathrm{par}} = - \varepsilon_d - \frac{1}{2}
\varepsilon' + \frac{1}{2} \varepsilon_h 
\end{equation}
for steps oriented in $x$- or $y$-direction. With our standard
parameter set, we have $\gamma_{\mathrm{diag}} = \sqrt{2} <
\gamma_{\mathrm{par}} = 3/2$. Therefore, at $T = 0$ steps oriented at
an angle of 45$^{o}$ to the coordinate axes are favoured
energetically. We expect, that this holds also at finite temperature
as long as the surface is $c(2 \times 2)$ reconstructed.

Diagonal steps in the simple cubic lattice of our model correspond to
steps oriented in the [100] or [010] direction on the zinc-blende
lattice. Indeed, investigations of CdTe(001) by means of STM
microscopy \cite{megm98,mem99} have shown, that this is the
preferential orientation of steps.

\begin{figure}
\begin{center}
\begin{picture}(100,45)(0,0)
\put(5,0){\resizebox{0.45\textwidth}{!}{\includegraphics{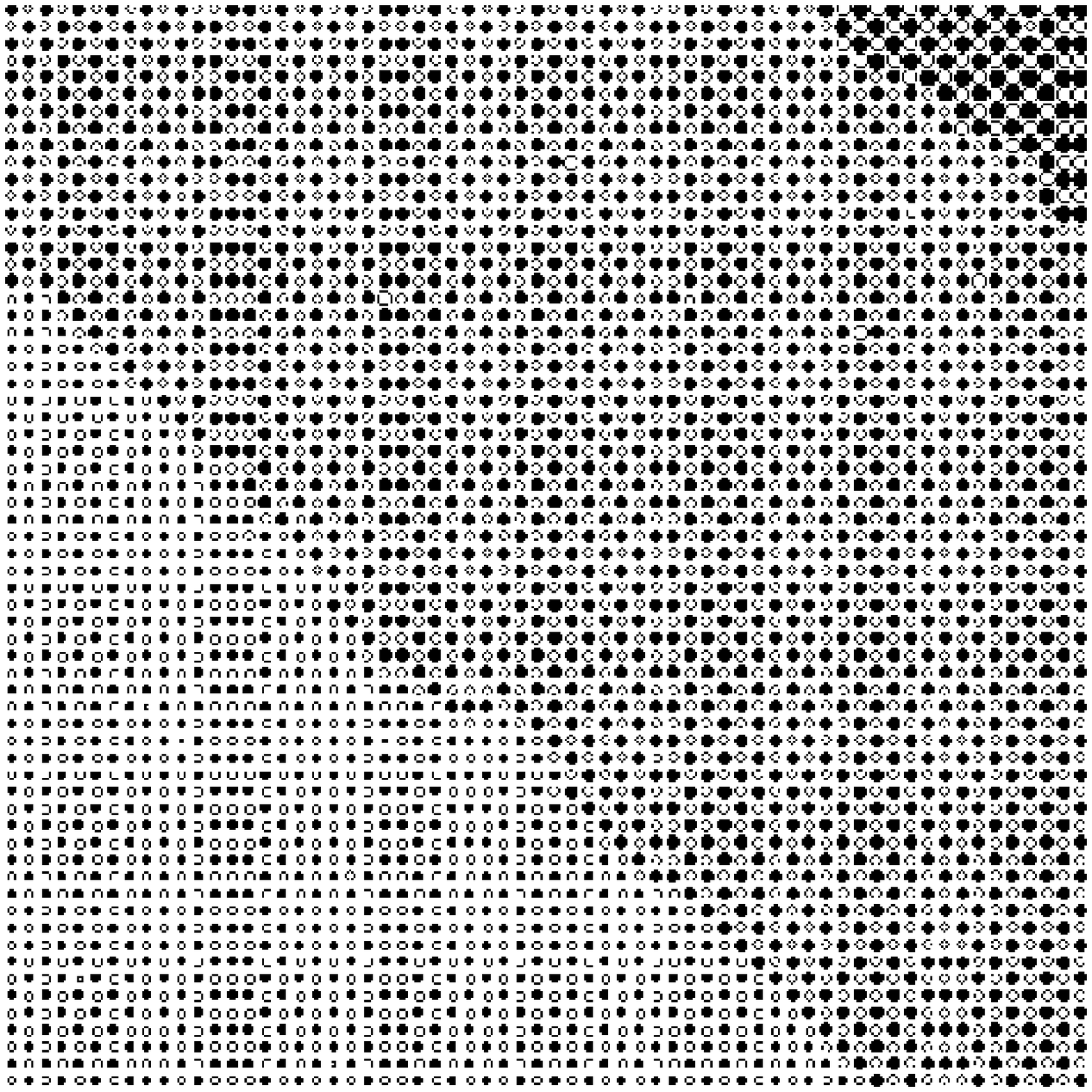}}}
\put(1,43){(a)}
\put(55,0){\resizebox{0.45\textwidth}{!}{\includegraphics{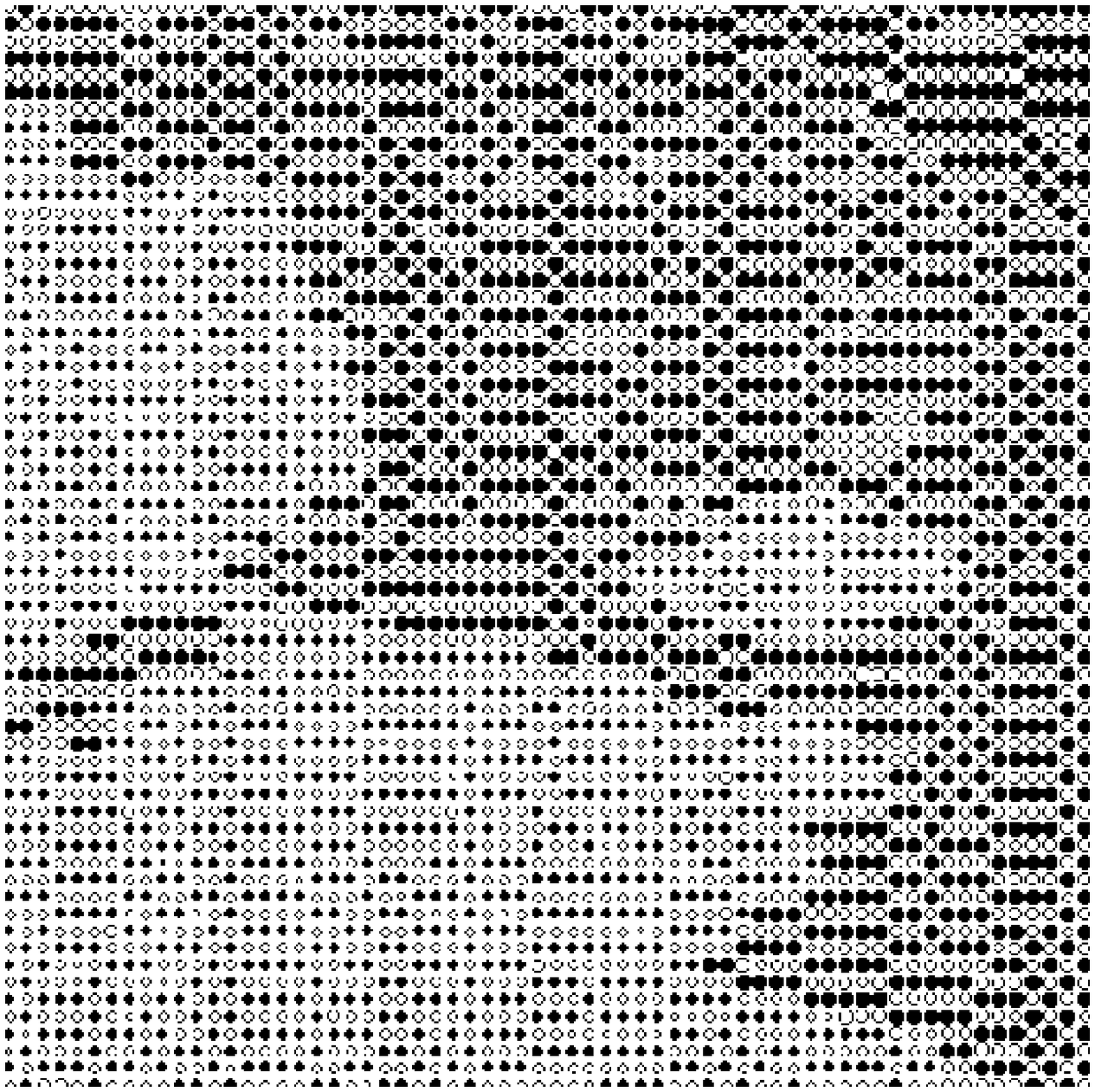}}}
\put(51,43){(b)}
\end{picture}
\end{center}
\caption{Vicinal surfaces sublimating in step flow mode. The images
show sections of $64\times 64$ lattice constants of surfaces at the
end of the simulation runs shown in figure \ref{rekfig2a}. Panel (a):
$T = 0.4$ (below the reordering), Panel (b): $T = 0.6$ (above the
reordering). \label{rekfigstuf}}
\end{figure}
In the following, we present the results of simulations of vicinal
surfaces with diagonal steps. We find that the oscillations in the
correlations disappear for terrace widths smaller than $\approx 50$
lattice constants, where sublimation proceeds in step-flow mode. Then,
the correlations become stationary apart from statistical fluctuations.
\begin{figure}[htb]
\botbase{\resizebox{0.48\textwidth}{!}{\rotatebox{270}{\includegraphics{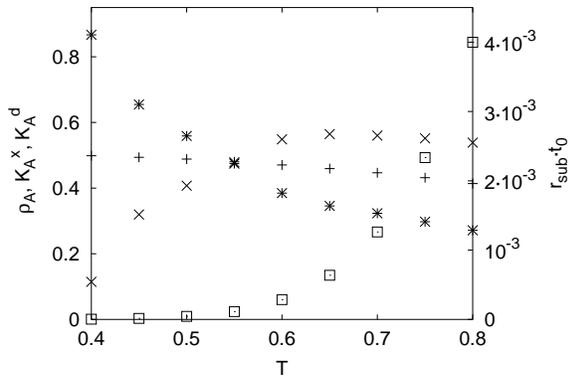}}}}
\hfill
\botbasebox{0.47\textwidth}{\caption{\A{}-coverage $\rho_{A}$ ($+$),
correlations $K_{A}^{x}$ ($\times$), $K_{A}^{d}$ ($\plustimes$) and
sublimation rate $r_{\mathrm{sub}}$ ($\boxdot$) in the stationary state
of the sublimation of vicinal surfaces under vacuum. The system 
size is $N = 128$. The steps are
oriented in the diagonal direction, the terrace width is $32 \sqrt{2}$
lattice constants. The data points show time averages of the
investigated quantities. $3\cdot 10^4 t_0$ time has been
simulated both for relaxation and for measurement.\label{rekfig2a}}}
\end{figure}

Figure \ref{rekfig2a} shows the $T$-dependence of $\rho_{A}$,
$K_{A}^{d}$, $K_{A}^{x}$ and the sublimation rate $r_{\mathrm{sub}}$ in this
stationary state at a step distance of $32\sqrt{2}$ lattice
constants. In the investigated temperature range $\rho_{A}$ decreases
only slightly from $\rho_{A} = 0.498$ at $T = 0.4$ to $\rho_{A} =
0.41$ at $T = 0.8$. Clearly, at $T = 0.55$ there is a transition from
a $c(2\times2)$ configuration at low temperature to a high temperature
regime where the $(2\times1)$ ordering dominates. This is reminiscent
of the behaviour of the correlations in the vicinity of the
order-disorder phase transition of the planar lattice gas (compare
with figure \ref{dreismuconst}). Figure \ref{rekfigstuf} shows typical
configurations of the surface at $T = 0.4$ which is below the
reordering (figure \ref{rekfigstuf}a) and at $T = 0.6$ above the
reordering (figure \ref{rekfigstuf}b).

Comparing the results of our simulations of the sublimating surface
with the anisotropic lattice gas introduced in section \ref{dreizwei}
we can check in how far the properties of the \A{} atoms on the surface
can be understood within an equilibrium theory of a flat surface. For
simplicity, in the following we denote the planar lattice gas as ``2D
model'' and the solid-on-solid model as ``3D model''. We investigate
the 2D model at the value of $\rho_A$ which is measured on the surface
of the 3D model sublimating in step-flow mode. We perform canonical
Monte Carlo simulations where the number of \A{} atoms is
fixed. Additionally, we consider results of transfer matrix
calculations where the chemical potential $\mu$ is adjusted to obtain
the desired \A{} coverage. Of course, the values of the parameters
$\varepsilon_x = -1.97$ and $\varepsilon_d = -1$ are identical to
those used in the 3D model.
\begin{figure}
\begin{picture}(100, 33)(0, 0)
\put(0,33){\resizebox{0.48\textwidth}{!}{\rotatebox{270}{\includegraphics{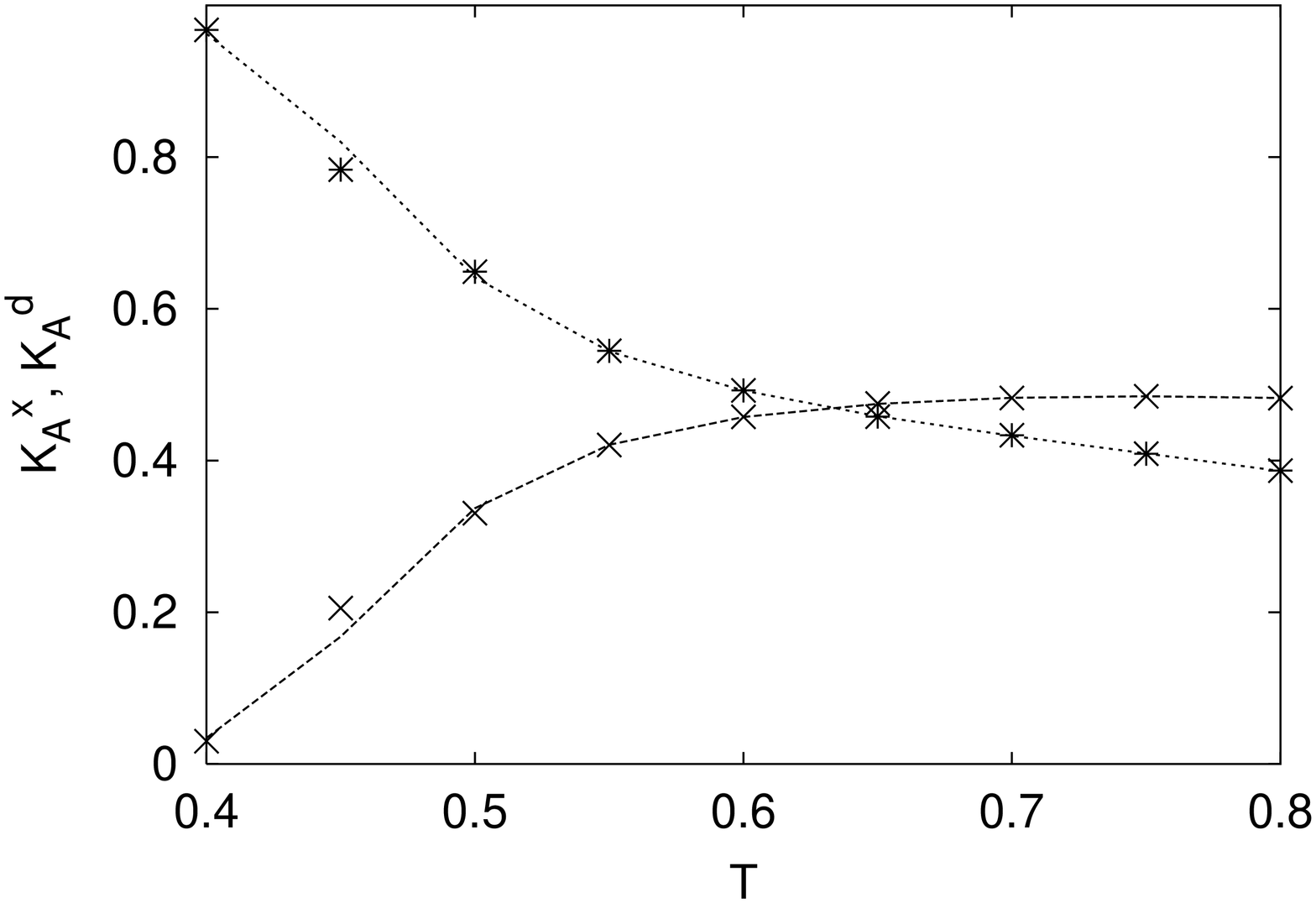}}}}
\put(50,33){\resizebox{0.48\textwidth}{!}{\rotatebox{270}{\includegraphics{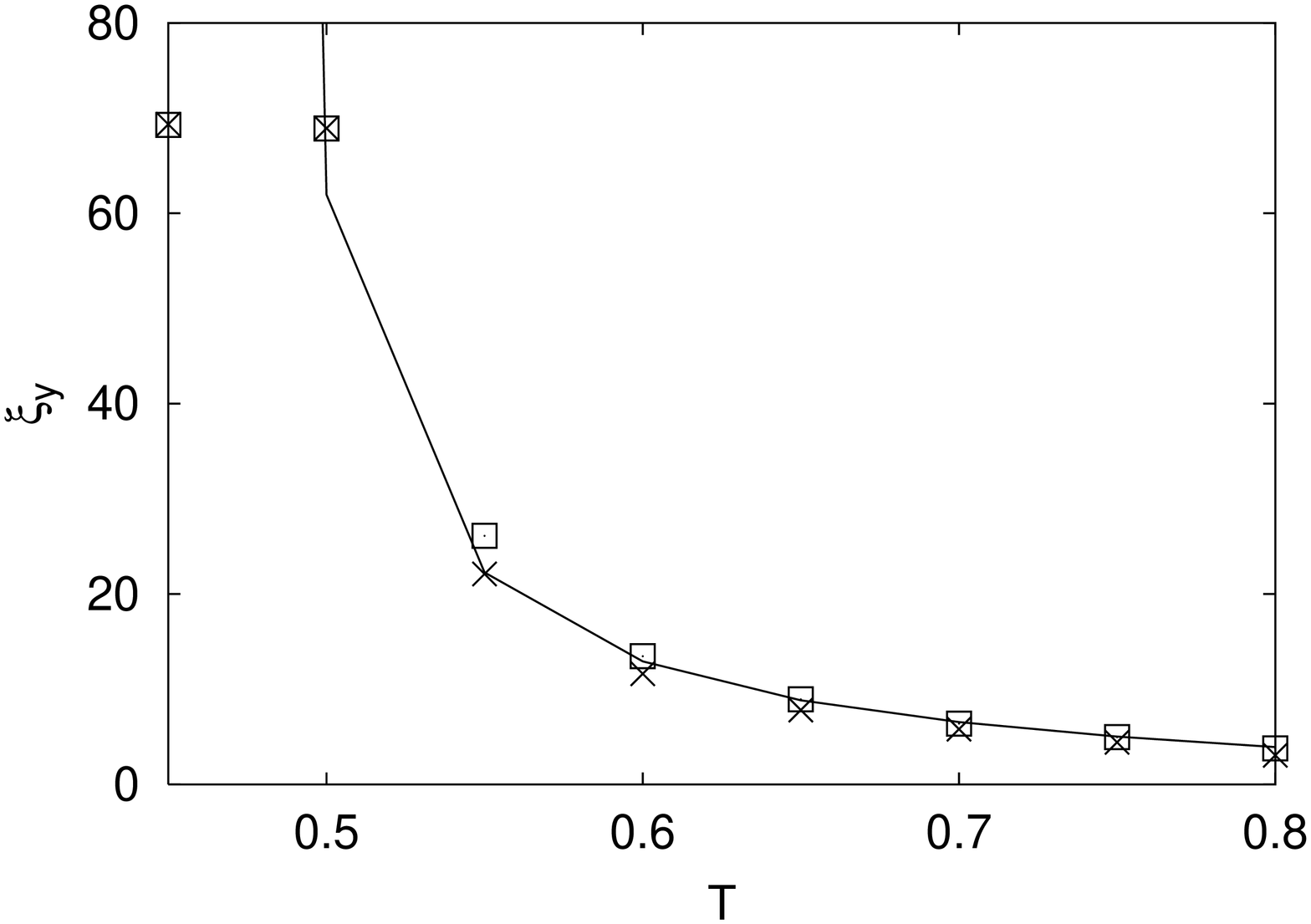}}}}
\put(0,30){(a)}
\put(50,30){(b)}
\end{picture}
\caption{Panel (a): Correlations in the planar lattice gas at the
value of $\rho_A$ which is found on a sublimating surface.  The dashed
(dotted) line shows $K_A^x$ ($K_A^d$) obtained from a transfer matrix
calculation at a strip width $L = 16$. Additionally, we show the
results of canonical Monte Carlo simulations ($\times$: $K_A^x$,
$\plustimes$: $K_A^d$). The system size is $N = 128$. 
$10^4 \cdot N^2$ Monte Carlo steps have been performed
both for equilibration and for measurement. Panel (b): Correlation length in the $y$-direction
measured on the sublimating surface ($\times$) and in an equilibrium
Monte Carlo simulation of the planar lattice gas ($\boxdot$). The
solid line shows results of a transfer matrix calculation. These data
have been measuered in the simulations shown in figure \ref{rekfig2a}
and in panel (a).}
\label{rek2dverg1}
\end{figure}
Figure \ref{rek2dverg1} shows the equilibrium values of $K_A^x$ and
$K_A^d$ in the 2D model. The simulation results agree well with the
results of the transfer matrix calculation. 
Qualitatively, the behaviour of the 2D model resembles that of the 3D
model. There is a crossover between a dominant $c(2\times 2)$
arrangement at low temperature and a high temperature regime where the
\A{} atoms preferentially arrange in rows. However, there is no
quantitative aggreement. In the 2D model, the reordering occurs at a
higher temperature. Additionally, at all investigated temperatures the
$c(2\times 2)$ reconstructed fraction of the surface is greater than
in the 3D model.

We determine the temperature $T_c$ at which the 2D model leaves the
long range ordered phase (I) and enters the disordered phase
(II). Monte Carlo simulations and transfer matrix calculations yield
$T_c \approx 0.45$ for the phase transition of the infinite
system. However, in our simulations of the 3D model we find that
$M_{A}^{(2\times 2)} = 0$ apart from statistical fluctuations even at
$T = 0.35$. Thus, there is no long range order of the \A{} atoms on the
surface of the sublimating crystal.

An inspection of surface snapshots of the 3D model (figure
\ref{rekfigstuf}a) at low temperature shows, that there are several
$c(2\times 2)$ reconstructed domains. In each domain, the \A{} atoms
occupy one of the two sublattices which correspond to positive and
negative values of $M_{A}^{(2\times 2)}$, respectively. Domains of
both orientations cover equal areas such that on average
$M_{A}^{(2\times 2)} = 0$. The domains have an elongated shape where
the domain size in the $y$-direction is much greater than that in the
$x$-direction. Frequently one encounters domains which run through the
whole system. Thus, there are long domain walls parallel to the
$y$-direction which consist of a row of pairs of \A{} atoms on
neighbouring sites in the $x$-direction. Of course, these pairs
contribute to $K_A^x$. Consequently, the formation of domain walls on
sublimating surfaces explains both the absence of long range order and
the greater values of $K_A^x$ compared to the 2D model.

We investigate correlation lengths in the 2D and in the 3D model. This
is done by fitting an Ornstein-Zernike law (equation \ref{ornstein})
to $C(\vec{x})$ which is measured in the Monte Carlo simulations. 

To check whether this method yields correct results we compare the
correlation length in the $y$-direction $\xi_y$ in the 2D model with a
transfer matrix calculation where $\xi_y$ is determined from the
eigenvalues according to equation \ref{korrelationslaenge}. We find,
that $C((0, y)^\top)$ follows an Ornstein-Zernike law for distances
greater than 15 lattice constants. The results are shown in figure
\ref{rek2dverg1}b. At high temperature, Monte Carlo data and results
of the transfer matrix calculation agree well. Clearly, in a finite
system with periodic boundary conditions, the correlation length
cannot exceed one half of the system size. Therefore, at low
temperature where the transfer matrix calculations yield very large
$\xi_y$, the Monte Carlo results saturate at values close to
64. Consequently, our fitting procedure is reasonable for correlation
lengths which are small compared to the system size.

We find, that in the 3D model $C((0, y)^\top)$ agrees well with an
Ornstein-Zernike law although this is strictly valid only in thermal
equilibrium. Interestingly, we find values of $\xi_y$ which agree well
with the equilibrium results of the 2D model (figure
\ref{rek2dverg1}b).

Since $C((x, 0)^\top)$ depends strongly on which of $K_A^x$ and
$K_A^d$ is greater, we expect a different behaviour of correlations in
the $x$-direction in the 2D and in the 3D model. Indeed, this is the
case. In general, correlation lengths $\xi_x$ in the $x$-direction are
small. Consequently, on large lengthscales where a fit to an
Ornstein-Zernike law would make sense, $C((x, 0)^\top)$ is
indistinguishable from its asymptotical value. Therefore, we present
only qualitative results. At all temperatures, $\xi_x$ in the 3D model
is significantly greater than in the 2D model. Probably this is a
consequence of the rows of \A{} atoms parallel to the $x$-direction
which are characteristic of the $(2\times 1)$ arrangement.
 
\section{The surface under an external particle flux \label{rekpflux}}

To get insight into the behaviour of a material in an MBE environment,
it is important to understand how the structure of the surface changes
if it is exposed to a particle beam. Additionally, these
investigations give us insight into the behaviour of the surface at
\A{} coverages which differ from the value found under vacuum
conditions.
\begin{figure}[htp]
\begin{picture}(100, 70)(0, 0)
\put(0,70){\resizebox{0.48\textwidth}{!}{\rotatebox{270}{\includegraphics{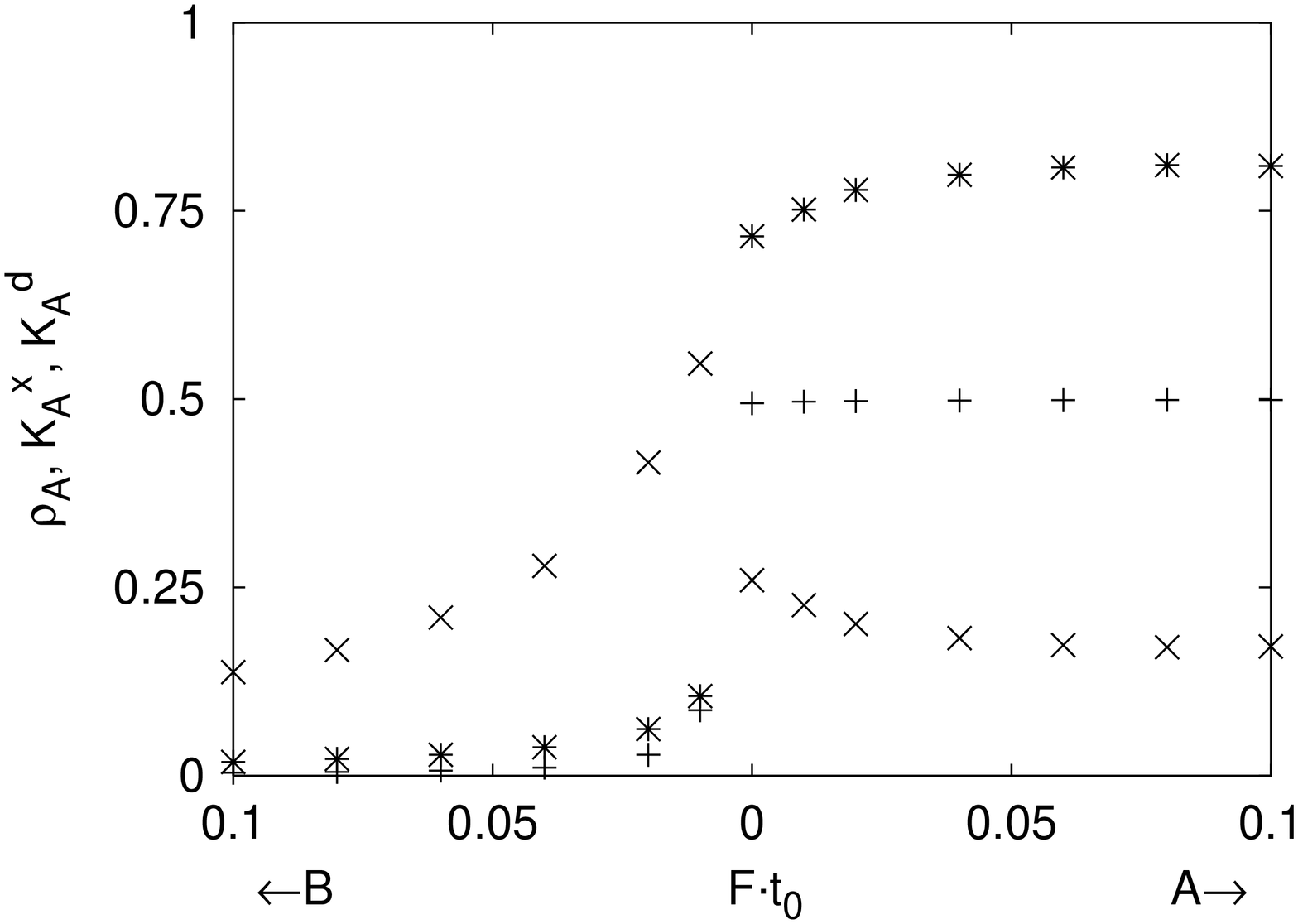}}}}
\put(50,70){\resizebox{0.48\textwidth}{!}{\rotatebox{270}{\includegraphics{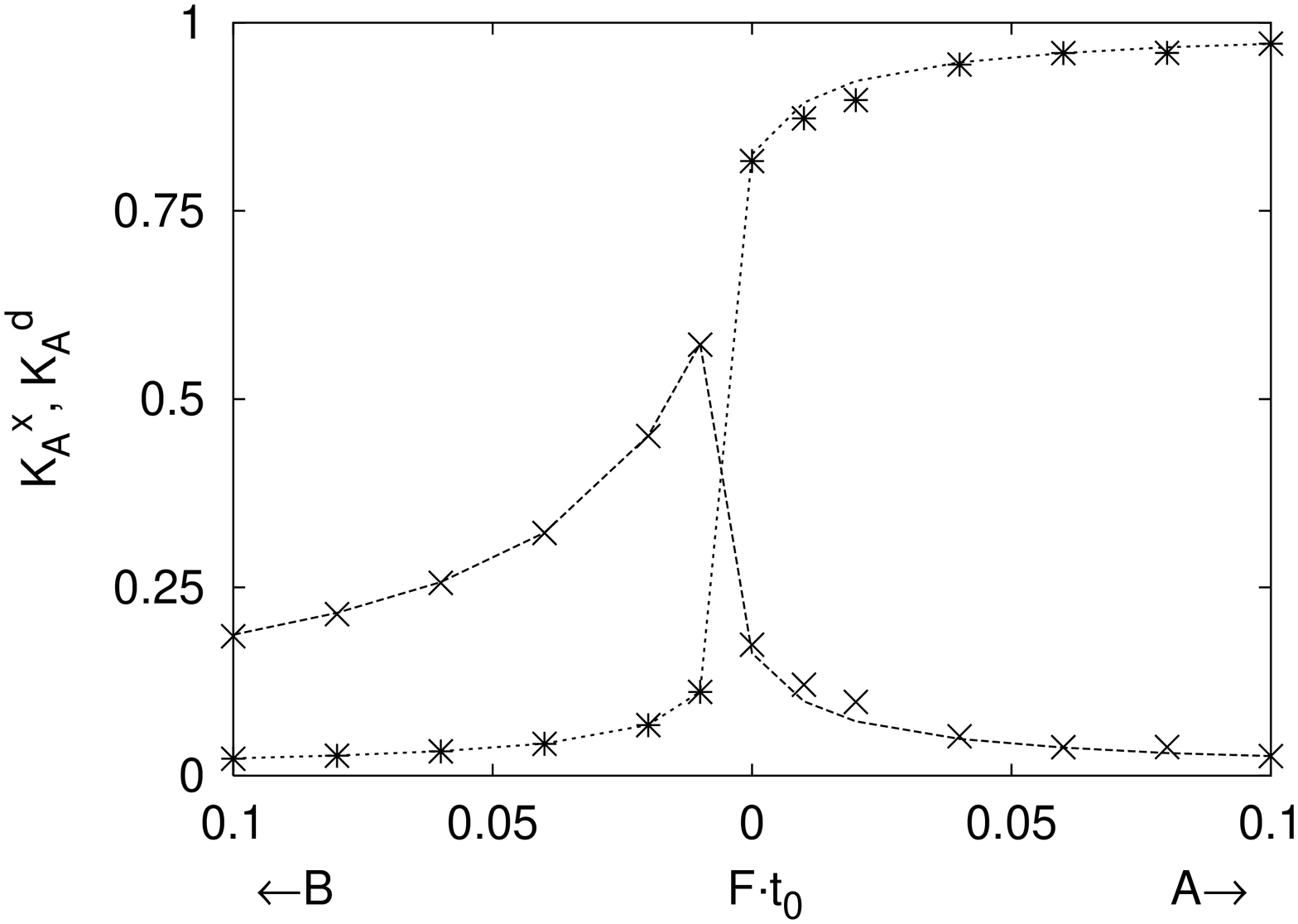}}}}
\put(0,33){\resizebox{0.48\textwidth}{!}{\rotatebox{270}{\includegraphics{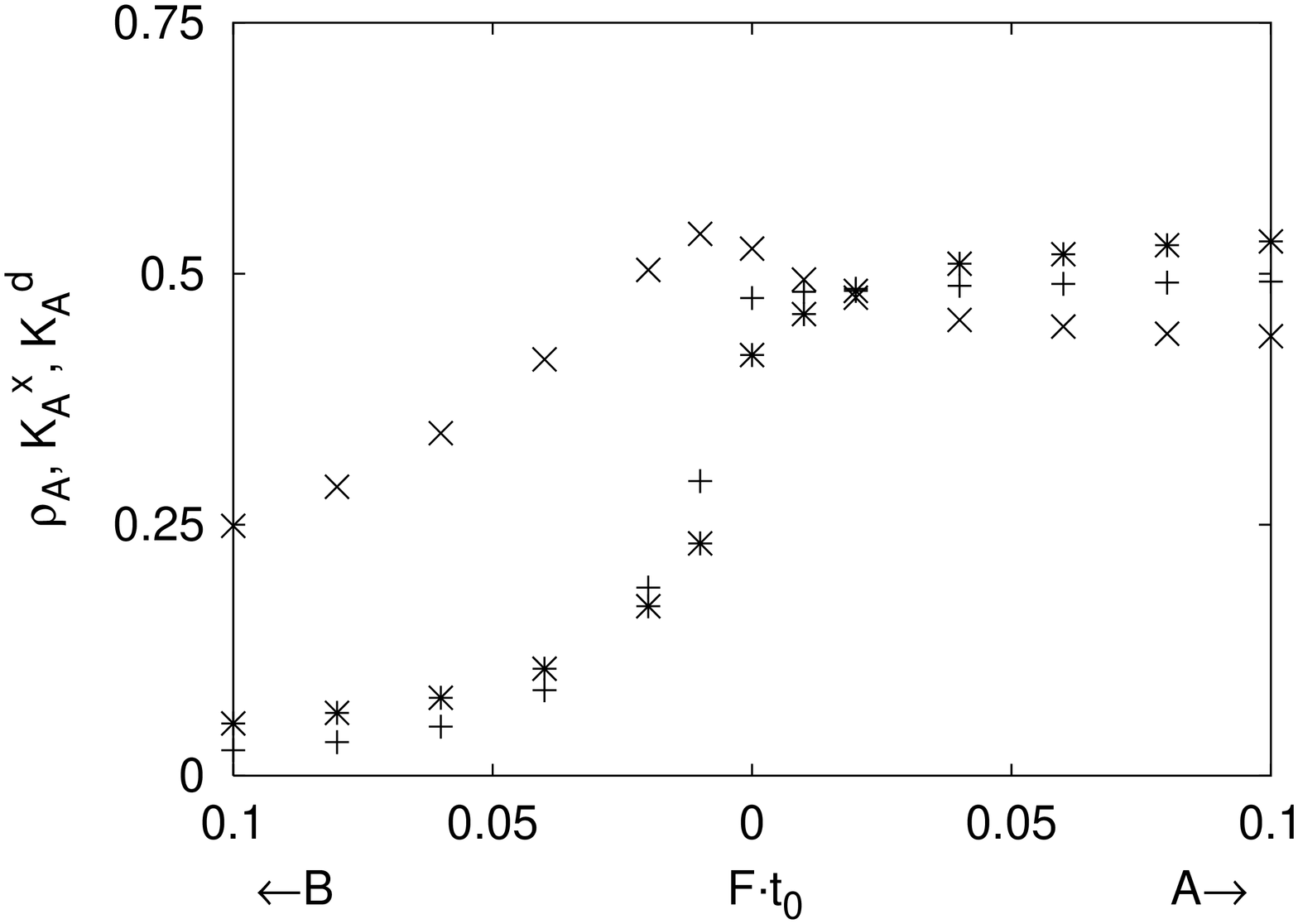}}}}
\put(50,33){\resizebox{0.48\textwidth}{!}{\rotatebox{270}{\includegraphics{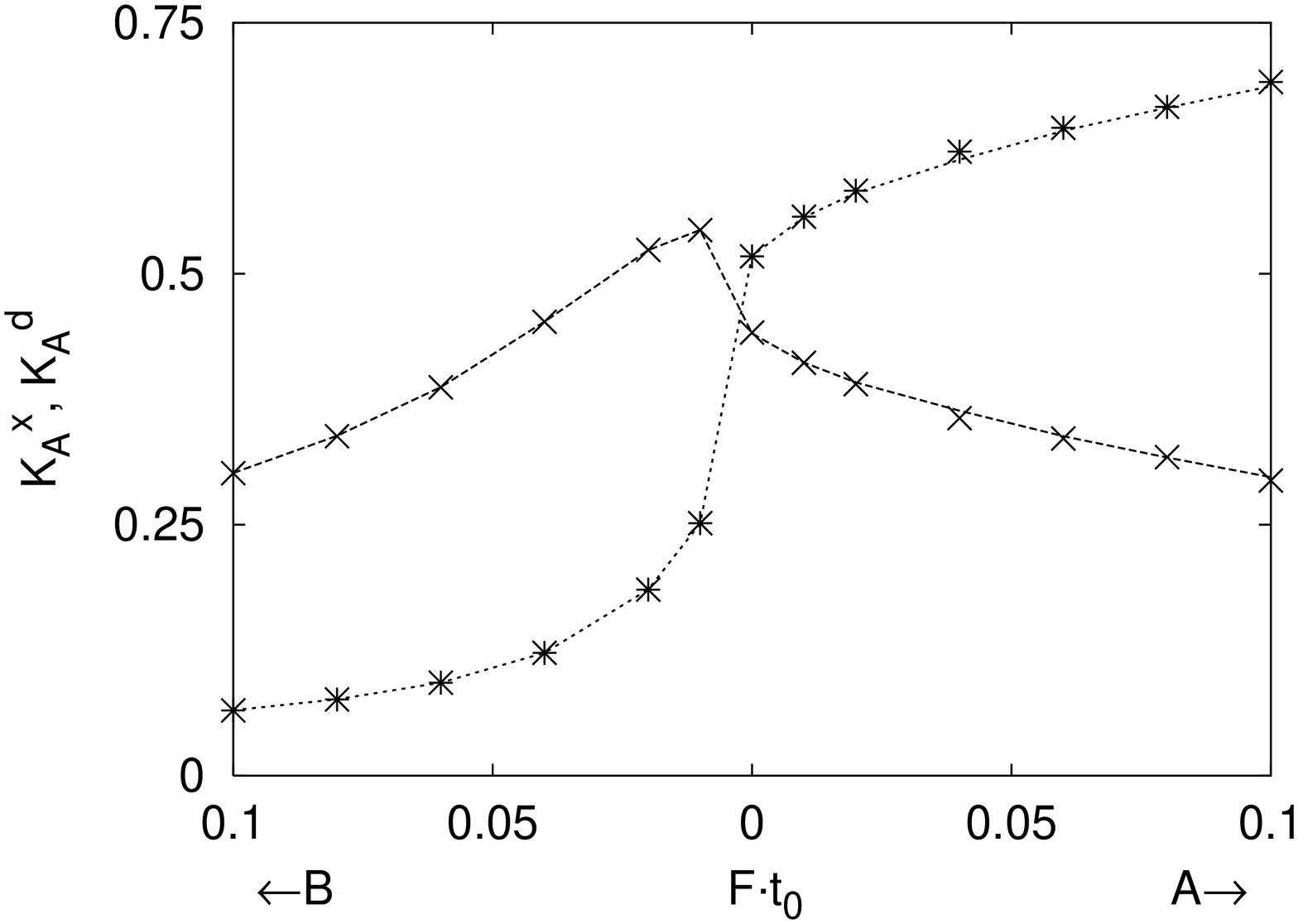}}}}
\put(0,67){(a)}
\put(50,67){(b)}
\put(0,30){(c)}
\put(50,30){(d)}
\end{picture}
\caption{\A{} coverage $\rho_{A}$ ($+$) and correlations $K_{A}^{x}$
($\times$), $K_{A}^{d}$ ($\plustimes$) in the stationary state of vicinal
surfaces under an external particle flux and equilibrium values in the
planar lattice gas at the \A{} coverage measured on the crystal surface.
Panel (a): Behaviour of the 3D model at $T = 0.45$. The steps are
oriented in the diagonal direction, the terrace width is $32 \sqrt{2}$
lattice constants. Panel (b): Results of the 2D model at the value of
$\rho_{A}$ measured on the sublimating surface at the flux indicated
at the $x$-axis. The symbols denote results of Monte Carlo
simulations, the lines show values determined by means of a transfer
matrix calculation at a strip width $L = 16$.  Panels (c) and (d): 3D
model (c) and 2D model (d) at $T = 0.57$. All other parameters are
identical to those used in (a) and (b). Note that an \A{}-flux applied
to the surface of the 3D model leads to a re-entry into the $c(2\times
2)$ reconstructed configuration at a temperature above the reordering
under vacuum.  In all Monte Carlo simulations shown, the system size
is $128 \times 128$ lattice constants. In the equilibrium simulations,
$10^4 \cdot N^2$ Monte Carlo steps have been performed both for
equilibration and for measurement. In the majority of the simulations
of the sublimating surface, time averages over $3\cdot 10^4 t_0$,
following a time interval of the same length for relaxation have been
performed. Since the relaxation of surfaces under an \A{} flux at $T =
0.45$ is very slow, a longer time interval of $10^6 t_0$ has been used
in these simulation runs.
\label{rekfig2b}}
\end{figure}

We simulate vicinal surfaces which are exposed to a flux of pure \A{} or
\B{}. Panels (a) and (c) of figure \ref{rekfig2b} show the quantities
$\rho_{A}$, $K_{A}^{x}$ and $K_{A}^{d}$ as functions of the particle
flux $F$ at a temperature $T = 0.45$ which is below the
$c(2\times2)$-$(2\times1)$ reordering under vacuum, and at $T = 0.57$
which is slightly above the transition.

An \A{} flux increases $\rho_A$. This leads to an increase of
$K_{A}^{d}$ such that under a large \A{} flux the \A{} atoms arrange
preferentially in a $c(2\times 2)$ pattern even at temperatures above
the reordering under vacuum (figure \ref{rekfig2b}c). Applying \B{}
fluxes, we can regulate $\rho_{A}$ to values smaller than those close
to 0.5 observed under vacuum. We find, that a decrease of $\rho_A$
makes the remaining \A{} atoms arrange preferentially in the rows of the
$(2\times 1)$ reconstruction. 

In figure \ref{rekfig2b}b,d we show the correlations $K_A^x$ and
$K_A^d$ which are found in the 2D model at the value of $\rho_A$ which
is measured in the 3D model. A comparison of these results with the
behaviour of the 3D model shows, that the qualitative similarity and
the quantitative differences between the 2D model and the 3D model
which have been observed under vacuum are preserved in a wide range of
\A{} coverages. There is no long range order of the \A{} atoms even at
high \A{} flux and low temperature. Instead, there are domains which are
elongated in the $y$-direction. This yields $M_{A}^{(2\times 2)} = 0$
and values of $K_A^x$ which are greater than those found in the 2d
model.  At low \A{} coverages we find that in the 3D model both $K_A^x$
and $K_A^d$ are slightly smaller than in the 2D model.  An
investigation of $C(\vec{x})$ at surfaces under a particle flux shows,
that correlation lengths in the $y$-direction in the 2D model and in
the 3D model agree quantitatively, while there are significant
differences in $\xi_x$.

\section{Conclusions \label{rekconclusions}}

In summary, our simulations reproduce the experimental results for the
reconstruction of Cd terminated CdTe(001) surfaces
qualitatively. Under vacuum, there is a crossover from a $c(2\times
2)$ reconstruction at low temperature to a high temperature phase with
a dominant $(2\times 1)$ arrangement. An external flux of \A{} atoms
stabilizes the $c(2\times 2)$ arrangement at temperatures above the
reordering in vacuum. We find that correlation lengths in the
$y$-direction are much greater than those in the $x$-direction. This
agrees with experimental results reported in \cite{n98,ntss00}.  

We have shown that the reconstruction order of \A{} atoms on a vicinal
surface sublimating in step flow mode does not change qualitatively
compared to a planar lattice gas in thermal equilibrium. In
particular, the behaviour of the short range correlations $K_A^x$ and
$K_A^d$ in the 3D model is quite similar to that found in the 2D
model. This shows, that the investigation of comparatively simple
models like those introduced in chapter \ref{dreikapitel} is an
appropriate tool to get insight into the behaviour of reconstructed
surfaces under MBE conditions.

The nonequilibrium conditions of sublimation induce
characteristic deviations of the arrangement of \A{} atoms on the
surface from the equilibrium configuration. Perhaps the most important
one is the absence of long range order at low $T$ and high
$\rho_A$. This is reminiscent of the behaviour of the growth model of
Chin and den Nijs \cite{cn01}. 

Our findings can be understood within the following picture of
sublimation: There is a permanent creation and annihilation of
reconstructed surface. As the terminating layer of the crystal
evaporates, the layer below is uncovered and becomes the new
surface. In this process, a given atom stays on the surface for a
typical time $\tau_{\mathrm{rel}} = 1/r_{\mathrm{sub}}$ before it is
desorbed. During this time, the arrangement of the atoms evolves
towards the equilibrium configuration. However, it is insufficient for
a complete equilibration. Therefore, there are characteristic defects
in the reconstruction order of \A{} atoms. 

At low temperature and high \A{} coverage, there is a division of the
surface in domains. Most of the domain walls run in the
$y$-direction. This can be understood from the fact that the formation
of such a domain wall requires only a small energy of $\Delta E = | 2
\varepsilon_d - \varepsilon_x|/2 = 0.015$ per unit length. Thus, there
is only a comparatively small driving force which makes the domains
coarse. Similarly, the small energy difference between a $c(2\times
2)$ order and a $(2\times 1)$ arrangement might be responsible for the
high values of $K_A^x$ at high temperature. At small $\rho_A$, both
$K_A^d$ and $K_A^x$ are smaller than in thermal equilibrium. This
indicates, that in the 3D model there is a higher fraction of single
\A{} atoms than in the 2D model. This may be understood from the fact
that $\tau_{\mathrm{rel}}$ is not sufficient for some diffusing
adatoms to meet a binding partner.

On the other hand, the good agreement between the correlation lengths
in the $y$-direction measured in the 2D model and in the 3D model
shows, that important aspects of the statistical mechanics of the
lattice gas in thermal equilibrium are preserved in step flow
sublimation. 

\chapter{Simulation of CdTe \label{zinkblendenmodell}}

In this chapter, we extend the model of a binary compound with surface
reconstructions presented in chapter \ref{rekkapitel} to make it a more
realistic model of CdTe. We simulate the zinc-blende lattice
of II-VI semiconductor crystals instead of a simple cubic lattice.  To
be meaningful, a fully three-dimensional modelling of the CdTe crystal
requires a detailed understanding of the interactions of the atoms on
{\em all} possible surface orientations and the correct treatment of
vacancies and non-stoichiometric crystals. Such a model is beyond the
scope of this investigation which aims at an understanding of the
behaviour of reconstructed $(001)$ surfaces. However, as we will show,
a physically relevant model is not possible within the limits of the
solid-on-solid condition.

To find a reasonable compromise between the complexity of a fully
three-dimensional model and the limits of a strict solid-on-solid
model, we permit the occupation of Cd sites by Te atoms and the
binding of Te atoms to single Cd atoms {\em on the surface}, while the
incorporation of point defects into the {\em bulk} of the crystal is
forbidden. Then we can treat those atoms which stay in the correct
sublattice and have at least two chemical bonds to atoms in the layer
below in a solid-on-solid manner, while the weakly bound Te atoms are
considered separately.

First, we study the behaviour of this model in vacuum and under a
flux of pure Cd or pure Te. In particular, we show that a reasonable
choice of model parameters yields semi-quantitative agreement with
experiments. Then, we perform simulations of atomic layer epitaxy and
compare our results with those of experimental investigations of CdTe
by means of this growth technique.

\section{The zinc-blende lattice \label{zbsos}}

In this section, we present a solid-on-solid representation of the
zinc-blende lattice. A sketch of its unit cell is shown in figure 
\ref{zbeinheitszell}. 
\begin{figure}
\botbase{\includegraphics{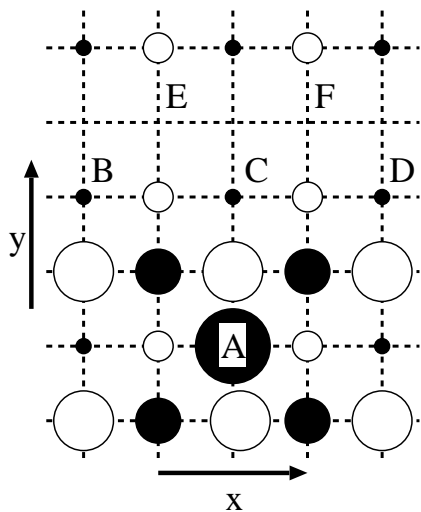}} \hfill
\botbasebox{0.6\textwidth}{\caption{The zinc-blende lattice lattice
seen from the $[001]$ direction.  The figure shows a Te terminated
surface with a step in the x-direction.  Cd atoms are shown as filled
circles, Te atoms as open circles. The size of the atoms corresponds
to their $z$-coordinate: the higher the atom, the larger the
circle.\label{zbsosfig1}}}
\end{figure}
We use cartesian coordinates where the $x$-axis points in the $[110]$
direction, the $y$-axis points in the $[1\overline{1}0]$ direction and
the $z$-axis points in the $[001]$ direction which is perpendicular to
the (001) surface. The origin of the coordinate system is at the
centre of a Te atom. If the zinc-blende lattice is projected onto the
$x$-$y$ plane in the $z$-direction, the image positions of all atoms
lie on a square lattice (figure \ref{zbsosfig1}).  Measuring the
lattice constant in appropriate units, the images of Te atoms are at
postitions $(2 i, 2 j)$ and $(2 i + 1, 2 j + 1)$ in the $x$-$y$ plane
where $i, j \in \mathbb{Z}$.  The images of Cd atoms are at positions
$(2 i, 2 j + 1)$ and $(2 i + 1, 2 j)$ with integer $i$, $j$.  This
square lattice is used as the basis of our solid-on-solid
representation of the zinc-blende $(001)$ surface, which is described
by a two-dimensional $2 L \times 2 N$ array $h$ of integers. $h_{x,y}$
is the maximal $z$-coordinate of the column of atoms whose images are
at $(x, y)$, measured in units of the spacing between a Cd and a Te
layer. In the following, for simplicity we will loosely denote any
atom whose image is at $(x, y)$ as ``atom at $(x, y)$'' and $h_{x,y}$
as ``height of the atom at $(x, y)$''.

An atom in the zinc-blende lattice may have at most 4 binding
partners.  Since the solid-on-solid condition forbids overhangs and
vacancies, each atom must have at least two bonds to atoms of the
opposite species in the layer below.  Only atoms with exactly two
bonds may be {\em removed} from the crystal without violating the
solid-on-solid condition: An atom with 4 bonds is inside the bulk of
the crystal. If we remove it, we create a vacancy. If we remove an
atom with three bonds the atom in the layer above which is bound to it
becomes an overhang. Atoms with only one bond form an overhang and do
not exist in a solid-on-solid model. In the following, we will denote
atoms which may be removed from the surface without violating the
solid-on-solid condition as {\em mobile}. Similarly, an atom can be
{\em deposited} only at a site where it has exactly two bonds. The
deposition of an atom with only one bond creates an overhang. Empty
sites where the deposition of an atom creates three or four bonds can
exist only in the presence of overhangs or vacancies, which is not the
case in a solid-on-solid model. In consequence, a {\em diffusion
process} where an atom is removed from one site and deposited at
another cannot change the number of chemical bonds in the system.

A Cd atom at $(x, y)$ is bound to Te atoms at $(x, y \pm 1)$ in the
layer below and can be bound to Te atoms at $(x \pm 1, y)$ in the
layer above.  The atom is mobile, if there are no bonds to Te atoms in
the layer above, which is the case if $h_{x, y} > h_{x \pm 1, y}$.
The neighbouring Cd atoms of a Te atom at $(x', y')$ are at $(x' \pm
1, y')$ in the layer below and at $(x', y'\pm 1)$ in the layer above.
Thus, the condition for mobility of a Te atom is $h_{x', y'} > h_{x',
y' \pm 1}$.  

An atom can be deposited at a site if the solid-on-solid condition is
fulfilled and the atomic species matches the character of the site. At
a Cd site at $(x, y)$, the solid-on-solid condition demands that
$h_{x, y-1} = h_{x, y+1}$ and $h_{x, y} < h_{x, y \pm 1}$.  A Te site
at $(x', y')$ can be occupied if $h_{x'-1,y'} = h_{x'+1, y'}$ and
$h_{x',y'} < h_{x'\pm 1, y'}$.  The deposition of an atom at $(x, y)$
is performed by setting $h_{x,y} \leftarrow h_{x, y} + 4$. Similarly,
setting $h_{x, y} \leftarrow h_{x, y} - 4$ removes the atom at $(x,
y)$.

As discussed in section \ref{dreikapitel}, the positions of the atoms
in {\em one half-layer} of the zinc-blende lattice lie on a square
lattice.  However, this is not identical to the square lattice
structure of the solid-on-solid representation of the crystal lattice,
as it is in the simple cubic model presented in chapter
\ref{rekkapitel}. The distance between nearest neighbours in the
square lattice of one half-layer (denoted as {\em lateral neighbours}
in the following) is {\em twice} the distance between neighbouring
images of atoms in the projection of the whole crystal onto the
$x$-$y$ plane.

\subsection{Desorption and adsorption \label{zbadsdes}}

Due to the solid-on-solid condition, only mobile atoms may desorb. We
consider the desorption of single atoms from the surface, neglecting
the fact that in reality Te probably desorbs as a binary molecule from
the CdTe surface. Similarly, we consider only the adsorption of
single atoms via an incorporation process (section
\ref{depositionintro}): First, the species of the atom and the landing
site are chosen randomly. Then, we search for sites where the
depositon of this atom is possible within an incorporation radius
$r_{i}$ around the landing site.  The atom is deposited at the lowest
site which has been found in that search. If two or more sites of
equal height have been found, one of these sites is chosen
randomly. If no appropriate site is found, the atom is reflected.

\subsection{Diffusion \label{zbdiffusion}}

We consider an atom diffusing on a flat surface, for example a Cd atom
on a Te terminated surface. For an atom at $\vec{x}_{0}$, the nearest
sites that can be occupied by an atom of its species are the lateral
neighbour sites at $\vec{x}_{0} + \Delta \vec{x}$, where the hopping
vector $\Delta \vec{x}$ is an element of the set $\{ (0, 2), (0, -2),
(2, 0), (-2, 0) \}$.  Similar to the model discussed in section
\ref{rekkapitel} we consider also diffusion to lateral diagonal
neigbour sites, i.e.\ $\Delta \vec{x} \in \{(2, 2), (2, -2), (-2, 2),
(-2, -2)\}$.

However, a particle that can perform only these diffusion hops is not
able to cross a step.  We discuss this for the example of the Cd atom
at site A in figure \ref{zbsosfig1} diffusing downhill at a step in
$x$-direction.  It is not possible to deposit the atom at either of
the sites B, C and D, since the condition $h_{x, y-1} = h_{x, y+1}$ is
not fulfilled there.  Similar considerations hold for other step
orientations, diffusion uphill and for the diffusion of Te atoms.
Thus there is an infinite Schwoebel barrier in a model in
which only these diffusion hops are considered. Since we do not want
to restrict ourselves in this special case, we introduce additional
hopping vectors for diffusion. In figure \ref{zbsosfig1}, the nearest
sites to site A at the lower side of the step that can be occupied by
a Cd atom are E and F, which yield $\Delta \vec{x} = (1, 3)$ and
$\Delta \vec{x} = (-1, 3)$.  Extending these considerations to
different step orientations, we obtain the set $\{(3, 1), (3, -1),
(-3, -1), (-3, 1), (1, 3), (-1, 3), (1, -3), (-1, -3)\}$ of hopping
vectors which are sufficient to permit the crossing of steps of one
monolayer height difference oriented in arbitrary direction.  To
permit the crossing of steps higher than one monolayer, we would have
to introduce additional hopping vectors. However, such high steps
rarely occur in our simulations.  Therefore, it is a reasonable
approximation to neglect diffusion accross high steps in order to
reduce the complexity of the model.  In summary, there are 16 hopping
vectors for the diffusion of an atom. However, due to the conditions
that must be fulfilled to permit the deposition of the atom at the
landing site, at most 8 diffusion hops are allowed for a single mobile
atom.
 
\section{The Hamiltonian of the solid-on-solid model}

The basic idea behind our model of the CdTe$(001)$ surface is similar
to that of the simple cubic crystal presented in chapter
\ref{rekkapitel}: besides the Cd-Te bond, there is an isotropic
interaction between Cd atoms and between Te atoms on lateral nearest
neighbour sites in the bulk. The interaction of Te atoms on the
surface is identical to that of Te atoms in the bulk, i.e.\ we neglect
the formation of Te dimers. The interaction of Cd atoms, however,
depends strongly on whether they are mobile or incorporated into the
crystal. Mobile Cd atoms interact with their lateral nearest neigbours
and lateral diagonal neighbours with the interactions of the
lattice gas model introduced in chapter \ref{dreikapitel}.

\subsection{The Cd-Te bond}

There is a chemical bond between Cd and Te atoms on nearest neighbour
sites.  We assume that the total energy of these bonds is proportional
to their number with an energy $\varepsilon_{c}$ per bond. Since the
strength of the interaction between atoms is a quickly decreasing
function of their distance, we expect this energy to be the greatest
part of the total binding energy of the crystal.  To remove a mobile
atom from the surface, two Cd-Te bonds have to be broken. This yields
an energy barrier $\varepsilon_v = -2 \varepsilon_{c}$ for the desorption of
an atom in addition to contributions from other, weaker interactions.

As discussed in section \ref{zbsos}, a diffusion process cannot change
the total number of nearest neighbour bonds. Consequently, the
strength of the Cd-Te bond does not influence the energy difference
between the starting and the final state of the atom. In our model,
the energy barrier a diffusing adatom must overcome is a separate
parameter. It is smaller than the energy required to break up its
bonds. Thus, the rates of diffusion processes are not a function of
$\varepsilon_{c}$.

\subsection{Interactions with lateral neighbours}

Interactions with a longer range than nearest neighbour interactions
are required to describe surface reconstructions and to ensure that
the equilibrium surface at low temperature is flat. 

It is a reasonable approximation to consider only interactions between
particles in the same layer. In a solid-on-solid model, the
environment of an atom in the vertical direction is always the
same. Therefore, interactions of longer range with atoms in this
direction will be always present as long as the atom remains on the
surface. In case of desorption they may be treated as an additional
contribution to $\varepsilon_c$.

We introduce an attractive interaction between atoms which are
incorporated in the crystal, i.e.\ atoms with more than two bonds. An
atom interacts with its lateral nearest neighbours and the strength of
the interaction is isotropic.  However, we allow a different strength
for the attraction between Cd atoms ($\varepsilon_{Cd,b}$) and Te atoms
($\varepsilon_{Te, b}$).

We assume that the interaction between Te atoms is independent on
whether they are incorporated in the crystal or mobile atoms on the
surface and do not consider the dimerization of surface Te atoms. This
simplification reduces the complexity of the model considerably.
Mobile Cd atoms interact via the anisotropic interactions
$\varepsilon_x$, $\varepsilon_d$ and a hardcore repulsion between particles
on lateral neighbour sites in the $y$-direction which were introduced
in chapter \ref{dreikapitel}. Additionally, we introduce an
interaction $\varepsilon_{Cd, h}$ between mobile and incorporated Cd
atoms which we assume to be isotropic.

In summary, the Hamiltonian of the solid-on-solid model is
\begin{equation}
H_{SOS} = \varepsilon_c n_c + \underbrace{ \varepsilon_{Cd, b} n_{Cd, b} +
\varepsilon_{Te, b} n_{Te, b} + \varepsilon_h n_{Cd, h} + \varepsilon_x n_{Cd,
x} + \varepsilon_d n_{Cd, d} }_{H_{lat}}
\label{zbsoshamilton}
\end{equation}
Here, $n_c$ is the number of chemical bonds beween Cd and Te atoms on
nearest neighbour sites. $n_{Cd, b}$ and $n_{Te, b}$ are the numbers
of pairs of incorporated Cd and Te atoms on lateral nearest
neighbour sites. $n_{Cd, h}$ is the number of mobile Cd atoms which
are lateral nearest neighbours of incorporated Cd atoms. $n_{Cd, x}$
and $n_{Cd, d}$ are the numbers of pairs of mobile Cd atoms on lateral
nearest neighbour sites in the $x$-direction and on lateral diagonal
neighbour sites, respectively. In addition to these interactions we
impose the constraint that the number $n_{Cd,y}$ of mobile Cd atoms on
nearest neighbour sites in the $y$-direction must always be zero.

\section{Beyond the solid-on-solid condition \label{zbteskapitel}}

In the present state of our model, there is a violation of the
ergodicity condition similar to the one discussed in section
\ref{rekmodel}. Consider the surface configuration shown in figure
\ref{zbsosfig3}a. Energetically, it is favourable to shift the small
island by one lattice constant into the negative $y$-direction to
obtain the configuration shown in figure \ref{zbsosfig3}e.  However,
it is not possible to construct a sequence of allowed single particle
moves to do so.
\begin{figure}
\begin{center}
\includegraphics{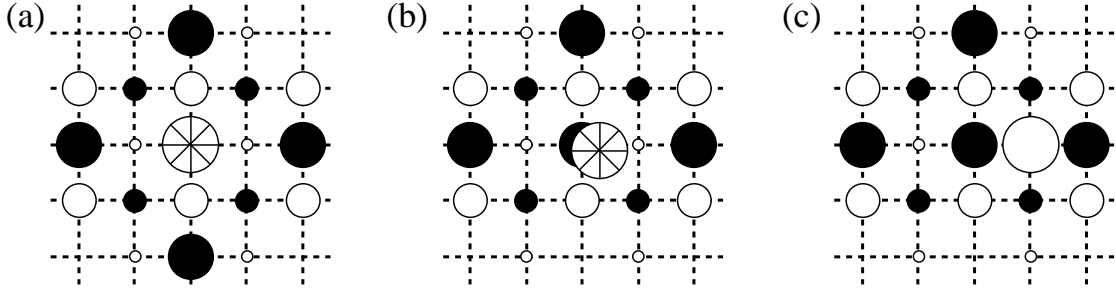}
\end{center}
\caption{Deposition of a Te atom on a \cdtxt{} reconstructed surface via
a Te$^*$ state. The Te$^*$ atom is shown as an open circle with a star
inside. \label{zbsosfig2}}
\end{figure}
\begin{figure}
\includegraphics{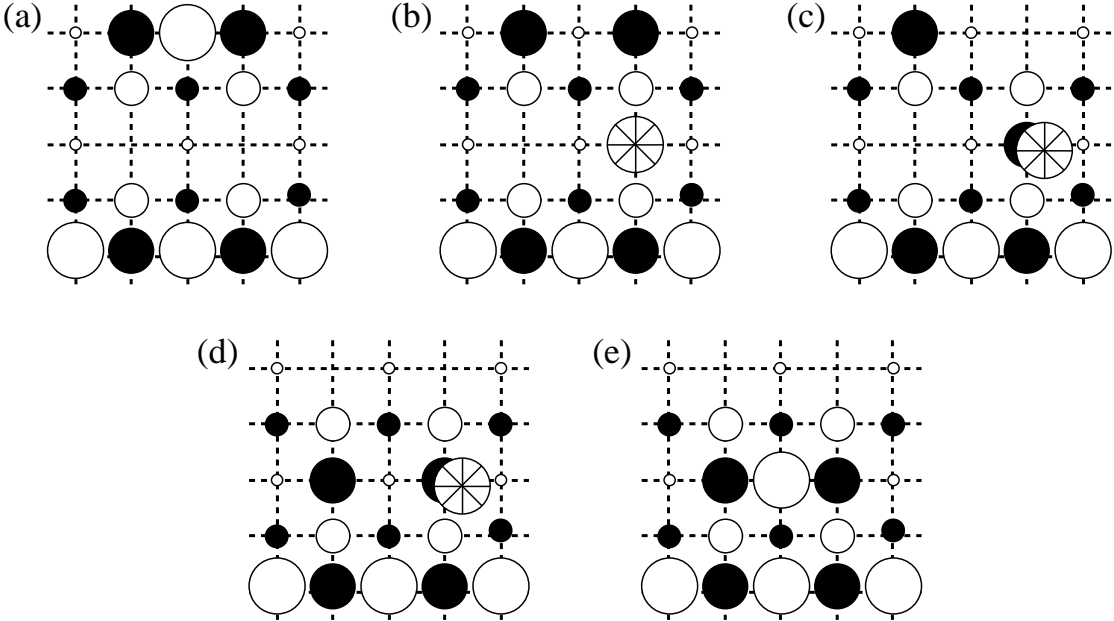}
\caption{Coalescence of islands. The Te atom on top of the small island 
stays temporarily in a Te$^*$ state. \label{zbsosfig3}}
\end{figure}
In chapter \ref{rekkapitel} we introduced the diffusion of \AB{}
molecules with \B{} atoms on top to solve an analogous problem. The
counterpart of this process on the zinc-blende lattice would be the
diffusion of a cluster of two Cd atoms and one Te atom. Then, the
small island in figure \ref{zbsosfig3} could diffuse as a
whole. However, this is not a general solution to the ergodicity
problem. It is not possible to remove such a cluster from the edge of
a greater Te terminated island without violating the solid-on-solid
condition. There is at least one additional Te atom in the top layer
of the island which is bound to a Cd atom in the cluster. If we remove
the cluster, this atom becomes an overhang. 

Additionally, in the present form of our model it is not possible to
deposit an atom on a perfectly \cdtxt{} reconstructed surface. There
are no sites which can be occupied by a Te atom without violating the
solid-on-solid condition. The hardcore repulsion between mobile Cd
atoms in the $y$-direction forbids a completion of the topmost Cd
layer.

There is a solution to both of these problems which has the advantage
of being consistent with an experimental observation made on CdTe:
under a strong flux of Tellurium one finds surfaces with a Te coverage
of $3/2$ and a $(2 \times 1)$ reconstruction (figure
\ref{cdtephasdschematisch}). This is explained by the formation of Te
trimers which arrange in rows in the $x$-direction.  A trimer is
created if a Cd site on the surface is occupied by a Te atom. We
consider this arrangement in our model. Additionally, we allow Te
atoms which have only {\em one} bond to a Cd atom.  In the following,
we denote Te atoms in one of these configurations which both violate
the solid-on-solid condition as Te$^*$ atoms. It is not necessary to
give up a solid-on-solid model in favour of a fully three-dimensional
model to consider Te$^*$ states if Te$^*$ atoms may exist on the
surface of the crystal but the incorporation of point defects into the
crystal is forbidden. We describe the crystal by two two-dimensional
$2 L \times 2 N$ arrays of integers, $h$ and $t$. $h$ represents those
atoms which fulfil the solid-on-solid condition in the manner
described in section \ref{zbsos}. $t$ represents the Te$^*$ atoms. A
Te$^*$ atom at $(x, y)$ corresponds to $t_{x, y} = 1$, otherwise
$t_{x, y} = 0$.

The rules for Te$^*$ atoms are the following:
\begin{enumerate}
\item Te$^*$ atoms are allowed on sites with a mobile Cd atom (Te
atoms with one bond to a Cd atom) and on sites where the deposition
of a Cd atom would not violate the solid-on-solid condition (middle Te
atom in a trimer). The coordinates of these sites correspond to Cd
sites in the solid-on-solid model. Each site may be occupied by at
most one Te$^*$ atom.
\item Te$^*$ atoms on different sites do not interact. The binding
energy $\varepsilon^{*}$ of a Te$^*$ is independent on whether it is
bound to two Te atoms or to one Cd atom.
\item Two laterally neighbouring sites in the $y$-direction may be
occupied by two Cd atoms, if at least one of the sites is occupied by
a Te* atom.  Thus, a Te$^*$ atom neutralizes the hardcore repulsion
between laterally neighbouring Cd atoms in the $y$-direction.
\item Atoms on the solid-on-solid surface which are mobile may diffuse
or desorb even if they are binding partners of Te$^*$ atoms.  If a
Te$^*$ atom is on a site which may not be occupied by a Te$^*$ atom
any more after such a process (according to rule (1)), it is desorbed.
Of course, the activation energy for the desorption of Te$^*$ atoms
has to be considered in the calculation of the rate of the
process. Te$^*$ atoms do not follow the motion of their binding
partners.
\item Both the desorption of Te$^*$ atoms and the deposition of Te
atoms from an external particle beam in Te$^*$ states are possible. If
a Te atom is landing on the surface, in the incorporation process
Te$^*$ states within $r_i$ are considered in addition to states in the
solid-on-solid lattice. Te$^*$ atoms may diffuse on the surface.  We
permit the same hopping vectors as for the diffusion of atoms on the
solid-on-solid lattice.
\item Te atoms may change between states in the solid-on-solid lattice
and Te$^*$ states. This is a diffusion process where the Te atom hops
between Te sites in the solid-on-solid lattice and Te$^*$ sites, which
have the coordinates of Cd sites in the solid-on-solid model. We
consider the set
$$ 
\begin{array}{l} \{ (1, 0), (-1, 0), \\ (1, 2), (-1, 2), (1,
-2), (-1, -2), (2, 1), (2, -1), (-2, 1), (-2, -1), \\ (3, 0), (-3, 0),
(0, 3), (0, -3) \} \end{array}
$$ 
of hopping vectors for this process.
\end{enumerate}
Thus, the Hamiltonian of our model is
\begin{equation}
H = H_{lat} + \varepsilon_c n_c + \varepsilon^{*} n^{*}. 
\label{zbhamilton}
\end{equation}
Here, $H_{lat}$ is the contribution of the lateral interactions to the
energy of the system. It has been defined in equation
\ref{zbsoshamilton}.  $n_{c}$ is the number of Cd-Te bonds in the
solid-on-solid lattice.  $n^{*}$ is the number of Te$^*$
atoms. Additionally, we consider only surface configurations where the
number $n_{Cd,y}'$ of pairs of mobile Cd atoms without a Te$^*$ atom
on lateral neighbour sites in the $y$-direction is zero.

Now we show examples of the role of Te$^*$ states in the deposition of
a Te atom on a Cd terminated surface and the coalescence of islands.
Figure \ref{zbsosfig2} shows the deposition of a Te atom on a \cdtxt{}
reconstructed surface. First, the atom is deposited in a Te$^*$ state
in one of the gaps of the \cdtxt{} reconstruction (figure
\ref{zbsosfig2}a). According to rule (3), one of the neighbouring Cd
atoms can diffuse into the gap (figure \ref{zbsosfig2}b). Following
rule (4), now the Te$^*$ atom is bound to that Cd atom. This process
creates a site in the solid-on-solid lattice which can be occupied by
a Te atom. In the last step, the Te$^*$ atom diffuses to that site
(figure \ref{zbsosfig2}c) following rule (6).  Figure \ref{zbsosfig3}
shows the coalescence of islands. The Te atom on top of the small
island stays temporarily in a Te$^*$ state. Then, the two Cd atoms can
diffuse to the step edge. Finally, the Te atom is deposited on the
site in the solid-on-solid lattice provided by the Cd atoms.

The assumption of noninteracting Te$^*$ atoms (rule (2)) is a
simplification which has been introduced to make an investigation of
this model computationally feasible. Considering interactions between
the quickly diffusing Te$^*$ atoms would lead to a considerable
slowing down of the simulations. This simplification is justified if
the density of Te$^*$ is so low that the probability to find a partner
to interact with is small. We will show that interactions between
Te$^*$ atoms should be considered to describe the behaviour of a
surface under a strong flux of pure Te at low temperature.

A Te$^*$ atom which is bound to a Cd atom has three dangling bonds.
As discussed in section \ref{reconstintro}, these bonds are negatively
charged. It is plausible to assume that this compensates the
energetically unfavourable concentration of positively charged Cd
dangling bonds on a pair of neighbouring Cd atoms in the
$y$-direction. This might justify rule (3) which has been introduced
to make the landing of Te atoms on a perfectly \cdtxt{} reconstructed
surface possible. Without rule (3), it would be impossible for a Cd
atom to perform a diffusion process which provides a site in the
solid-on-solid lattice that can be occupied by a Te atom.

In the set of hopping vectors introduced in rule (6), we have
considered diffusion to those sites which have a smaller distance to
the starting site than the final sites of a diffusion process within
the same sublattice. These hopping vectors permit the crossing of
steps with a height of one monolayer. We have omitted the hopping
vectors $(0, 1)$ and $(0, -1)$, since due to the topology of the
zinc-blende lattice it is never possible to deposit the atom on the
final site of such a diffusion process.

\section{The dynamics of adatoms \label{zbdynamik}}

In our model, we assume Arrhenius rates (equation
\ref{arrheniusgesetz}) for diffusion and desorption processes. To
calculate the activation energy we assume the energy of the
transition state between the starting and the final configuration of
the system to be equal to the greater of the energies of both
configurations plus an additional barrier height. Since atoms on the
solid-on-solid surface are bound stronger than Te$^*$ atoms, it is
reasonable to assume that the barrier height for diffusion of atoms on
the solid-on-solid surface is greater than that for diffusion of
Te$^*$ atoms.
\begin{figure}
\begin{center}
\includegraphics{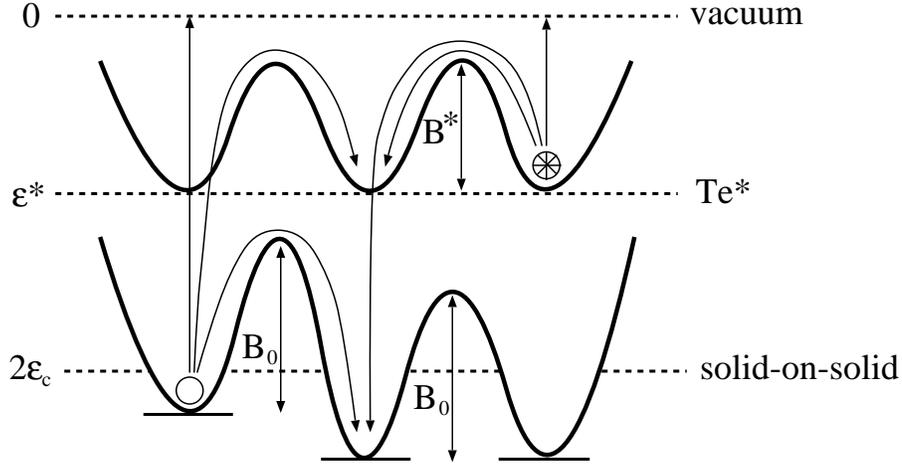} 
\end{center}
\caption{The potential energy surface of
adatoms \label{zbfig4}}
\end{figure}
In our model, we use the simplest choice for the energy barriers of
all processes that meets this condition. There is an equal barrier
height $B_{0}$ for all diffusion hops of particles without lateral
nearest neighbours and a barrier height $B^{*}$ for the diffusion of
Te$^*$ atoms. The barrier heights for the other processes are chosen
such that the barriers for all moves of one particle which do not
increase the energy of the system are equal and detailed balance is
fulfilled for diffusion processes.

This principle yields the following activation energies for diffusion
on the solid-on-solid surface ($E_{s \rightarrow s}$), diffusion of
Te$^*$ atoms ($E_{* \rightarrow *}$), diffusion of a Te atom from a
solid-on-solid state to a Te$^*$ state ($E_{s \rightarrow *}$),
diffusion of a Te$^*$ atom into a solid-on-solid state ($E_{*
\rightarrow s}$), desorption of an atom on the solid-on-solid surface
($E_{s \rightarrow v}$) and desorption of a Te$^*$ atom ($E_{*
\rightarrow v}$):
\begin{eqnarray}
E_{s \rightarrow s} & = & \max \{ B_0, B_0 + \Delta H \}
\label{zbdiffrate} \\
E_{* \rightarrow *} & = & \max \{ B^*, B^* + \Delta H \} \\ E_{s
\rightarrow *} & = & \max \{ B_0, B^* + \Delta H \} \\ E_{*
\rightarrow s} & = & \max \{ B^*, B_0 + \Delta H \} \\ E_{s
\rightarrow v} & = & \max \{ B_0, \Delta H \} \\ E_{* \rightarrow v} &
= & \max \{ B^*, \Delta H \}.
\end{eqnarray}
Figure \ref{zbfig4} shows a sketch of the energy barriers in our
model.

In the following, we will focus on the parameter set $r_i = 2$,
$\varepsilon_x = -1.95 |\varepsilon_d|$, $\varepsilon_h = 0$, $\varepsilon_{Cd,b}
= \varepsilon_{Te,b} = -0.8 |\varepsilon_d|$, $\varepsilon_c = -6.5
|\varepsilon_d|$, $\varepsilon^{*} = -3 |\varepsilon_d|$, $B_0 = 9
|\varepsilon_d|$ and $B^{*} = 2 |\varepsilon_d|$. Since there is an
attractive interaction between mobile Cd atoms on lateral diagonal
neighbour sites, $\varepsilon_d$ is necessarily negative. This choice of
parameters is guided by the following principles: $\varepsilon_x = -1.95
|\varepsilon_d|$ ensures that \cdtxt{} is energetically favourable compared
to \cdtxo{} and the difference between the surface energies of both
reconstructions is small compared to the total surface energy. The
parameters $\varepsilon_{Cd,b}$, $\varepsilon_{Te,b}$ and $\varepsilon_h$ have
been chosen such that we obtain a \cdtxt{} reconstruction under vacuum
at low temperature. Our model of a flat surface in thermal equilibrium
(chapter \ref{dreikapitel}) indicates that $|\varepsilon_d|$ is slightly
smaller than $0.1 \,\mathrm{eV}$. On the other hand, we expect the
activation energy $2 \varepsilon_c$ which is required for the desorption
of a single adatom to be on the same order of magnitude as the
macroscopic activation energy of sublimation. Experimentally, values
between $0.96 \,\mathrm{eV}$ and $1.95 \,\mathrm{eV}$ have been measured
\cite{ct97,nskts00,ntss00} depending on particle species and mode of
sublimation. Thus, $\varepsilon_c = -6.5 |\varepsilon_d|$ seems to be a good
guess. An adsorption of Te on a \cdtxt{} reconstructed surface via a
Te$^{*}$ state is possible only if Te$^{*}$ atoms are not desorbed too
fast such that $\varepsilon^*$ must not be too small. The barrier for
diffusion of a Te atom on the solid-on-solid surface should be smaller
than that for diffusion into a Te$^*$ state such that $B_0 < |2
\varepsilon_c - \varepsilon^*|$. Finally, the diffusion barrier for a Te$^*$
atom must be smaller than that for desorption of Te$^*$ which implies
$B^* < |\varepsilon^*|$. The prefactor of the Arrhenius rates has been
set to $\nu = 10^{12}/\mathrm{s}$ which has turned out to be a good choice for
CdTe \cite{nskts00,schi99}.

The value of $|\varepsilon_d|$ sets the energy scale of the model
which needs not to be specified if temperature is expressed by means
of the dimensionless parameter $\Theta = kT/|\varepsilon_d|$.  If we
set $k = |\varepsilon_d| = 1$ like in the previous chapters, we have
$\Theta = T$. In this chapter, we will determine $|\varepsilon_d|$ in
physical units by identifying the temperature of the \cdtxt{}-\cdtxo{}
transition with the experimental value. As we will show, this yields
activation energies for sublimation which are on a reasonable order of
magnitude.

\section{Characterization of the surface \label{zbcharakterisier}}

In principle, the methods applied for the characterization of the
surface of this model follow the ideas introduced in chapters
\ref{dreikapitel} and \ref{rekkapitel}. Note however, that there is an
important difference between the simple cubic lattice and the
zinc-blende lattice: In our representation of the zinc-blende lattice,
the distance between lateral nearest neighbours in one half-layer is
two unit lengths. The positions of atoms in each CdTe monolayer are
shifted by one unit length both in the $x$- and in the $y$-direction
compared to the atoms in the layer below or above. Thus, the pattern
of a surface reconstruction cannot be continued across a step as it is
possible in the simple cubic lattice. Consequently, it is not possible
to define the order parameter $M_{Cd}^{(2\times 2)}$ of the whole
system in analogy to the planar lattice gas. Similarly, this effect
hinders a calculation of correlation lengths with the method
introduced in section \ref{rekcharakterisier}. Therefore, we restrict
ourselves to an investigation of the short-range correlations
$K_{Cd}^{x}$ and $K_{Cd}^{d}$.

To calculate these, we define a $2 L \times 2 N$ array
of integers $c$, where $c_{i,j} = 1$ if there is a {\em mobile} Cd
atom at $(i, j)$ and zero otherwise. In terms of these variables, we
have
\begin{eqnarray}
K_{Cd}^{d} & := & \frac{1}{2\rho_{Cd} N L} \sum_{i,j=1}^{2L,2N} c_{i, j} \left( c_{i+2, j+2} + c_{i+2, j-2} \right) \\
K_{Cd}^{x} & := & \frac{1}{\rho_{Cd} N L} \sum_{i,j=1}^{2L,2N} c_{i,j} c_{i+2,j}
\end{eqnarray}
where $\rho_{Cd} = (L N)^{-1} \sum_{i,j=1}^{2L,2N} c_{i,j}$ is the density of
mobile Cd atoms on the surface. 

\section{Sublimation \label{zbsub}}

Qualitatively, the behaviour of a surface under vacuum is quite
similar to that observed in our model of a simple cubic crystal with
surface reconstructions presented in section \ref{rekkapitel}.

\begin{figure}[htb]
\botbase{\resizebox{0.48\textwidth}{!}{\rotatebox{270}{\includegraphics{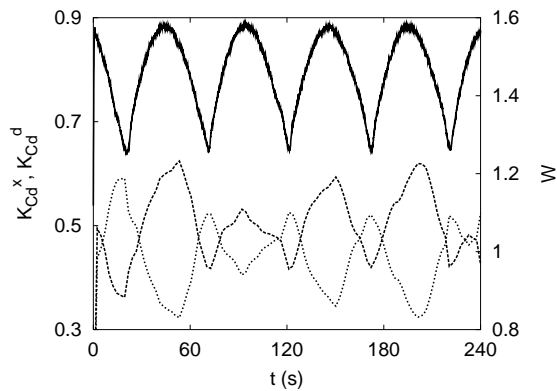}}}}
\hfill
\botbasebox{0.5\textwidth}{\caption{Surface width $W$(solid) and
correlations $K_{Cd}^{x}$ (dashed), $K_{Cd}^{d}$ (dotted) during the
sublimation of a flat surface at $\Theta = 0.525$ which corresponds to
$T = 525 \,\mathrm{K}$ in physical units. For clarity of plotting, the curves of
the correlations have been smoothed; $W$ shows the natural
fluctuations. The system size is $L = N = 64$.\label{zblayersub}}}
\end{figure}

On a flat surface sublimating in layer-by-layer mode, the correlations
$K_{Cd}^{x}$ and $K_{Cd}^{d}$ oscillate (figure
\ref{zblayersub}). Like in the simple cubic model, island boundaries
induce a preferential arrangement of Cd atoms in rows such that there
is a maximum of $K_{Cd}^{x}$ whenever the surface width $W$ is large
and a maximum of $K_{Cd}^{d}$ whenever the surface is flat.

\begin{figure}[htb]
\begin{picture}(100, 33)(0, 0)
\put(0,33){\resizebox{0.48\textwidth}{!}{\rotatebox{270}{\includegraphics{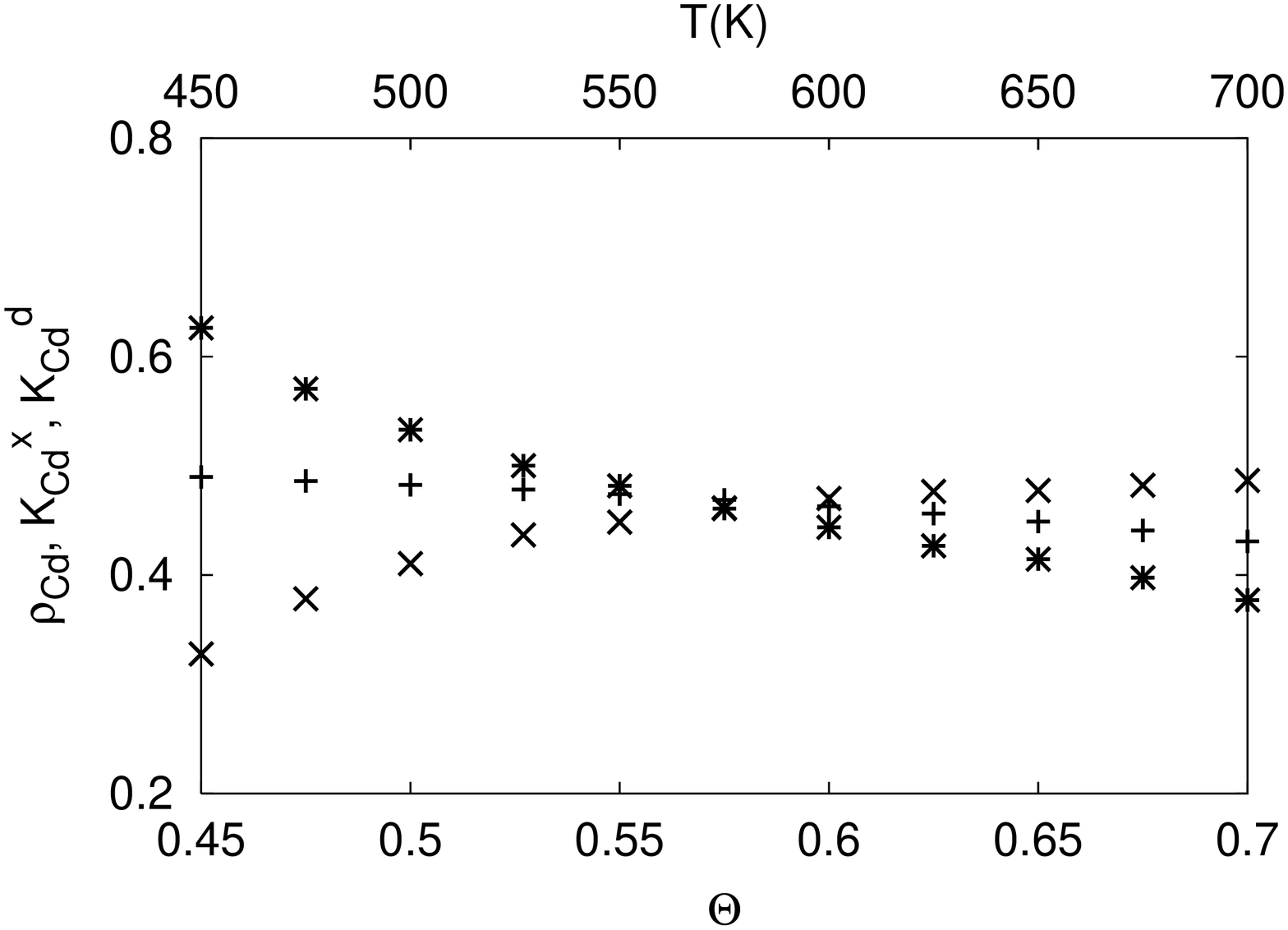}}}}
\put(50,33){\resizebox{0.48\textwidth}{!}{\rotatebox{270}{\includegraphics{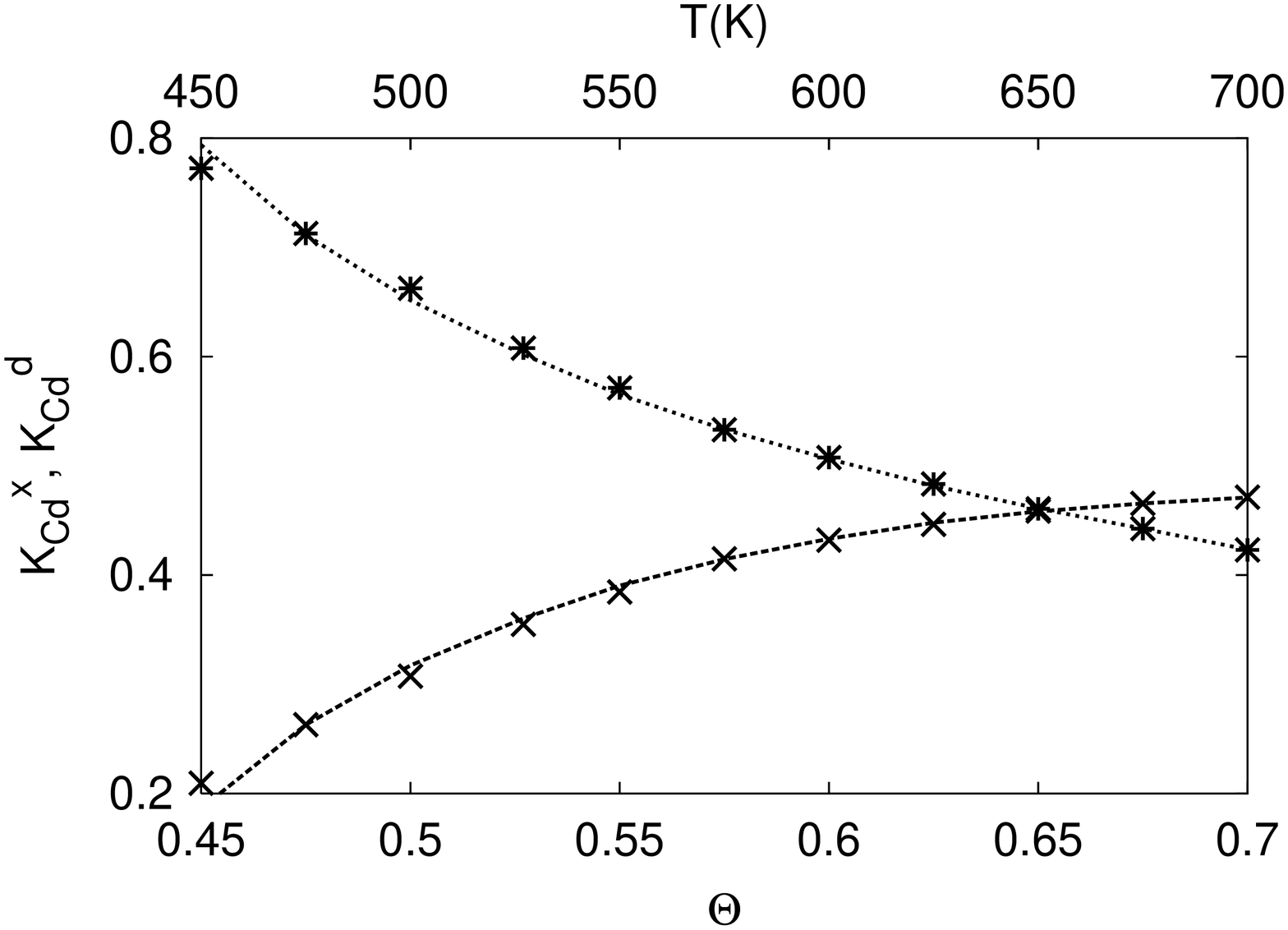}}}}
\put(0,30){(a)}
\put(50,30){(b)}
\end{picture}
\caption{Panel (a): Stationary values of $\rho_{Cd}$ ($+$),
$K_{Cd}^{x}$ ($\times$) and $K_{Cd}^{d}$ ($\plustimes$) in step flow
sublimation. Shown are time averages of these quantities which have
been measured during the sublimation of 5 monolayers. Additionally,
averages over 3 independent simulation runs have been performed. Panel
(b): Equilibrium values of these quantities in the planar lattice gas
in thermal equilibrium. The symbols show results of canonical Monte
Carlo simulations at the values of $\rho_{Cd}$ which are found on
sublimating surfaces. $10^4 \cdot L N$ events have been performed both
for equilibration and for measurement. The lines show results of a
transfer matrix calculation at a strip width $L = 16$. The chemical
potential $\mu$ has been adjusted to obtain the desired value of
$\rho_{Cd}$. In both Monte Carlo simulations, the system size is $L =
N = 64$.
\label{zbstufensub}}
\end{figure}
Figure \ref{zbstufensub}a shows the correlations $K_{Cd}^{x}$,
$K_{Cd}^{d}$ and the Cd coverage $\rho_{Cd}$ as functions of
temperature in the stationary state of step flow sublimation. The
steps are parallel to the $[100]$ direction and have a distance of 16
lattice constants. In the investigated temperature range, $\rho_{Cd}$
is close to 0.5. It decreases slightly from $\rho_{Cd} = 0.49$ at
$\Theta = 0.45$ to $\rho_{Cd} = 0.43$ at $\Theta = 0.7$. At low
temperature, the Cd atoms arrange preferentially in a \cdtxt{} order. At
high temperature, the Cd atoms prefer to arrange in rows which are
characteristic of the \cdtxo{} reconstruction. The crossover between
both regimes where $K_{Cd}^{x} = K_{Cd}^{d}$ is at $\Theta = 0.57$.

We compare the values of $K_{Cd}^{x}$ and $K_{Cd}^{d}$ which are found
on the sublimating surface with those measured in the planar lattice
gas introduced in section \ref{dreizwei} at the same temperature and
Cd coverage. We find, that the nonequilibrium conditions of
sublimation enhance the tendency of Cd atoms to arrange in rows. At
all temperatures, in step flow sublimation we find greater (smaller)
values of $K_{Cd}^{x}$ ($K_{Cd}^{d}$) than on a flat surface in
thermal equilibrium. An inspection of surface snapshots shows, that at
low temperature the \cdtxt{} reconstructed surface is divided into
domains which are elongated in the $y$-direction. Thus, there is no
long range order in the lattice gas of Cd atoms on one terrace. These
observations parallel our findings in the simple cubic model.
Consequently, the ideas about reconstruction order on sublimating
surfaces developed in section \ref{rekkapitel} are valid also in the
more complicated model presented in this chapter.

We set the energy scale of our model by identifying $\Theta = 0.57$
where $K_{Cd}^{x} = K_{Cd}^{d}$ with the experimental value of
$\approx$ 570 K \cite{ct97,ntss00} for the temperature of the
transition between \cdtxt{} and \cdtxo{}.  Then, we have $T = \Theta \cdot
1000 \,\mathrm{K}$ which corresponds to $\varepsilon_d = - 0.0862 \,\mathrm{eV}$.  

Given the parameter set in physical units, we check its physical
relevance by comparing activation energies of sublimation processes
with the corresponding experimental values. A Te terminated surface
decays in a temperature-dependent time interval $\tau$ which
corresponds to a sublimation rate $r_{\mathrm{Te}} = 1/\tau$. The
measurement of sublimation rates in stoichiometrical evaporation is
done straightforward by counting the number of desorbed particles. It
is convenient to plot the logarithm of the sublimation rate versus
$1/T$. In this representation which is denoted as {\em Arrhenius
plot}, the data points lie on a straight line if the dependence of the
sublimation rate on temperature is given by an Arrhenius law.

\begin{figure}[htb]
\botbase{\resizebox{0.48\textwidth}{!}{\rotatebox{270}{\includegraphics{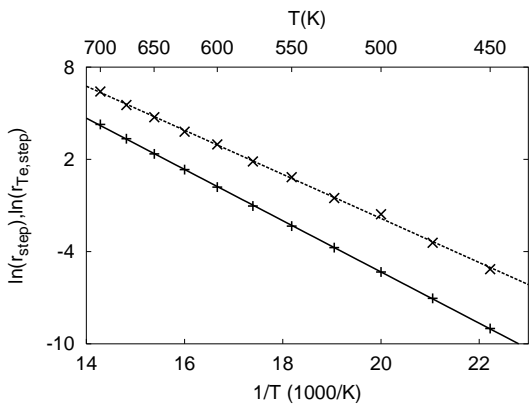}}}}
\hfill
\botbasebox{0.5\textwidth}{\caption{Arrhenius plot of the sublimation
rate in step-flow mode ($+$) and the rate at which the initial
Te terminated surface decays ($\times$). These data have been obtained
in the simulation runs shown in figure \ref{zbstufensub}a. The lines
show fits with an Arrhenius law.\label{zbarrhenius}}}
\end{figure}

Figure \ref{zbarrhenius} shows that this is the case both for
$r_{\mathrm{Te,step}}$ and for the rate $r_{\mathrm{step}}$ at which
CdTe evaporates stoichiometrically in step flow mode. The
corresponding activation energies are $E_{\mathrm{Te,step}} = 1.23
\,\mathrm{eV}$ and $E_{\mathrm{step}} = 1.43
\,\mathrm{eV}$. Experimentally, values of $E_{\mathrm{Te,step}} = 0.96
\,\mathrm{eV}$ \cite{tdbev94} and $E_{\mathrm{step}} = 1.54
\,\mathrm{eV}$ \cite{ntss00} have been found. In layer-by-layer
sublimation, the agreement of the data with an Arrhenius law is as
good as in step flow sublimation.  We find an activation energy
$E_{\mathrm{layer}} = 1.50 \,\mathrm{eV}$ which is smaller than the
experimental value $1.94 \,\mathrm{eV}$ \cite{nskts00}. The activation
energy for the decay of the Te-terminated surface in layer-by-layer
sublimation is $E_{\mathrm{Te,layer}} = 1.35
\,\mathrm{eV}$. Surprisingly the difference between
$E_{\mathrm{step}}$ and $E_{\mathrm{layer}}$ is quite small. From the
quality of the fits of simulation data with an Arrhenius law, we
estimate the accuracy of the values of the activation energies to
about $\pm 0.02 \,\mathrm{eV}$. For the attempt frequencies, we obtain
values on the order of magnitude of $10^{12}/\mathrm{s}$. In general,
the deviation of our values of the activation energies from the
experimental results is less than 30\%. This shows that our model
parameters are at least in a physically reasonable region of the
parameter space.

At present, an optimization of model parameters to obtain better
agreement with experiments does not make much sense. First, the
experimental data are insufficient for a systematical fit of the nine
parameters in our model and there are no values of diffusion barriers
available from ab initio or at least semi-empirical
calculations. Therefore, any parameter set will be partly based on
estimates and physical reasoning. Second, our model still contains
several important simplifications like a neglecting of Te dimerization
and interactions between Te$^*$ atoms. Additionally, we have made a
comparatively simple ansatz for the potential energy
surface. Therefore, a perfect agreement with experiments is not
expected.

\section{Surfaces under an external particle flux \label{zbflux}}

\begin{figure}[htb]
\begin{picture}(100, 33)(0, 0)
\put(0,33){\resizebox{0.48\textwidth}{!}{\rotatebox{270}{\includegraphics{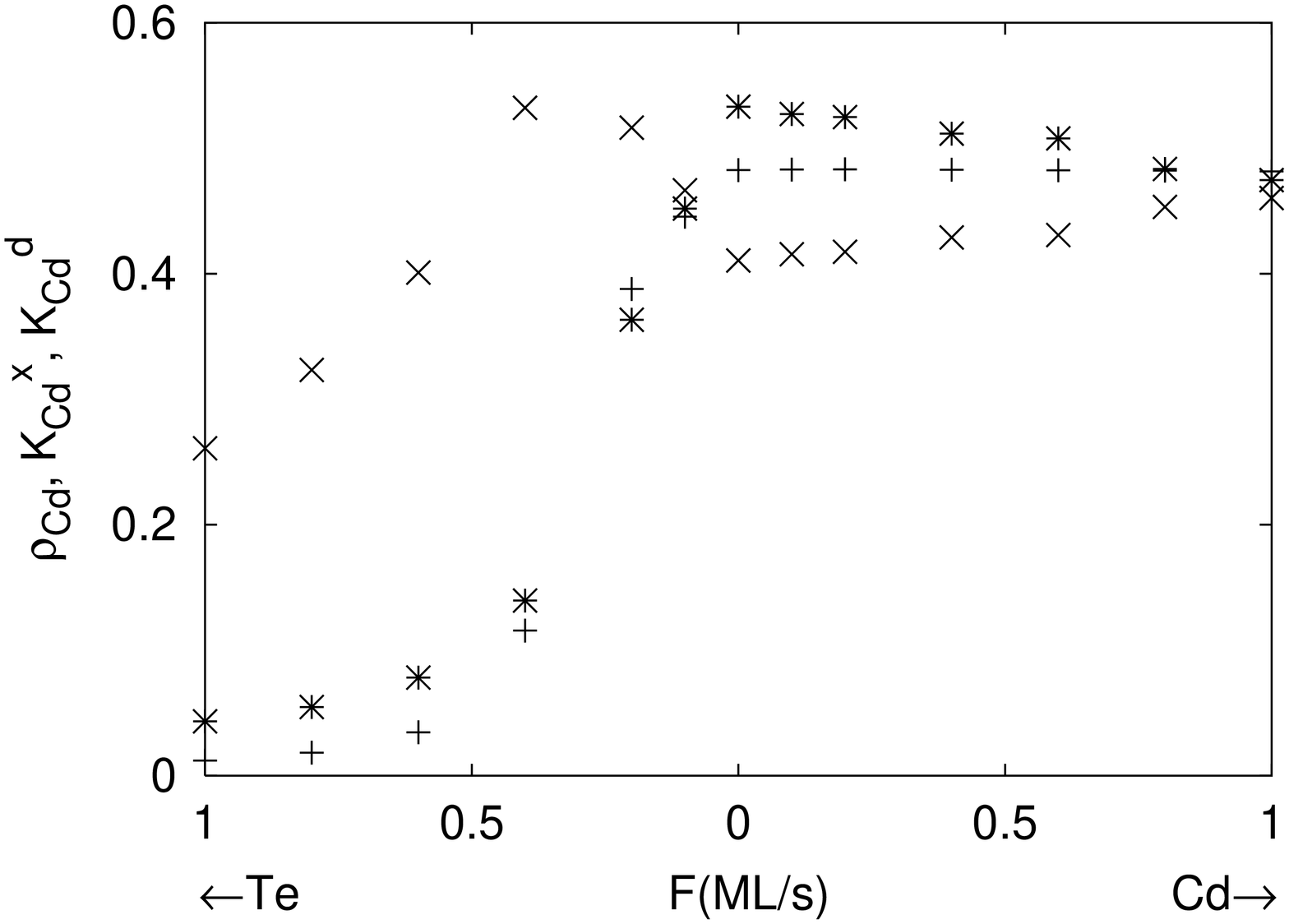}}}}
\put(50,33){\resizebox{0.48\textwidth}{!}{\rotatebox{270}{\includegraphics{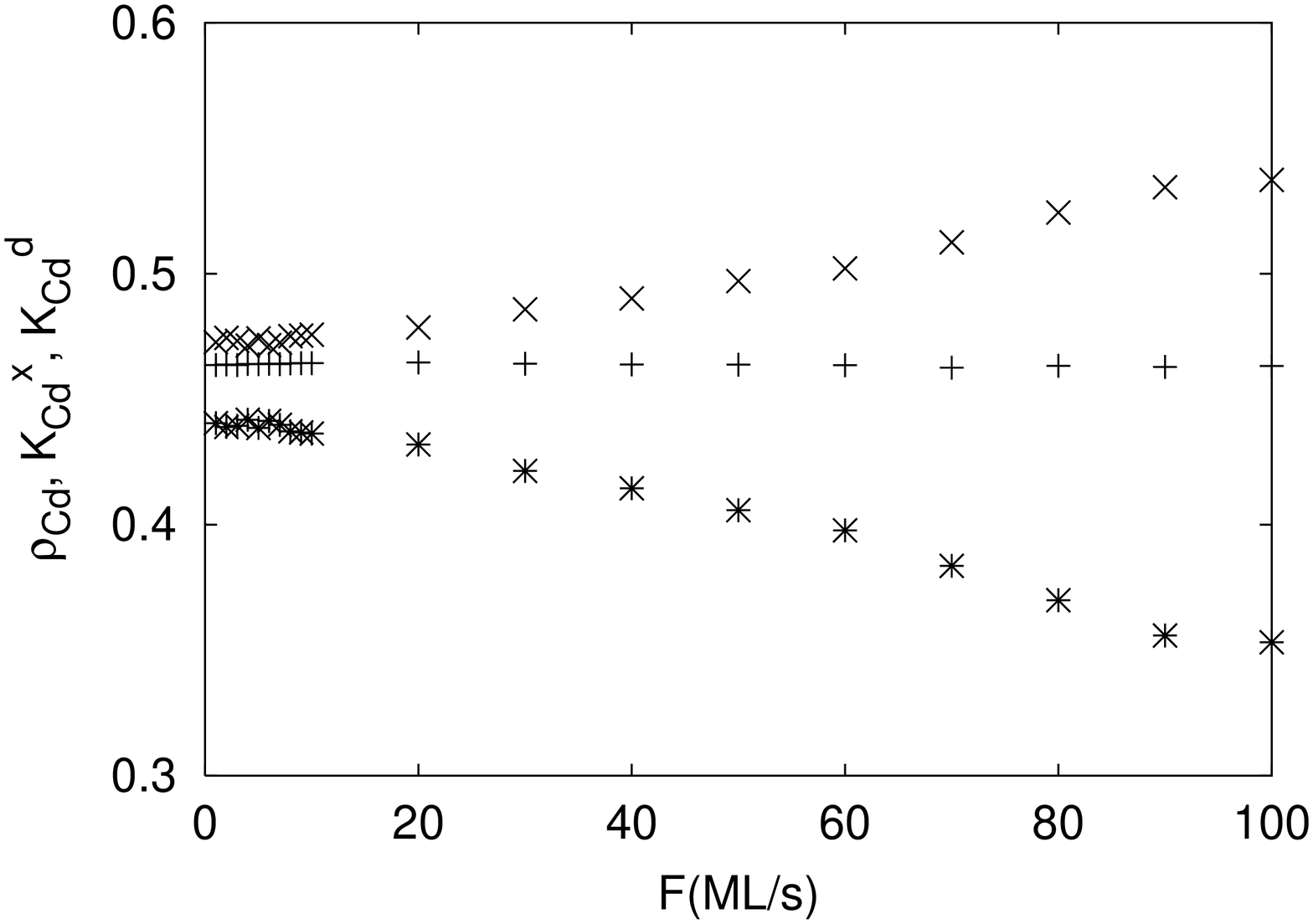}}}}
\put(0,30){(a)}
\put(50,30){(b)}
\end{picture}
\caption{$\rho_{Cd}$ ($+$), $K_{Cd}^{x}$ ($\times$) and $K_{Cd}^{d}$
($\plustimes$) on stepped surfaces with a terrace width of 16 lattice
constants under an external particle flux. The system size is $L = N =
64$. Panel (a): $T = 500 \,\mathrm{K}$. The simulation shows time
averages over $140\,\mathrm{s}$ of simulated time after a relaxation
phase of $140\,\mathrm{s}$. Panel (b): $T = 600 \,\mathrm{K}$. The
surface is exposed to a flux of pure Cd. Time averages have been
performed over $7\,\mathrm{s}$ of simulated time after a relaxation
phase of equal length. \label{zbfluxgraf}}
\end{figure}

Figure \ref{zbfluxgraf}a shows the Cd coverage and the correlations
$K_{Cd}^{x}$, $K_{Cd}^{d}$ on a stepped surface which is exposed to a
flux of pure Cd or pure Te at $T = 500 \,\mathrm{K}$. Under vacuum, the Cd atoms
arrange preferentially in a \cdtxt{} pattern at this temperature. A Te
flux decreases the Cd coverage of the surface. The remaining Cd atoms
arrange preferentially in rows. Thus, the behaviour of our model at
low $\rho_{Cd}$ agrees qualitatively with that of the planar lattice
gas in thermal equilibrium. Of course, this is what we expect in
analogy to the simple cubic model introduced in section
\ref{rekkapitel}.

However, a Cd flux does not increase the tendency of the Cd atoms to
arrange in a checkerboard pattern as it is the case for the \A{} atoms
in the simple cubic model. For Cd fluxes smaller than $\approx 0.6$
ML/s, the correlations $K_{Cd}^{x}$ and $K_{Cd}^{d}$ remain almost
constant. At greater Cd flux, $K_{Cd}^{x}$ increases and $K_{Cd}^{d}$
decreases. A similar behaviour is found at $T = 600 \,\mathrm{K}$ where
$K_{Cd}^{x} > K_{Cd}^{d}$ under vacuum (figure
\ref{zbfluxgraf}b). For Cd fluxes smaller than 10 ML/s, $K_{Cd}^{x}$ and
$K_{Cd}^{d}$ remain close to their vacuum values. A greater Cd flux
leads to an increase of $K_{Cd}^{x}$.

Surprisingly, there is no significant difference in the Cd coverage
between a surface in vacuum and a surface which is exposed to an
external Cd flux. If we determine the equilibrium values of
$K_{Cd}^{x}$ and $K_{Cd}^{d}$ in the two-dimensional lattice gas at
the values of $\rho_{Cd}$ which are measured on a surface under a Cd
flux, we obtain values which are almost identical to those found at
the $\rho_{Cd}$ of a surface in vacuum. This agrees qualitatively with
the result that on a surface of a three-dimensional crystal which is
exposed to a small Cd flux $K_{Cd}^{x}$ and $K_{Cd}^{d}$ are close to
the values which are measured in vacuum. However, the increase of
$K_{Cd}^{x}$ under a large Cd flux cannot be understood within the
framework of a qualitative agreement of our model with the
two-dimensional lattice gas.

On the zinc-blende lattice, Cd atoms can be deposited only at sites
where they are bound to two Te atoms in the layer below. This implies,
that the maximum possible density of mobile Cd atoms on a stepped
surface is smaller than 0.5. The exact value depends on the shape of
the steps. Therefore, the fact that we measure $\rho_{Cd} < 0.5$ does
not imply that there are many sites where an additional Cd atom can be
deposited. If the density of such sites is small, an additional Cd
flux cannot increase $\rho_{Cd}$ further such that there is no reason
for the mobile Cd atoms on the surface to arrange preferentially in a
checkerboard pattern. This might explain why the behaviour of our
model deviates from experiments which show that a Cd flux stabilizes
the \cdtxt{} reconstruction at high temperature.

We expect a different behaviour if the dimerization of Te atoms in the
terminating layer is considered. In a dimer, there is a chemical bond
between two Te atoms which might also be the binding partners of a Cd
atom in the layer above. Consequently, dimers are sites where a Cd
atom from a particle beam can be deposited. The binding energy of a
dimer stabilizes the Te atoms with respect to desorption. Therefore,
we expect that there are dimers on a surface sublimating under vacuum
at high temperature if the binding energy of dimers is large
enough. Such a surface is probably similar to the disordered phase of
the planar lattice gas with Te dimerization introduced in chapter
\ref{dreikapitel} (figure \ref{dreifig2}e). Under a Cd flux, Cd atoms
are incorporated into the dimers which leads to an increase of
$\rho_{Cd}$. This makes the Cd atoms arrange preferentially in a
\cdtxt{} pattern.

\begin{figure}
\begin{center}
\begin{picture}(100,45)(0,0)
\put(5,0){\resizebox{0.45\textwidth}{!}{\includegraphics{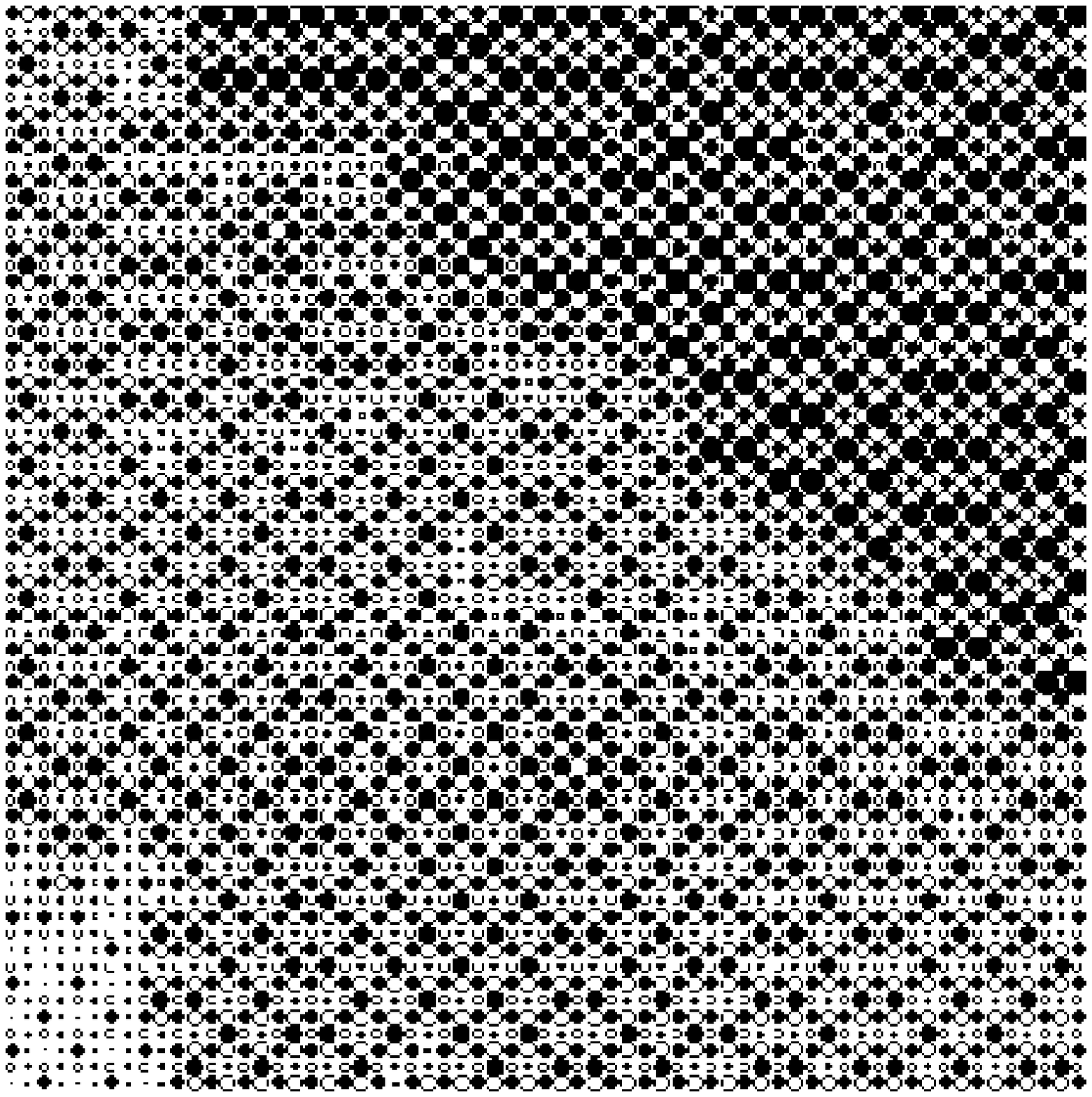}}}
\put(1,43){(a)}
\put(55,0){\resizebox{0.45\textwidth}{!}{\includegraphics{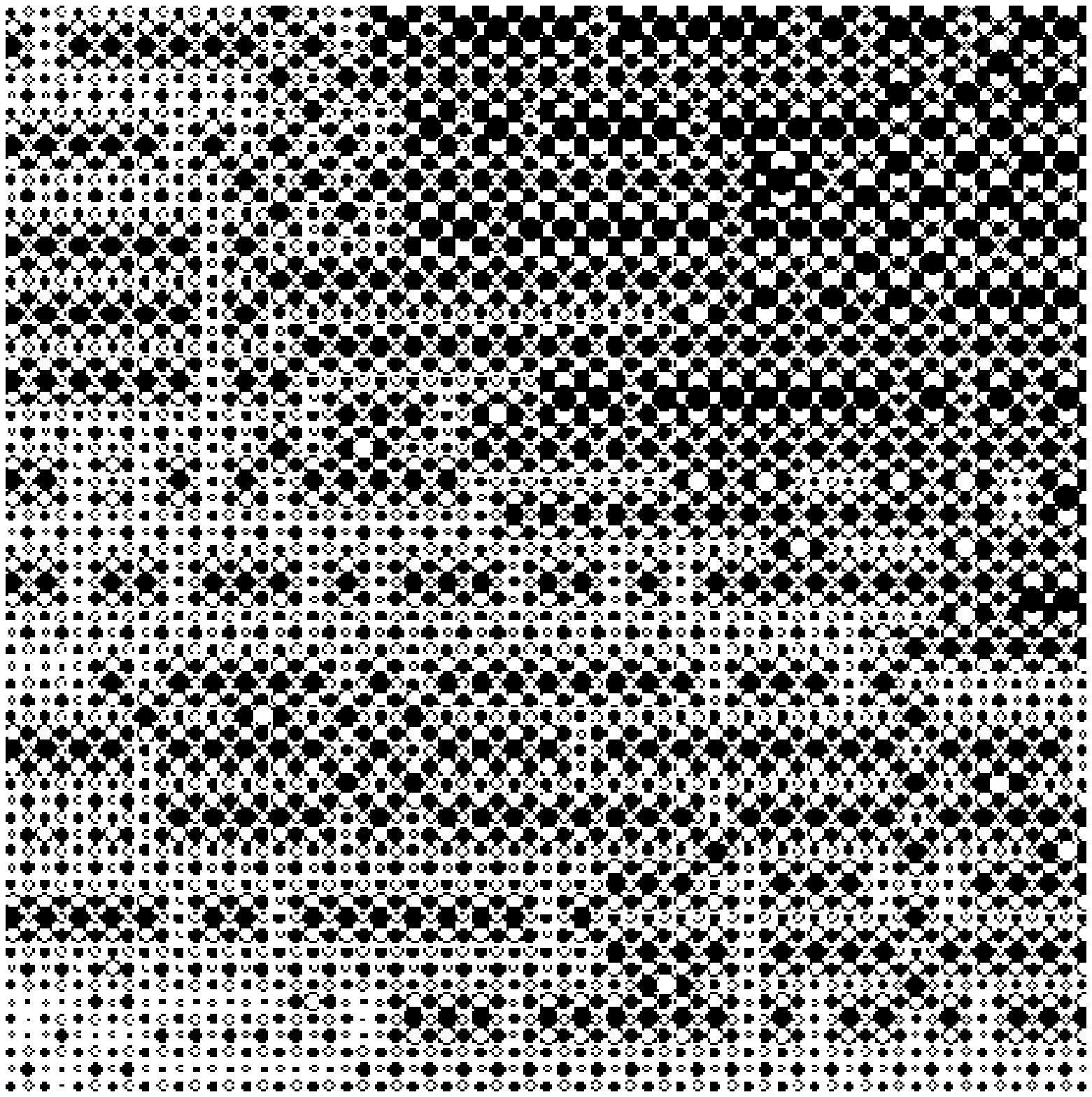}}}
\put(51,43){(b)}
\end{picture}
\end{center}
\caption{Sections of $16\sqrt{2} \times 16\sqrt{2}$ lattice constants
of surfaces which have been exposed to a flux of pure Cd for 280\,s at
$T = 500 \,\mathrm{K}$. Panel (a): $F =
0.6\,\mathrm{ML}/\mathrm{s}$. Panel (b): $F = 10
\,\mathrm{ML}/\mathrm{s}$. Cd atoms are shown as filled circles, Te
atoms as open circles. Greater circles denote greater values of
$h_{i,j}$. \label{zbfluxbild}}
\end{figure}

The morphology of a surface under a small Cd flux is similar to that
of a surface in vacuum. An example is shown in figure
\ref{zbfluxbild}a. 
At large Cd flux, the morphology changes. In figure \ref{zbfluxbild}b,
we show a snapshot of a surface which is exposed to a Cd flux of 10
ML/s at $T = 500 \,\mathrm{K}$. The terraces are not flat. Instead, there is a
network of grooves with a depth of one CdTe monolayer which run across
the surface. The edges of the grooves are $(111)$, $(1\overline{1}1)$,
$(\overline{1}11)$ and $(\overline{1}\overline{1}1)$ facets. Between
the grooves, there are ridges which consist of long rows of Cd atoms
which are parallel to the $x$-direction. These rows yield the
extremely large values of $K_{Cd}^{x} \approx 0.8$ which are measured
on such a surface.  Consequently, the increase of $K_{Cd}^{x}$ on
surfaces under a strong flux of Cd is induced by a facetting of the
crystal surface.

At low temperature, the increase of $K_{Cd}^{x}$ which encompanies the
formation of grooves and ridges occurs at a smaller flux (0.6 ML/s at
$500 \,\mathrm{K}$) than at high temperature (10 ML/s at $600
\,\mathrm{K}$). This indicates that under a small flux facetting is
suppressed by the thermally activated diffusion of particles which
smoothens the surface. A quantitative estimate confirms this
consideration. We assume that the onset of facetting occurs at a fixed
ratio between the Cd flux and the rate at which adatoms diffuse. In
our model, the latter is given by an Arrhenius law with the energy
barrier defined in equation \ref{zbdiffrate}. Attractive interactions
between particles in one layer are small compared to $B_0$. Therefore,
$B_0$ is a good approximation for the typical energy barrier of a
diffusing atom. This yields a ratio of $20$ for the fluxes where
facetting begins at $T = 600 \,\mathrm{K}$ and $T = 500 \,\mathrm{K}$
which agress well with the results of our simulations.

The facetting of surfaces of II-VI compounds under an excess of one
constituent has also been observed in experiments. In \cite{wewr00} it
is reported that the (001) surface of ZnSe is facetting if it is
exposed to a mixed particle beam with a large ratio of Zn to Se. The
critical ratio between Zn and Se increases with temperature.

\section{Atomic layer epitaxy \label{zbalekapitel}}

Atomic layer epitaxy (ALE) is used frequently in the fabrication of
semiconductor heterostructures and has been applied successfully to
determine the stoichiometry of the surface reconstructions of CdTe and
other II-VI semiconductors. Thus, it provides a scenario for the
investigation of our model in a situation present in technical
applications and permits a direct comparison of computer simulations
with experiments. As explained in the introductory section
\ref{aleintro}, the basic idea of ALE is to obtain self-regulated
growth by alternate deposition of pure Cd and Te. In the absence of
reconstructions, one would expect the formation of complete
half-layers of Cd and Te during deposition of the elements such that
in each cycle one CdTe monolayer is grown. However, this is not
necessarily the case in the presence of surface reconstructions where
the number of atoms in the terminating layer of the crystal is
different from that in a layer inside the bulk.

In CdTe experimental investigations of the ALE growth rate $v_{g}$,
i.e.\ the number of CdTe monolayers deposited per cycle have shown
that it decreases in subsequent plateaus as temperature is increased
\cite{ct97,dbt96,fs90}. Within large temperature intervals, $v_{g}$ is
almost constant. At temperatures below $\approx 510 \,\mathrm{K}$,
CdTe grows at a speed of 1\,ML/cycle. In the temperature interval $530
\,\mathrm{K} \leq T \leq 640 \,\mathrm{K}$ growth rates which are
slightly lower than 0.5\,ML/cycle have been found. At $T > 700
\,\mathrm{K}$, ALE growth is not possible at all.

\begin{figure}[htb]
\botbase{\resizebox{0.48\textwidth}{!}{\rotatebox{270}{\includegraphics{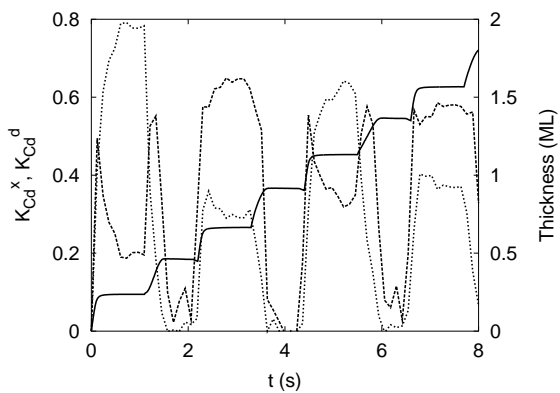}}}}
\hfill \botbasebox{0.5\textwidth}{\caption{Thickness of the deposited
layer (solid), $K_{Cd}^{x}$ (dashed) and $K_{Cd}^{d}$ (dotted) as
functions of time in atomic layer epitaxy at $T = 500
\,\mathrm{K}$. Pulse time and dead time are $t_p = 1\,\mathrm{s}$ and
$t_d = 0.1\,\mathrm{s}$. The particle flux during the pulses is 5
ML/s. The system size is $L = N = 128$. For clarity of plotting, the
curves of $K_{Cd}^{x}$ and $K_{Cd}^{d}$ have been
smoothed.\label{zbalebild}}}
\end{figure}
\begin{figure}
\begin{picture}(100,143)(0,0)
\put(4,98){\botbase{\resizebox{0.45\textwidth}{!}{\includegraphics{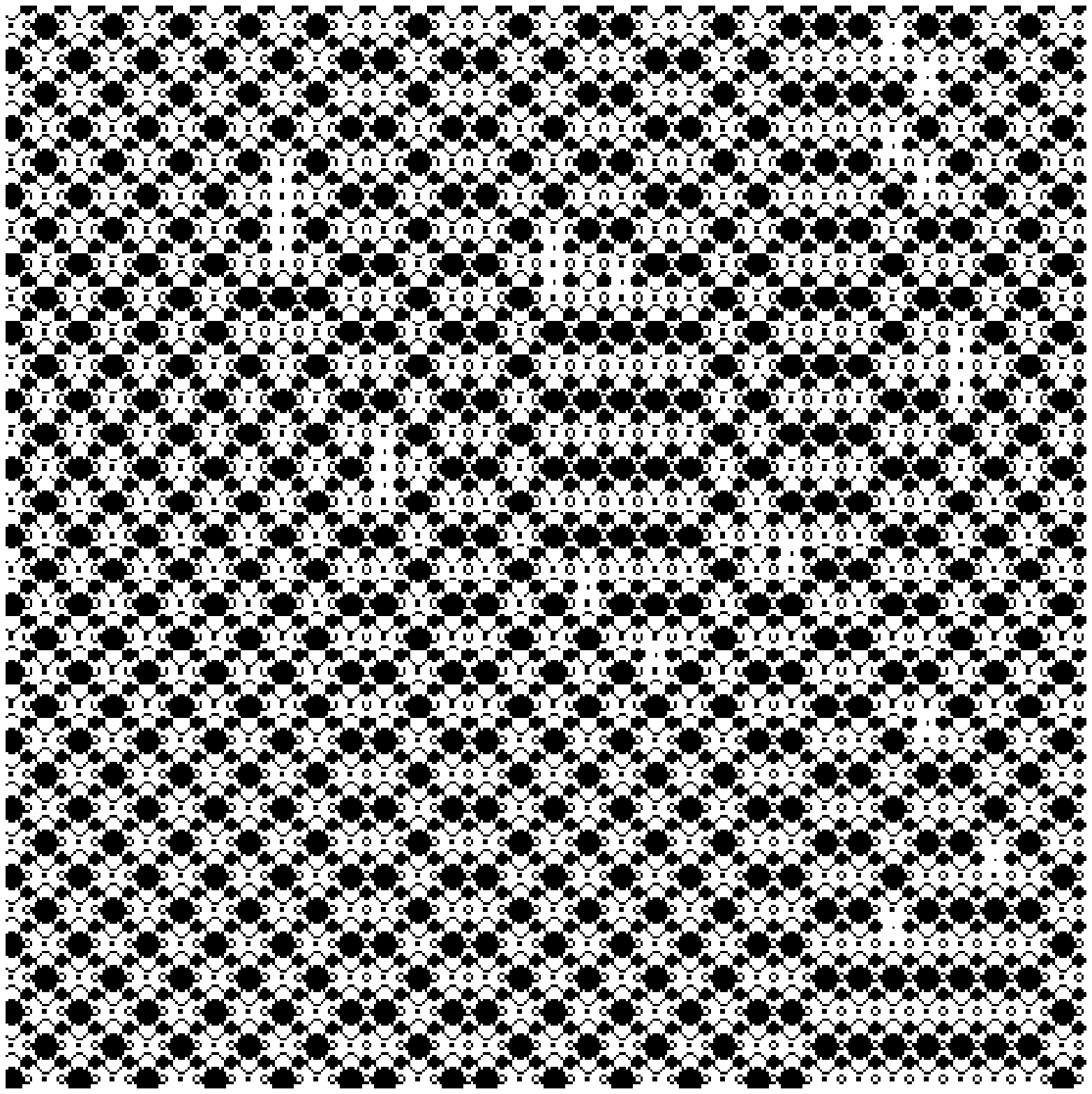}}}}
\put(1,141){(a)}
\put(54,98){\botbase{\resizebox{0.45\textwidth}{!}{\includegraphics{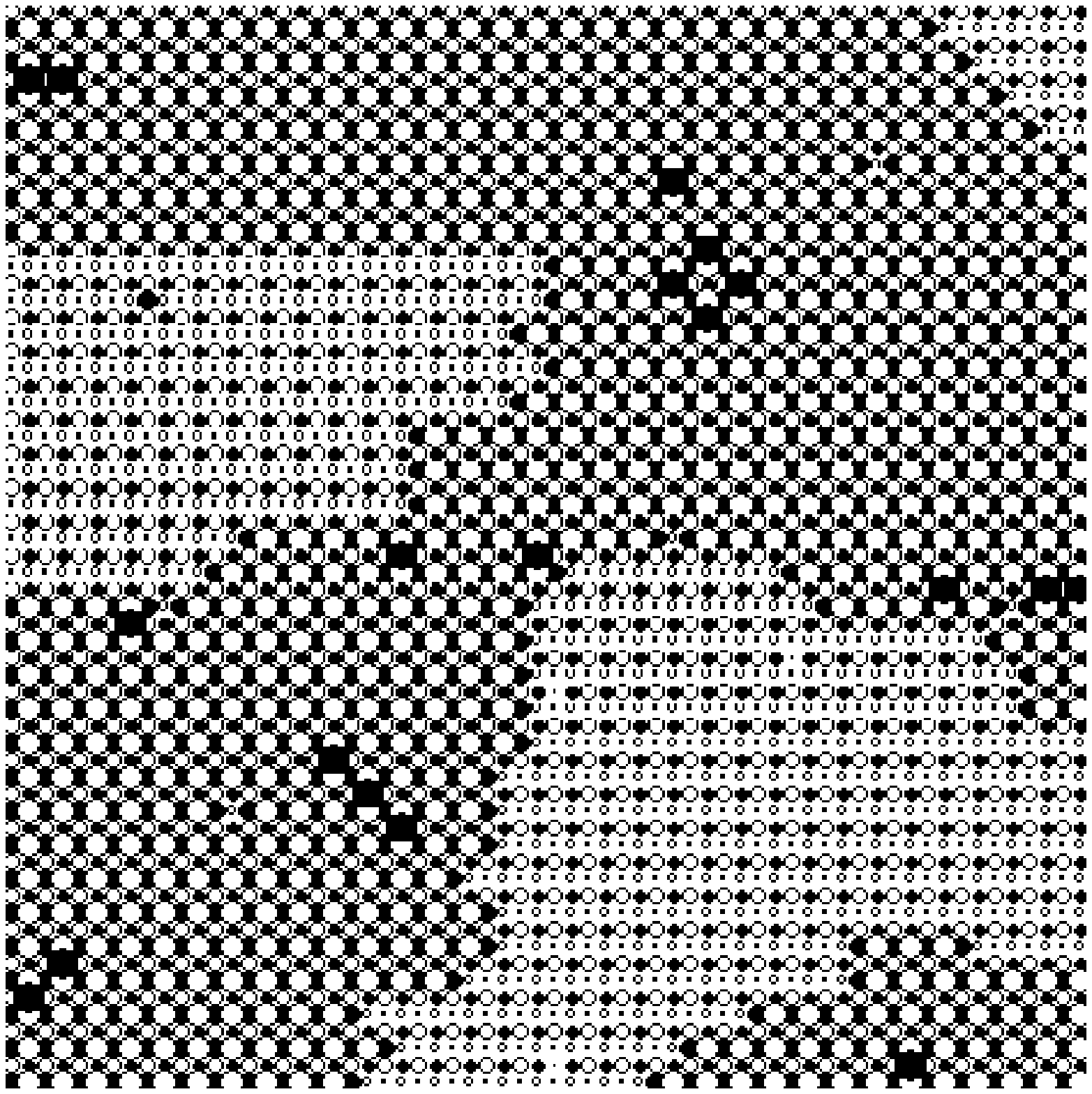}}}}
\put(51,141){(b)}
\put(4,49){\botbase{\resizebox{0.45\textwidth}{!}{\includegraphics{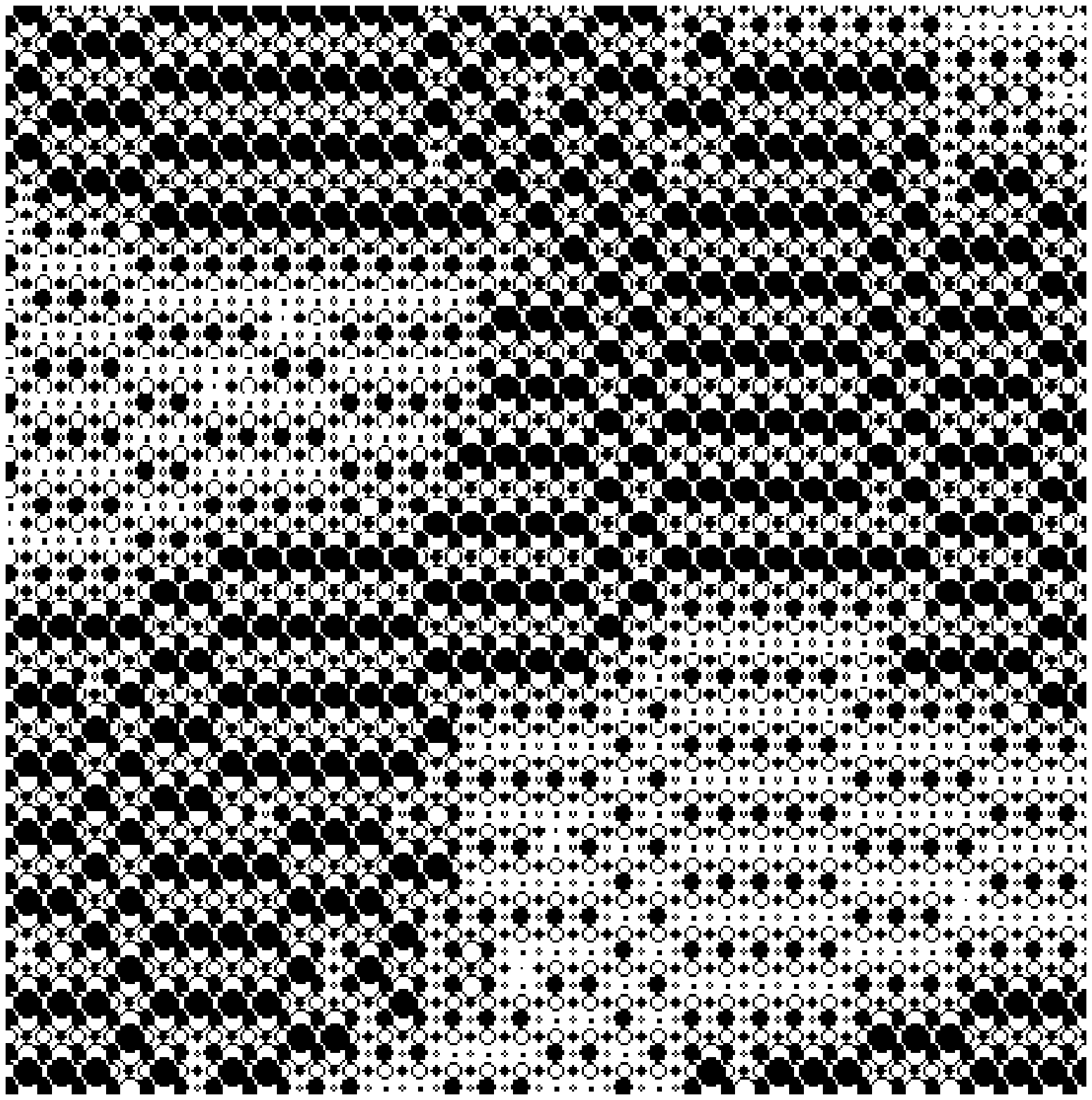}}}}
\put(1,92){(c)}
\put(54,49){\botbase{\resizebox{0.45\textwidth}{!}{\includegraphics{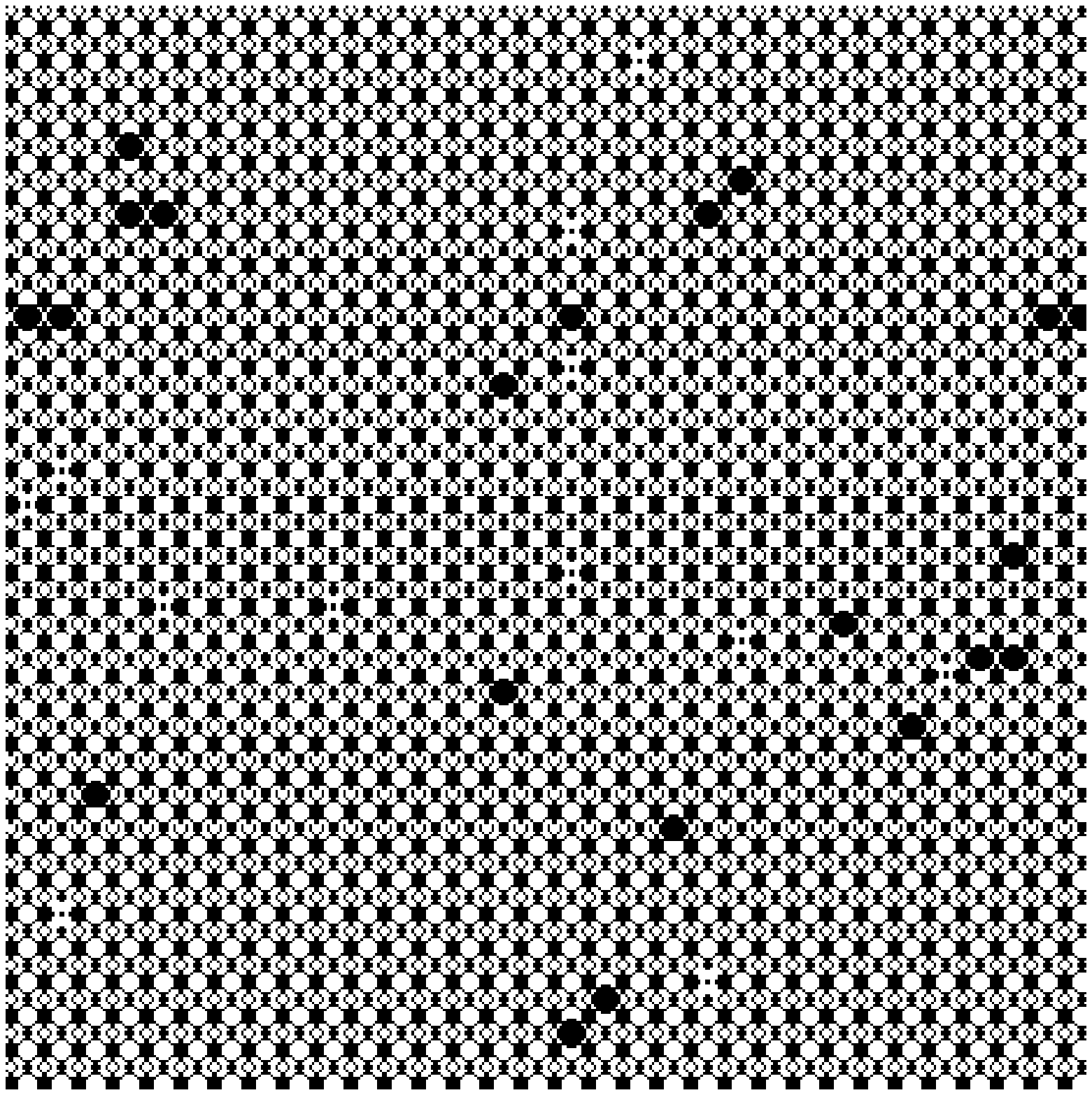}}}}
\put(51,92){(d)}
\put(4,0){\botbase{\resizebox{0.45\textwidth}{!}{\includegraphics{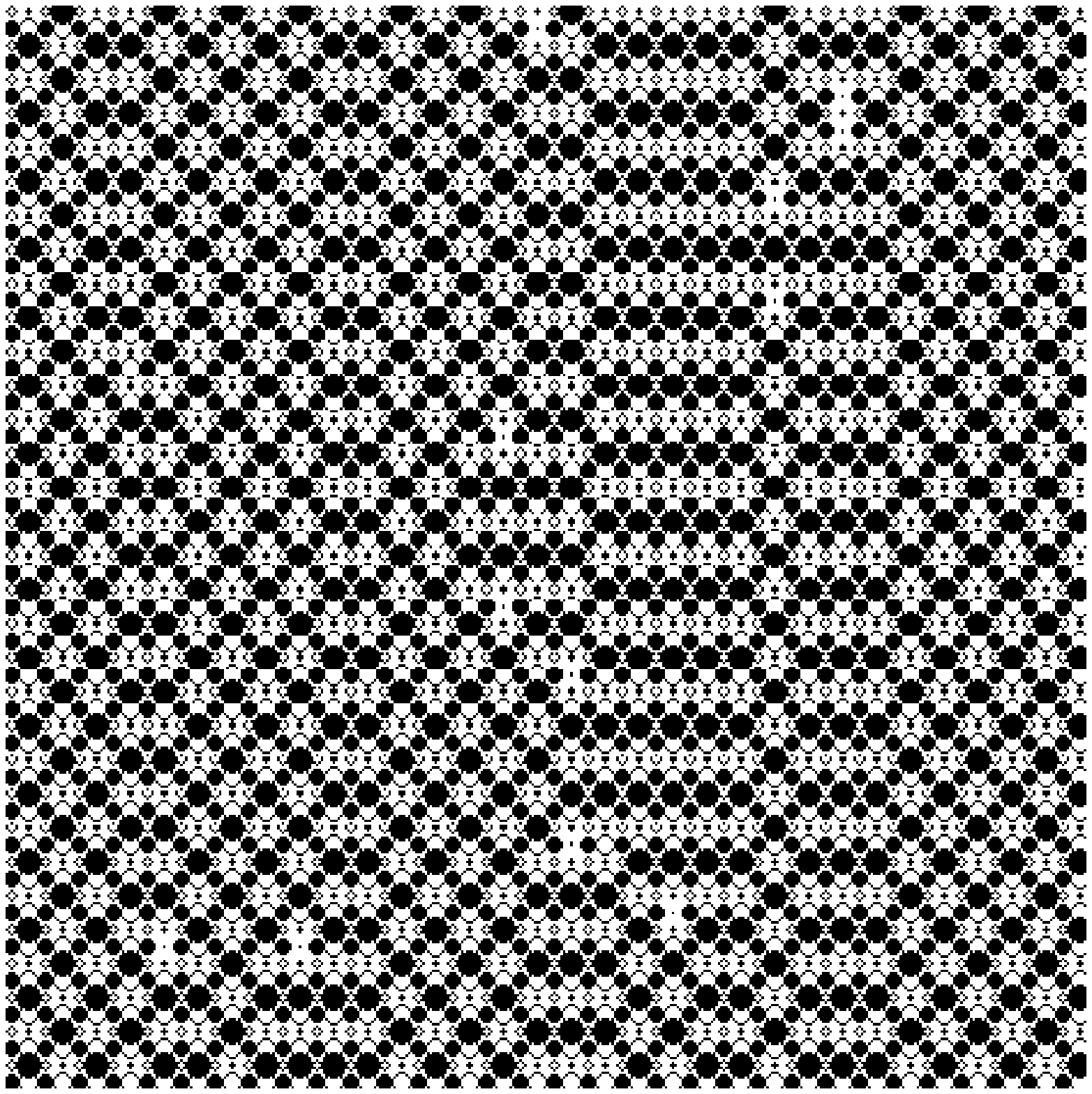}}}}
\put(1,43){(e)}
\put(54,0){\botbasebox{0.45\textwidth}{\caption{Sections of $16 \sqrt{2} \times 16
\sqrt{2}$ lattice constants of the surface during the simulation run
shown in figure \ref{zbalebild}. Panel (a): $t = 1.1 \,\mathrm{s}$. Panel (b): $t
= 2.2 \,\mathrm{s}$. Panel (c): $t = 3.3 \,\mathrm{s}$. Panel (d): $t = 4.4 \,\mathrm{s}$. Panel (e):
$t = 5.5 \,\mathrm{s}$. Cd atoms are shown as filled circles, Te atoms as open
circles. Note the alternation of a flat surface with a surface covered
with islands in subsequent cycles.\label{zbalesnapshot}}}}
\end{picture}
\end{figure}

Since experiments \cite{vadt96} indicate that ALE growth proceeds in
layer-by-layer mode, we focus on simulations of flat surfaces.  Figure
\ref{zbalebild}a shows the evolution of $\rho_{Cd}$, $K_{Cd}^{x}$ and
$K_{Cd}^{d}$ during ALE at $T = 500 \,\mathrm{K}$. The initial
configuration is a flat Te terminated surface. In the first phase, a
Cd flux of 5\,ML/s is applied for a pulse time $t_p = 1 \,\mathrm{s}$,
followed by a dead time $t_d = 0.1 \,\mathrm{s}$. In the second phase,
Te is deposited with the same flux and identical parameters $t_p$ and
$t_d$. This cycle is repeated permanently. Snapshots of the surface
during ALE growth are shown in figure \ref{zbalesnapshot}.

The first Cd pulse creates a Cd terminated surface with $\rho_{Cd}
\approx 1/2$ (figure \ref{zbalesnapshot}a). Since $T$ is below the
\cdtxt{}-\cdtxo{} transition, the dominant reconstruction is
\cdtxt{}. At the onset of the Te phase, $\rho_{Cd}$ decreases
rapidly. This leads to an increase of $K_{Cd}^{x}$ since at low Cd
coverages it is energetically favourable for the remaining Cd atoms to
arrange in rows. At the end of the cycle, approximately 50\% of the
surface are covered with Te terminated islands (figure
\ref{zbalesnapshot}b). The formation of a closed layer is impossible
since only one half of the Cd atoms needed for a complete monolayer
are present on the surface. The following Cd phase deposits half a
monolayer of Cd again (figure \ref{zbalesnapshot}c). However, now the
reconstruction is preferentially \cdtxo{}. This is due to the
influence of the island edges on the reconstruction and analogous to
the maximum of $K_{Cd}^{x}$ we observe in layer-by-layer sublimation
after half a monolayer has desorbed. During the following Te phase, Cd
atoms from the terminating layer of the islands and newly deposited Te
atoms diffuse into the gaps between the islands which leads to the
formation of an almost complete Te terminated surface (figure
\ref{zbalesnapshot}d). In the next Cd phase, a \cdtxt{} reconstructed
flat surface is created and the whole process repeats (figure
\ref{zbalesnapshot}e). Thus, at the end of the odd cycles, the surface
is rough, i.e.\ covered with islands, whereas at the end of the even
cycles the surface is atomically flat. In our simulations the
alternation of a rough and a flat surface can be traced during 3-4
cycles. There are two effects which keep it from repeating
infinitely. First, we obtain a growth rate of 0.45\,ML/cycle which is
smaller than the ideal value of 0.5\,ML/cycle. Second, the diffusion of
particles from the islands into the gaps is not complete. Both effects
hinder a perfect closure of a monolayer at the end of the even cycles
which leads to a damping of the characteristic behaviour.

These observations closely resemble the ideas developed to explain
CdTe growth at a speed of $\approx$ 0.5\,ML/cycle in
\cite{dbt96}. 
\begin{figure}
\botbase{\resizebox{0.48\textwidth}{!}{\rotatebox{270}{\includegraphics{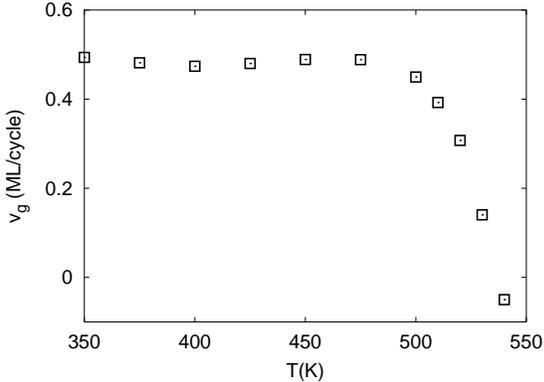}}}}
 \hfill
\botbasebox{0.5\textwidth}{\caption{Temperature dependence of the
growth rate in ALE on a flat surface. $v_g$ has been determined from
the increase of $\left< h_{i,j} \right>$ in cycles 2--4. Apart from
$T$, all parameters are identical to the simulation shown in figure
\ref{zbalebild}.\label{zbalerate}}}
\end{figure}
Figure \ref{zbalerate} shows $v_{g}$ as a function of temperature. For
$350 \,\mathrm{K} \leq T \leq 500 \,\mathrm{K}$, growth proceeds at a
speed of about 0.5\,ML/cycle. If temperature is increased further,
$v_{g}$ decreases quickly. Thus, our model is also capable of
reproducing the experimental observation of plateaus in $v_{g}(T)$.

There are two differences between our computer simulations and
experimental results. First, in experiments the plateau of ALE growth
at 0.5\,ML/cycle extends to considerably higher temperatures than in
our model. Second, our model does not reproduce growth at a speed of
one monolayer per cycle at low temperature.

An inspection of the surface shows, that the breakdown of ALE at high
temperature is mainly due to the quick evaporation of Te atoms during
the Te phase. On the contrary, surfaces under a Cd flux appear to be
comparatively stable. A simple consideration confirms this picture. We
expect a surface under a flux $F$ of the particle species which
terminates the reconstruction to be stable if the sublimation rate
$r_{\mathrm{sub}}$ is smaller than $F$. If sublimation is an Arrhenius
activated process with macroscopic activation energy
$E_{\mathrm{sub}}$ and attempt frequency $\nu_{\mathrm{sub}}$, the
temperature $T_{\mathrm{inst}}$ where $r_{\mathrm{sub}} = F$ is given
by
\begin{equation}
T_{\mathrm{inst}} = - \frac{E_{\mathrm{sub}}}{k \ln
\left(F/\nu_{\mathrm{sub}} \right)}.
\label{zbstabil}
\end{equation} 
Inserting the activation energy and the corresponding prefactor for
the decay of a flat Te-terminated surface, we obtain
$T_{\mathrm{inst}} = 570 \,\mathrm{K}$ which is close to the temperature $540 \,\mathrm{K}$
at which $v_g$ becomes zero. The experimental value of
$T_{\mathrm{inst}} = 700 \,\mathrm{K}$ \cite{fs90} corresponds to an activation
energy of $\approx 1.6 \,\mathrm{eV}$\footnote{Since $T_{\mathrm{inst}}$ depends
on $\ln(F/\nu_{\mathrm{sub}})$, variations of particle flux and
attempt frequency within a physically reasonable range have small
effects.}. Thus, the deviation from our value of $1.35 \,\mathrm{eV}$ is about
20\%. While this deviation might be cured by a tuning of model
parameters, our considerations indicate an inconsistency between
different experiments. Clearly, the fact that ALE is possible at
temperatures above 600 K is incompatible with an activation energy as
low as $0.96 \,\mathrm{eV}$ for the decay of the Te terminated surface which has
been measured in \cite{tdbev94} unless an unphysically low attempt
frequency $\sim 10^7 /\mathrm{s}$ is assumed.

The observation of a growth rate of 1\,ML/cycle has been explained with
the formation of a surface reconstruction with a Te coverage of $3/2$
in the Te phase. Here, one half of the potential Cd sites on the
surface are occupied by Te$^*$ atoms. In the following Cd phase, these
particles are bound to newly deposited Cd atoms where they provide
sites which can be occupied by additional Cd atoms. Thus, in each
cycle a complete CdTe monolayer is deposited in spite of the hard
repulsion between Cd atoms on neighbouring sites in the $y$-direction.

This effect can occur only if Te$^*$ atoms on a Te terminated surface
are bound strong enough such that the rate of desorption of Te$^*$ is
smaller than the particle flux. Since there is no interaction between
Te$^*$ atoms in our model, macroscopical attempt frequency and
activation energy are identical to the microscopical model
parameters. Inserting these in equation \ref{zbstabil}, we find that a
layer of Te$^*$ atoms on the surface is stable at $T < 116 \,\mathrm{K}$ which is
far below ALE growth temperatures. Simulations confirm this
result. Conversely, the experimental ALE data and equation
\ref{zbstabil} yield an activation energy of $1.28 \,\mathrm{eV}$ for the decay
of a surface terminated by Te trimers. This high value indicates that
there are attractive interactions between trimers. On the other hand,
there must be also repulsive interactions which stop the Te coverage
from becoming greater than $3/2$.  Thus, a consistent description of
ALE growth at comparatively low temperature probably requires the
introduction of anisotropic interactions between Te trimers.

For completeness, we have also performed simulations of ALE on a
stepped surface. At a terrace width of 16 lattice constants ALE growth
proceeds in step flow mode. We find, that the growth rate $v_g$ is
identical to that found in ALE on a flat surface. However, there is no
alternation between a rough and a flat surface in subsequent
cycles. The deposited particles are captured by step edges and no
islands are created. Therefore, we find a dominance of a \cdtxt{}
arrangement in all Cd phases.

\section{Summary and discussion}

In summary, we have presented a model of growth and sublimation of
CdTe(001) which might be an important step towards a realistic model
of this surface. We have introduced an efficient representation of the
(001) surface of the zinc-blende lattice. On the one hand, the great
advantage of a solid-on-solid model to allow for a two-dimensional
representation of a three-dimensional crystal is preserved. On the
other hand, it is possible to consider overhangs and atoms in the
wrong sublattice as long as these are not incorporated into the
crystal. The reconstructions of the Cd terminated surface are
considered in a way which is inspired by the ideas developed in
sections \ref{dreikapitel} and \ref{rekkapitel}. Simulations of this
model show, that a physically reasonable parameter set yields a rough
quantitative agreement with experiments.

To the best of our knowledge, we have performed the first simulations
of atomic layer epitaxy. Our results show that it is possible to
investigate the influence of surface reconstructions on this important
growth process by means of Monte Carlo simulations. Additionally, a
comparison of the ALE growth rate as a function of temperature between
simulations and experiments should be useful to fit model parameters.

However, there are still some open questions. (1) What is the role of
Te dimers on a non-equilibrium crystal surface? The lattice gas model
of a flat CdTe(001) surface in thermal equilibrium (chapter
\ref{dreikapitel}) suggests that Te dimerization might be important,
in particular at low $\rho_{Cd}$. As discussed in section
\ref{zbflux}, the dimerization of Te atoms might lower the Cd coverage
of surfaces sublimating under vacuum at high temperature. This should
lead to an increase of $\rho_{Cd}$ and $K_{Cd}^{d}$ and to a decrease
of $K_{Cd}^{x}$ under an external Cd flux. Experiments indicate that
the Cd coverage under vacuum is indeed smaller than the values which
we find in our simulations. In \cite{ntss00} $\rho_{Cd} \approx 0.35$
has been measured at the temperature of the crossover between dominant
\cdtxt{} and dominant \cdtxo{}. As discussed in section
\ref{dreiexpcmp}, this would correspond to a coexistence of a \cdtxt{}
and a \tetxo{} phase on the surface in vacuum. (2) Our model does not
reproduce ALE growth at a speed of one ML/cycle at low temperature. As
discussed in section \ref{zbalekapitel}, an understanding of this
feature requires the consideration of interactions between Te
trimers. Experiments have shown, that trimers arrange in rows parallel
to the $x$-direction yielding a $(2\times 1)$
reconstruction\cite{ct97,tdbev94}. There are no trimers on nearest
neighbour sites in the $y$-direction which can be understood from the
electron counting rule (section \ref{reconstintro}). This suggests a
modelling by means of an attractive interaction between trimers on
neighbouring sites in the $x$-direction and a hard repulsion between
trimers on neighbouring sites in the $y$-direction. However, it is
questionable whether the simulation of such a model is computationally
feasible. Since Te$^*$ atoms are comparatively weakly bound, the
diffusion of Te$^*$ atoms is fast. This leads to a dramatical slowing
down of the simulations if there is a high density of Te$^*$ on the
surface. Itoh et.\ al.\ \cite{ibajjv98,itoh99} faced a similar problem
in the simulation of epitaxial growth of GaAs(001) where there is a
high density of weakly bound As$_2$ molecules. They solved it by
treating all As$_2$ states as one homogeneous particle reservoir
without distinguishing between different arrangements of the
molecules. In our model, a similar treatment of the Te$^*$ atoms
should be possible. The basic idea is to assume that the diffusion of
Te$^*$ atoms is so fast that the probability of a Te$^*$ atom to
interact with a given atom on the solid-on-solid surface does not
depend on its history. Then, it is a reasonable approximation to
assume that there is an equal density of Te$^*$ atoms above each site
on the solid-on-solid surface. Interactions between Te$^*$ atoms might
be treated in a mean field manner by assuming that the binding energy
of a Te$^*$ atom depends on the total number of Te$^*$ atoms in the
reservoir.

\chapter{Singularity spectra and dynamic scale invariance of
kinetically rough surfaces \label{waveletkapitel}}

The past decade has raised much theoretical research in the subject of
{\em kinetic roughening} of surfaces during growth. The investigation
of this effect promises deep insight into statistical physics far from
thermal equilibrium, see e.g.\ \cite{bs95} for an overview. As
discussed in section \ref{kineticroughenintro}, this process creates
surfaces which have fractal properties on lengthscales smaller than
the correlation length $\xi(t)$.

We focus on a solid-on-solid model with a simple cubic lattice
structure and nearest neighbour interactions only. The surface is
represented by a two-dimensional array $h$ of integers $h_{x,y}$ which
denote the height of the column of atoms at $(x, y)$. The rates of
thermally activated processes are given by the Arrhenius law
\ref{arrheniusgesetz}. Energy barriers are determined by the number of
bonds which must be broken to remove the particle from its initial
site. 

Atoms may diffuse to nearest neighbour sites. The rate of such a
process is $\nu \exp(-(\varepsilon_{b} + n \varepsilon_{n})/(k T))$,
where $\varepsilon_{b}$ and $\varepsilon_{n}$ are the binding energies
of a particle to the substrate and to its $n$ nearest neighbours in
the same layer. $\nu$ is the attempt frequency, and $k T$ has its
usual meaning. On each site, new particles are deposited with a rate
$F$. There is no incorporation process. In contrast to earlier
investigations of similar models \cite{dslkg96}, we permit the {\em
desorption} of particles from the surface with rates $\nu
\exp(-(\varepsilon_{v} + n \varepsilon_{n})/(kT))$, where
$\varepsilon_{v} > \varepsilon_{b}$.  Simulations of this model are
performed by means of the continuous time algorithm which is
introduced in appendix \ref{mcappendix}.  Clearly, such a simple model
neglects many effects which are present in real materials. On the
other hand, it captures the essential features of processes which
occur on unreconstructed surfaces in the absence of a Schwoebel
barrier. Therefore, it is adequate to obtain insight into fundamental
properties of the roughening process which are not specific to a
particular material.

Recently, Arn\'{e}odo et.\ al.\ \cite{adr99,adr00,dra00,rad99} have
suggested a method for the investigation of fractal and multifractal
surfaces which uses the continuous wavelet transform. This is superior
to the structure function approach which has to date solely been used
in the analysis of models of epitaxial growth. In this chapter, we use
the wavelet method for a precise characterization of kinetically rough
surfaces. In particular, we investigate the influence of desorption on
the surface morphology. We show that our model fulfils the dynamic
scale invariance which was introduced in section
\ref{introdynamicscale}. We relate the dynamic exponents $\alpha$, $\beta$
with the static scaling properties of the multifractal surface. We
conclude with some remarks on the question of universality of scaling
exponents.

\section{Anomalous scaling and multiscaling}

Originally, it was believed that the scaling properties of kinetically
rough surfaces can be understood within the framework of Family-Vicsek
scaling. Then, the properties of the surface are invariant under the
dynamic scale transformation \ref{dynamicscalingintro}. On
lengthscales smaller than $\xi(t)$ the surface is self-affine. Its
Hurst exponent equals the dynamic exponent $\alpha$. Under these
conditions, the scaling properties of the surface are characterized
completely by the two exponents $\alpha$ and $\beta = \alpha/z$. These
are assumed to be universal, i.e.\ independent of model parameters.

Different methods have been suggested to measure $\alpha$ and $\beta$
in computer simulations.  Frequently, $\beta$ is measured from the
increase of the surface width with time, $W(t) \propto t^\beta$
(equation \ref{Wdynamicscaling}). Since $\alpha$ characterizes both
the static scaling properties of the surface at fixed time and the
dynamic scaling properties, there are two fundamentally different
approaches to determine its value. The {\em static} approach
investigates the height-height correlation function $G(\vec{l}, t)$ at
fixed time $t$. For small $l := |\vec{l}|$, from equation
\ref{selbstaffing} we expect it to increase $\sim l^{2 \alpha}$. The
{\em dynamic}\footnote{In the literature (e.g.\
\cite{l99,lr96,lrc97,rlr00}) the static and the dynamic approach are
frequently denoted as ``local'' and ``global'' approach,
respectively.} approach analyzes the dependence of the {\em surface
width} $W_{\mathrm{sat}}$ in the saturation regime on the system size
$N$: According to equation \ref{wasympdynamic}, we have
$W_{\mathrm{sat}}(N) \sim N^{\alpha}$. An alternative method uses
equation \ref{gdynamikscaling}: $\alpha$ and $z$ are chosen such that
the curves of $G(\vec{l}, t)/l^{2 \alpha}$ versus $l/t^{1/z}$ collapse
on a unique function $u$ within a large range of $t$ and $l$.

However, a careful analysis of simulation data
\cite{bbjkvz92,dslkg96,dsp97,k94} has shown that several models of
epitaxial growth show significant deviations from this simple
picture. First, frequently one obtains significantly different 
exponents from the static and the dynamic approach. This
phenomenon is commonly denoted as {\em anomalous scaling}. Second, a
powerlaw behaviour of $G(\vec{l}, t)$ for small $|\vec{l}|$ does not
necessarily imply that the surface is self-affine. The standard
approach to test for self-affinity considers the height-height
correlation functions of order $q$
\begin{equation}
\Gamma (q, \vec{l}, t) := \left< \left| f(\vec{x}, t) - f(\vec{x} +
\vec{l},t) \right|^{q} \right>_{\vec{x}},   
\label{wavegg}
\end{equation} 
which are generalizations of $G(\vec{l}, t) \equiv \Gamma(2, \vec{l},
t)$. Here, $f(\vec{x}, t)$ is the reduced surface height $h(\vec{x},t)
- \left< h(\vec{x}, t) \right>$.  On a self-affine surface we expect
$\Gamma(q, \vec{l}, t) \sim l^{\gamma_q q}$ for small $l$, where
$\gamma_q = \alpha$ for all $q$. However, sometimes one encounters the
phenomenon of {\em multiscaling}: the initial powerlaw behaviour of
$\Gamma(q, \vec{l}, t)$ yields a hierarchy of $q$-dependent exponents
$\gamma_q$.

These observations can be interpreted within the mathematical
framework of {\em multifractality}: The {\em H\"{o}lder exponent}
\cite{adr99,bma92,bs95,mba93} $\chi(\vec{x_{0}})$ of a function $f$ at
$\vec{x}_{0}$ is defined as the largest exponent such that there
exist a polynomial of order $n < \chi(\vec{x}_{0})$ and a constant $C$
which yield $|f(\vec{x}) - P_{n}(\vec{x} - \vec{x}_{0})| \leq C
|\vec{x} - \vec{x}_{0}|^{\chi(\vec{x}_{0})}$ in the neighbourhood of
$\vec{x}_{0}$.
The H\"{o}lder exponent characterizes {\em local} symmetry properties
of the surface. In the vicinity of $\vec{x}_0$, we have
$$ f(\vec{x}_0 + b \vec{l}, t) - f(\vec{x}_0, t) \sim 
b^{\chi(\vec{x}_0)} \left[ f(\vec{x}_0 + \vec{l}, t) - f(\vec{x}_0, t)
\right]. $$
Thus, the H\"{o}lder exponent is a local counterpart of the Hurst 
exponent: a self-affine function with Hurst exponent $H$ has 
$\chi(\vec{x}) = H$ everywhere. 
However, in the case of a {\em multiaffine} function different points
$\vec{x}$ might be characterized by different H\"{o}lder exponents.
In this case, the surface consists of interwoven fractal sets of
points where there is one particular H\"{o}lder exponent. 
This general case is characterized by the {\em singularity spectrum}
$D(\chi)$, which denotes the Hausdorff dimension of the set of points
where $\chi$ is the H\"{o}lder exponent of $f$.

\section{The wavelet approach to multifractality \label{waveletdef}}

There is a deep analogy between multifractality and thermodynamics
\cite{bs95,s90,v92}, where the H\"{o}lder exponents play the role of
energy, the singularity spectrum corresponds to entropy, and $q$ plays
the role of inverse temperature.  So, theoretically $D(\chi)$ might be
calculated via a Legendre transform of $\gamma_q$: $D(\chi) =
\mbox{min}_{q} (q \chi - q \gamma_q + 2) $ \cite{adr99,bma92,mba93}, a
method which has been called {\em structure function} (SF)
approach. However, its practical application raises fundamental
difficulties: First, to obtain the complete singularity spectrum, one
needs $\gamma_q$ for positive {\em and} negative $q$. But as
$|f(\vec{x}, t) - f(\vec{x} + \vec{l}, t)|$ might become zero,
$\Gamma(q, \vec{x}, t)$ is in principle undefined for $q <
0$. Therefore, only the left, ascending part of $D(\chi)$ is accessible
to this method.  Additionally, the results of the SF method can easily
be corrupted by polynomial trends in $f(\vec{x})$ \cite{bma92}.  It
might be due to these difficulties, that---to our knowledge---no
attempt to determine the singularity spectrum of {\em growing
surfaces} from $\gamma_q$ has ever been made. On the basis of
simulations of several simple growth models, it has been argued that
the $\gamma_q$ might collapse onto a single $\gamma$ in the limit $t
\rightarrow \infty $, which characterizes the asymptotic universality
class of the model \cite{dslkg96,dsp97,dkdds97}. This would imply that
multiscaling is a finite size effect. However, there is no physical
reason why this should be the case. A precise measurement of $D(\chi)$
will help to test this hypothesis. 

To this end, we follow the strategy suggested by Arn\'{e}odo et.\ al.\
\cite{adr99,bma92,wave}, which circumvents the problems of the SF
approach and permits a reliable measurement of the complete
singularity spectrum.  Mathematically, the wavelet transform of a
function $f(\vec{x})$ of two variables is defined as its convolution
with the complex conjugate of the wavelet $\psi(\vec{x})$ which is {\em
dilated} with the {\em scale} $a$ and rotated by an angle $\theta$
\cite{wave}:
\begin{equation}
T_{\psi}[f] (\vec{b}, \theta, a) = C_{\psi}^{-1/2} a^{-2} \int d^{2}x
\; \psi^{*}(a^{-1} \mathbf{R_{-\theta}} (\vec{x} - \vec{b}) ) f(\vec{x}). 
\label{wavewgen}
\end{equation} 
Here $\mathbf{R_{\theta}}$ is the usual 2-dimensional rotation matrix,
and $C_{\psi} = (2 \pi)^{2} \int d^{2}k |\vec{k}|^{-2}
|\hat{\psi}(\vec{k})|^{2}$ is a normalization constant, whose
existence requires square integrability of the wavelet $\psi(\vec{x})$
in fourier space. Apart from this constraint, the wavelet can (in
principle) be an arbitrary complex-valued function.  Introducing the
wavelet $\psi_{\delta}(\vec{x}) = \delta(\vec{x}) - \delta(\vec{x} +
\vec{n})$, where $\vec{n}$ is an arbitrary unit vector, one obtains
easily
\begin{eqnarray}
T_{\psi_{\delta}}[f](\vec{b}, \theta, a) &=& C_{\psi_{\delta}}^{-1/2}
\left[ f(\vec{b}) - f(\vec{b} + a \mathbf{R_{\theta}} \vec{n}) \right]
\nonumber \\
& \Rightarrow & \int d^{2}b \left| T_{\psi_{\delta}}[f](\vec{b},
\theta, a)\right|^{q} \propto \Gamma(q, a \mathbf{R_{\theta}} \vec{n}).
\label{wavetg}
\end{eqnarray} 
Consequently, a calculation of the moments of the wavelet transform of
the surface yields the SF approach as a special case. To avoid its
weaknesses, two major improvements are necessary.

First, we use a class of wavelets with a greater number of {\em
vanishing moments} $n_{\vec{\Psi}}$ than
$\psi_{\delta}(\vec{x})$. Then, the wavelet is orthogonal to any
polynomial of degree less than $n_{\vec{\Psi}}$. This increases the
range of accessible H\"older exponents and improves the insensitivity
to polynomial trends in $f(\vec{x})$. We introduce a two-component
version of the wavelet transform
\begin{equation}
\vec{T}_{\vec{\Psi}}[f] (\vec{b}, a) = \frac{1}{a^{2}} \int d^{2}x \left(
\begin{array}{c}
\Psi_{1}(a^{-1} (\vec{x} - \vec{b})) \\
\Psi_{2}(a^{-1} (\vec{x} - \vec{b}))
\end{array}
\right)
f(\vec{x}) \; , 
\label{wavewspec}
\end{equation}
where the components $\Psi_1$, $\Psi_2$ of the analyzing wavelet
$\vec{\Psi}$ are defined as partial derivatives of a radially
symmetrical convolution function $\Phi(\vec{x})$: $\Psi_{1}(\vec{x}) =
\partial \Phi / \partial x$, $\Psi_{2}(\vec{x}) = \partial \Phi /
\partial y$. Then $\vec{T}_{\vec{\Psi}}[f](\vec{b}, a)$ can be written
as the gradient of $f(\vec{x})$, smoothed with a filter $\Phi$ with
respect to $\vec{b}$.  This definition becomes a special case of
equation \ref{wavewgen}, when multiplied with $\vec{n}_{\theta} =
(\cos(\theta), \sin(\theta))^{\top}$, yet allows for an easier
numerical computation\footnote{For simplicity, the irrelevant constant
$C_{\Psi}$ has been omitted.}.  For example, $\Phi$ can be a gaussian,
where $n_{\vec{\Psi}} = 1$, or $\Phi_{1}(\vec{x}) = (2 - \vec{x}^{2})
\exp(-\vec{x}^{2}/2)$ which has two vanishing moments.

Second, the integration over $\vec{b}$ in equation \ref{wavetg} is
undefined for $q < 0$, since the wavelet coefficients might become
zero. The basic idea is to replace it with a discrete summation over
an appropriate partition of the wavelet transform which takes on 
nonzero values only, but preserves the relevant information on the
H\"older regularity of $f(\vec{x})$. In the following, we give a
brief outline of the rather involved algorithm and refer the reader to
\cite{adr99,bma92,wave} for more details and a mathematical proof.

The {\em wavelet transform modulus maxima} (WTMM) are defined as local
maxima of the modulus $M_{\vec{\Psi}}[f](\vec{b}, a) :=
|\vec{T}_{\vec{\Psi}}[f](\vec{b}, a)|$ in the direction of
$\vec{T}_{\vec{\Psi}}[f](\vec{b}, a)$ for fixed $a$. 
If $\Phi(\vec{x})$ is the gaussian function, the computation
of their positions is equivalent to the detection of edges in image
processing \cite{mh92,mz92}.  
Similarly, for arbitrary wavelets the WTMM lie on
connected curves which trace structures of size $\sim a$ on the
surface. The strength of each curve is characterized by the {\em maximal}
value of $M_{\vec{\Psi}}[f](\vec{b}, a)$ along the line, the so-called
{\em wavelet transform modulus maxima maximum} (WTMMM) \cite{adr99}.
While proceeding from large to small $a$, successively smaller
structures are resolved. Connecting the WTMMM at different scales
yields the set $\mathcal{L}$ of maxima lines $l$, which lead to the
locations of the singularities of $f(\vec{x})$ in the limit $a
\rightarrow 0$.  The partition functions
\begin{equation}
Z(q, a) = \sum_{l \in {\mathcal L}(a)} \left(
\sup_{(\vec{b}, a') \in l, 
a' \leq  a} M_{\vec{\Psi}}[f](\vec{b}, a')
\right)^{q} \sim a^{\tau(q)} \quad \mbox{for} \quad a \rightarrow 0
\label{wavezsum}
\end{equation}
are defined on the subset $\mathcal{L}(a)$ of lines which cross the
scale $a$.  From the analogy between the multifractal formalism and
thermodynamics, $D(\chi)$ is calculated via a Legendre transform of the
exponents $\tau(q)$ which characterize the scaling behaviour of $Z(q,
a)$ on small scales $a$: $D(\chi) = \mbox{min}_{q} (q \chi - \tau(q))$.
Additionally, $\tau(q)$ itself has a physical meaning for some $q$:
$-\tau(0)$ is the fractal dimension of the set of points where
$h(\vec{x}) < \infty$, while the fractal dimension of the surface
$f(\vec{x})$ itself equals $\mbox{max} (2, 1 - \tau(1))$.

\section{Results}

In our simulations, we choose the parameters $\nu = 10^{12}\,\mathrm{s}$, 
$\varepsilon_{b} = 0.9\,\mathrm{eV}$, 
$\varepsilon_{n} = 0.25\,\mathrm{eV}$, and a temperature 
$T = 450\,\mathrm{K}$.  To
study the influence of desorption, we consider three models with
different activation energies $\varepsilon_{v}$: in model A desorption
is forbidden, i.e.\ $\varepsilon_{v} = \infty$. Models B and C have
$\varepsilon_{v} = 1.1 \; \,\mathrm{eV}$ and $\varepsilon_{v} = 1.0 \;
\,\mathrm{eV}$.  We simulate the deposition of $2 \cdot 10^{4}$
monolayers at a growth rate of one monolayer per second on a lattice
of $N \times N$ unit cells using periodic boundary conditions, our
standard value being $N = 512$.
\begin{figure}
\figpanel{\resizebox{0.3\textwidth}{!}{\includegraphics{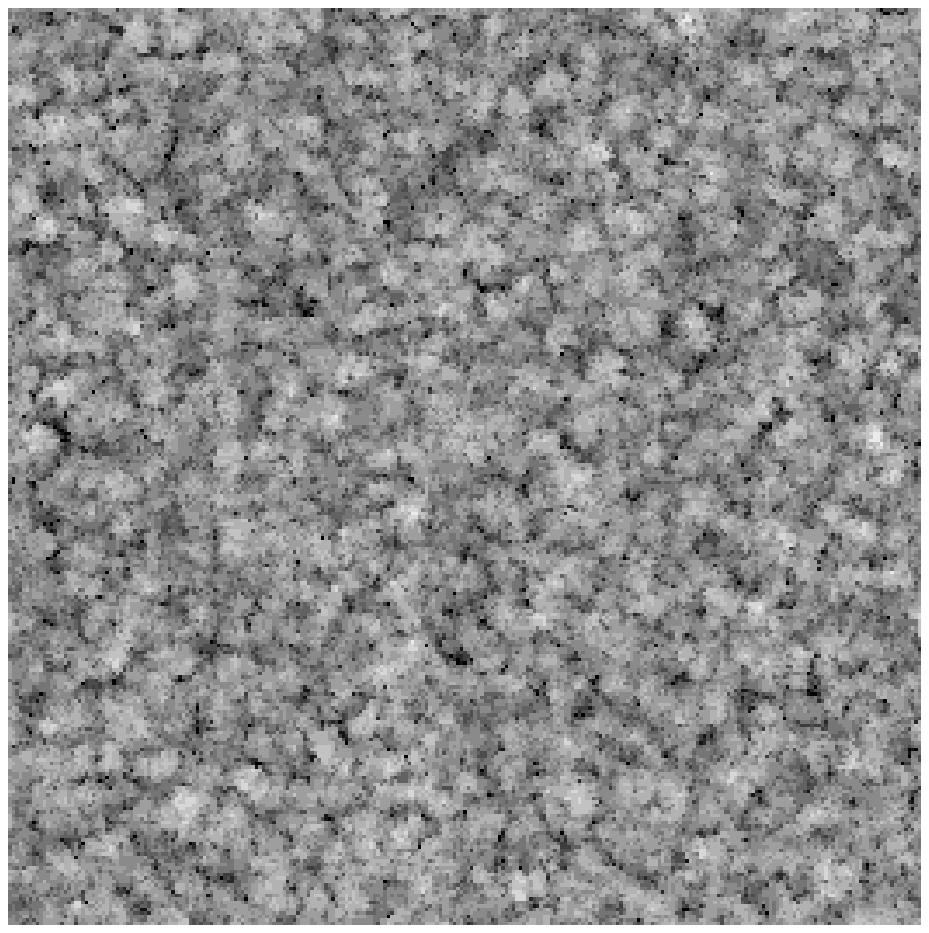}}}{$t
= 200\,\mathrm{s}$} \hfill
\figpanel{\resizebox{0.3\textwidth}{!}{\includegraphics{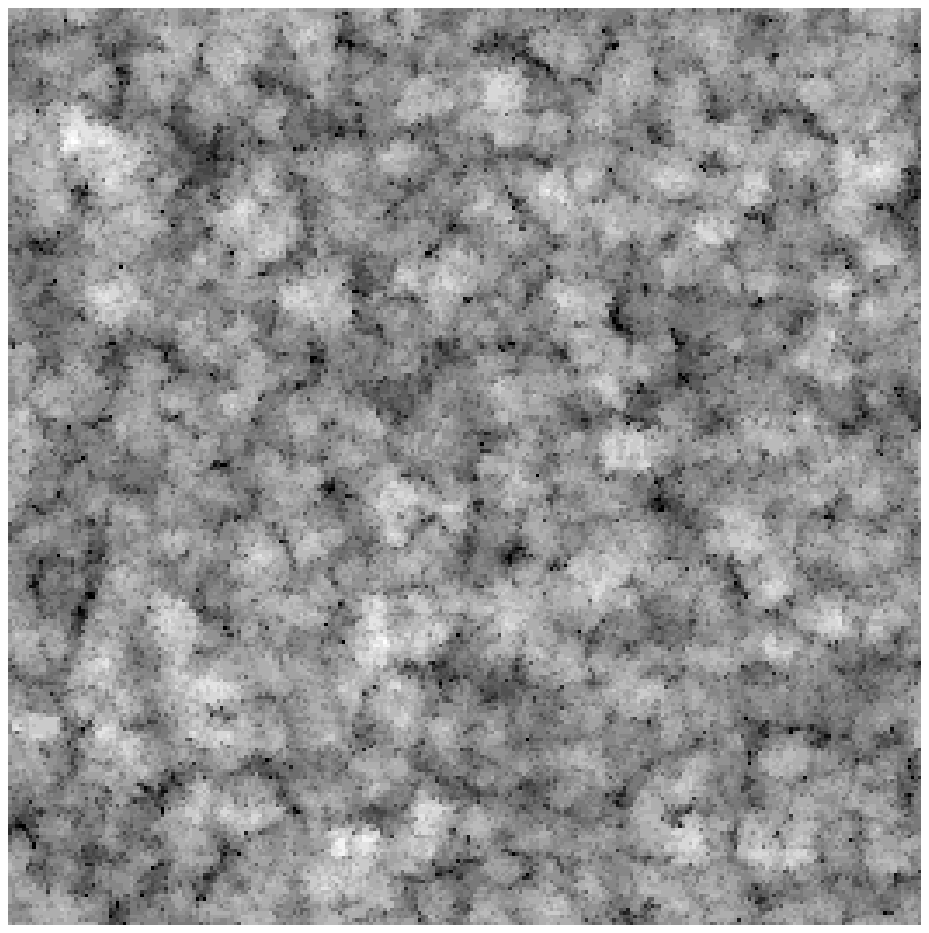}}}{$t
= 2000\,\mathrm{s}$} \hfill
\figpanel{\resizebox{0.3\textwidth}{!}{\includegraphics{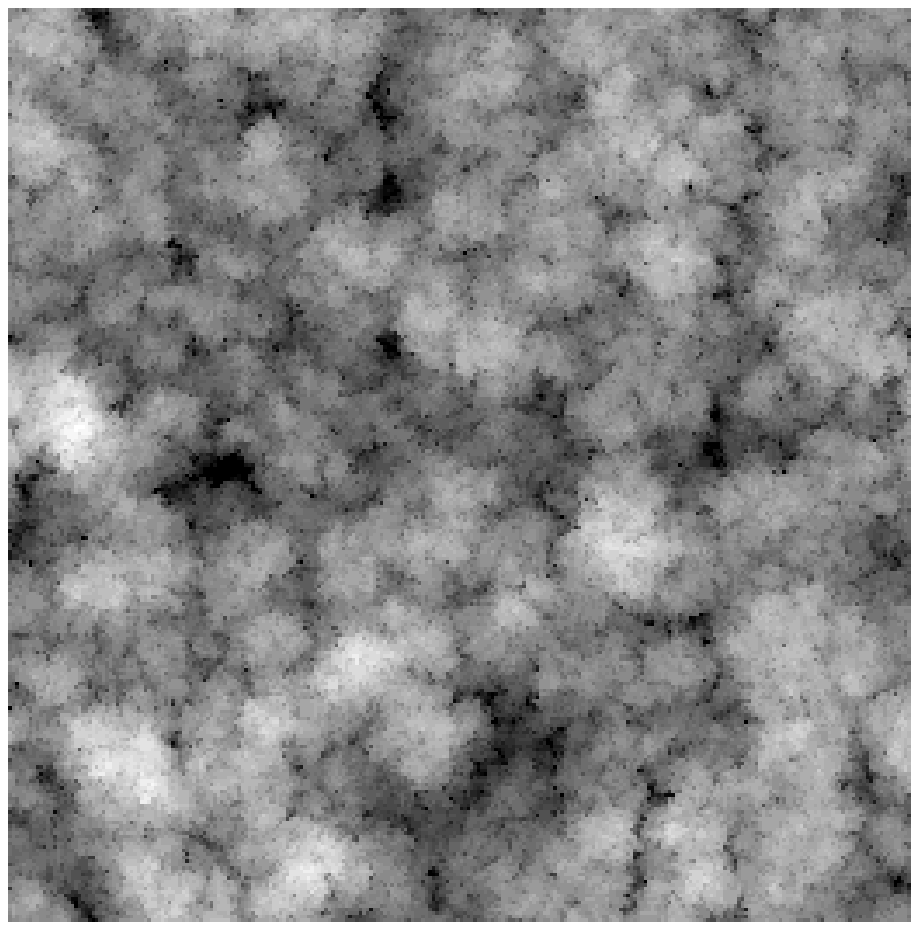}}}{$t
= 20000\,\mathrm{s}$}
\caption{Evolution of the surface in a simulation run of model
A. Brighter shades of grey denote greater surface
heights. \label{wavesnapshot}}
\end{figure}
To check for finite size effects, we have also simulated $N = 256$.
In all presented results averages over 6 independent simulation runs
have been performed. Snaphots of the surface in a
simulation of model A are shown in figure
\ref{wavesnapshot}. Visually, surfaces of models B and C look quite
similar.

\subsection{Singularity spectra}

\begin{figure}
\begin{picture}(100, 33)(0, 0)
\put(0,33){\resizebox{0.48\textwidth}{!}{\rotatebox{270}{\includegraphics{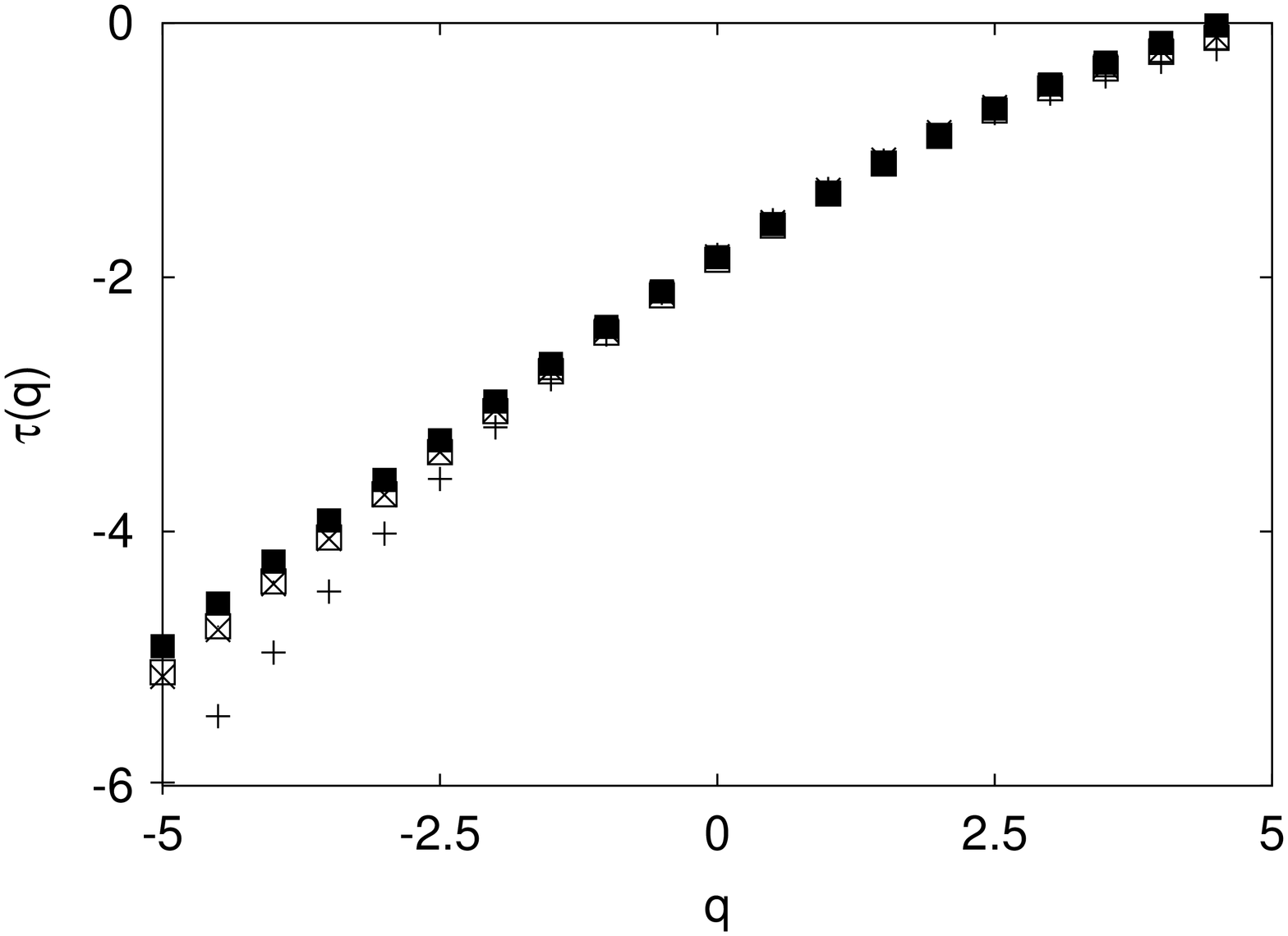}}}}
\put(50,33){\resizebox{0.48\textwidth}{!}{\rotatebox{270}{\includegraphics{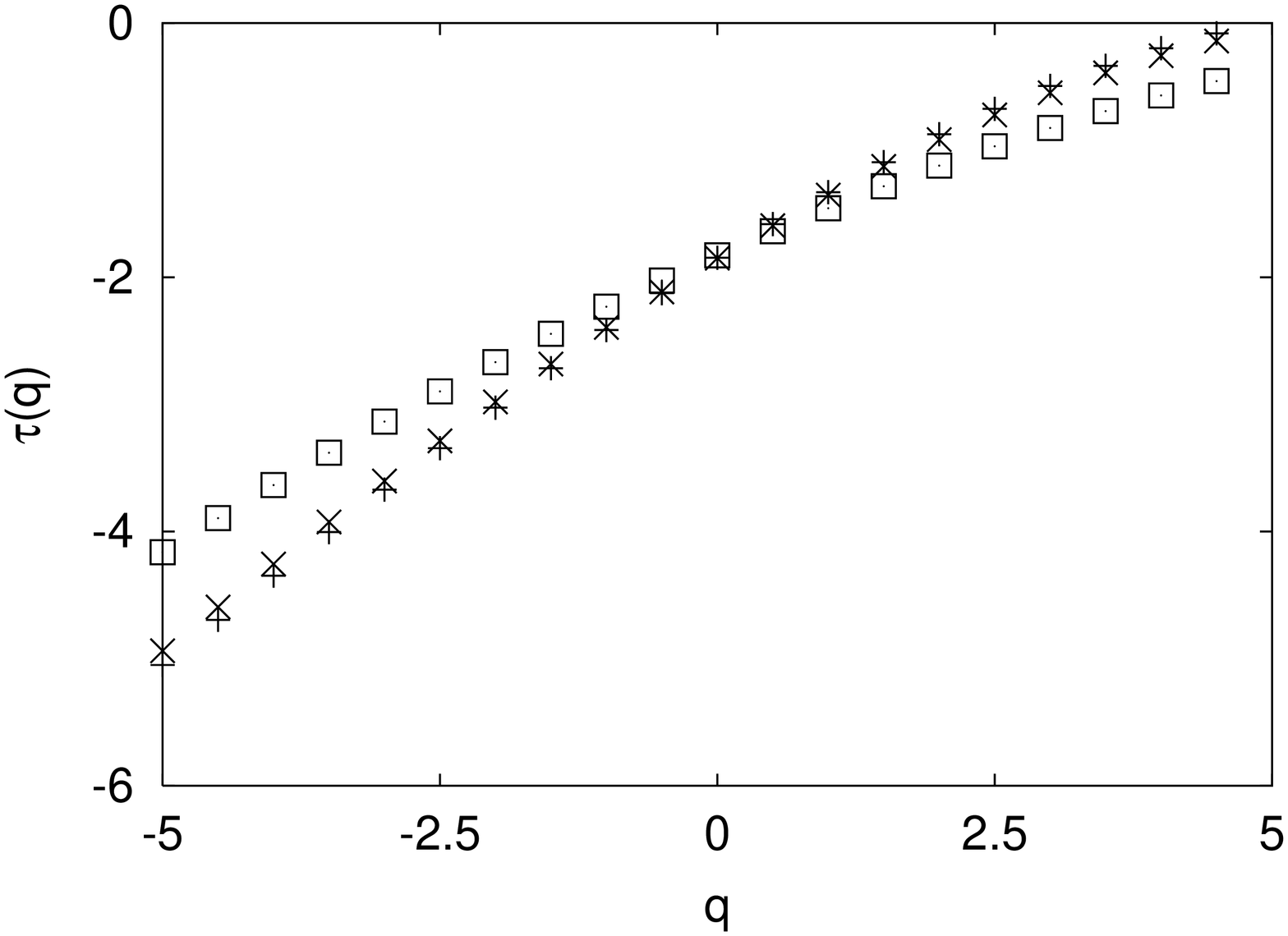}}}}
\put(0,30){(a)}
\put(50,30){(b)}
\end{picture}
\caption{Panel (a): $\tau(q)$ in model A as obtained from
investigations with wavelets derived from the filters $\Phi_0$ ($+$),
$\Phi_1$ ($\times$), $\Phi_2$ ($\boxdot$) and $\Phi_3$
($\blacksquare$). Note the deviations of the data obtained with the
gaussian function $\Phi_{0}$.  Panel (b): $\tau(q)$ for models A
($+$), B($\times$) and C ($\boxdot$).  All data have been obtained
from surfaces after $2 \cdot 10^{4}\,\mathrm{s}$ of simulated
time. Sizes of errorbars are on the order of symbol sizes.}
\label{waveeins}
\end{figure}
First, we have checked our results for artifacts resulting from
properties of the analyzing wavelet rather than the analyzed
surface by using different convolution functions $\Phi_{n}$:
$\Phi_{0}$ is the gaussian function, $\Phi_{n}, n \geq 1$ are products
of gaussians and polynomials which have been chosen in a way that the
first $n$ moments vanish. Then, the analyzing wavelets have
$n_{\vec{\Psi}_{n}} = n + 1$ vanishing moments.  We find (figure
\ref{waveeins}a), that the $\tau(q)$-curve obtained with $\Phi_{0}$
deviates significantly from those obtained with $\Phi_{1}$, $\Phi_{2}$
and $\Phi_{3}$.  The latter agree apart from small differences which
are mainly due to the discrete sampling of the wavelet in the
numerical implementation of the algorithm.  This is explained by the
theoretical result \cite{mba93} that $d \tau(q) / dq = n_{\vec{\Psi}}$
for $q < q_{\mathrm{crit.}} < 0$ if the number of vanishing moments of the
analyzing wavelet is too small.  Consequently, the agreement of the
other curves proves their physical relevance.

Figure \ref{waveeins}b shows averages of $\tau(q)$ curves obtained with
the convolution functions $\Phi_{1}$, $\Phi_{2}$, $\Phi_{3}$ from
surfaces after $2 \cdot 10^{4} \,\mathrm{s}$ of growth on an initially flat
substrate.  For all our models, their nonlinear behaviour reflects the
multiaffine surface morphology. From the fact that these curves are
reproduced within statistical errors in simulations with $N = 256$ we
conclude that finite size effects can be neglected.  Clearly,
desorption reduces the slope of $\tau(q)$, although only a small
fraction of the incoming particles is desorbed: $0.18 \%$ in model B
and $2.57 \%$ in model C with slightly higher values at earlier times.
\begin{figure}[tbh]
\begin{picture}(100, 33)(0, 0)
\put(0,33){\resizebox{0.48\textwidth}{!}{\rotatebox{270}{\includegraphics{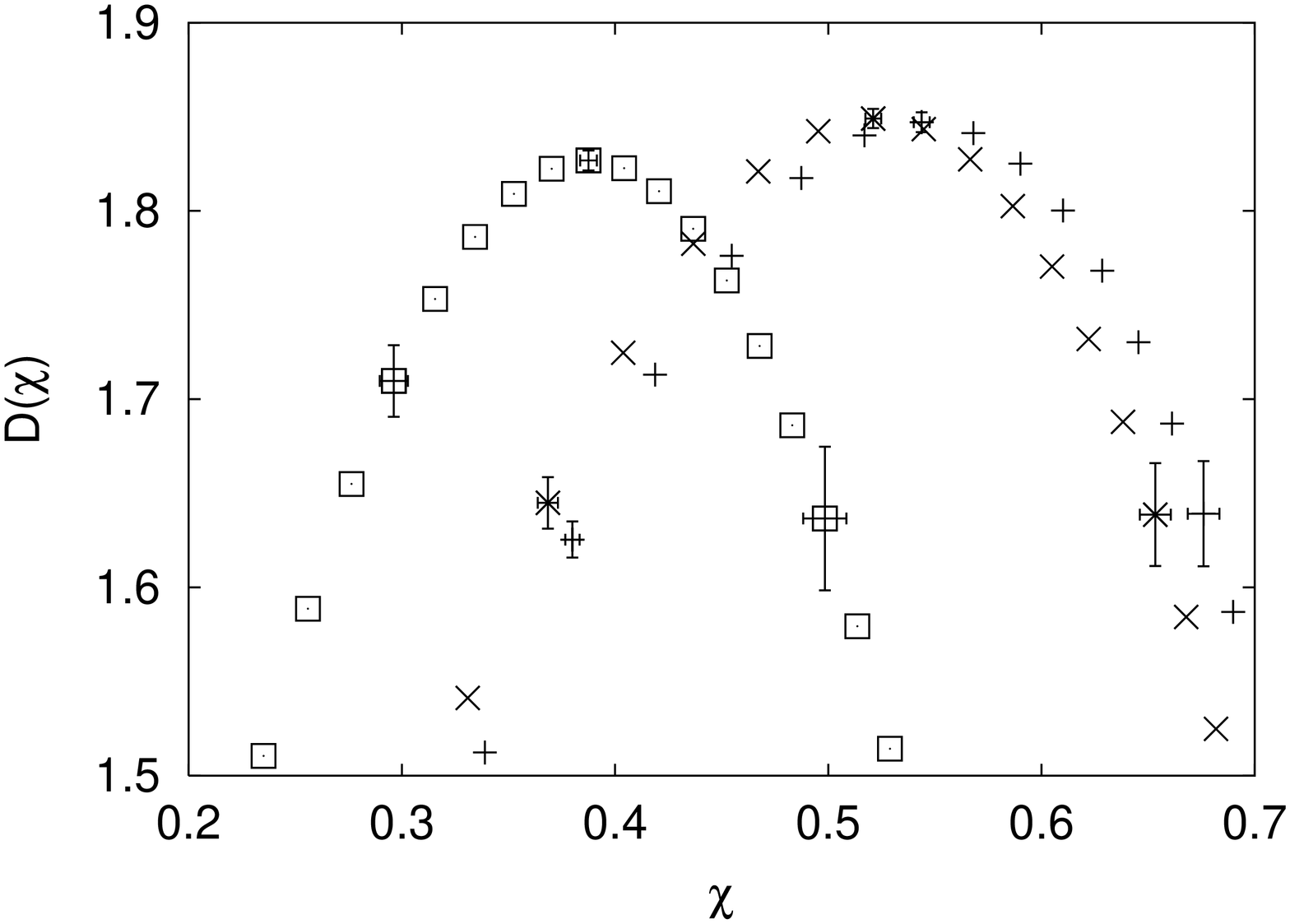}}}}
\put(50,33){\resizebox{0.48\textwidth}{!}{\rotatebox{270}{\includegraphics{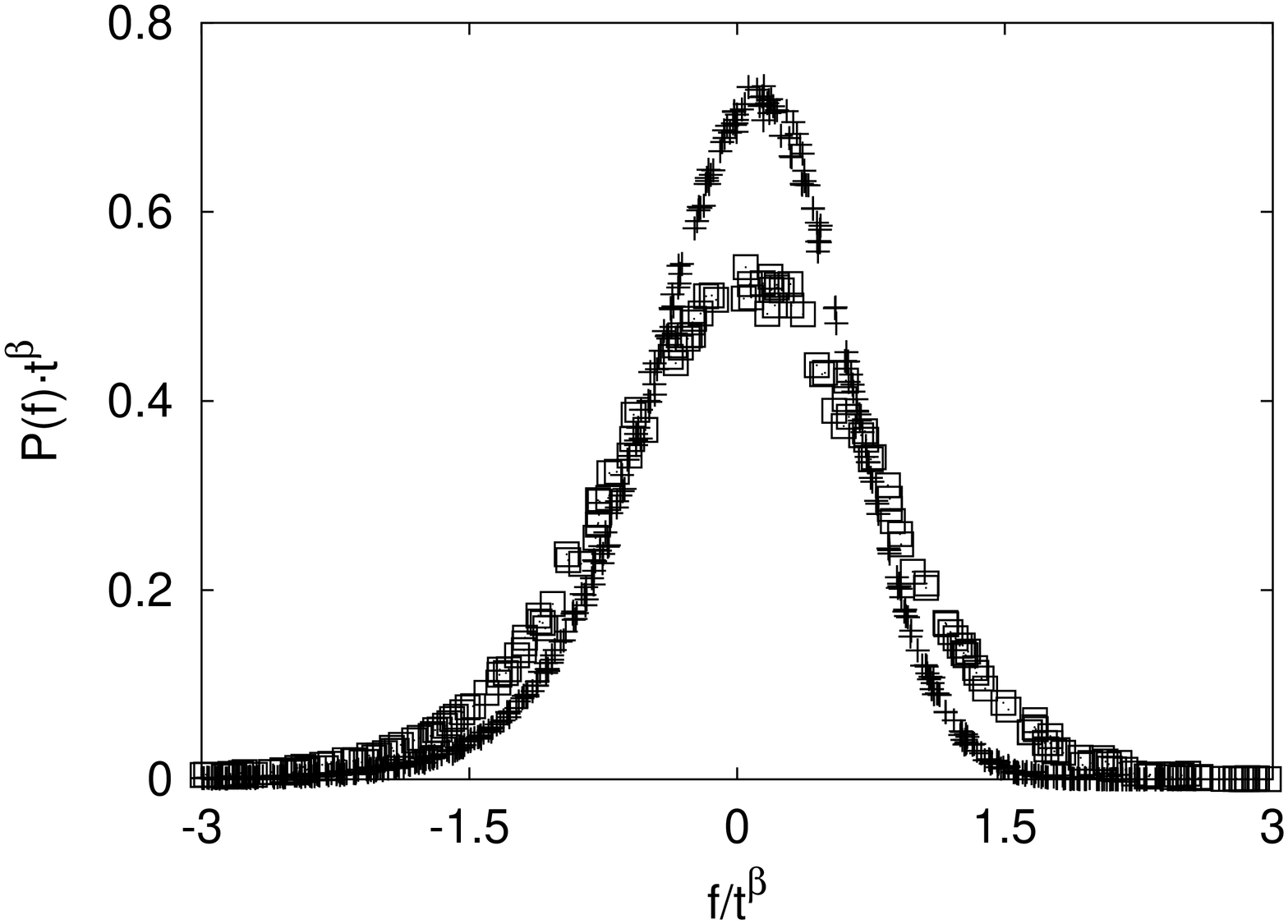}}}}
\put(0,30){(a)}
\put(50,30){(b)}
\end{picture}
\caption{Panel (a): Singularity spectra obtained from a Legendre
transform of the data in figure \ref{waveeins}b. Panel (b): Data
collapse of rescaled PDFs of surface heights for times between $150
\,\mathrm{s}$ and $2\cdot 10^{4} \,\mathrm{s}$ (model A) respectively
$150 \,\mathrm{s}$ and $7500 \,\mathrm{s}$ (model C). The
corresponding scaling exponents are summarized in table
\ref{wavetabelle}. Time is measured in seconds. In both panels, the
symbol $+$ represents results of model A, $\times$ corresponds to
model B and $\boxdot$ denotes results of model C. Model B 
which is not shown in panel (b) yields a
data collapse of similar quality as model A.
\label{wavezwei}}
\end{figure}
The corresponding singularity spectra are shown in figure
\ref{wavezwei}a. They have a typical shape whose descending part seems
to be symmetrical to the ascending part and which changes at most
slightly, while the whole spectra are shifted towards smaller
H\"{o}lder exponents as desorption becomes more important.  We
emphasize that we find no evidence for a {\em time dependence} of the
singularity spectra within the range $9700\,\mathrm{s} \leq t \leq 2
\cdot 10^{4} \,\mathrm{s}$, such that our results do {\em not} support
the idea of an asymptotic regime which is governed by Family-Vicsek
scaling. However, the accessible time range of computer simulations is
limited, so we cannot finally disprove the existence of such a regime.

\subsection{The relation between static and dynamic scaling properties}

In Family-Vicsek scaling, we have $H = \alpha$ such that the scaling
exponent of a static surface is identical to one of the dynamic
exponents. The multifractal formalism has replaced the unique Hurst
exponent $H$ of a self-affine surface with a wide spectrum of
H\"{o}lder exponents. This raises the question whether the analogy
between static and dynamic scaling behaviour holds also in the
presence of multiscaling such that there is a distribution of dynamic
scaling exponents. Alternatively, it is possible that multiscaling
affects only the properties of static surfaces while dynamic scaling
is governed by two unique exponents $\alpha$ and $\beta$.

To answer this question, we need a reliable method to decide whether
the properties of the surface are invariant under the transformation
\ref{dynamicscalingintro}. To this end, we investigate the probability
distribution function (PDF) $P_f(f, t)$ of the reduced surface
height. Invariance under dynamic scaling implies that $P_f(b^\alpha f,
b^z t) d(b^\alpha f) = P_{f}(f, t) df$. This is fulfilled if
\begin{equation}
P_f(f, t) = \frac{1}{t^{\beta}} \ p_f \left( \frac{f}{t^{\beta}} \right).   
\label{wavebkollaps}
\end{equation}
Consequently, within a large time range the rescaled PDFs 
$P_f t^{\beta}$ should collapse onto a
single function $p_f$ if they are plotted as a function of
$f/t^{\beta}$.  Additionally, we consider
the PDF $P_{\Delta f}(\Delta f, \vec{l}, t)$ of the height increment
at a distance $\vec{l}$, $\Delta f := f(\vec{x}, t) - f(\vec{x} +
\vec{l}, t)$.
Similar to equation \ref{wavebkollaps}, we obtain
\begin{equation}
P_{\Delta f}(\Delta f, l, t) = \frac{1}{t^\beta} \ p_{\Delta f} \left(
\frac{\Delta f}{t^\beta}, \frac{\vec{l}}{t^{1/z}} \right). 
\label{wavepdeltafkollaps}
\end{equation}
Therefore, in a large time range one should obtain a data collapse if
$t^\beta P_{\Delta f}$ is plotted versus $\Delta f /t^{\beta}$ and the
ratio $\vec{l}/t^{1/z}$ is kept fixed. Surface width and height-height
correlation functions of arbitrary order are moments of $P_f$ and
$P_{\Delta f}$, respectively. Consequently, our analysis of the PDFs
considers all the information which is contained in $W(t)$ and
$\Gamma(q, \vec{l}, t)$ for arbitrary $q$.

\begin{figure}
\begin{center}
\begin{picture}(100,35)(0,0)
\put(0,34){\resizebox{0.48\textwidth}{!}{\rotatebox{270}{\includegraphics{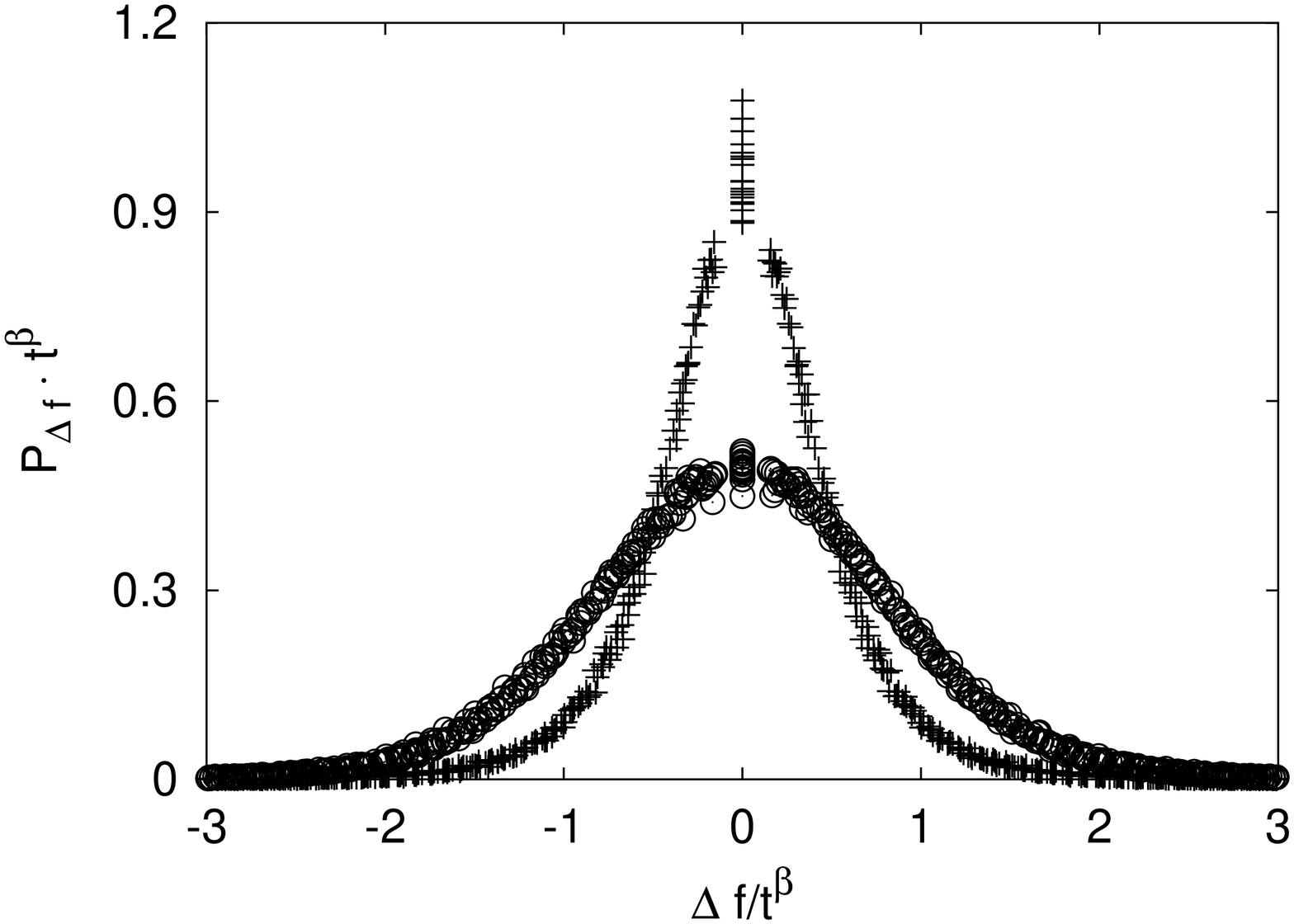}}}}
\put(0,31){(a)}
\put(52,34){\resizebox{0.48\textwidth}{!}{\rotatebox{270}{\includegraphics{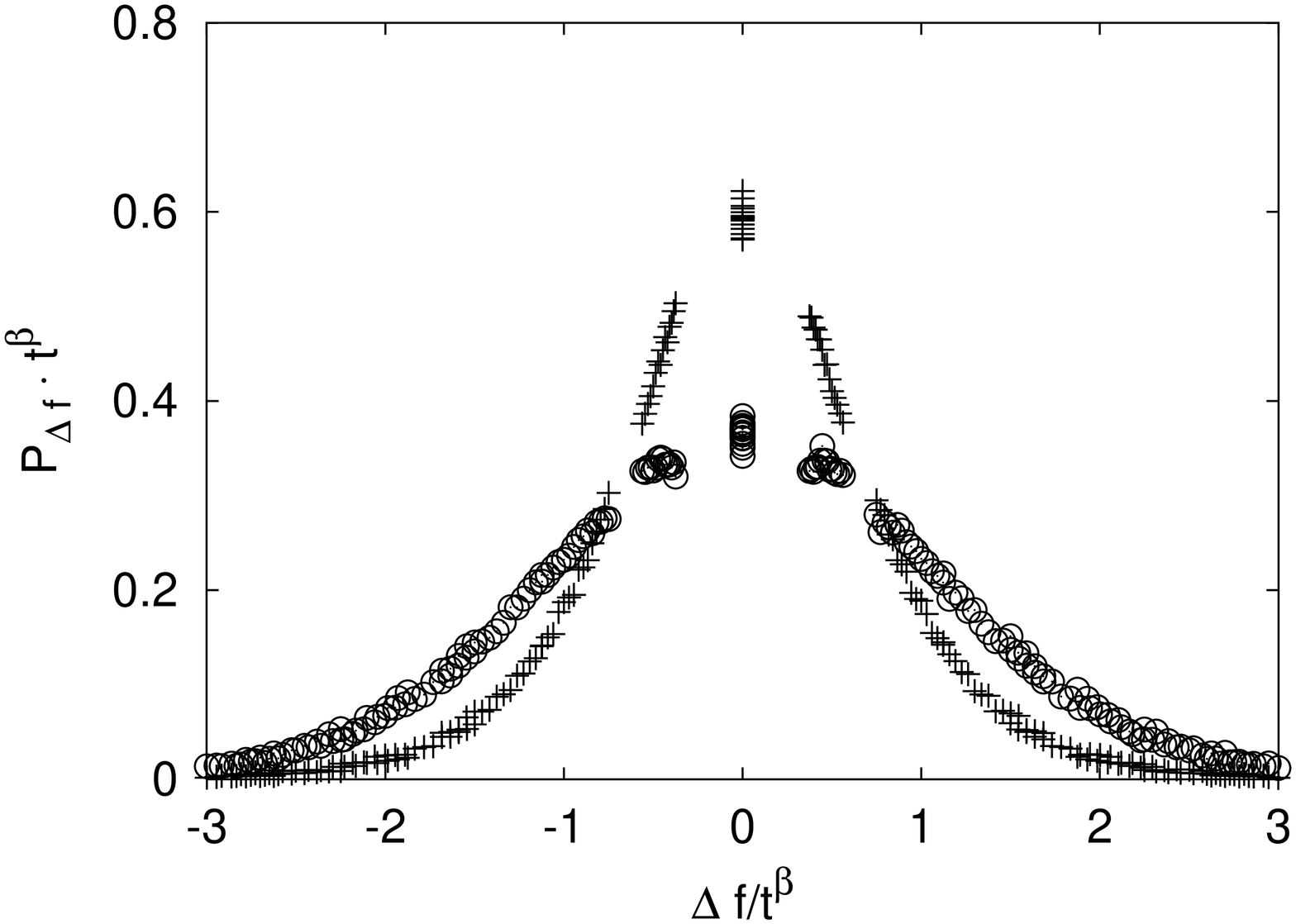}}}}
\put(52,31){(b)}
\end{picture}
\end{center}
\caption{Data collapse of $P_{\Delta f}$. Panel (a) shows data from
model A in the time range $150 \,\mathrm{s} \leq t \leq 2\cdot 10^4
\,\mathrm{s}$ at $l/t^{1/z} = 0.33$ ($+$) and $l/t^{1/z} = 8.42$
($\odot$). In panel (b), we show results for model C within the time
range $150\,\mathrm{s} \leq t \leq 7500 \,\mathrm{s}$ at $l/t^{1/z} =
0.63$ ($+$) and $l/t^{1/z} = 15.1$ ($\odot$). The data collapse is
good apart from small lattice effects at $\Delta f = 0$. Model B
yields results similar to model A. In all results shown, the vector
$\vec{l}$ is parallel to the $x$-direction. \label{wavepdeltaf}}
\end{figure}
To obtain values of the dynamic exponents $\alpha$ and $\beta$, we
apply standard methods which have been reported in the literature. We
measure $\beta$ from the increase of the surface width with time,
which follows a power law for $t \geq 150 \,\mathrm{s}$ in models A
and B and for $ 150 \,\mathrm{s} < t < 7500 \,\mathrm{s}$ in model C
which then starts to approach the final {\em saturation regime}. The
dynamic exponents $\alpha$ and $z$ have been obtained from the data
collapse of the scaled height-height correlation function $G( \vec{l},
t)$. The results are summarized in table \ref{wavetabelle}.

In figure \ref{wavezwei}b, we show the data collapse of $P_f$ in
models A and C which is of high quality. Model B yields similar
results. Examples of rescaled $P_{\Delta f}$ are shown in figure
\ref{wavepdeltaf}. The data collapse is good for $\Delta f \neq
0$. However, the probability that the height difference between points
at a distance $l/t^{1/z}$ is zero does not follow the behaviour which
is expected from the 
assumption of dynamic scale invariance. This deviation is a
finite size effect. Dynamic scale invariance is a continuous symmetry
property. On the other hand, due to the discrete lattice structure,
differences between surface heights can assume only integer
values. This makes it impossible to fulfil dynamic scale invariance
exactly, since the transformation \ref{dynamicscalingintro} maps
integer height differences to noninteger height differences. If we
approximate a real random number $r$ with an integer $i$, on average
the relative error $|i - r|/r$ is greater for smaller $r$.  Therefore,
it is plausible that the deviations of the empirical $P_{\Delta
f}(\Delta f, \vec{l}, t)$ from the theoretical result in equation
\ref{wavepdeltafkollaps} are greater for small $\Delta f$. However,
in the limit of late time and lengthscales which are large compared to
the lattice constant, corrections due to the lattice structure can be
neglected.  Consequently, there is no indication that there is a
violation of dynamic scale invariance apart from small finite size
corrections.  This parallels the result of Krug \cite{k94} that there
is an unique exponent $\beta$ in the one-dimensional Das
Sarma-Tamborenea model.

Finally, we relate the static scaling properties of the surface which
have been determined precisely by means of the WTMM method with the
dynamic scaling properties. Our results (table \ref{wavetabelle})
show, that the dynamic scaling exponent $\alpha$ and the value $\chi_m$
of the H\"{o}lder exponent which {\em maximizes} $D(\chi)$ agree within
errorbars. This empirical finding can be explained with a saddle point
argument. We calculate the standard deviation of the surface in a
square of size $L \times L$,
\begin{equation}
w(L, t)^{2} = \frac{1}{L^{2}} \int_0^L \int_0^L d^{2} x f(\vec{x},
t)^{2} = \frac{1}{L^{2}} \int d\tilde{\chi} \underbrace{ \int_0^L
\int_0^L d^{2}x \; \delta(\tilde{\chi} - \chi(\vec{x})) f(\vec{x},
t)^{2}}_{I(\tilde{\chi})} \; .
\label{lokalglobalgleichgleichung}
\end{equation}
Since $I(\tilde{\chi})$ grows like $L^{D(\tilde{\chi})}$ with $L$, for
large $L$ the integral over $\tilde{\chi}$ will be dominated by
$I(\chi_{m})$.  Thus, $w(L, t)$ is governed by the subset of points
which has the greatest fractal dimension. Consequently, in a
multiaffine surface $w(L, t)$ is identical to that of a self-affine
surface with Hurst exponent $\chi_m$.  In particular, for $L <
\xi(t)$ we have $w(L, t) \sim L^{\chi_m}$ (equation
\ref{selbstaffinw}). On the other hand, dynamic scale invariance
implies $w(L, t) \sim L^{\alpha}$ for $L < \xi(t)$. Consequently,
$\chi_m = \alpha$.

The conventional picture of anomalous scaling \cite{l99,lr96,lrc97}
notes the difference between a ``global $\alpha$'' (the dynamic
exponent) and a ``local $\alpha$'' which is determined from
the power-law behaviour of $G(\vec{l}, t)$ for small $l$ and, in our
notation, is termed $\gamma_2$.  Within the multifractal formalism,
the local $\alpha$ simply corresponds to some H\"{o}lder exponent on
the ascending part of the singularity spectrum. In contrast to the
dynamic scaling exponents, it is not related to any symmetry property
of the model itself. Therefore, we expect the dynamic exponents
$\alpha$ and $z$ to be most significant to characterize the scaling
properties of a growth model.  

\section{Conclusions}

\begin{table}
\begin{center}
\begin{tabular}[t]{|l|c|c|c|c|c|c|}
\hline
Model & $p_{d}$ & $\beta$       &$\chi_{m}$        &$\alpha$& $z$ & $D_{f}$ \\ \hline 
A     & 0       &0.19 $\pm$ 0.01&0.54 $\pm$ 0.01& 0.55       & 2.9     & 2.32 $\pm$ 0.01 \\
B     & 0.18 \% &0.17 $\pm$ 0.01&0.52 $\pm$ 0.01& 0.51       & 3.3     & 2.35 $\pm$ 0.01 \\
C     & 2.57 \% &0.11 $\pm$ 0.01&0.38 $\pm$ 0.01& 0.39       & 3.5     & 2.45 $\pm$ 0.02 \\
\hline
\end{tabular}
\end{center}
\caption{Simulation results: $p_{d}$ is the fraction of particles
which desorbs, $\chi_{m}$ the H\"{o}lder exponent which maximizes
$D(\chi)$, $\alpha$, $\beta$ and $z$ are the dynamic scaling
exponents, and $D_{f}$ is the fractal dimension of the surface. }
\label{wavetabelle}
\end{table}
Our results support the following picture of the scaling behaviour of
kinetically roughening surfaces: Dynamic scale invariance is fulfilled
on time- and lengthscales where lattice effects can be
neglected. Surfaces at fixed time have multifractal properties which
are characterized by a wide spectrum of H\"{o}lder exponents. The
H\"{o}lder exponent which maximizes $D(\chi)$ is identical to the
dynamic exponent $\alpha$. There is no indication that the singularity
spectra narrow with time. Consequently, it is plausible that the
multifractal properties of the surface are preserved in the limit of
infinite time and there is no transition to Family-Vicsek scaling.

Our discussion of equation \ref{lokalglobalgleichgleichung} indicates
that this is the most general scaling behaviour given the following
conditions are fulfilled: (1) The statistical properties of the model
are invariant under the dynamic scale transformation
\ref{dynamicscalingintro}. (2) The surface is {\em scale-free} such
that the statistical properties of sections of the surface of size $L$
are independent of $L$ whenever $L < \xi(t)$. Then, $\xi(t)$ is the
only relevant lengthscale of the surface. This picture contains
Family-Vicsek scaling as the special case where there is only one
H\"{o}lder exponent.

According to L\'{o}pez \cite{lopezmail}, there are continuum models of
epitaxial growth which exhibit anomalous scaling but not multiscaling.
In particular, this is the case for the Mullins equation and the
Lai-Das Sarma equation. Since the singularity spectrum is the Legendre
transform of $\gamma_q$, in this case $D(\chi)$ has a single peak at
one characteristic H\"{o}lder exponent which equals the exponent
$\gamma_2$. Thus, the static scaling properties which are measured on
short lengthscales (equation \ref{wavezsum}) are not related to the
dynamic scaling properties. This implies necessarily, that the
dynamic properties are determined by large structures which
appear smooth on small lengthscales. Consequently, the surface is not
{\em scale-free}. Instead, we expect that there is a second
lengthscale $\Xi(t) < \xi(t)$. On lengthscales $\ll \Xi(t)$, the
surface appears self-affine with a Hurst exponent $H = \gamma_2$. The
dynamic properties are determined by structures on lengthscales
between $\Xi(t)$ and $\xi(t)$. Note, that dynamic scale invariance of
the model implies $\Xi(t) \propto \xi(t)$.

Table \ref{wavetabelle} summarizes our quantitative results. Model A
without desorption reviews the results in \cite{dslkg96}, which have
been obtained with slightly different activation energies on smaller
systems and shorter timescales. Models B and C show, that desorption
is an important process, which, although it affects only a small
fraction of the adsorbed particles, must not be neglected, since it
alters the scaling properties of the surfaces by reducing both
$\alpha$ and $\beta$.  Since the dynamic exponents depend continuously
on the desorption rate, the paradigm of a few universality classes
characterized by a {\em small} number of exponents which are {\em
independent} of details of the model is {\em not} adequate to catch
the features of kinetic roughening in our computer simulations. Since
the time range of computer simulations is limited by the available CPU
time, our results cannot finally disprove that there is a crossover to
a different, possibly universal scaling behaviour after the deposition
of a number of monolayers which is large compared to our $2\cdot
10^4$. However, if this should indeed be the case, it is improbable
that the asymptotic scaling behaviour might ever be observed on
experimentally relevant timescales of a few hours of growth.

\chapter{The influence of the crystal lattice on coarsening in 
unstable epitaxial growth \label{coarsekapitel}}

In spite of considerable efforts
\cite{kscm99,mg00,rk96,s98,spz97,t98,tsv97}, see e.g.\ \cite{pgmpv00} for an
overview, a thorough theoretical understanding of the coarsening
process of mounded surfaces is still lacking. In this chapter we
investigate the problem of growth in the presence of a strong
Schwoebel barrier which hinders interlayer transport. As discussed in
section \ref{intromound}, this leads to an instability of the flat
surface. During an initial transient, {\em mounds} form on the surface
which start to merge as soon as their flanks have assumed the stable
slope. It is generally accepted that this asymptotic coarsening regime
fulfils the dynamic scale symmetry which has been introduced in
section \ref{introdynamicscale}.

It was first pointed out by Siegert et.\ al.\ \cite{s98,spz97} that
lattice symmetries may play an important role in the coarsening
process. These authors investigated continuum equations, using an
analogy between coarsening and a phase ordering process which has
recently gained popularity \cite{ks95}: areas of constant slope should
correspond to domains of a constant order parameter. They derived
scaling exponents $\alpha = 1, \beta = 1/3$ on surfaces with a
triangular symmetry, and $\beta = 1/(3 \sqrt{2})\approx 0.24$ for
generic cubic surfaces, while $\beta = 1/3$ requires a fine-tuning of
parameters. Moldovan and Golubovic \cite{mg00} applied similar methods
to surfaces with hexagonal symmetry and found $\beta = 1/3$.
   
However, Monte Carlo simulations have raised doubts on these
predictions, since they yield $\beta \simeq 1/3$ on cubic surfaces for
a great range of parameters, see e.g.\
\cite{amar99,bkks99,skbk99,ssbk99}. In this chapter, we adress
the following questions: (1) What are the {\em mesoscopic} processes
which make the mounds coarse? (2) How does the coarsening process
depend on the crystal lattice and its symmetries? (3) Do these
results support a deeper analogy between coarsening and phase
ordering?

\section{Effective single particle simulations}

To answer these questions, we perform computer simulations of growth
on (001) surfaces of the simple cubic (sc) lattice, the simple
hexagonal lattice (sh), the body-centred cubic (bcc) lattice and the
(0001) surface of the hexagonal close packing (hcp). This is done
under solid-on-solid conditions, i.e.\ the effects of overhangs or
dislocations are neglected. Then, the simple lattices can be
represented by a square (sc) and a triangular mesh (sh) of integers
respectively, which denote the height $h(\vec{x})$ of the surface. We
build the bcc (hcp) lattice out of two intersecting sc (sh)
sublattices, which contain the even and the odd heights,
respectively\footnote{For simplicity, we assume the spacing between
the layers to be one unit length. Our algorithm depends only on the {\em
topology} of the lattice.}. Here, an adatom in a {\em stable}
configuration is bound to 4 (3) neighbours below in the other
sublattice, which will be denoted as {\em vertical neighbours} in the
following. Particles with fewer vertical neighbours form overhangs
which are forbidden by the solid-on-solid condition. This is
physically reasonable, since such particles are only weakly bound and
therefore these configurations will be unstable. Additionally,
particles may have neighbours in the {\em lateral} direction which are
in the same sublattice.

The investigation of the coarsening process requires a fast algorithm
which allows for the simulation of the deposition of thick films on
comparatively large systems. Kinetical Monte Carlo techniques which
consider the moves of many particles on the surface simultaneously are
too slow for our purpose even if the efficient continuous time
algorithms introduced in appendix \ref{mcappendix} are used. Instead,
we simulate the moves of a {\em single} particle from deposition until
an immobile state is reached \cite{bkks99,k99,skbk99}.

An adatom impinges on a randomly chosen lattice site. As discussed in
section \ref{depositionintro}, it is reasonable to assume that a newly
deposited adatom undergoes an incorporation process. In our model, the
particle funnels downhill \cite{e91,estp90,yhd98} to the lowest
(vertical) neighbour site. On bcc and hcp this is repeated until it
reaches a {\em stable site} as defined above. On the simple lattices
this is a site $\vec{x}$ where all nearest neighbour sites have a
height $\geq h(\vec{x})$.

Then, the adatom diffuses on the surface. If the particle has no
lateral neighbours, one of its neighbour sites is chosen at random. On
the simple lattices, the particle moves to this site only if its
height does not change, i.e.\ we introduce an infinite Schwoebel
barrier. On the bcc or hcp lattice, the particle is moved to the
neighbour site if it is stable. This condition implies an infinite
Schwoebel barrier, too. Similar to our model of the zinc-blende
lattice introduced in chapter \ref{zinkblendenmodell}, interlayer
diffusion would require the consideration of additional hopping
vectors. 
 
As discussed in section \ref{epitaxintro}, from time to time adatoms
which diffuse around on a flat surface collide and nucleate a new
island. The typical distance $l_i$ between neighbouring islands
depends on the ratio of the adatom diffusion constant $D$ and the
particle flux $F$. In our effective single particle dynamics, we
consider this effect by limiting the maximal number of diffusion steps
of a particle. If an atom has performed $l_d^2$ diffusion steps,
diffusion stops and the particle becomes an island nucleus which will
capture subsequent particles. The {\em diffusion length} $l_d$ fixes
the number of islands in the first layer which is identical to the
initial number of mounds. In the later stages of growth, the typical
terrace width is much smaller than $l_{d}$ and nucleation occurs only
at the top terraces of the mounds.

As soon as a particle has a lateral neighbour, it is bound to it.
This process is irreversible in the sense that we forbid diffusion
processes which reduce the number of bonds like the detachment from a
step edge. In section \ref{epitaxintro} we have shown that this is a
reasonable approximation if the rates of these processes are much
smaller than $F$. However, the adatom may diffuse along the edge.
After $l_{k}^{2}$ steps, or if the particle has reached a kink site,
it is fixed to the surface. If not stated otherwise, diffusion around
corners is allowed. Analogous to planar diffusion, $l_{k}$ determines
the typical distance of nucleation events in the effectively
one-dimensional step edge diffusion, see \cite{vptw92} for a
discussion.

In this model, we measure time $t$ in units of the time needed to
deposit one monolayer (ML). This algorithm may be programmed very efficiently
and is at least one order of magnitude faster than full diffusion
Monte Carlo algorithms.

The two essential simplifications in our model are (a) the effective
representation of nucleation events in a single particle dynamics and
(b) the consideration of irreversible processes, i.e.\ infinite energy
barriers for detachment and downward diffusion at edges. For the
simple cubic lattice, the model was studied in
\cite{bkks99,skbk99,ssbk99} with particular emphasis on the role of
step edge diffusion. In \cite{ssbk99} a full diffusion model is
considered for comparison which includes detachment as well as finite
Schwoebel barriers and displays the same asymptotic coarsening
exponents as measured in the simplified model.

In all our simulations we choose $l_{d} = 15$ which controls the
initial island distance. We simulate a variety of different values of
the step edge diffusion length $l_{k}$ in the range between $l_{k} =
1$ and $l_{k} = 20$. The simulations are performed on a square of $N
\times N$ lattice constants using periodic boundary conditions (bcc
and sc) and a regular hexagon with edges of length $M$ (hcp and sh)
using helical boundary conditions, our standard values being $N = 512$
and $M = 300$. For every parameter set we have performed 7
independent simulation runs.

\section{Surface morphologies}

\begin{figure}
\figpanel{\hspace{-1mm}\resizebox{0.33\textwidth}{!}{\includegraphics{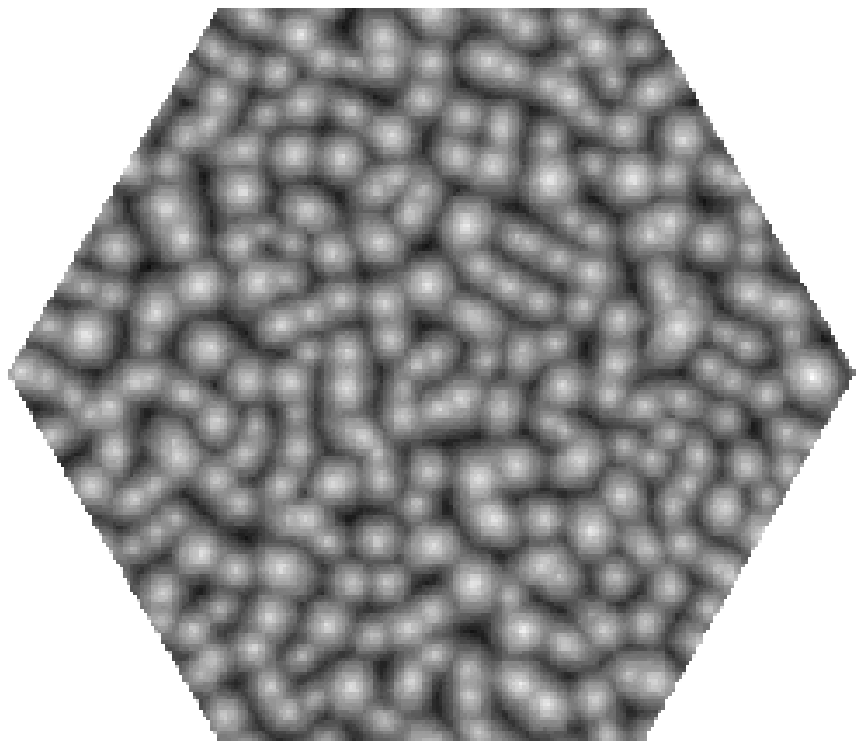}}\hspace{-1.5mm}}{$t=
200\,\mathrm{ML}$} \hfill
\figpanel{\hspace{-1.5mm}\resizebox{0.33\textwidth}{!}{\includegraphics{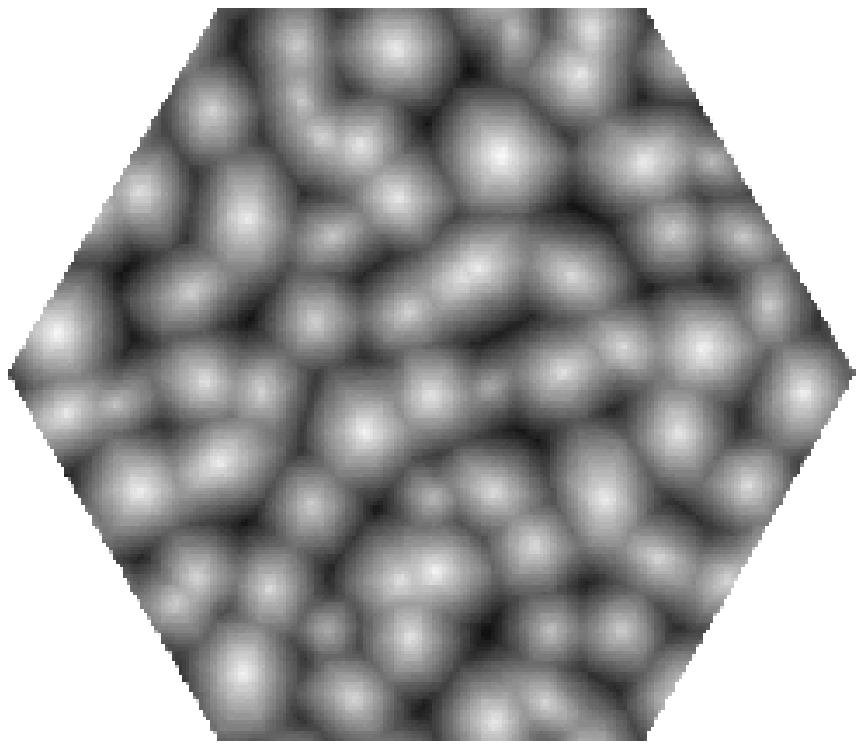}}\hspace{-1.5mm}}{$t
= 2000\,\mathrm{ML}$} \hfill
\figpanel{\hspace{-1.5mm}\resizebox{0.33\textwidth}{!}{\includegraphics{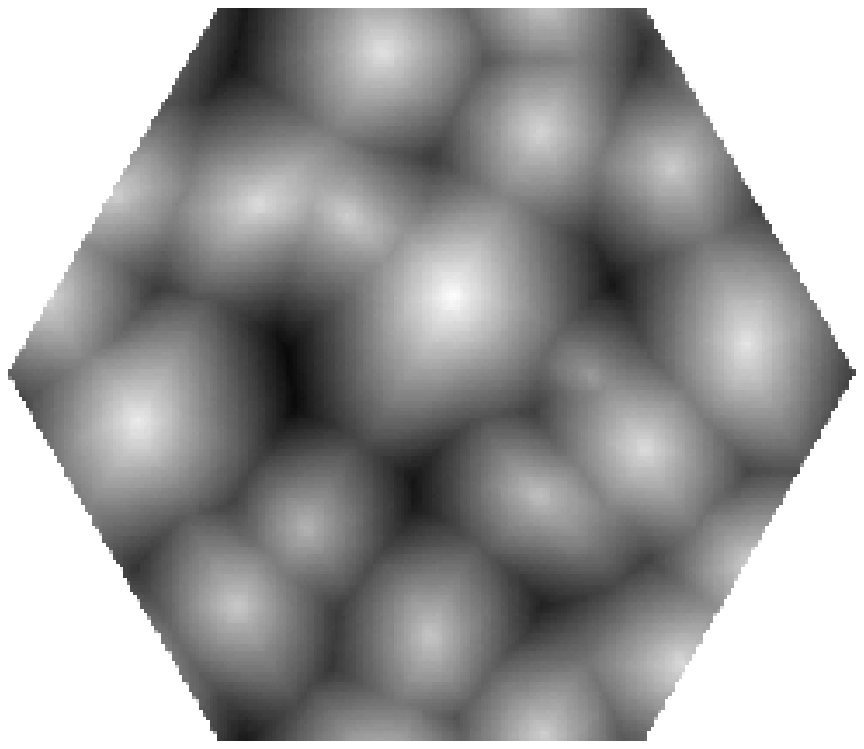}}\hspace{-1.5mm}}{$t
= 20000\,\mathrm{ML}$}
\caption{Coarsening of mounds on the hcp(0001) surface at $M = 300$,
$l_d = 15$ and $l_k = 20$. Brighter shades of grey correspond to
greater values of the surface height. \label{coarsefigb}}
\end{figure}

\begin{figure}
\begin{center}
\begin{picture}(100,49.5)(0,0)
\put(5,5){\resizebox{0.45\textwidth}{!}{\includegraphics{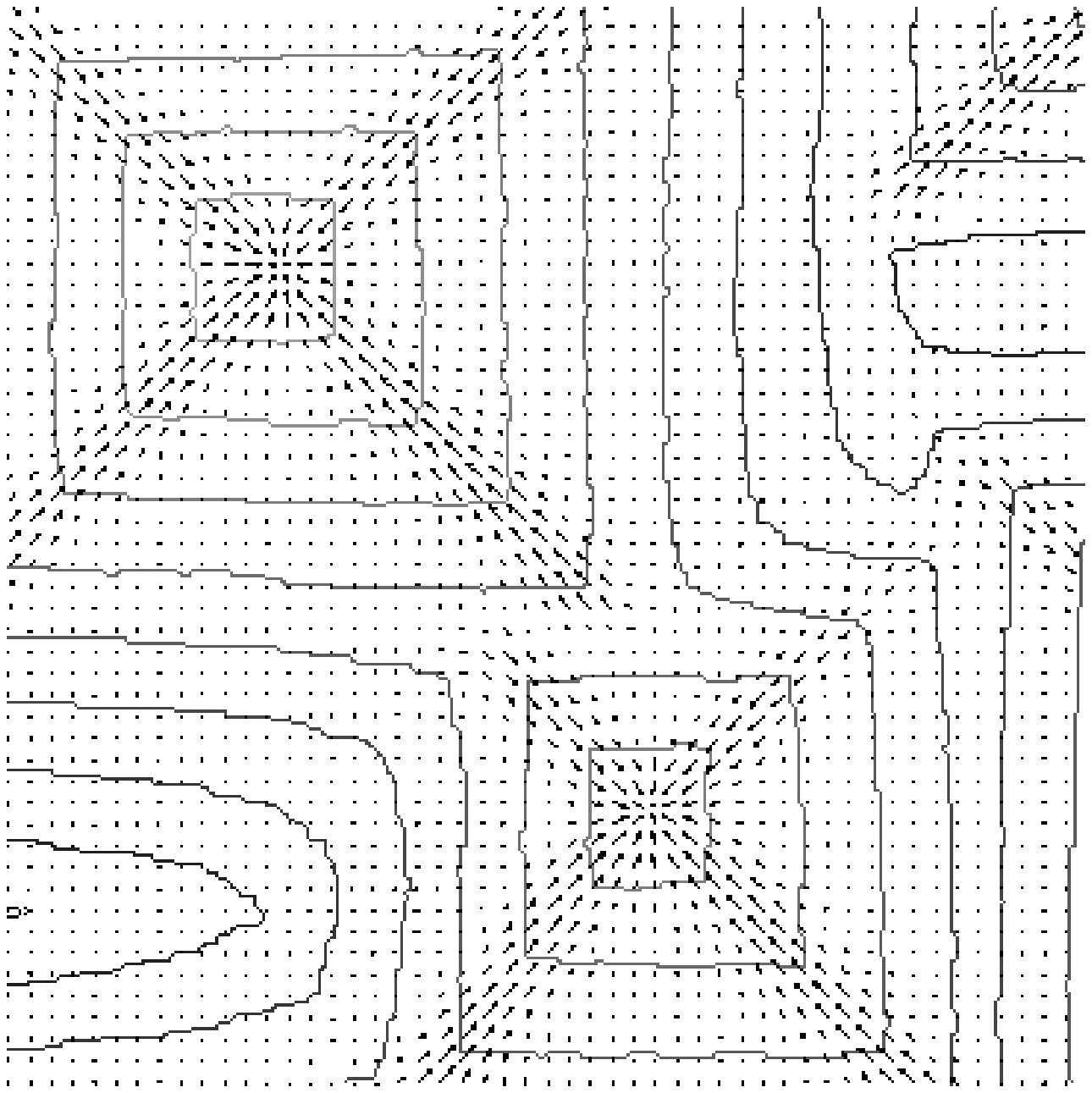}}}
\put(1,47.5){(a)}
\put(55,5){\resizebox{0.45\textwidth}{!}{\includegraphics{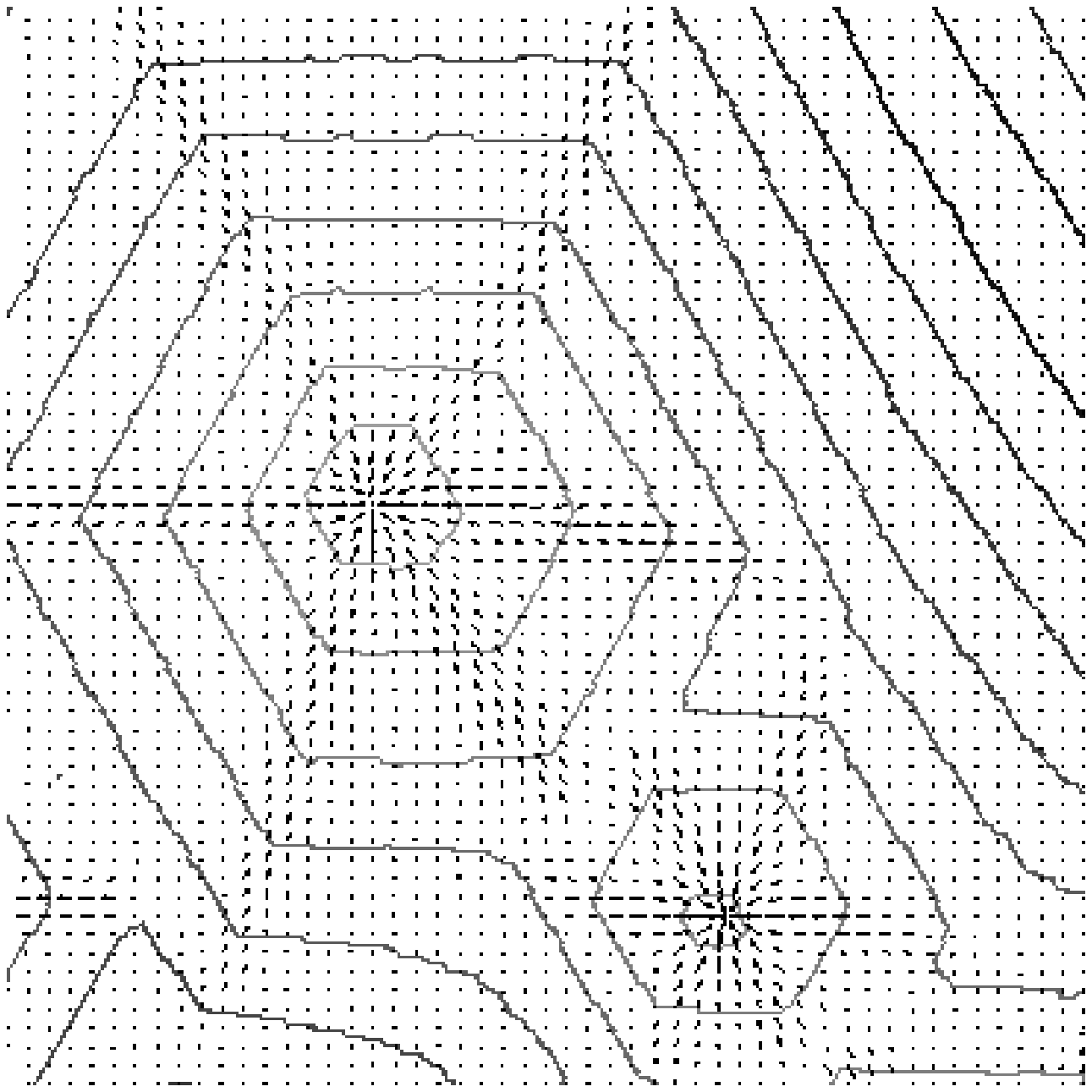}}}
\put(51,47.5){(b)}
\put(40,3){\vector(1,0){20}}
\put(47.5,0){$[100]$}
\end{picture}
\end{center}
\caption{Contour plots of surfaces of the simple lattices at $t =
2\cdot 10^4\,\mathrm{ML}$. Brighter contours denote greater surface
heights. The arrows show the average diffusion current on the surface
during the deposition of additional 200\,ML. The figures show sections
of $250 \times 250$ lattice constants of systems of size $N = 512$
(simple cubic, panel (a)) and $M = 300$ (simple hexagonal, panel
(b)). Diffusion length and step edge diffusion length are $l_d = 15$
and $l_k = 20$, respectively. \label{coarsescurrents}}
\end{figure}

During the first (about $100$) monolayers of growth on an initially
flat surface islands nucleate on which mounds build up and take on
their stable slope. Then, the asymptotical coarsening regime starts.
As an example, in figure \ref{coarsefigb} we show the evolution of the
hcp surface at $l_k = 20$. 
As expected, on the simple lattices the mounds obtain regular shapes
which are determined by the symmetry of the surface: square pyramids
on sc, hexagonal ones on sh (figure \ref{coarsescurrents}). On bcc
and hcp however, {\em rounded} corners as well as sharp corners can be
found (figures \ref{coarsefigb}, \ref{coarsefig1} and
\ref{coarsefig2}).

\begin{figure}
\botbase{
\begin{picture}(48,48)(2,0)
\put(5,6){\resizebox{0.45\textwidth}{!}{\includegraphics{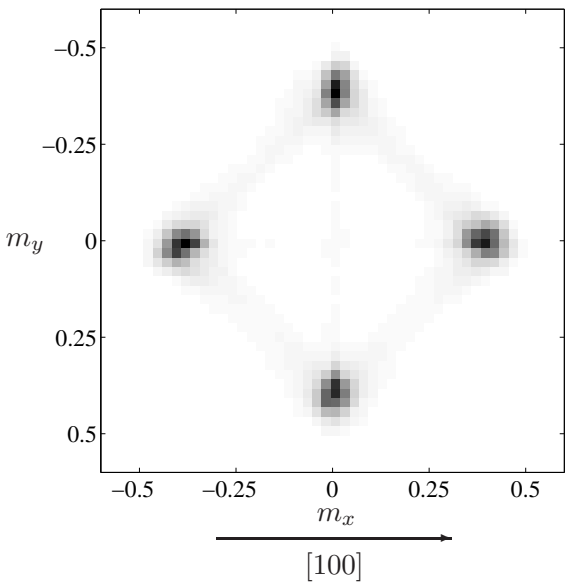}}}
\put(2,28){$m_{y}$}
\put(28.5,4.5){$m_{x}$}
\put(20,3){\vector(1,0){20}}
\put(27.5,0){$[100]$}
\end{picture}
}
\hfill
\botbasebox{0.45\textwidth}{\caption{Histogram of the slopes on simple cubic
surfaces ($N = 512$, $l_d = 15$, $l_k = 20$, $t = 2\cdot 10^4$
ML). High probabilities are drawn dark. Each of the four dark spots
corresponds to the slope of one pyramid edge.\label{coarsehcub}}}
\end{figure}

\begin{figure}
\begin{center}
\begin{picture}(100,49)(0,0)
\put(5,5){\resizebox{0.45\textwidth}{0.45\textwidth}{\includegraphics{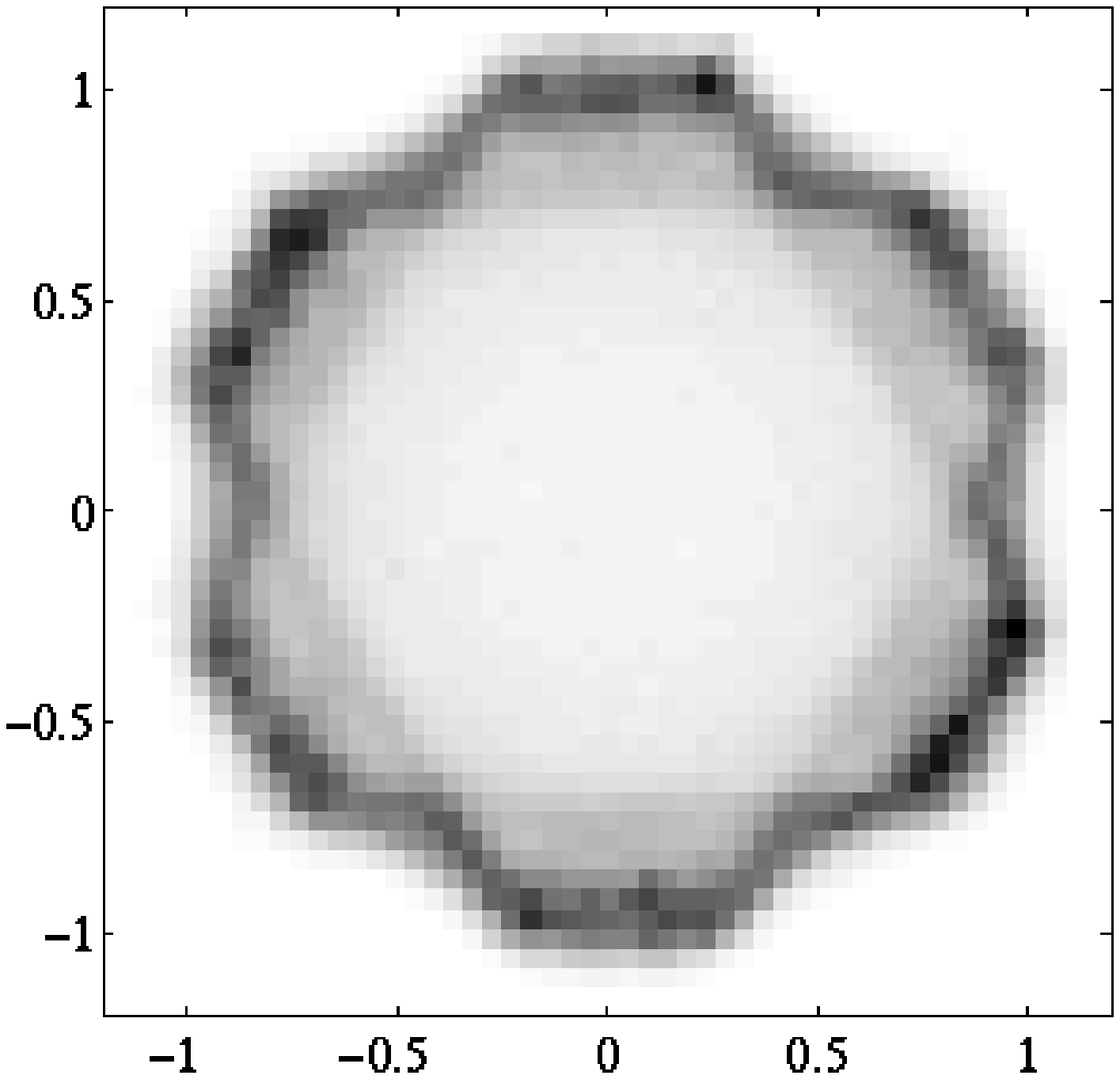}}}
\put(1,47){(a)}
\put(2,28){$m_{y}$}
\put(28.5,3.5){$m_{x}$}
\put(55,5){\resizebox{0.45\textwidth}{0.45\textwidth}{\includegraphics{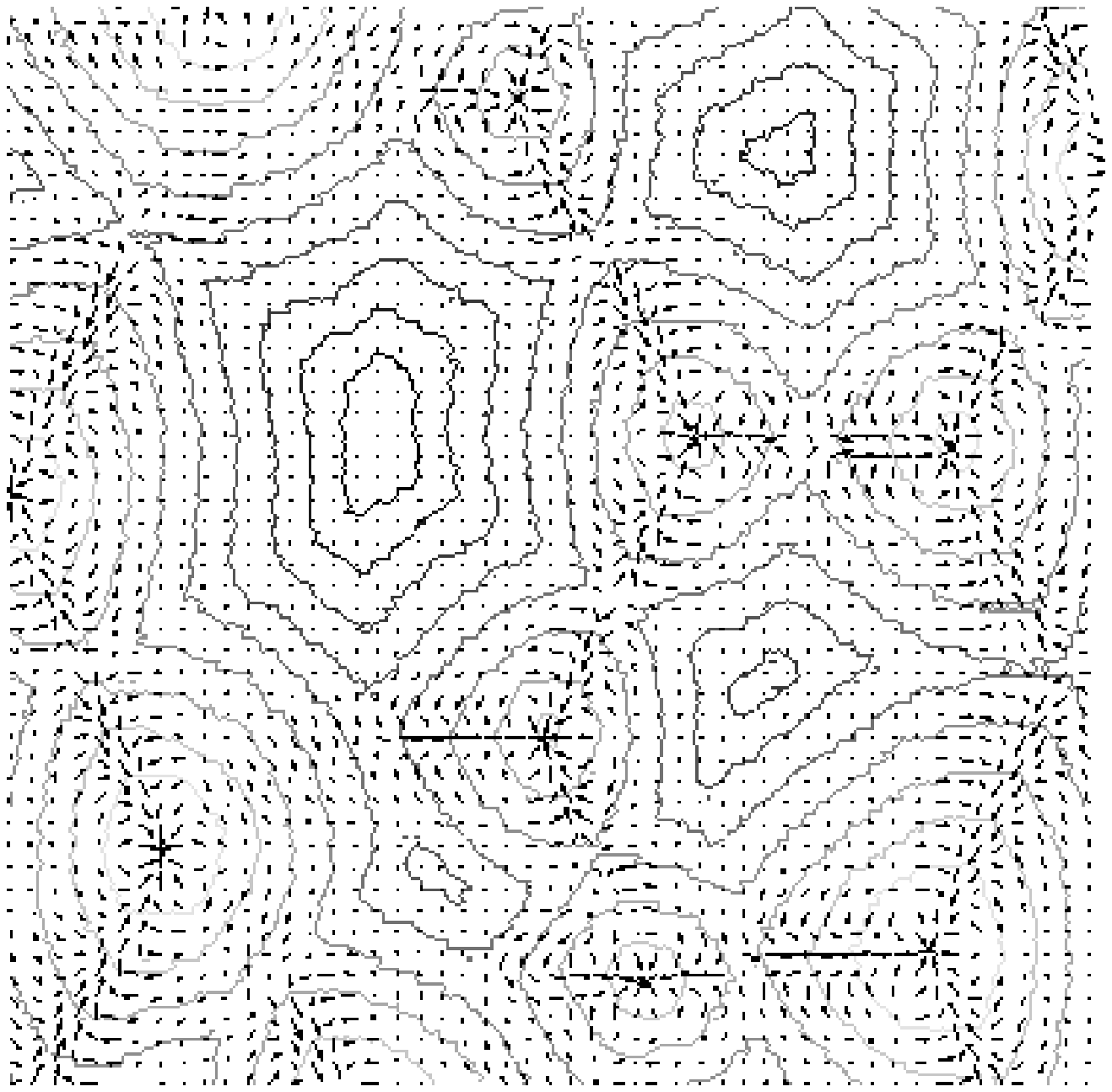}}}
\put(51,47){(b)}
\put(40,3){\vector(1,0){20}}
\put(47.5,0){$[100]$}
\end{picture}
\end{center} 
\caption{Simulation results on hcp ($M = 300$, $l_{d} = 15$, $l_{k} =
10$, $t = 20000\,\mathrm{ML}$). Panel (a): Histogram of slopes
(density plot). High probabilities are drawn dark. Panel (b): Contour
plot of a part of size $250 \times 250$ lattice constants of the
surface. High levels are plotted in light grey. The arrows show the
diffusion current as measured during the deposition of additional $200
\,\mathrm{ML}$.}
\label{coarsefig1}
\end{figure}
Here one finds two fundamentally different types of mounds: On the one
hand there are mounds, where {\em all} corners are rounded with
octagonal shapes on bcc (figure \ref{coarsefig2}b), and 12-cornered
ones on hcp.  On the other hand, one observes a {\em breaking} of the
{\em symmetry} given by the lattice: approximately triangular mounds
on hcp (figure \ref{coarsefig1}b) and oval shapes on bcc (figure
\ref{coarsefig2}a), where every second corner is sharp while
the others are rounded. We have observed that in the process of
coarsening, on all lattice structures mounds merge preferentially at
the {\em corners}, while mounds touching at the flanks are a {\em
metastable} configuration. On bcc and hcp, the metastable
configuration consists of mounds with completely rounded
corners. Mounds whose shapes break the lattice symmetry are always in
the course of merging with a neighbour they touch with a sharp corner.
Note that similar configurations were observed in a full diffusion
simulation of bcc(001) growth \cite{amar99}.

\begin{figure}
\begin{center}
\begin{picture}(100,49)(0,0)
\put(5,5){\resizebox{0.45\textwidth}{0.45\textwidth}{\includegraphics{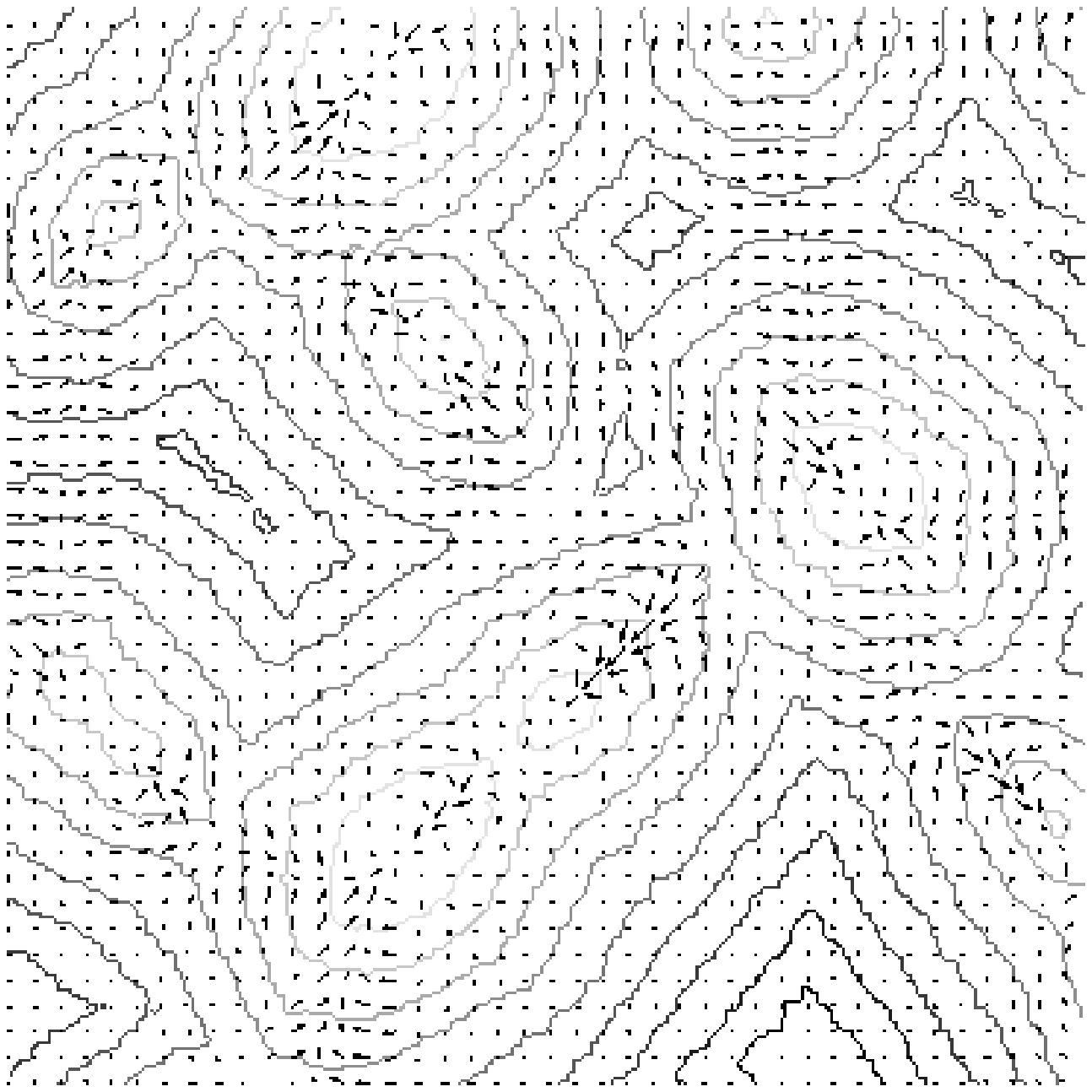}}}
\put(1,47){(a)}
\put(55,5){\resizebox{0.45\textwidth}{0.45\textwidth}{\includegraphics{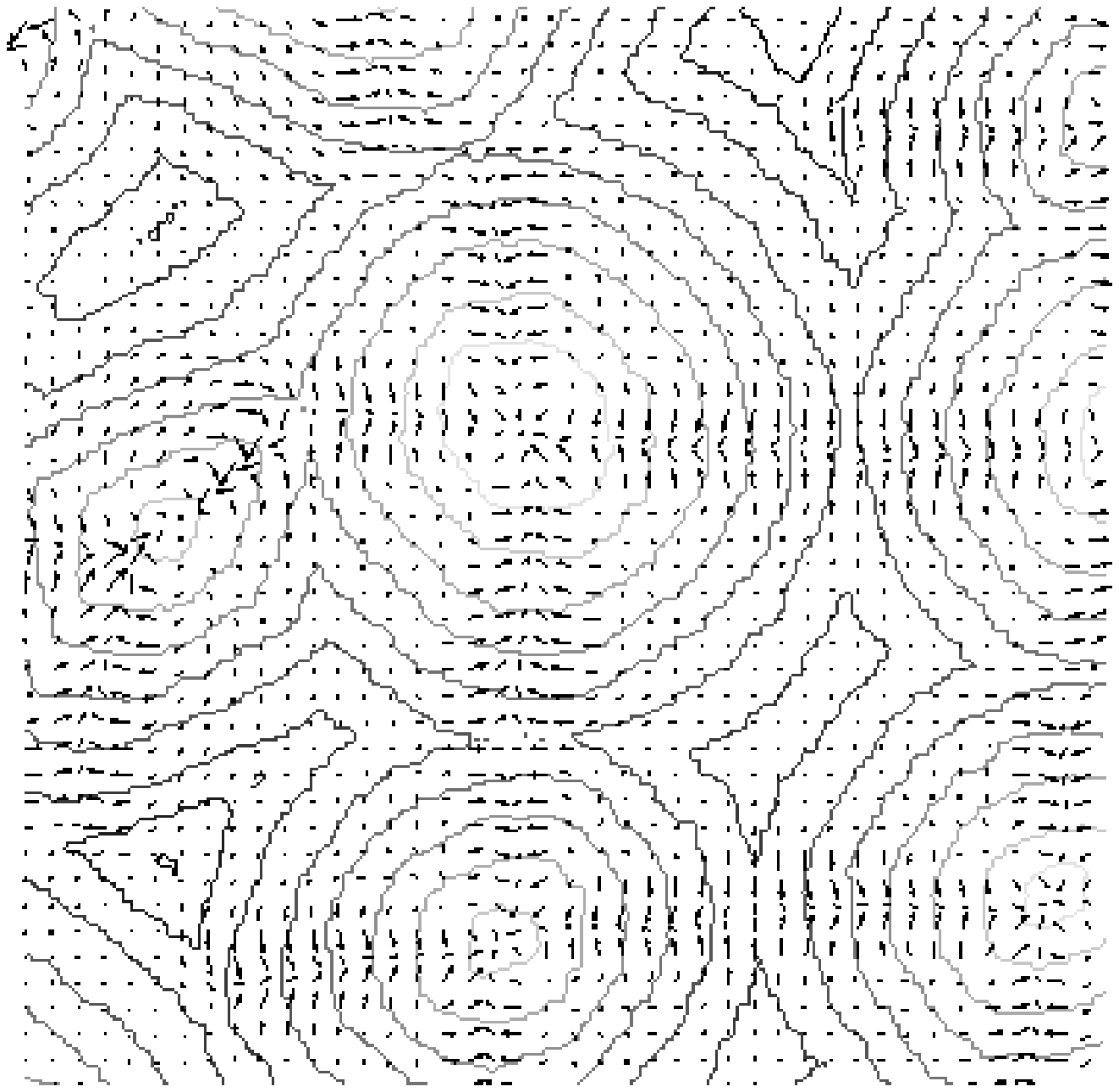}}}
\put(51,47){(b)}
\put(40,3){\vector(1,0){20}}
\put(47.5,0){$[100]$}
\end{picture}
\end{center} 
\caption{Contour plots of parts of size $250 \times 250$ lattice
constants of bcc surfaces ($N = 512$, $l_{d} = 15$, $l_{k} = 10$, $t =
2\cdot 10^4 \,\mathrm{ML}$). Panel (a): section dominated by quickly
coarsening mounds breaking the surface symmetry. Panel (b): metastable
configuration.
\label{coarsefig2}}
\end{figure}
To characterize the morphology quantitatively, we investigate the
slopes $\vec{m} = \nabla h$ of the surface. On average, $|\vec{m}|$
obtains a value which can be estimated by a simple argument
\cite{psk93}: The Schwoebel effect yields an uphill current due to the
preferential attachment of particles to steps from the lower terrace
which is compensated by a downhill current due to funneling. If we
assume that a particle reaches its final height after at most one
incorporation step, the sc and the sh surface select $|\vec{m}| =
1/2$, while bcc and hcp select $|\vec{m}| = 1$.  In practice, the
selected slopes are slightly smaller, since the maximal number of
incorporation steps in the funneling process is unlimited.  Since a
direct computation of the numerical gradient of simulated surfaces
fails due to the discrete heights, we first apply a gaussian filter
with variance $\sigma = 4$.  This value has turned out to be a good
compromise: it removes the atomistic structure but preserves the
shape of the mounds.  On sc and sh, we find a regular square and
hexagon of slightly broadened peaks, respectively, which correspond to
the slope of the flanks of the mounds (figure \ref{coarsehcub}). This
is what one expects at pyramidal mounds.

On bcc and hcp, however, we obtain a more complicated pattern. Figure
\ref{coarsefig1}a shows a histogram of the slopes on hcp surfaces: due
to the round shapes of the mounds, one finds a pronounced maximum of
the probability density function in every direction of $\vec{m}$. The
most probable value of $|\vec{m}|$ has a directional dependence: it is
{\em minimal} in the lattice directions ($0.88 \pm 0.02$), and {\em
maximal} ($0.99 \pm 0.02$) in the intermediary directions.
Consequently, the rounded corners are {\em steeper} than the flanks.
This is possible because the rounding occurs at the base of the
mounds, whereas close to the tips corners are typically sharp,
cf. figure \ref{coarsefig2}.  On bcc the results are equivalent, apart
from the different symmetry.

\section{Diffusion currents}

In order to gain deeper insight into the coarsening process, we
investigate the {\em material transport} on mesoscopic lengthscales by
tracing the motion of particles on the surface. On all lattice
structures, we observe a strong uphill current at the corners of
mounds which is a geometrical effect: an adatom attaching at the
corner is transported over a mean distance $l_{k}$, namely with equal
probability along either of the two flanks of the mound. This results
in a net current $\vec{j}$ in the direction of the bisector of the
angle at the corner, which is directed uphill. A simple calculation
yields
$$ |\vec{j}| = r_{a} l_{k} \cos \frac{\phi}{2} $$ if there is a
straight step edge of length $\gg l_{k}$, which is the case on the
simple lattices (figure \ref{coarsescurrents}). Here, $r_{a}$ is the
rate at which particles attach to a step edge. $\phi$ is the angle
between the two steps which meet at a corner. We have $\phi = \pi/2$
on surfaces with cubic symmetry and $\phi = 2 \pi /3$ on hexagonal
surfaces.

On the bcc and the hcp lattice however, material transport along step
edges is severely restricted due to the rounded corners. There,
particles diffusing along the edge are typically {\em reflected}.
This can be understood from the fact that the slope of the rounded
corners is high. Consequently, there are sites next to a step where an
additional atom would have less than 4 (on bcc) or 3 (on hcp) vertical
neighbours. Since we consider only diffusion to stable sites, an atom
cannot cross such sites.

In the metastable configuration, this leads to a current towards the
middle of the smooth step edges at the flanks of the mounds which have
an extension $\approx l_k$ (figure \ref{coarsefig2}). This current has
a non-vanishing {\em uphill} component due to the Schwoebel effect. On
the rounded corners on the contrary, one observes a weak {\em
downhill} current. This is explained by the observation of steep
gradients (figure \ref{coarsefig1}a), at which the downhill current
due to funneling dominates the uphill current of the Schwoebel
effect. On these surfaces the stable slope at which the currents
cancel is {\em not} assumed {\em locally}, but only in the {\em
global} average, which results in spatial current patterns on the
surface.  An analogous behaviour is observed at the symmetry-breaking,
merging mounds (figures \ref{coarsefig1}b, \ref{coarsefig2}a). Here,
the smooth edges are in the proximity of the sharp corners. Particles
are reflected at the rounded corners, too. This yields a net current
towards the sharp corners which {\em compensates} the geometrically
induced uphill current exactly if the smooth edges have a length
$l_{k}$. This fact explains the broken symmetry of these mounds: if
there is any process which transports matter towards sharp corners,
this material is not completely diffusing inward as it would be the case
on pyramidal mounds but is attached to the corner and makes it
overgrow its neighbours. Merging of mounds, however, {\em is} such a
process: consider a terrace which goes around two mounds touching
each other at the corners. Due to its curvature, in this contact zone
there is a high concentration of kink sites which make it a {\em
sink} for diffusing particles.
 
\section{Theoretical estimation of scaling exponents}

Our observation that mounds merge perferentially in positions with
touching corners gives strong evidence that the current which fills
the gap between two mounds plays a dominant role in the process of
coarsening, at least in the case of large $l_{k}$ where it is
efficient. If this is the case, then a simple consideration yields the
scaling exponents: Due to step edge diffusion, adatoms are transported
over a typical distance $l_{k}$. If $l_{k}$ is noticeably greater than
a few lattice constants, this is much more than the average terrace
width. Then, material transport via diffusion on terraces is small
compared to transport via step edge diffusion and can be neglected.
Our investigation of the diffusion currents shows, that the current
into the gap is significant on a few atomic layers in the vicinity of
the contact point only. In consequence, this current is {\em
independent} of the size of the mounds if the latter is much greater
than $l_{k}$. Since the volume of the gap is proportional to $L^{3}$,
if $L$ is the typical distance between the mounds, it will take a time
$t \sim L^{3}$ to fill it. In consequence, $L \sim t^{1/3}
\Longrightarrow z = 3$.  Since $\alpha = 1$ in the presence of slope
selection, this yields $\beta = 1/3$. In contrast to the effects
discussed in \cite{s98,spz97} these considerations are completely
independent of lattice symmetries. In \cite{t98,tsv97} a similar
argument has been applied to the case of coarsening in the absence of
dominant step edge diffusion processes (``bond energy driven
coarsening''), where material is transported mainly by terrace
diffusion. This yields $\beta = 1/4$. The same exponent is obtained,
if coarsening is exclusively due to fluctuations in the particle beam
(``noise assisted coarsening'').  We expect, that these processes
dominate step edge diffusion in the limit of small $l_{k}$ which leads
to a transient from $\beta = 1/4$ to $\beta = 1/3$ if $l_{k}$ is
increased.

\section{Measurement of scaling exponents}

We apply a variety of methods to measure the scaling exponents. This
is important as a consistency check and to eliminate systematical
errors which might be intrinsic to some methods.  According to
equation \ref{Wdynamicscaling}, $\beta$ can be obtained from the
increase of the {\em surface width} $W(t) \sim t^{\beta}$ \cite{bs95}.
This power law behaviour may be corrupted by the presence of noise on
the surface profile, an additional contribution to $W(t)$ which has
been called {\em intrinsic width} \cite{bs95,kw88}. Similarly, the
{\em number of mounds} $n_{m}$ in the system will decrease like
$t^{-2/z}$, if this scaling hypothesis holds. We measure $n_{m}$ by
counting the number of {\em top terraces} in the system. Since a
single particle is counted as a top terrace, this method may be
misleading if particles nucleate on terraces at the flanks of
mounds. A method which is widely used in the literature uses the
fourier transform $\hat{f}(\vec{k})$ of the reduced surface height
$f(\vec{x})$: The {\em structure factor} $S(\vec{k}) :=
\left<\hat{f}(\vec{k}) \hat{f}(-\vec{k}) \right>$ is maximal at
nonzero wavenumbers $|\vec{k}_{m}| = 2 \pi / l_{m}$, if there are
structures at a typical lengthscale $l_{m}$ on the surface. Since
$l_{m} \sim t^{1/z}$, $|\vec{k}_{m}|$ will decay $\sim t^{-1/z}$.  In
practice, a direct search for the maximum often fails due to noise
effects; one avoids this problem by calculating the averages
$k_{m}^{(p)} := \left( \sum_{\vec{k}} |\vec{k}| S(\vec{k})^{p} \right)
/ \sum_{\vec{k}}S(\vec{k})^{p}$.  However, the choice of the correct
power $p$ is a bit arbitrary.

\subsection{The wavelet method}

To avoid these difficulties, we propose a method which uses the
continuous wavelet transform which was introduced in chapter
\ref{waveletkapitel}. Due to the easier numerical computation, we employ
the two-component version \ref{wavewspec}. 

The basic idea of this transformation is to convolve $h(\vec{x})$ with
the wavelet which is {\em dilated} with the scale $a$. As already
noted in section \ref{waveletdef}, in a wavelet transform where the
radially symmetrical filter $\Phi(\vec{x})$ is a gaussian, the search
for the wavelet transform modulus maxima (WTMM) at scale $a$ is
equivalent to the detection of edges of structures of size $\sim a$
\cite{mh92,mz92}.  This makes it a natural tool to search for the
typical size $a_{t}$ of mounds on the surface. Therefore, in the
remainder of this chapter we use this particular wavelet.
\begin{figure}
\figpanel{\resizebox{0.3\textwidth}{!}{\includegraphics{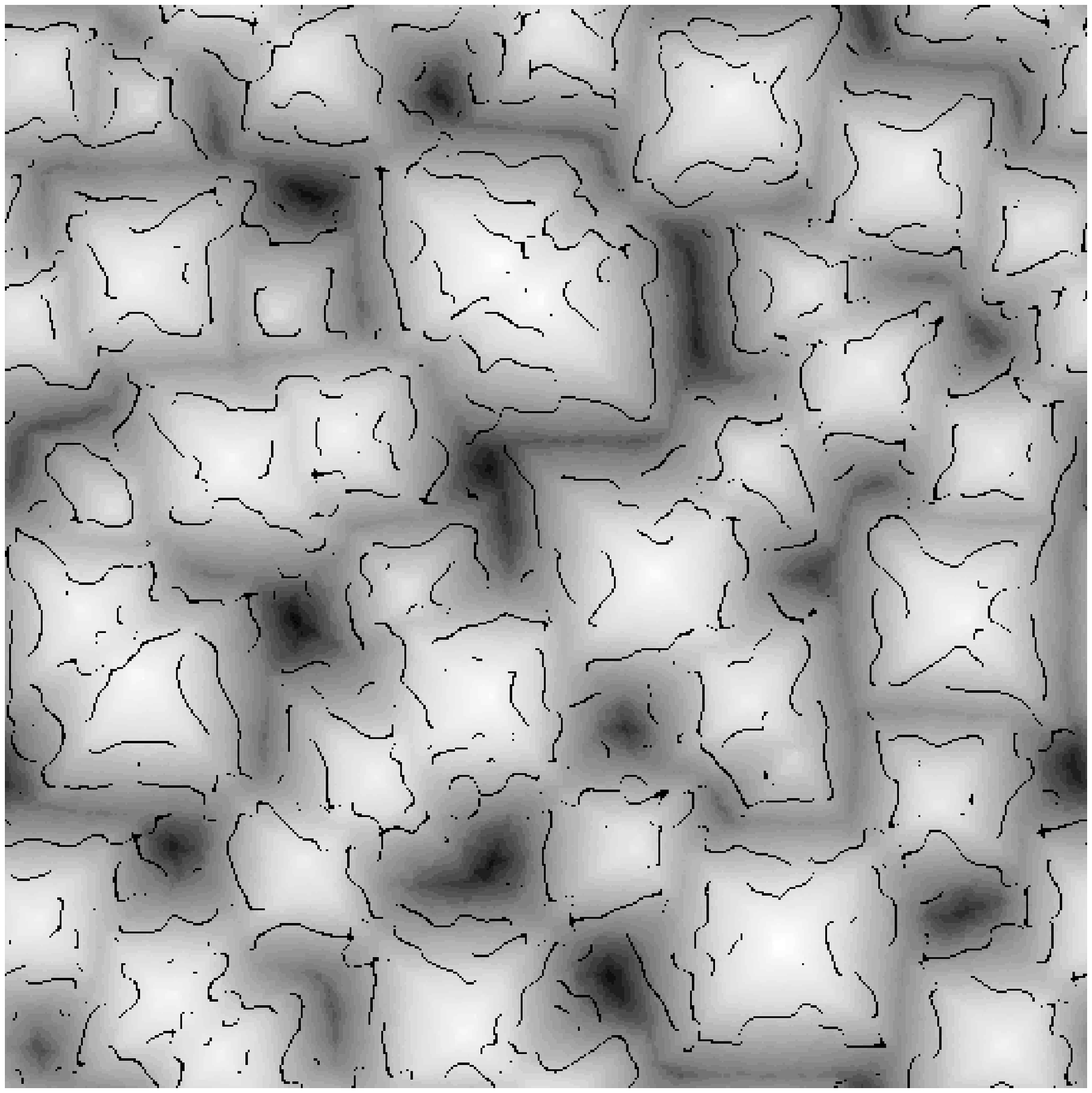}}}{$a
= 5$} 
\hfill
\figpanel{\resizebox{0.3\textwidth}{!}{\includegraphics{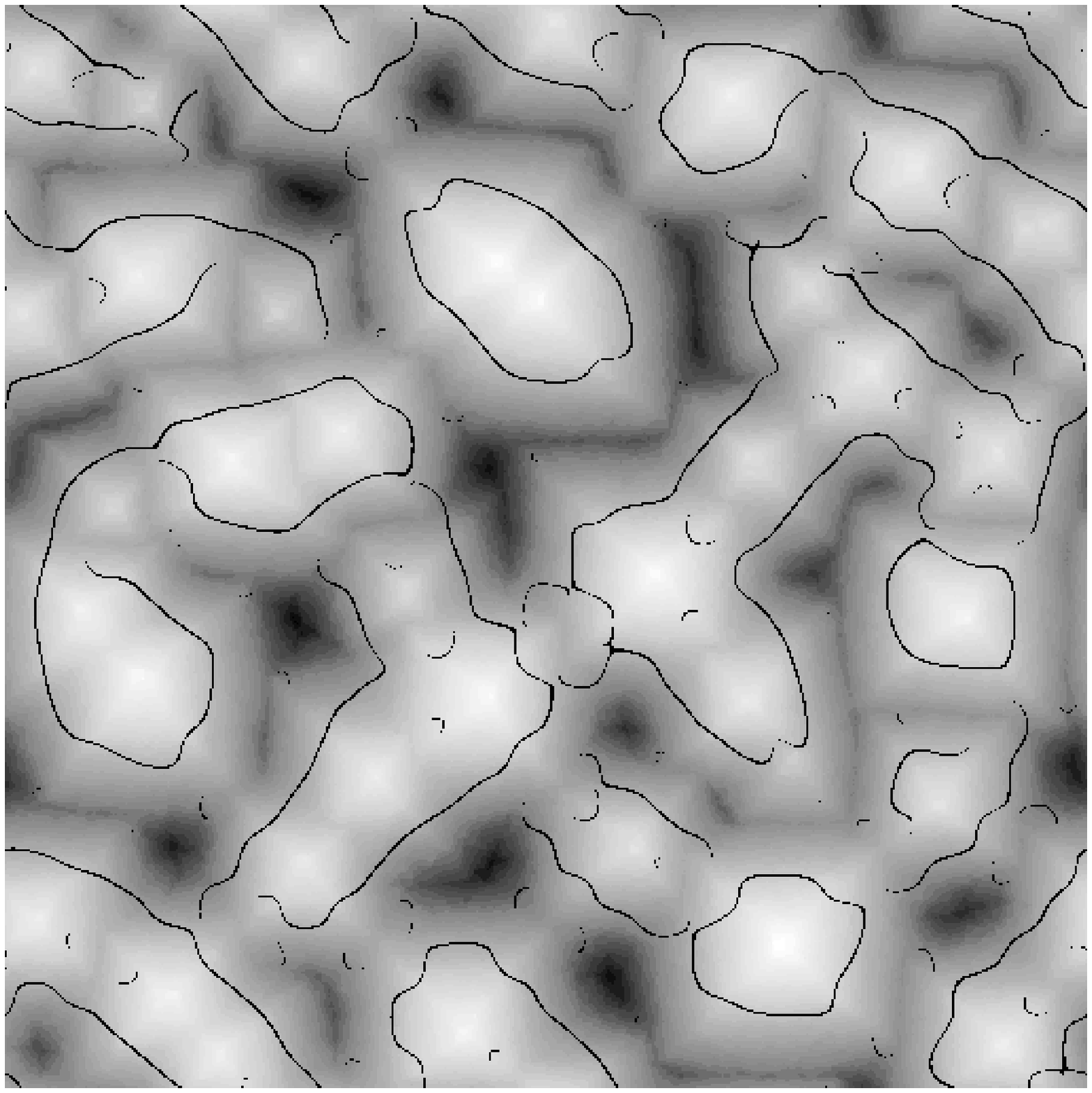}}}{$a
= a_m = 17$} 
\hfill
\figpanel{\resizebox{0.3\textwidth}{!}{\includegraphics{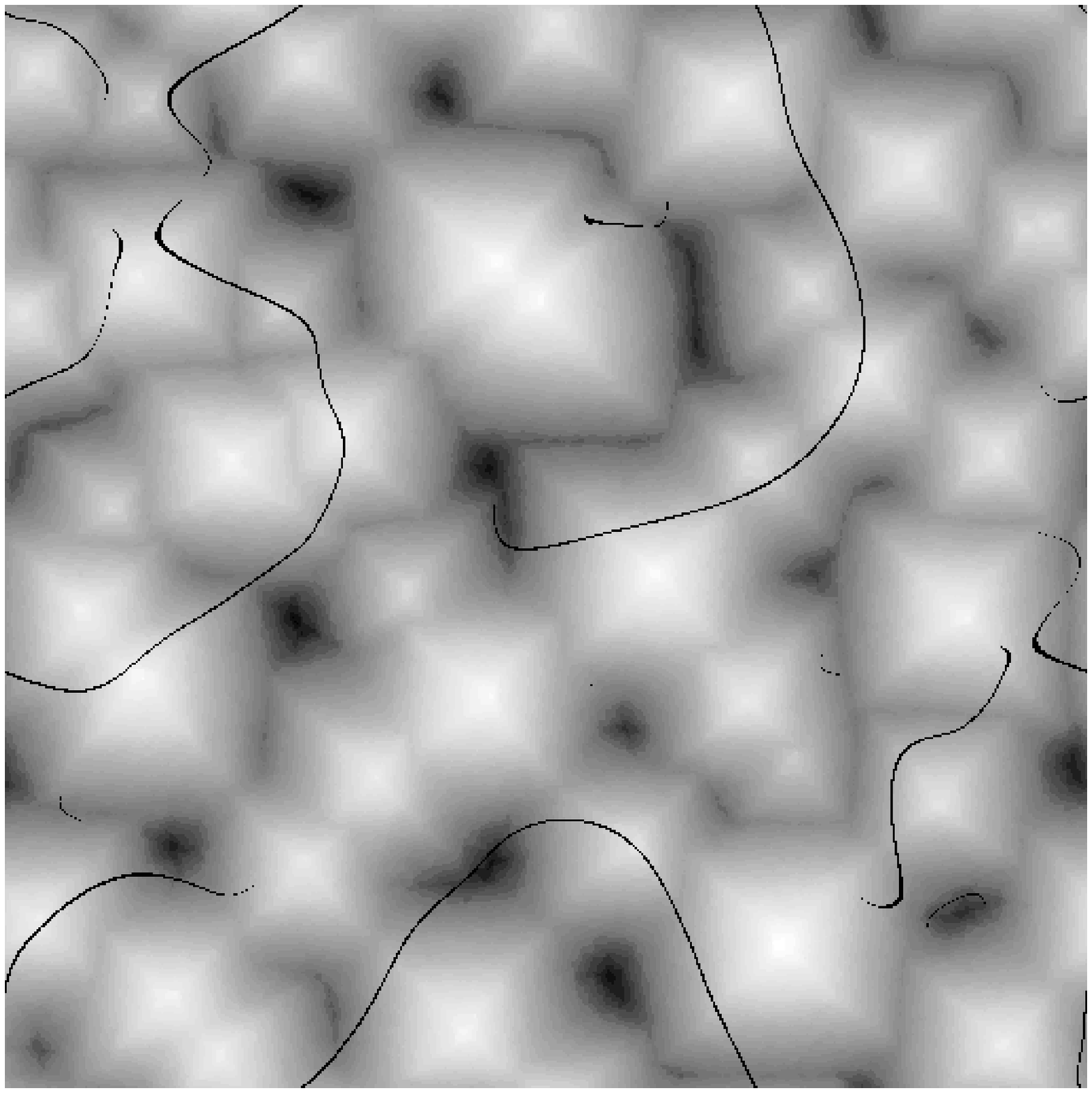}}}{$a
= 80$}
\caption{WTMM at the surface of a simple cubic crystal ($N = 512$,
$l_d = 15$, $l_k = 8$, $t = 3400$ ML) at different scales $a$. The
WTMM are shown in black. At $a = a_m$, the WTMM lines trace the
contours of the mounds. At smaller $a$, the WTMM lines enclose details
of the surface at the flanks of the mounds, whereas at large $a$ the
WTMM seem to be independent of the mounds.
\label{coarsewtmmlines}}
\end{figure}
We investigate the average of the WTMM on the scale $a$
\begin{equation}
\Omega(a) := \left< M_{\vec{\Psi}}[h](\vec{b}, a) \right>_{\vec{b} = 
\mbox{WTMM}}.
\end{equation} 
This function has a pronounced maximum at a value $a_{m}$ which is the
scale that contains the most relevant contributions to the surface
morphology. On mounded surfaces, we expect $a_m \propto a_t$.  As an
example, $\Omega(a)$ of the surface of a simple cubic crystal is shown
in figure \ref{coarsewma}. Figure \ref{coarsewtmmlines} shows the WTMM
of the same surface at different scales including $a = a_m =
17$. Clearly, at $a = a_m$ the WTMM lines trace the contours of the
mounds. On the contrary, at the small scale $a = 5$, the search for
the WTMM yields structures at the flanks of the mounds which are small
compared to the mounds themselves. The positions of the WTMM lines at
$a = 80$ which is much greater than $a_m$ are not related to the
mounds at all. On other lattice structures and at different time, we
obtain similar results. Consequently, $a_m$ is indeed a measure for
the {\em lateral} size of the mounds. Since the convolution of the
surface with the wavelet $\vec{\Psi}_0$ is the gradient of the surface
smoothed with the gaussian filter $\Phi_0$, $\Omega(a_m)$ is a measure
for the {\em height} of the mounds.  During the coarsening process, we
have $a_m \sim t^{1/z}$ and $\Omega(a_m) \sim t^\beta$.
\begin{figure}[htb]
\botbase{\resizebox{0.48\textwidth}{!}{\rotatebox{270}{\includegraphics{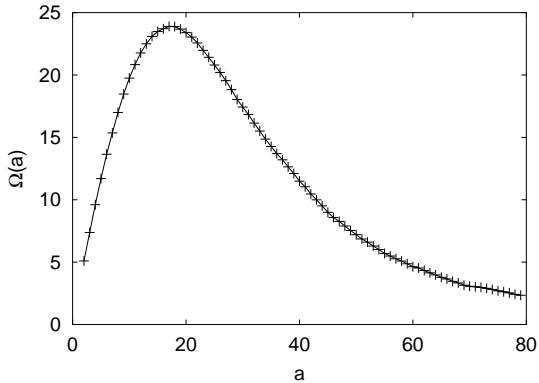}}}}
\hfill
\botbasebox{0.48\textwidth}{\caption{$\Omega(a)$ for the surface shown in figure
\ref{coarsewtmmlines}. There is a pronounced maximum at $a = a_m = 17$
where the WTMM lines enclose the mounds.\label{coarsewma}}}
\end{figure}

This procedure has several advantages compared to the standard methods
used in the literature: (1) The detection of the WTMM leads to an
efficient suppression of noise and therefore eliminates the (noisy)
intrinsic width. (2) Since only the most important lengthscale is
considered, the results may not be corrupted by details of the surface
morphology on small lengthscales. (3) Appropriate wavelets are well
localized both in real space and in Fourier space. Therefore, this
analysis combines the advantages of real space techniques, like the
counting of mounds, and Fourier techniques, like the calculation of
$k_m^{p}$. We have tested this method on various artificial test
surfaces and found a clear superiority to the standard methods,
especially in the presence of noise which should make it interesting
for experimental investigations.

\subsection{Results}

In our simulations, the asymptotic scaling behaviour is obtained after
a long initial transient of about $100$ monolayers. Since after this
transient only a comparatively small number of mounds is left on the
surface, the measurement of exponents is complicated by finite size
effects: due to the periodicity of the system, there is a {\em
self-interaction} of the currents on the mounds, if their size is no
more small compared to the system size.  We measure $\alpha \approx 1$
for all $l_{k} > 1$, a consequence of slope selection.  At $l_{k} = 1$
however, we measure smaller values which {\em do} depend on the
crystal lattice: $\alpha = 0.85$ on bcc and $\alpha = 0.94$ on hcp. A
similar effect was observed in \cite{bkks99,skbk99} on the simple
cubic lattice.  We remark that our results are independent of the
method which was applied to measure the exponents.
 
\begin{figure}[htb]
\botbase{\resizebox{0.48\textwidth}{!}{\rotatebox{270}{\includegraphics{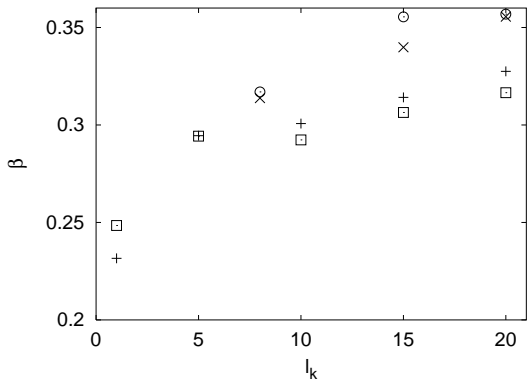}}}}
\hfill \botbasebox{0.48\textwidth}{\caption{$\beta$, as obtained from
the wavelet method, plotted as a function of the step edge diffusion
length. $+$:\ bcc lattice, $\times$:\ sc, $\boxdot$:\ hcp, $\odot$:\
sh. 
Results from measurements of $\beta$ from the surface width are
identical within error bars which are on the order of magnitude of the symbol 
size. These have been estimated from the
logarithmic fits; we expect the true errors to be larger due to
systematical deviations.\label{coarsefig3}}}
\end{figure}
In figure \ref{coarsefig3}, $\beta$ is plotted as a function of
$l_{k}$.  Clearly, its value does {\em not} depend systematically on
the {\em symmetry} of the crystal surface. We find no significant
deviations between $\beta$ on bcc and on hcp (sc and sh). As already
observed in \cite{bkks99,skbk99}, there is a dependence of $\beta$ on
the step edge diffusion length. One obtains values $\approx 1/4$ at
$l_{k} = 1$ and higher values at greater $l_{k}$. The values obtained
at large $l_{k}$ are compatible with $\beta = 1/3$. As indicated
above, we explain this transient as a competition between different
coarsening mechanisms. At large $l_{k}$ the merging of mounds should
proceed mainly by the step edge diffusion driven transport of material
into the gap, at small $l_{k}$ it is dominated by noise assisted
coarsening and/or bond energy driven coarsening. This competition
might also explain why the exponents measured on bcc and hcp are still
slightly smaller than those obtained on the simple lattices even at
values of $l_{k}$ as large as $20$, since the symmetry breaking of the
mounds on these lattices halves the number of corners which are active
in the coarsening process.

All the simulation results reported above have been obtained in
simulations with unhindered step edge diffusion. To further
corroborate our ideas of the coarsening process, we have also
performed simulations with a corner diffusion barrier. A particle may
perform at most $l_{k}^{2}$ diffusion hops along a straight step edge,
but diffusion around corners is forbidden. Then, we observe {\em no
symmetry breaking} of the mounds, which now obtain regular polygonal
shapes on all lattice structures.  These are rotated by an angle of
$45$ degrees against the lattice directions on cubic surfaces and $30$
degrees on hexagonal ones. Due to the absence of aligned step edges,
material transport by step edge diffusion is severely restricted under
these conditions. Thus, we measure only small diffusion currents and
observe no pronounced long-range order in the flux lines. Here, we
find $\alpha \approx 1$, $\beta \approx 1/4$ on {\em all} lattices,
{\em independent} of the step edge diffusion length. This slow
coarsening in absence of the characteristic features of step edge
diffusion driven coarsening---currents on mesoscopic lengthscales and
symmetry breaking on bcc and hcp---strongly supports the important
role of this mechanism for fast coarsening with $\beta = 1/3$.

\section{Conclusions}

In summary, we have presented a detailed investigation of the
coarsening process in epitaxial growth. Our most important finding is
that the crystal lattice has a considerable influence on the
morphology of the growing surface which is by no means restricted to
trivial symmetry effects. In spite of these differences, in the limit
of large step edge diffusion lengths the mounds coarse according to
power laws with universal exponents. We obtain $\beta \approx 1/3$ in
the case of unhindered step edge diffusion. In the case of restricted
step edge diffusion, $\beta \approx 1/4$ is observed as it is for
unhindered step edge diffusion with a small value of $l_{k}$.

Amar \cite{amar99} has obtained values of the coarsening exponents of
bcc(001) surfaces which agree with our results within errorbars. He
performed standard kinetical Monte Carlo simulations which consider
the motion of many particles on the surface simultaneously. However,
with this method only the simulation of the deposition of at most 300
ML was possible, and the initial formation of mounds took up to 100
monolayers. Our effective single particle dynamics, on the contrary,
allows for the simulation of $2\cdot 10^4$ ML such that coarsening can
be investigated over two decades in time. Thus, our simulations give
strong evidence that the mesured values of $\alpha$ and $\beta$ are
indeed the dynamic exponents of the asymptotic regime. Conversely, the
agreement of the results proves that the simplifications made in the
effective single particle dynamics do not alter the values of the
coarsening exponents.

Our results contradict previous studies of the coarsening process
using continuum equations \cite{mg00,rk96,s98,spz97} which predict
slower coarsening on surfaces with a cubic symmetry. In these
publications, equations have been studied which are invariant under
the transformation $h(\vec{x}) \rightarrow -h(\vec{x})$. This symmetry
reflects itself also in the up-down-symmetry of the solutions of the
equations. Clearly, our simulation results {\em break} this symmetry
and the arguments of \cite{rk96,s98,spz97} cannot apply here.  In
figures \ref{coarsescurrents}, \ref{coarsefig1}b and \ref{coarsefig2}
the contours of the mounds and those of the valleys are clearly
distinct. This is also true for the {\em currents}, especially the
step edge diffusion current which dominates the coarsening behaviour
in the limit of large $l_{k}$ and unhindered step edge diffusion and
determines the scaling exponents.  We conclude, that the breaking of
the up-down symmetry is a central feature of unstable epitaxial
growth. Its precise effect on the coarsening dynamics clearly deserves
further attention.

However, we find it difficult to interpret our result in the context
of an analogy to phase ordering processes, where the local slope of
the surface corresponds to an order parameter. In this picture, the
importance of the breaking of up-down symmetry leads to the
conclusion, that now the stability of a domain wall between areas of
equal slope is a complicated function of both the {\em orientation} of
the wall and the {\em order parameters} on both sides of it.
Additionally, in contrast to the simple lattices where there are only
4 (sc) respectively 6 (sh) stable slopes, on bcc and hcp one would
have to deal with a {\em continuous} order parameter.

In any case, the behaviour of our models which implement the
microscopic processes on the surface is governed by much more complex
rules than those implemented in all the continuum models we are aware
of. Differences are by no means restricted to some details, but have a
fundamental influence on the behaviour of the system on large
lengthscales.
  
\appendix

\chapter{Transfer matrices and finite size scaling \label{tmappendix}}

In principle, the transfer matrix method \cite{cf72,t99} is a
technique to calculate the free energy of {\em one-dimensional}
lattice systems exactly in the thermodynamic limit. However, a system
of arbitrary dimension can be treated formally as a one-dimensional
system if it is decomposed in slices along an arbitrary
direction. Each slice is then treated as one site in an infinite chain
of slices. Since the transfer matrix method can consider only a {\em
finite}, however large, number of possible states per site, the size
of the slices must be finite. Therefore, in general we can investigate
only semi-infinite systems which are finite in all directions but
one. Only in some special cases, like the two-dimensional Ising model
in the absence of a magnetic field, the limit of a real infinite
system can be performed exactly. Otherwise, in two or more dimensions
we must apply finite size scaling techniques to extrapolate to the
thermodynamic limit from results obtained for semi-infinite systems.

\section{The transfer matrix method}

Consider a chain of $N$ identical slices of a system of arbitrary
dimension. Let $\mathcal{M}$ denote the number of sites per
slice. Each slice $j$ can be in one of $\tilde{n}$ different states
$X_j$.  We assume that there are interactions only between
neighbouring slices. Interactions of longer but finite range can be
considered by increasing the thickness of the slices. Then, the energy
of the system can be written as
\begin{equation}
H = \tilde{H}(X_N, X_1) + \sum_{j = 1}^{N-1} \tilde{H}(X_j, X_{j+1})
\label{hamiltonzeilen}
\end{equation}
where we have assumed periodic boundary conditions. For simplicity, we
assume without loss of generality that $\tilde{H}(X, Y) = \tilde{H}(Y,
X)$. It is always possible to rearrange the Hamiltonian of the system
such that this property is fulfilled. The partition function is
\begin{equation}
Z = \sum_{\{X_j\}_{j=1}^{N}} 
\exp \left[ -\beta \tilde{H}(X_N, X_1) \right] \prod_{j=1}^{N-1}
\exp \left[ -\beta \tilde{H}(X_j, X_{j+1}) \right]. 
\end{equation} 
To calculate it, we map each of the $\tilde{n}$ possible
configurations of a slice to one of $\tilde{n}$ orthonormal vectors,
which span up a $\tilde{n}$-dimensional Hilbert space:
\begin{equation}
X \longleftrightarrow \ket{X} \; \; \mbox{such that} \; \;
\sprod{X}{Y} = \left\{
\begin{array}{ll}
1 & \mbox{if}\;  X \equiv Y \\
0 & \mbox{else}
\end{array}
\right. . 
\end{equation}
In the base of these vectors we define the $\tilde{n} \times
\tilde{n}$ transfer matrix
\begin{equation}
T = \sum_{\ket{X}, \ket{Y}} \ket{X} \exp \left[ -\beta \tilde{H}(X, Y)
\right] \bra{Y},
\label{transfermatrixdefinition}
\end{equation}
which has the property that the trace of $T^N$ is equal to the
partition function:
\begin{eqnarray*}
\mbox{Tr}\left[T^N\right] & = & \sum_{\ket{X_1}} \bra{X_1} T^N \ket{X_1} 
\\
&=& \sum_{\ket{X_1},...,\ket{X_N}} \bra{X_1} T \ket{X_2} \bra{X_2} T \ket{X_3}
\cdots \bra{X_N} T \ket{X_1}\\ 
&=& \sum_{\ket{X_1},...,\ket{X_N}} 
\exp \left[ -\beta \tilde{H}(X_N, X_1) \right]
\prod_{j=1}^{N-1}
\exp \left[ -\beta \tilde{H}(X_j, X_{j+1}) \right] \\
&=& Z 
\end{eqnarray*}
Consequently, the partition function can be expressed in terms of the
eigenvalues $\lambda_i$ of $T$. Since $\tilde{H}(X,Y) =
\tilde{H}(Y,X)$, $T$ is symmetric and all $\lambda_i$ are real. We
have
\begin{equation}
Z = \mbox{Tr} \left[ T^N \right] = \sum_{i=1}^{\tilde{n}} \lambda_i^N.
\end{equation}
This identity can be used to calculate the free energy per site $f$
in the limit of infinite $N$. The theorem of Perron-Frobenius states
that the eigenvalue with the greatest absolute value is positive and
nondegenerate.  In the following, we will assume that the eigenvalues
are sorted in descending order such that $\lambda_1 > \lambda_2 \geq
... \geq \lambda_{\tilde{n}}$.  Then,
\begin{equation}
f = -\frac{1}{\beta N \mathcal{M}} \ln Z = - \frac{1}{\beta N
\mathcal{M}} \ln \left[ \lambda_1^N \left( 1 + \sum_{i=2}^{\tilde{n}}
\left(\frac{\lambda_i}{\lambda_1} \right)^N \right) \right]
\stackrel{N \rightarrow \infty}{\longrightarrow} -\frac{1}{\beta
\mathcal{M}} \ln \lambda_1.
\label{freieenergietm}
\end{equation} 
Thus, we can calculate the free energy of a semi-infinite system by
calculating the greatest eigenvalue of the $\tilde{n} \times
\tilde{n}$-dimensional transfer matrix $T$. Usually this is done
numerically since $\tilde{n}$ increases exponentially fast with the
number of sites per slice.

\subsection{Thermal averages} 
The standard method in statistical physics to calculate the thermal
average $\left< q \right>$ of an arbitrary intensive quantity $q$ uses
a field $h$ which couples to $q$. Then,
\begin{equation}
\left< q \right> = \lim_{h \rightarrow 0} \frac{\partial f_q}{\partial h}
\label{mittelstandard}
\end{equation}
where $f_q$ is the free energy of the system with the modified
Hamiltonian $H_q = H + h N \mathcal{M} q$.  Of course, this method can
be applied straightforwardly in the transfer matrix formalism by
performing the differentiation with respect to $h$ numerically.

There is however an alternative which can be applied whenever $q$ is
well defined as a function $q(X_k)$ of the state of a single slice
$k$.  Then, since all slices are identical, the thermal average of $q$
is the same in all slices and identical to the thermal average of $q$
in the whole system.  In a chain of $N$ slices with periodic boundary
conditions
\begin{eqnarray}
\left< q \right> &=& \frac{1}{Z} \sum_{\{X_i\}_{i=1}^{N}} q(X_k) \exp
\left[ -\beta \tilde{H}(X_N, X_1)\right] \prod_{i =1}^{N-1} \exp
\left[ -\beta \tilde{H} (X_i, X_{i+1}) \right] \nonumber \\ 
&=& \frac{1}{Z} \sum_{\ket{X_1}, \ket{X_k}} \bra{X_1} T^{k-1}
\ket{X_k} q(X_k) \bra{X_k} T^{N-k+1} \ket{X_1}.
\end{eqnarray}
The transfer matrix $T$ can be written as $ T = \sum_{i=1}^{\tilde{n}}
\ket{\lambda_i} \lambda_i \bra{\lambda_i}, $ where $\ket{\lambda_i}$
is the normalized eigenvector to the eigenvalue $\lambda_i$.  Further,
we introduce the matrix
\begin{equation}
\hat{q} = \sum_{\ket{X}} \ket{X} q(X) \bra{X}
\end{equation}
which will be denoted as the operator of $q$ in the following. Then,
\begin{equation}
\left< q \right> = \frac{1}{Z} \sum_{i=1}^{\tilde{n}} \lambda_i^N
\bra{\lambda_i} \hat{q} \ket{\lambda_i} \stackrel{N \rightarrow
\infty}{\longrightarrow} \bra{\lambda_1} \hat{q} \ket{\lambda_1},
\label{mittelwerttm}
\end{equation}
where we have used that $\sprod{\lambda_i}{\lambda_j} = \delta_{i,j}$
and $\sum_{\ket{X_1}} \ket{X_1} \bra{X_1} = 1$. Thus, the thermal
average of $q$ can be calculated easily from the normalized
eigenvector $\ket{\lambda_1}$.

In particular, the probability $P(X)$ to find a slice in state $X$ is
the thermal average of $n_X$ which is defined as one, if the slice is
in state $X$ and zero otherwise.  Its operator is $\hat{n}_X = \ket{X}
\bra{X}$.  Using equation \ref{mittelwerttm}, we find
\begin{equation}
P(X) = \sprod{\lambda_1}{X} \sprod{X}{\lambda_1} = \left|
\sprod{\lambda_1}{X} \right|^2.
\end{equation}
The square of every component of the normalized eigenvector to the
greatest eigenvalue of $T$, represented in the base of the vectors
$\ket{X}$, is the probability to find a slice in a particular state.

\subsection{The correlation length}

If we measure an arbitrary quantity $q$ simultaneously in two
different slices, due to the interactions between the slices these
values are not statistically independent.  This fact is measured
by the correlation function
\begin{equation}
g_q(|l-k|) := \left< q(X_k) q(X_l) \right> - \left< q \right>^2. 
\end{equation}
Nonzero values indicate correlations between $q(X_k)$ and $q(X_l)$.
The thermal average of $q(X_k) q(X_l)$ in a chain of $N$ slices can be
expressed in terms of the transfer matrix:
\begin{eqnarray}
\left< q(X_k) q(X_l) \right> &=& \frac{1}{Z} \sum_{\{X_i\}_{i=1}^{N}}
q(X_k) q(X_l) \exp \left[ -\beta \tilde{H}(X_N, X_1) \right]
\prod_{i=1}^{N-1} \exp \left[ -\beta \tilde{H}(X_i, X_{i+1}) \right]
\nonumber \\ & = & \frac{1}{Z} \sum_{\ket{X_1}} \bra{X_1} T^{k-1}
\hat{q} T^{l-k} \hat{q} T^{N-l+1} \ket{X_1}.
\end{eqnarray}
For simplicity, we have assumed that $l > k$. We continue by
expressing the transfer matrix in terms of its eigenvectors and
eigenvalues and performing the limit $N \rightarrow \infty$.  We find
\begin{eqnarray}
\left< q(X_k) q(X_l) \right> &=& \frac{1}{Z} \sum_{i,j=1}^{\tilde{n}}
\lambda_i^{k-1} \bra{\lambda_i} \hat{q} \ket{\lambda_j}
\lambda_j^{l-k} \bra{\lambda_j} \hat{q} \ket{\lambda_i}
\lambda_i^{N-l+1} \nonumber \\ 
&\stackrel{N \rightarrow
\infty}{\longrightarrow}& \bra{\lambda_1} \hat{q} \ket{\lambda_1}^{2}
+ \sum_{j=2}^{\tilde{n}} \left(\frac{\lambda_j}{\lambda_1}
\right)^{l-k} \left| \bra{\lambda_1} \hat{q} \ket{\lambda_j} \right|^2
\label{quhkaell}
\end{eqnarray} 
The first term in equation \ref{quhkaell} yields $\left< q
\right>^2$. If $l-k \gg 1$, the sum is dominated by the term
where the {\em absolute} value of $\lambda_j$ is maximal and the
matrix element $\bra{\lambda_1} \hat{q} \ket{\lambda_j} \neq 0$.  In
the following, we denote this eigenvalue as $\lambda_m$ and the
corresponding eigenvector as $\ket{\lambda_m}$. This yields the
following asymptotics for the correlation function:
\begin{equation}
g_q(|l-k|) \sim \left| \bra{\lambda_1} \hat{q} \ket{\lambda_m}
\right|^2 \left( \frac{\lambda_m}{\lambda_1} \right)^{|l-k|}
\end{equation}
If $\lambda_m$ is positive, $g_{q}(|l-k|)$ decays exponentially. On
the other hand, if $\lambda_m$ is negative, $g_q(|l-k|)$ oscillates
with an exponentially decaying amplitude between positive and negative
values.  This indicates that the values of $q$ in neighbouring slices
are anticorrelated.  In both cases, the amplitude decays on a typical
lengthscale
\begin{equation}
\xi = \left[ \ln \left( \left| \frac{\lambda_1}{\lambda_m} \right|
\right)\right]^{-1}
\label{korrelationslaenge}
\end{equation}
which is called {\em correlation length}. 

\subsection{Symmetries of the Hamiltonian and vanishing matrix elements}

The transfer matrix formalism is based on the assumption that the
Hamiltonian of the investigated system is invariant under translations
of the system in the infinite direction. However, frequently both the
Hamiltonian and the operator of a quantity $q$ fulfil additional
symmetry properties. These can be used to predict which of the matrix
elements $\bra{\lambda_i} \hat{q} \ket{\lambda_j}$ are zero.  Let
there be an operator $\hat{U}$ which maps each configuration of a
slice uniquely to another one. We assume that the energy of the
system is invariant under this transformation. Since the transfer
matrix has the same symmetry properties as the Hamiltonian, we have
\begin{equation}
\hat{U}^\top \hat{U} = 1 \label{symmeins}\\
\end{equation}
\begin{equation}
\hat{U} T \hat{U}^\top = \hat{U}^\top T \hat{U} = T \label{symmzwei}.
\end{equation}
Further, we assume that the operator $\hat{q}$ is either symmetric or
antisymmetric under the transformation $\hat{U}$ such that
\begin{equation}
\hat{U} \hat{q} \hat{U}^\top = \chi_{\hat{q}} \hat{q}\; \;
\mbox{where} \; \; \chi_{\hat{q}} \in \{-1, 1\}.
\label{symmq}
\end{equation}
Applying equations \ref{symmeins} and \ref{symmzwei} to the product of
the transfer marix with its eigenvector $\ket{\lambda_i}$ we find
$$
\hat{U} T \ket{\lambda_i} = \hat{U} T \hat{U}^\top \hat{U}
\ket{\lambda_i} = T \hat{U} \ket{\lambda_i} = \lambda_i \hat{U}
\ket{\lambda_i}.
$$
Consequently, $\hat{U} \ket{\lambda_i}$ is an eigenvector to the
eigenvalue $\lambda_i$ of the transfer matrix.  In case of a
non-degenerate eigenvalue $\lambda_i$, this condition implies
\begin{equation}
\hat{U} \ket{\lambda_i} = \chi_i \ket{\lambda_i} \;\; \mbox{where}
\;\; \chi_i \in \{ -1, 1\},
\label{symmeigenvektor}
\end{equation}
since due to equation \ref{symmeins} the multiplication with $\hat{U}$
leaves the norm of $\ket{\lambda_i}$ invariant.  If $\lambda_i$ is
degenerate, the eigenvectors of $T$ are not uniquely determined.
However, it is always possible to construct a set of orthonormal
eigenvectors such that equation \ref{symmeigenvektor} is
fulfilled. Vectors with $\chi_i = 1$ are {\em symmetric} under the
transformation $\hat{U}$. $\chi_i = -1$ characterizes an {\em
antisymmetric} vector.  Using equations \ref{symmeins}, \ref{symmq}
and \ref{symmeigenvektor}, we find
\begin{equation}
\bra{\lambda_i} \hat{q} \ket{\lambda_j} = \bra{\lambda_i} \hat{U}^\top
\hat{U} \hat{q} \hat{U}^\top \hat{U} \ket{\lambda_j} = \chi_i \chi_j
\chi_{\hat{q}} \bra{\lambda_i} \hat{q} \ket{\lambda_j}.
\end{equation} 
For a nonzero matrix element $\bra{\lambda_i} \hat{q}
\ket{\lambda_j}$, this equation can be fulfilled only if $\chi_i
\chi_j \chi_{\hat{q}} = 1$.  Consequently, the matrix elements of a
{\em symmetric} operator $\hat{q}$, where $\chi_{\hat{q}} = 1$ vanish
whenever the eigenvectors $\ket{\lambda_i}$ and $\ket{\lambda_j}$ have
different symmetry properties under the transformation
$\hat{U}$. Conversely, for an {\em antisymmetric} operator
($\chi_{\hat{q}} = -1$) all matrix elements between eigenvectors of
equal symmetry must be zero.

\subsection{Application to two-dimensional lattice systems 
\label{tmgittergas}}

\begin{figure}
\includegraphics{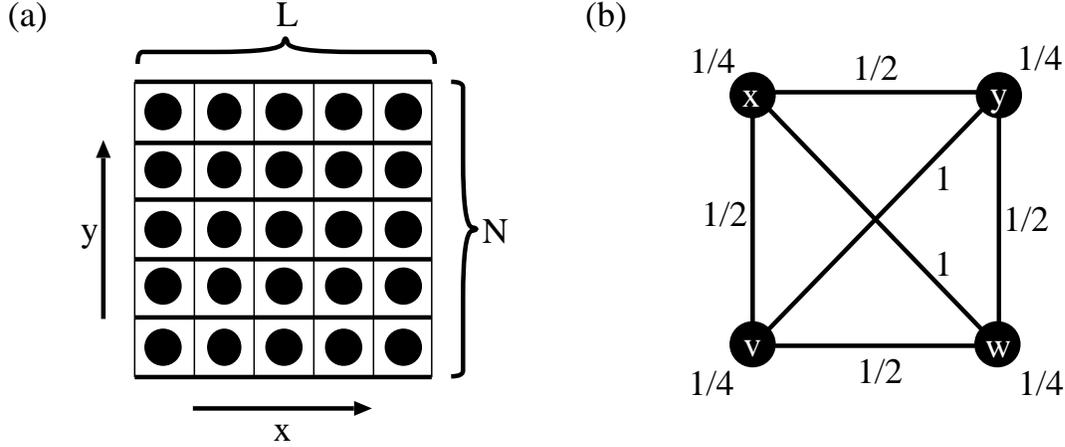}
\caption{Sketch of a square lattice system as discussed in section
\ref{tmgittergas}.  Panel (a) shows the decomposition of the system in
lines with $L$ sites each. Panel (b) shows the unit cell of the
lattice. Sites are shown as black circles. Possible interactions are
symbolized by lines. Note, that each of the sites belongs to four unit
cells.  The nearest neighbour bonds belong to two unit cells.
Therefore, in the calculation of $h(v, w, x, y)$, the energies of
these objects have to be weighted with the factors shown in the figure
to avoid multiple counting of energies in equation
\ref{hamiltonquadratgitter}.
\label{einheitszelletm}}
\end{figure}
Now we apply this formalism to a special class of models which are
defined on a two-dimensional square lattice.  Each lattice site can be
in one of $n$ different states.  There are interactions only between
nearest neighbour and diagonal neighbour sites.  The models discussed
in section \ref{dreikapitel} belong to this class.  We consider a
rectangular system with an extension of $L$ lattice constants in the
$x$-direction and $N$ lattice constants in the $y$-direction as shown
in figure \ref{einheitszelletm}a.  The boundary conditions are assumed
to be periodic.

To investigate the limit $N \rightarrow \infty$ of an infinite strip
with the transfer matrix method, we decompose this system into slices
which consist of a single line of sites in the $x$-direction.  The
configuration $X$ of such a line is determined by the states of the
$L$ sites $(x_1, x_2, ..., x_L)$.  Since there are $n$ possibilities
for each of the $x_i$, we have $\tilde{n} = n^L$ different
configurations of a line.  The transfer matrix formalism requires a
mapping of these configurations to a set of vectors such that
$\sprod{X}{Y} = \delta_{x_1, y_1} \cdot \delta_{x_2, y_2}\cdots
\delta_{x_L, y_L}$ for any two vectors $\ket{X}$, $\ket{Y}$ which
represent configurations $X$, $Y$.  For this purpose, we introduce
real $n^L$-dimensional unit vectors
\begin{equation}
\ket{X} = \left(\delta_{1, \zeta(X)}, \delta_{2, \zeta(X)}, ...,
 \delta_{n^L, \zeta(X)} \right)^\top ,
\label{xzweidkonkret}
\end{equation}
where $\zeta(X)$ is a function which maps each configuration uniquely
to an integer $\in \{1, ..., n^L\}$. To define such a function, we
assume without loss of generality that the $x_i$ can have values from
$0$ to $n-1$. Then, we can read each configuration of a line as a
number $\in \{0, ..., n^L -1\}$ written in a positional notation with
radix $n$. Consequently,
\begin{equation}
\zeta(X) = 1 + \sum_{i=1}^{L} x_i n^{i-1}
\label{zetavonxkonkret}
\end{equation}
has the desired properties. It is convenient to use the notation
$\ket{X} \equiv \ket{x_1, ..., x_L}$ which shows the configuration of
the line that is represented by $\ket{X}$ directly. Note, that the
$x_i$ are {\em not} the components of $\ket{X}$!

The Hamiltonian of the system can be written in the form of equation
\ref{hamiltonzeilen}, where
\begin{equation}
\tilde{H}(X, Y) = h(x_L, x_1, y_L, y_1) + \sum_{i=1}^{L-1} h(x_i,
x_{i+1}, y_i, y_{i+1}).
\label{hamiltonquadratgitter}
\end{equation}
Here, $h(v, w, x, y)$ is the total energy of a unit cell as shown in
figure \ref{einheitszelletm}b.  Inserting equation
\ref{hamiltonquadratgitter} into equation
\ref{transfermatrixdefinition}, we obtain the transfer matrix
\begin{equation}
T = \sum_{\ket{X}, \ket{Y}} \ket{x_1, ..., x_L} t(x_L, x_1, y_L, y_1)
\prod_{i=1}^{L-1} t(x_i, x_{i+1}, y_i, y_{i+1}) \bra{y_1, ..., y_L},
\end{equation}
where we have introduced $t(v, w, x, y) = \exp(-\beta h(v, w, x, y))$.

We are interested in the extremal eigenvalues of $T$ and the
corresponding eigenvectors.  Since $T$ is a $n^L \times n^L$ matrix,
the number of its elements, $n^{2L}$, increases extremely fast with
$L$.  Therefore, any approach where the transfer matrix is stored in
computer memory is limited to very small strip widths $L$. However,
there are algorithms for the computation of extremal eigenvalues which
do not need the matrix itself but only a method to compute the product
of the investigated matrix with an arbitrary column vector. One of
these is the Lanczos algorithm which we use in this work. We refer the
reader to the numerical literature \cite{sb93} for a description of this
method.  In this context, it is important to note that the transfer
matrix can be written as a product of $L$ matrices which are extremely
sparse \cite{na78,t99}:
\begin{equation}
T = T_L T_{L-1} \cdots T_1
\end{equation}
The $n^{L+2} \times n^{L}$ matrix
\begin{equation}
T_1 = \sum_{\ket{X}} \sum_{y_1, y_{L+2}} \ket{y_1, x_1, ..., x_L,
y_{L+2}} t(y_{L+2}, y_1, x_L, x_1) \bra{x_1, ..., x_L}
\end{equation}
maps the $n^L$-dimensional space which is spanned up by the vectors
$\{\ket{x_1, ..., x_L}\}$ to a $n^{L+2}$-dimensional space.  An
orthonormal base which spans up this space can be constructed using a
mapping of the configurations of a line of $L+2$ sites to vectors
similar to that defined in equations \ref{xzweidkonkret} and
\ref{zetavonxkonkret}. The matrix elements are then defined in terms
of the configuration both of the line of $L$ sites and the line of
$L+2$ sites. Note, that $T_1$ has only $n^2 \cdot n^L$ nonzero
elements. 
Similarly, we have $L-2$ matrices of dimension $n^{L+2} 
\times n^{L+2}$
\begin{equation}
T_i = \sum_{\ket{X}} \sum_{y_i} \ket{x_1, ..., x_{i-1}, y_i, x_{i+1},
..., x_{L+2}} t(x_{i-1}, y_i, x_i, x_{i+1}) \bra{x_1, ..., x_{L+2}}
\end{equation}
where $i = 2, ..., L-1$. Each of these matrices has $n \cdot n^{L+2}$
nonzero elements.  Finally, we have the $n^L \times n^{L+2}$ matrix
\begin{equation}
T_L = \sum_{\ket{X}} \ket{x_1, ..., x_{L-1},x_{L+2}} t(x_{L-1},
x_{L+2}, x_2, x_{L+1}) \bra{x_1, ..., x_{L+2}}
\end{equation}
which has $n^{L+2}$ nonzero elements. It is easy but lengthy to verify
that the product of these matrices yields the transfer matrix.  Using
a sparse representation, we need to keep only the nonzero elements to
store the $T_i$ in computer memory.  The $T_i$ altogether have
$n^{L+2}\left[ 2 + \left( L-2 \right) n \right]$ nonzero elements,
which is significantly less than the $n^{2L}$ elements of $T$ itself.
Since the product of the transfer marix with an arbitrary vector
$\ket{v}$ can be performed by successively multiplying $\ket{v}$ with
the $T_i$, we can save a great amount of memory and CPU time by using
this factorization.  On our computer equipment, we can deal with strip
widths up to $L=18$ for $n=2$ and $L=10$ for $n=3$.
		
\section{Finite size scaling}

Using the transfer matrix formalism, the thermodynamic properties of
semi-infinite systems can be calculated exactly.  However, we are
interested in the thermodynamic limit of a system which is infinite in
{\em all} directions. In particular, we want to study the phase
transitions of the infinite system.  A phase transition is an abrupt
change of the order of the system in dependence of a parameter like
temperature. At continuous transitions and first order transitions
which are investigated in this thesis, the correlation length diverges
to infinity. Additionally, there is a singularity in the free energy.
From equation \ref{korrelationslaenge}, it follows that a diverging
correlation length requires a degenerate greatest eigenvalue of the
transfer matrix.  Consequently, there are no phase transitions in
semi-infinite systems since there the greatest eigenvalue of the
transfer matrix is always nondegenerate. On the other hand, it is
reasonable to assume that we can draw conclusions on the infinite
system from results obtained for a semi-infinite system with
sufficiently large slices. The theory of finite size scaling describes
how this can be done. The procedure is different for continuous phase
transitions and first order transitions.

\subsection{Continuous phase transitions
\label{finitesizekontinuierlich}}

In continuous phase transitions, the order parameter $m$ which
measures the order of the system decreases continuously to zero as the
critical temperature $T_c$ is approached from below. There is no
latent heat and no interface tension between the low temperature phase
and the high temperature phase. The description of this kind of
transition is based on the observation that the system is self-similar
at the critical point, i.e.\ the statistical features are invariant
under rescaling. The method which is presented in this section has
been suggested by Nightingale \cite{n82,na78} as ``phenomenological
renormalization group theory''.

We introduce the parameter $\epsilon := T - T_c$\footnote{In
principle, the temperature might be replaced by any other parameter
which can be tuned to reach the critical point, e.g.\ a chemical
potential.}.  Additionally, we consider an ordering field $h$ which
couples to the order parameter $m$. Then, in the infinite system the
critical point is at $\epsilon = h = 0$. For simplicity, we assume an
equal system size $L$ in all finite directions. We divide the system
into hypercubes with a linear dimension of $b$ lattice sites. For
small enough $\epsilon, h$ we have $b \ll \xi$. We map each block to a
single site in a system of size $L/b$ which might be done e.g.\ by a
majority decision among the sites in the block. Following
renormalization group theory, the free energy of the interactions
between the blocks can be approximated by the free energy of the
rescaled system at a different temperature $\epsilon_b$ and field
$h_b$.  Then, the total free energy per block is
\begin{equation}
b^d f(\epsilon, h, 1/L) = b^d g_b(\epsilon, h, 1/L) + f(\epsilon_b, h_b, b/L).
\label{freieenergiekadanoff}
\end{equation}
$d$ is the dimension of the system. The first term on the right hand
side of this equation is the free energy of the interactions between
the sites in the block. Since $b$ is finite, it is analytical even at
the critical point where $\epsilon = h = 0$ and $L = \infty$.  All
free energies have been expressed in terms of free energy densities
per site. It is convenient to define the free energy as a function of
the {\em inverse} system size. Then, at the critical point all
arguments of $f$ are zero.  Clearly, the correlation length in the
rescaled system is by a factor $1/b$ smaller than in the original
system such that
\begin{equation}
\xi(\epsilon, h, 1/L) = b \xi(\epsilon_b, h_b, b/L).
\label{korrelationslaengekadanoff}
\end{equation}

A simple consideration yields the functional dependence of
$\epsilon_b$ and $h_b$ on $\epsilon$ and $h$. We iterate the rescaling
procedure by dividing the {\em rescaled} system into blocks of size
$b'$. Clearly, this should yield the same result as a division of the
original system into blocks of size $b b'$:
$\epsilon_{b'}(\epsilon_b(\epsilon)) = \epsilon_{b' b}(\epsilon)$ and
$h_{b'}(h_b(h)) = h_{b' b}(h)$. This group property is fulfilled if
\begin{equation}
\begin{array}{ccc}
\epsilon_b(\epsilon) &=& b^{y_T} \epsilon \\
h_b(h) &=& b^{y_h} h 
\end{array}
\label{skalenexponentenkadanoff}
\end{equation}
with scaling exponents $y_T$ and $y_h$. These exponents characterize
the {\em universality class} of the model. Note, that formally the
inverse system size enters the free energy and the correlation length
like a field which drives the system away from the phase
transition. The corresponding scaling exponent is one.

One can show \cite{na78} that the free energy can be decomposed into a
singular and a regular part $f = f_{\mathrm{sing}} +
f_{\mathrm{reg}}$. $f_{\mathrm{reg}}$ is analytical even at the
critical point. Close to the phase transition, the behaviour of the
system is dominated by the singular part $f_{\mathrm{sing}}$ which
fulfils
\begin{equation}
b^d f_{\mathrm{sing}}(\epsilon, h, 1/L) = f_{\mathrm{sing}}(b^{y_T}
\epsilon, b^{y_h} h, b/L).
\label{freieenergieskalierung}
\end{equation} 
The second derivatives of this equation with respect to $\epsilon$ and
$h$, respectively, yield similar scaling forms for the specific heat
$c_L$ and the susceptibility $\chi_L$ of $m$. From these, we obtain
how the singularities of $c_L$ and $\chi_L$ build up as $L$ goes to
infinity:
\begin{eqnarray}
c_L &\propto& f_{\epsilon \epsilon}(0, 0, 1/L) \sim L^{2 y_T - d} \\
\chi_L &\propto & f_{h h}(0, 0, 1/L) \sim L^{2 y_h - d}.
\end{eqnarray}
Using equations \ref{korrelationslaengekadanoff} and
\ref{skalenexponentenkadanoff}, we obtain the scaling behaviour of the
inverse correlation length $\kappa := \xi^{-1}$:
\begin{equation}
b \kappa (\epsilon, h, 1/L) = \kappa( b^{y_T} \epsilon, b^{y_h} h,
b/L).
\label{kappaskalierung}
\end{equation}
Consequently, at $\epsilon = h = 0$ the inverse correlation length,
its derivative with respect to temperature and its second derivative
with respect to the field $h$ behave like
\begin{eqnarray}
\kappa(0, 0, 1/L) &\sim& L^{-1} \label{kappaeins} \\
\kappa_{\epsilon}(0, 0, 1/L) &\sim & L^{y_T - 1} \label{kappazwei} \\
\kappa_{h h}(0, 0, 1/L) & \sim & L^{2 y_h - 1} \label{kappadrei}.
\end{eqnarray}
In equation \ref{kappadrei}, we have assumed $h \leftrightarrow -h$
symmetry of the investigated model such that the first derivative of
$\kappa$ with respect to $h$ vanishes.  These properties are
sufficient to determine the critical temperature and the scaling
exponents $y_T$ and $y_h$.  To this end, we measure the inverse
correlation lengths $\kappa^{(1)}$, $\kappa^{(2)}$ at zero field for
two different strip widths $L^{(1)}$ and $L^{(2)}$ as a function of
temperature.  Due to equation \ref{kappaeins}, $T_c$ is the
temperature where $\kappa^{(1)} L^{(1)} = \kappa^{(2)} L^{(2)}$.  Once
$T_c$ is known, $y_T$ and $y_h$ can be determined from equations
\ref{kappazwei} and \ref{kappadrei}. At $T_c$, we have
\begin{equation}
y_T = \ln \left(\frac{L^{(1)} \kappa^{(1)}_{\epsilon}}{L^{(2)}
\kappa^{(2)}_{\epsilon}}\right) \left[ \ln
\left(\frac{L^{(1)}}{L^{(2)}} \right) \right]^{-1} \; \; \; \mbox{and}
\; \; \; y_h = \frac{1}{2} \ln \left( \frac{L^{(1)} \kappa^{(1)}_{h
h}}{ L^{(2)} \kappa^{(2)}_{hh}} \right) \left[ \ln \left(
\frac{L^{(1)}}{L^{(2)}}\right) \right]^{-1}.
\end{equation}

\subsection{First order phase transitions \label{finitesizefirst}}

At a first order phase transition, there is a discontinuity both in
energy and the order parameter. The energy difference between the high
temperature phase and the low temperature phase is called {\em latent
heat}. Additionally, there is an {\em interface tension} between the
phases. Since the surface of a phase and its volume scale differently
if the rescaling procedure introduced in chapter
\ref{finitesizekontinuierlich} is applied, this interface energy
forbids self-similar properties of the system.  It is connected with
the phenomenon of {\em metastability}: The low (high) temperature
phase is insensitive to {\em small} perturbations even at temperatures
above (below) the transition temperature.  Due to the interface
tension, small droplets of the stable phase in the unstable phase are
unfavourable.  At the transition temperature, the system will
spontaneously order in one of the phases unless this is impossible due
to some constraint, e.g.\ a fixed particle number. In that case, the
coexisting phases are separated and exist in different parts of the
system.

The description of these phenomena requires an extension of the
concept of free energy. Conventionally, the free energy is defined as
a thermodynamic potential which characterizes the state of the {\em
system}. In the theory of phase equilibria \cite{reif}, the concept of
the free energy of a {\em phase} is used. In the thermodynamic limit,
the free energy of the system is then assumed to be the sum of the
free energies of all coexisting phases. In the following paragraph, we
discuss the connection of this idea with the transfer matrix
formalism.

\subsubsection{Phases and asymptotically degenerate eigenvalues}

Due to the dependence of the correlation length on the eigenvalues of
$T$ in a finite system (equation \ref{freieenergietm}), one might
expect that in the thermodynamic limit the greatest eigenvalue of the
transfer matrix becomes degenerate at a phase transition. However, the
transfer matrix of an infinite system is ill-defined. Since energy and
free energy are extensive, $\tilde{H}(X, Y)$ and the free energy per
slice grow proportional to the number of sites per slice,
$\mathcal{M}$. Therefore, some elements and eigenvalues of $T$ go to
infinity as $L \rightarrow \infty$. To avoid this mathematical
problem, we define the quantities
\begin{equation}
f_i := \frac{-1}{\beta \mathcal{M}} \ln |\lambda_i|\; \; \; \mbox{for}
\; \; \; i = 1, ..., \tilde{n}.
\end{equation}
The eigenvalues $\lambda_i$ are assumed to be sorted in descending
order. For $L \rightarrow \infty$, $f_1$ converges to the free energy
per site of the infinite system.  Consider a system at a phase
transition.  Quite generally, we assume that there are $\nu$ positive
eigenvalues $\lambda_1 > \lambda_2 \geq ... \geq \lambda_{\nu}$ and
$\mu$ negative eigenvalues $\lambda_{\tilde{n}} \leq
\lambda_{\tilde{n} - 1} \leq ... \leq \lambda_{\tilde{n} - \mu + 1}$
such that
\begin{equation}
\xi_i^{-1} := \ln \left( \left| \frac{\lambda_1}{\lambda_i} \right|
\right) = - \beta \mathcal{M} \left(f_1 - f_i \right) \stackrel{L
\rightarrow \infty}{\longrightarrow} 0
\label{pseudoentartung}
\end{equation} 
for $i \leq \nu$ and $i > \tilde{n} - \mu$. In the following, we will
denote these eigenvalues as ``asymptotically degenerate''. Each of the
$\xi_i$ might be the diverging correlation length of an appropriately
chosen quantity.

We investigate the influence of a perturbing field $h$ which couples
to the extensive quantity $N \mathcal{M} q$.  $q$ is an
arbitrary intensive quantity the operator of which is $\hat{q}$. This
field drives the system away from the phase transition at $h=0$. In
the remainder of this paragraph it is assumed that $|h|$ is so small
that terms of order $h^2$ and higher orders in $h$ can be
neglected. Variables which refer to the perturbed system are marked
with a prime. The transfer matrix is
\begin{eqnarray}
T' &=& \sum_{\ket{X}, \ket{Y}} \ket{X} \exp \left[ - \beta \left( \tilde{H}(X, Y) 
- \frac{h \mathcal{M}}{2} (q(X) + q(Y)) \right) \right] \bra{Y} \nonumber \\
& = & \sum_{\ket{X}, \ket{Y}} \ket{X} \left[ 1 + \frac{\beta h \mathcal{M}}{2} 
\left( q(X) + q(Y) \right) \right] \exp \left[ - \beta \tilde{H}(X, Y) \right] 
\bra{Y} \nonumber \\
& = & T + \frac{\beta h \mathcal{M}}{2} \left( \hat{q} T + T \hat{q} \right)
\label{tmgestoert}.
\end{eqnarray}
According to perturbation theory, for small $L$ the greatest eigenvalue of 
$T'$ is $\lambda_1' = \bra{\lambda_1} T' \ket{\lambda_1} + \mathcal{O}(h^2)$. 
However, for large $L$ the gap between the greatest eigenvalue of $T$ and the 
asymptotically degenerate eigenvalues becomes smaller. 
For any small but finite $h$, there will be some value of $L$ where the 
perturbation induced by $h$ is comparable to the separation between the 
eigenvalues. In this case, the eigenvector $\ket{\lambda_1'}$ to the greatest 
eigenvalue of $T'$ will be the linear combination of the eigenvectors to 
the asymptotically degenerate eigenvalues of $T$ which maximizes 
$\bra{\lambda_1'} T' \ket{\lambda_1'}$. We make the ansatz
\begin{equation}
\ket{\lambda_1'} = \sum_{k = 1}^{\nu + \mu} c_k \ket{\lambda_{z(k)}}
\label{eigenvektorgestoert}
\end{equation}
where $\sum_{k=1}^{\nu + \mu} c_k^{2} = 1$ and
\begin{equation}
z(k) = \left\{
\begin{array}{lll}
k & \mbox{if} & k < \nu \\
k + \tilde{n} - \nu - \mu & \mbox{else} & 
\end{array}
\right. 
\end{equation}
enumerates the asymptotically degenerate eigenvalues.  Using equations
\ref{eigenvektorgestoert} and \ref{tmgestoert}, we obtain
$$
\bra{\lambda_1'} T' \ket{\lambda_1'} = \lambda_1 \sum_{k,l = 1}^{\nu +
\mu} c_k c_l \left[ \frac{\lambda_{z(k)}}{\lambda_1} \delta_{k,l} +
\frac{\beta h \mathcal{M}}{2} \bra{\lambda_{z(l)}} \hat{q}
\ket{\lambda_{z(k)}} \left( \frac{\lambda_{z(l)}}{\lambda_1} +
\frac{\lambda_{z(k)}}{\lambda_1} \right) \right].
$$
Due to equation \ref{pseudoentartung}, we have $\lim_{L \rightarrow
\infty} \lambda_i/\lambda_1 = \mbox{sign}(\lambda_i)$ for all
asymptotically degenerate eigenvalues. Using this fact and equation
\ref{freieenergietm}, we obtain after some lengthy algebra that in the
limit $L \rightarrow \infty$
\begin{equation}
f_1' = \left\{
\begin{array}{ccc}
f_a = f_1 - h q_{\mathrm{max}} & \mbox{if} & h > 0 \\
f_b = f_1 - h q_{\mathrm{min}} & \mbox{if} & h < 0
\label{freieenergieentartet}
\end{array}
\right.
\end{equation}
where $q_{\mathrm{max}}$ ($q_{\mathrm{min}}$) is the greatest
(smallest) eigenvalue of the matrix
\begin{equation}
Q = \left(
\begin{array}{cc}
\begin{array}{ccc}
Q_{1,1} & \cdots & Q_{1 ,\nu} \\ \vdots & \ddots & \vdots \\ Q_{\nu,
1} & \cdots & Q_{\nu ,\nu}
\end{array} & 0 \\
0 & \begin{array}{ccc} 
Q_{\tilde{n} - \mu + 1, \tilde{n} - \mu +1} &
\cdots & Q_{\tilde{n} - \mu +1, \tilde{n}} \\ \vdots & \ddots & \vdots
\\ Q_{\tilde{n}, \tilde{n} - \mu + 1} & \cdots & Q_{\tilde{n},
\tilde{n}}
\end{array}
\end{array}
\right).
\label{grossematrix}
\end{equation}
Here, we have introduced $Q_{i,j} = \lim_{L \rightarrow \infty}
\bra{\lambda_i} \hat{q} \ket{\lambda_{j}}$. This result is reminiscent
of degenerate perturbation theory. Note, that there is no mixing
between positive and negative eigenvalues.  Using equation
\ref{mittelstandard}, we find that there might be different
expectation values of $q$ at the phase transition depending on whether
the limit $h \rightarrow 0$ is performed for a positive or a negative
field:
\begin{eqnarray}
\left< q_{a} \right> &=& \lim_{h \rightarrow 0+} \frac{\partial f_1'}{
\partial h} \; = \; q_{\mathrm{max}}
\label{limitqplus} \\
\left< q_{b} \right> &=& \lim_{h \rightarrow 0-} \frac{\partial f_1'}{
\partial h} \; = \; q_{\mathrm{min}} \label{limitqminus}. 
\end{eqnarray}
Clearly, this is what one would expect at a first order phase
transition where there is a discontinuity in $q$. Obviously, this is
the case if $q$ is an order parameter. On the contrary, at a
continuous phase transition we have $q_{\mathrm{max}} =
q_{\mathrm{min}}$ for an arbitrary quantity $q$. This implies that $Q$
has only one $\nu + \mu$ -fold degenerate eigenvalue.
\begin{figure}
\includegraphics{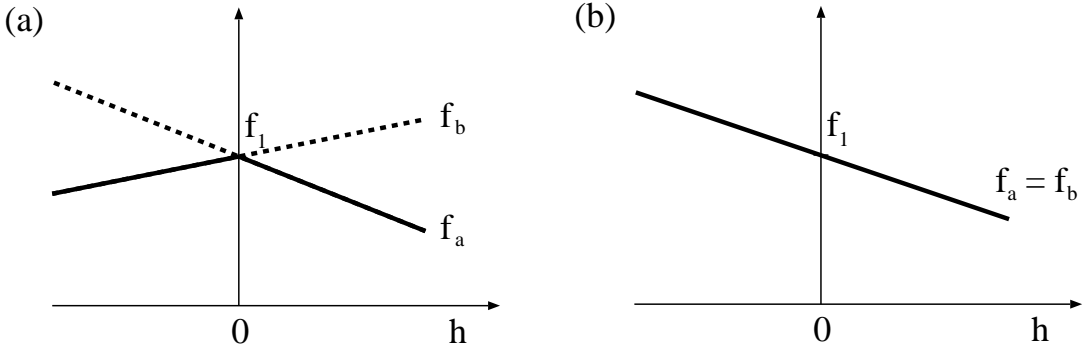}
\caption{Sketch of the behaviour of $f_1'$ at a phase transition where
$\nu + \mu = 2$ in the limit $L \rightarrow \infty$. Panel (a): First
order phase transition. The solid line shows the free energy of the
system. At the phase transition ($h = 0$), there is a discontinuity in
$\partial f_1' / \partial h$.  The dashed lines show the continuation
of the free energies of the phases into the regions of metastability.
Panel (b): Continuous phase transition. At $h = 0$, the first derivative of
$f_1'$ is continuous. \label{phasenuebergangunendlich}}
\end{figure}
Figure \ref{phasenuebergangunendlich} shows a sketch of the free
energy $f_1'$.  As expected, at a first order phase transition there is
a cusp in the free energy. At a continuous transition, the first
derivative of $f_1'$ with respect to $h$ is continuous.

Consider a system with a first order phase transition and $\nu + \mu =
2$ asymptotically degenerate eigenvalues. In the absence of
constraints, at $h = 0$ the symmetry of the system is spontaneously
broken. The system is either in a phase where $\left< q \right> =
q_{\mathrm{max}}$ (denoted as phase (a) in the following) or a phase
where $\left< q \right> = q_{\mathrm{min}}$ (termed phase (b)).  For
$h > 0$, phase (a) is thermodynamically stable, for $h < 0$ (b) is the
stable phase.  If the concept of the free energy of a {\em phase} can
be applied, $f_a$ should be the free energy of phase (a). Conversely,
the free energy of phase (b) should be $f_b$. We expect that both
phases are metastable in the region where they are thermodynamically
unstable. Since there is no singularity in the free energy of a phase
at the transition (in contrast to the free energy of the system), the
free energies of the phases in the metastable region should be given
by the continuation of $f_a$ and $f_b$.

Generalizing this idea, we find that in the vicinity of a first order
phase transition the free energies of the metastable phases are
functions of those eigenvalues of the transfer matrix which are
asymptotically degenerate at the phase transition.  The number of
these eigenvalues equals the number of coexisting phases at the
transition itself. For example, at a triple point we expect $\nu + \mu
= 3$.

\subsubsection{Application to semi-infinite systems}
At continuous transitions, the inverse system size enters the free
energy like a field which drives the system away from the phase
transition. According to Privman, Fisher and Schulman
\cite{pf83,ps82}, a similar analogy holds in the vicinity of a first
order transition.  They argue, that the $f_i$ of a semi-infinite
system can be calculated from measurends of an infinite system via a
perturbation theory similar to that presented in the preceeding
paragraph. The finiteness of the system size plays the role of a
perturbing field.  Consequently, the free energy per site in a
semi-infinite system is the smallest eigenvalue of a $(\nu + \mu)
\times (\nu + \mu)$- dimensional free energy matrix which has the same
structure as $Q$ in equation \ref{grossematrix}. Its diagonal elements
are the free energies per site of the phases which coexist at the
transition. The off-diagonal elements determine the finite size
corrections. Since there is no mixing between positive and negative
eigenvalues, the free energy matrix can be diagonalized blockwise.

\begin{figure}
\includegraphics{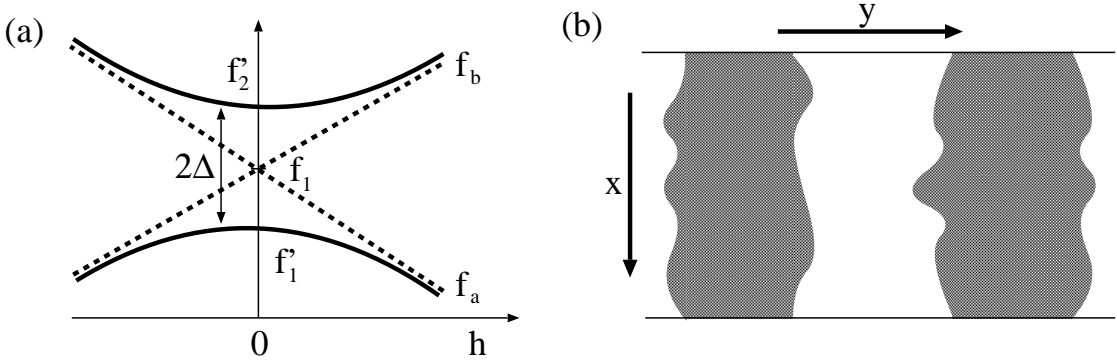}
\caption{ Panel (a): Schematical behaviour of the $f_i$ at a finite
strip width $L$. The dashed lines show the free energies $f_a$, $f_b$
of the phases of the infinite system.  Panel (b): Sketch of a
semi-infinite system at the phase transition of the infinite
system. White and grey areas symbolize two coexisting
phases. \label{semiinfinitefirstorder}}
\end{figure}

In the models which are discussed in this thesis, we have to deal with
lines of triple points, lines of points where four phases coexist and,
in the model of the CdTe(001) surface with Te dimerization, a point
where five phases coexist.  In our investigations, the system is
arranged such that the repulsive interactions are in the infinite
direction. Then, neighbouring slices are anticorrelated which implies
negative asymptotically degenerate eigenvalues.  We find, that
whenever the number of coexisting phases is $\leq 4$, there are at
most two positive and two negative eigenvalues which are
asymptotically degenerate.  In the following, we restrict ourselves to
this case. The point where 5 phases coexist can be found from the
intersection of lines of other phase transitions.  Due to the theorem
of Perron-Frobenius, $f_1$ is the smallest eigenvalue of the positive
block of the free energy matrix which has the form
\begin{equation}
\left(
\begin{array}{cc}
f_a & \Delta \\
\Delta & f_b
\end{array}
\right). 
\end{equation}
Its eigenvalues are 
\begin{equation}
f_{1, 2} = - \frac{1}{\beta \mathcal{M}} \ln \lambda_{1,2} =
\frac{1}{2} \left( f_a + f_b \right) \mp \frac{1}{2} \sqrt{ \left( f_a
- f_b \right)^2 + 4 \Delta^2 }.
\label{freieenergieprivman}
\end{equation}
At the phase transition, we have $f_a = f_b$. Then, the inverse
correlation length is $\xi^{-1} = \ln |\lambda_1 / \lambda_2| = 2
\beta \Delta$. Further away from the phase transition, where $| f_a -
f_b | \gg \Delta$, $f_1$ approaches to the minimum of $f_a$ and $f_b$
(figure \ref{semiinfinitefirstorder}a).

Since $\Delta$ is a function of the correlation length at the phase
transition, we can determine the functional form of the dependence of
the finite size corrections on $L$ from an estimation of $\xi$. In a
semi-infinite system, there is no spontaneous symmetry breaking at the
transition temperature of the infinite system. Instead, the system
consists of alternating domains of the coexisting phases. A sketch of
such a system is shown in figure \ref{semiinfinitefirstorder}b.
The correlation length is proportional to the typical distance between
two domain walls. Due to the interface tension $\sigma$, the formation
of such a domain wall costs a free energy $\alpha L \sigma$, where
$\alpha$ is a dimensionless number $\mathcal{O}(1)$. Consequently, the
probability for the formation of a domain wall is proportional to
$\exp(-\beta \alpha L \sigma)$. Therefore, we have
\begin{equation}
\xi \propto \exp \left( \beta \alpha L \sigma \right) \; \;
\Longrightarrow \; \; \Delta \propto \exp \left( - A L \right) \; \;
\mbox{where} \; \; A = \beta \alpha \sigma.
\label{auweia}
\end{equation}
Since the behaviour of all other quantities depends on the free
energy, an exponential decay of the finite size corrections should be
a generic feature of first order phase transitions.  However,
frequently one encounters the phenomenon of a {\em tricritical point}
at a temperature $T_{\mathrm{tri}}$.  At $T_{\mathrm{tri}}$, the
nature of the phase transition changes from first order at $T <
T_{\mathrm{tri}}$ to a continuous transition at $T >
T_{\mathrm{tri}}$. Conseqently, at $T_{\mathrm{tri}}$ we expect a
crossover from exponential finite size scaling to the linear scaling
of the continuous transition \cite{rkgk83}. Unfortunately, at feasible
values of $L$, this crossover is smeared out over a finite temperature
range which complicates finite size scaling in the vicinity of the
tricritical point.

To determine the locus of the phase transition, we consider an
intensive quantity $q$ and a field $h$ which couples to $N \mathcal{M}
q$.  We assume the phase transition to be at $h = 0$. Since equation
\ref{freieenergieprivman} holds also in the presence of a field, we
have
$$ \frac{\partial f_{1,2}}{\partial h} = \frac{1}{2} \left[
\frac{\partial f_a}{\partial h} + \frac{\partial f_b}{\partial h} \mp
\left( \frac{\partial f_a}{\partial h} - \frac{\partial f_b}{\partial
h} \right) \frac{f_a - f_b}{\sqrt{\left(f_a - f_b \right)^2 + 4
\Delta^2}} \right].
$$
At the transtion, we have $f_a = f_b$. Consequently, 
\begin{equation}
\left<q \right>_{1,2} := \lim_{h \rightarrow 0} \frac{\partial f_{1,
2}}{ \partial h} = \frac{1}{2} \left( \left. \frac{\partial
f_a}{\partial h} \right|_{h = 0} + \left. \frac{\partial f_b}{\partial
h} \right|_{h = 0} \right).
\label{mirbrummtderkopfvondemzeug}
\end{equation}
Using equation \ref{mittelwerttm}, we can calculate $\left< q
\right>_{1,2}$ from the eigenvectors $\ket{\lambda_1}$ and
$\ket{\lambda_2}$ of the transfer matrix. Then from equation
\ref{mirbrummtderkopfvondemzeug} at the phase transition we have
\begin{equation}
\bra{\lambda_1} \hat{q} \ket{\lambda_1} = \bra{\lambda_2} \hat{q} 
\ket{\lambda_2}.
\end{equation} 
Note, that this identity is {\em independent} of the order of the
phase transition and the functional form of the finite size
corrections. 

Once the locus of the transition is known, the values of $q$ in the
phases can be determined from the matrix elements $\bra{\lambda_i}
\hat{q} \ket{\lambda_j} =: Q_{i,j}^{(f)}$.  Consider the matrix
\begin{equation}
Q^{(f)} = \left(
\begin{array}{cc}
\begin{array}{ccc}
Q_{1,1}^{(f)} & \cdots & Q_{1 ,\nu}^{(f)} \\
\vdots & \ddots & \vdots \\
Q_{\nu, 1}^{(f)} & \cdots & Q_{\nu ,\nu}^{(f)} 
\end{array} & 0 \\
0 & \begin{array}{ccc}
Q_{\tilde{n} - \mu + 1, \tilde{n} - \mu +1}^{(f)} & \cdots & 
Q_{\tilde{n} - \mu +1, \tilde{n}}^{(f)} \\
\vdots & \ddots & \vdots \\
Q_{\tilde{n}, \tilde{n} - \mu + 1}^{(f)} & \cdots & 
Q_{\tilde{n}, \tilde{n}}^{(f)}
\end{array}
\end{array}
\right).
\end{equation}
Since $\lim_{L \rightarrow \infty} Q^{(f)} = Q$, the eigenvalues
$q_i^{(f)}$ of $Q^{(f)}$ converge to the eigenvalues $q_i$ of $Q$ in
the limit $L \rightarrow \infty$. Due to equation \ref{auweia}, we
expect
\begin{equation}
q_{i}^{(f)} = q_i + B \exp \left(-A L \right).
\end{equation} 
Therefore, $q_i$ and the parameters $A$, $B$ can be determined from
transfer matrix calculations with three different strip widths.
 
Methods similar to that presented in this section have been suggested
in the literature for the investigation of the spontaneous
magnetization \cite{h82} and latent heat \cite{h83} of the Ising model
and for the calculation of coverage discontinuities in interacting
hard square lattice gases \cite{ber86}. However, to our knowledge a
consistent presentation of this subject in the general case has not
been published yet.  

\chapter{Continuous time Monte Carlo simulations \label{mcappendix}}

A large number of Monte Carlo algorithms is based on the mathematical
concept of {\em Markov chains} \cite{nb99}. In all the simulations
presented in this thesis, such methods were used.  Consider a system
with a finite set $\mathcal{S}$ of possible states $s$.  A Markov
chain is a stochastic process in discrete timesteps which is
characterized completely by a set of transition probabilities $\{W_{s
\rightarrow s'} | s,s' \in \mathcal{S} \}$, where $W_{s \rightarrow
s'}$ is the conditional probability to find the system in state $s'$
at step $n+1$ given that it is in state $s$ at step $n$.

In equilibrium simulations one is interested in the calculation of
averages over an ensemble of states distributed according to some
probability distribution function $P(s)$. Of course, the most
important example is the Gibbs distribution
\begin{displaymath}
P(s) = \frac{\exp \left[ -H(s)/(k T) \right]}{
\sum_{s} \exp \left[ -H(s)/(kT) \right]},
\end{displaymath}
where $H(s)$ is the energy of the system in state $s$, $k$ is
Boltzmann's constant and $T$ is temperature.  The idea is to construct
a Markov chain, where {\em time averages} over this sequence of states
converge to {\em ensemble averages} over $P(s)$ in the limit of
infinite time.  The first condition for this to be the case is that
every state $s$ of the system with $P(s) > 0$ can be reached within a
finite number of steps from every initial configuration.  A dynamics
with this property is called {\em ergodic}.  The second condition is
that $P(s)$ is {\em stationary} under the dynamics.  Consider an
ensemble of Markov chains. Let $P_{n}(s)$ denote the probability of a
randomly chosen member of such an ensemble to be in state $s$ at time step
$n$.  The change of this probability in the next step is then given by
\begin{equation}
P_{n+1}(s) - P_{n}(s) = \sum_{s' \neq s} W_{s' \rightarrow s}
P_{n}(s') - \sum_{s' \neq s} W_{s \rightarrow s'} P_{n}(s).
\end{equation}
The first sum is the total probability to enter state $s$ in step $n$,
the second sum is the total probability to leave state $s$. A
distribution is stationary, if the left hand side of this equation is
zero. In particular, this is the case if the transition probabilities
fulfil a detailed balance condition
\begin{equation}
W_{s \rightarrow s'} P(s) = W_{s' \rightarrow s} P(s'). 
\end{equation} 
In case of a Gibbs distribution, this condition yields
\begin{equation}
\frac{W_{s \rightarrow s'}}{W_{s' \rightarrow s}} = \exp \left( 
- \frac{H(s) - H(s')}{kT} \right).
\label{balance} 
\end{equation}
Since there is no unique solution of equation \ref{balance}, there is
some freedom in the choice of transition probabilities in simulations
of thermal equilibrium.  In particular, there is no need to choose the
$W_{s \rightarrow s'}$ in agreement with the physical dynamics of the
system.

In the simulation of systems far from equilibrium like crystal growth
or sublimation one is interested in the evolution of the state of the
system with time. If we look at the state of the system on a
coarse-grained scale in discrete time intervals of appropriate length,
we can describe its evolution by a Markov process. The transition
probabilities depend on the chaotic microscopic motion of atoms
(section \ref{mobilityintro}).  Therefore, the $W_{s \rightarrow s'}$
must reflect the physics of the investigated system.  However, in the
absence of external forces any physical system will finally reach
thermal equilibrium. Therefore, any reasonable choice of the dynamics
of the {\em isolated} system must fulfil {\em at least} the conditions
of ergodicity and the stationarity of the Gibbs distribution. The
simplest way to ensure the latter condition is to make an ansatz for
the $W_{s \rightarrow s'}$ in agreement with the detailed balance
condition \ref{balance}.

It is favourable to express the transition probabilities in terms of
events which modify the system. Such events might be e.g.\ diffusion
hops of atoms on a crystal surface or spin flips in an Ising model. In
general, let there be a system where $\mathcal{N}$ events $i =
1,...,\mathcal{N}$ are possible. There is a {\em rate} $\rho_i(s)$ for
every event, which is a function of the actual state $s$ of the
system.  We introduce the transition probabilities
\begin{equation}
W_{s \rightarrow s'} = \left\{  
\begin{array}{ll}
\frac{\rho_{i(s \rightarrow s')}(s)}{\mathcal{N} \rho_{0}} & \mbox{if} \; 
s \neq s' \; \mbox{and} \; i(s \rightarrow s') \; \mbox{exists} \\
1 - \sum_{i = 1}^{\mathcal{N}} \frac{\rho_i(s)}{\mathcal{N} \rho_0} & \mbox{if} \; 
s = s' \\ 
0 & \mbox{otherwise}
\end{array}
\right.,
\label{markovprozess}
\end{equation}
where $\rho_0$ is a constant such that $\rho_0 \geq \mbox{max}_i \;
\rho_i $. Event $i(s \rightarrow s')$ changes the state of the system
from $s$ to $s'$.  These $W_{s \rightarrow s'}$ fulfil the conditions
$W_{s \rightarrow s'} \geq 0$ and $\sum_{s'} W_{s \rightarrow s'} = 1$
for probabilities. Detailed balance is fulfilled if $\rho_{i(s
\rightarrow s')}/\rho_{i(s' \rightarrow s)} = P(s)/P(s')$.

These transition probabilities describe the dynamics of a stochastic
system with independent events if the probability that two or more
events occur simultaneously in the physical time interval $\Delta t$
between subsequent Markov steps can be neglected. The probability of a
random event with rate $\rho_{i}(s)$ to happen in a time interval
$\Delta t$ is $\rho_{i}(s) \Delta t$.  Identifying this with the
corresponding transition probability we find
\begin{equation}
\Delta t = \frac{1}{\mathcal{N} \rho_0}.
\end{equation}
Clearly, the probability for two or more events in $ \Delta t$
decreases $\sim 1/\rho_0^2$ as $\rho_0$ is increased. In continuous
time algorithms, we can perform the limit $\rho_0 \rightarrow \infty$
exactly, such that there are no problems due to simultaneous events.

\section{Fundamental algorithms}

In the Markov chain defined in equation \ref{markovprozess}, there are
two different classes of steps: On the one hand, there are steps in
which the state of the system changes. In the following, we will
denote these as ``modifying steps''. On the other hand, there are
steps in which the state of the system remains the same (non-modifying
steps).  Especially at low temperature the probability $W_{s
\rightarrow s}$ of a non-modifying step usually is high. This leads to
a considerable slowing down of simple techniques like the Metropolis
algorithm which simulate the Markov chain directly.

{\em Continuous time algorithms} \cite{bkl75,m88,nb99} avoid this
problem. The basic idea is to jump over the non-modifying steps.  In
every iteration of the algorithm one event is selected randomly with
the conditional probability $\tilde{P}_i(s)$ that event $i$ is
performed in a Markov step {\em given} that this is a modifying step.
Then, the number $n_p$ of Markov steps which pass before the next
modifying step is drawn randomly according to the probability of
non-modifying steps.  This procedure yields the correct statistics of
the Markov chain, if quantities measured in the actual state of the
system are weighted with $n_p$ or, equivalently, with the
corresponding physical time interval.  According to equation
\ref{markovprozess}, the probability of a modifying step is
\begin{equation}
P_m(s) = 1 - W_{s \rightarrow s} = \frac{R(s)}{\mathcal{N} \rho_0} \; \; \; \mbox{where} \; 
\; \;  R(s) = \sum_{i = 1}^{\mathcal{N}} \rho_i(s). 
\label{peemm}
\end{equation}
Using this equation, we can calculate $\tilde{P}_i(s)$. The
probability that event $i$ is performed in one Markov step is
$$
P_i(s) = \frac{\rho_i(s)}{\mathcal{N} \rho_0} = P_m(s)
\frac{\rho_i(s)}{R(s)} = P_m(s) \tilde{P}_i(s).
$$ 
Consequently, 
\begin{equation}
\tilde{P}_i(s) = \frac{\rho_i(s)}{R(s)}.
\label{peischlange}
\end{equation}
The probability of a specific value of $n_p$ is the probability 
of a sequence of $n_p - 1$ non-modifying steps
followed by one modifying step which is 
\begin{equation}
P(n_p, s) = (1 - P_m(s))^{n_p -1} P_m(s).
\label{ennpe}
\end{equation} 
To perform the limit $\rho_0 \rightarrow \infty$, we express $n_p$ in
terms of the physical time interval $\tau = n_p \Delta t =
n_p/(\mathcal{N} \rho_0)$ which passes between subsequent modifying
steps.  Then, the probability distribution function of $\tau$ is
\begin{equation}
P(\tau, s) d\tau = \lim_{\rho_0 \rightarrow \infty} 
\left( 1 - \frac{R(s)}{\mathcal{N} \rho_0} \right)^{\mathcal{N} \rho_0 \tau} 
R(s) d\tau = R(s) \exp \left( - R(s) \tau \right) d\tau,
\label{tauverteilung}
\end{equation}
where we have used equations \ref{peemm} and \ref{ennpe}. 

\begin{algorithm} Continuous time  Monte Carlo simulation
\begin{enumerate}
\item 	Draw an event $i$ randomly with probability $\tilde{P}_i(s)$ and
	perform it. This changes the state $s$ of the system. 
\item 	Calculate the $\rho_i(s)$ and $R(s)$ in the new state of the system.
	If the event changes the state of the system only locally and 
	the range of interactions is short, $\rho_i(s)$ changes only for 
	a few $i$.  
\item	Draw a uniformly distributed random number $\chi \in \; ]0, 1]$ and 
	increase the system time by $\tau = - \ln \chi /R(s)$. 
	$\tau$ is a random number distributed according to equation 
	\ref{tauverteilung}. 
	In the calculation of thermal averages, weight the quantities measured in 
	the current state of the system with $\tau$. 
\item 	Go to step 1 unless some condition to terminate is fulfilled (e.g.\ 
	a maximum number of steps is done). 
\end{enumerate}
\label{ctmc}
\end{algorithm}
To apply this algorithm, we need fast methods to perform the random
selection of events in step 1 and the update of the rates in step 2.
We first introduce algorithms which perform both tasks in
$\mathcal{O}(\log_2 \mathcal{N})$ time \cite{bbs95}.  They can be
applied in the general case where there is a unique rate for every
event.
 
We use a {\em complete binary tree} \cite{knuth} to store the list of
the rates of all events.  Formally, a {\em binary tree} is defined as
a finite set of nodes that either is empty, or consists of a root node
and the elements of two disjoint binary trees which are called the
left and the right subtree of the root.
\begin{figure}
\begin{center}
\includegraphics{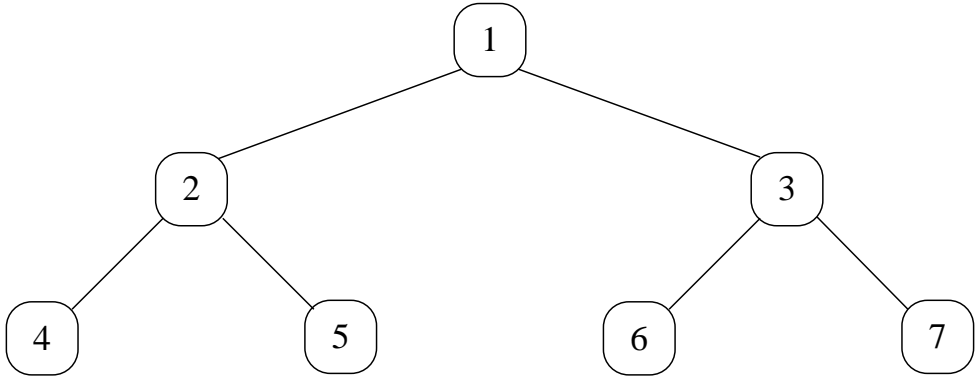}
\end{center}
\caption{Sketch of a complete binary tree. \label{baumbildchen}}
\end{figure}
The root nodes of the subtrees of a node are commonly denoted as its
``children''. Conversely, the node a node is child of is called its
``parent''.

Usually, a binary tree is drawn with the root node on top. Lines
connect every node with its children.  Such a sketch of a binary tree
is shown in figure \ref{baumbildchen}.  There is a simple way to
enumerate the nodes of a binary tree. The root node gets index $1$. In
general, node $k$ has the left child $2 k$ and the right child $2 k +
1$. Then, the parent of node $k$ is $\lfloor k/2 \rfloor$.  A binary
tree is called {\em complete} if this enumeration maps its nodes to a
set of consecutive integers. The tree shown in figure
\ref{baumbildchen} is such a complete binary tree. We use this
enumeration to store a complete binary tree in a linear array in
computer memory.

In our Monte Carlo simulations, we use a complete binary tree with
$\mathcal{N}$ nodes, where $\mathcal{N}$ is the maximum number of
events. Each node $k$ is attributed to a specific event.  There are
three real numbers for each node: $w_k = \rho_k(s)$ is the rate of
event $k$ in the current state of the system. $l_k$ is the sum of the
rates of all nodes in the left subtree of node $k$ and $r_k$ is the
sum of all rates in the right subtree of node $k$. Using these
quantities, the sum of all rates is $R(s) = w_1 + l_1 +
r_1$. Formally, we define $w_k = l_k = r_k = 0$ for all $k >
\mathcal{N}$.  Then, $l_k = w_{2k} + l_{2k} + r_{2k}$ and $r_k =
w_{2k+1} + l_{2k+1} + r_{2k+1}$.  The following algorithm selects a
random event by climbing down the tree. The index $k$ points to the
current node.
\begin{algorithm} Selection of events with probability 
$\tilde{P}_k(s) = w_k/(w_1 + l_1 + r_1)$
\begin{enumerate}
\item Draw a uniformly distributed random number 
	$\xi \in [0, w_1 + l_1 + r_1[$.  Set $k \leftarrow 1$ (Start
	at the root node).
\item	If $\xi < l_k$, then set $k \leftarrow 2k$, i.e.\ go to the
	left child. 
	Else, if $\xi < l_k + w_k$ return event $k$. In this case, the 
	algorithm 
	is done. 
	Else, go to the right child and adjust $\xi$ such that $0 \leq \xi < 
	r_k$: $k \leftarrow 2k+1$, 
	$\xi \leftarrow \xi - l_k - w_k$. 
\item	Go to step 2.
\end{enumerate}
\label{geteventbaum}
\end{algorithm}
The proof that algorithm \ref{geteventbaum} is correct is by
induction.  Whenever the algorithm enters node $k$ in step 2, the
actual value of $\xi$ is a uniformly distributed random number $\in
[0, w_k + l_k + r_k[$. This is ensured by the initialization in step
one and the adjustment in step 2. Therefore, the algorithm terminates
with probability $w_k/(w_k + l_k + r_k)$. The probabilities to go to
the left and the right child are $l_k/(w_k + l_k + r_k)$ and $r_k/(w_k
+ l_k + r_k)$, respectively.  In particular, the probability to
terminate at the root is $w_1/(w_1 + l_1 + r_1)$.  This is the basis
of induction. Node $k$ is a child of node $m = \lfloor k/2 \rfloor <
k$.  By induction, the algorithm terminates in node $m$ with the
correct probability, which can only be the case if the algorithm
enters node $m$ with probability $(w_m + l_m + r_m)/(w_1 + l_1 +
r_1)$.  The probability to go from node $m$ to node $k$ is $(w_k + l_k
+ r_k)/(w_m + l_m + r_m)$.  Therefore, the probability to terminate in
node $k$ is
$$
\frac{w_m + l_m + r_m}{w_1 + l_1 + r_1}\cdot \frac{w_k + l_k +
r_k}{w_m + l_m + r_m} \cdot \frac{w_k}{w_k + l_k + r_k} =
\frac{w_k}{w_1 + l_1 + r_1}. \; \; \; \; \; \mbox{q.e.d.}
$$ 
In each iteration where the algorithm does not terminate, one half of
the events in the subtrees of node $k$ are ruled out. Therefore, the
average running time of the algorithm is $\mathcal{O}(\log_2
\mathcal{N})$.

In step 2 of algorithm \ref{ctmc}, the tree must be updated to reflect
the changes in the $\rho_i(s)$. To do this for one event, we set $w_i
= \rho_i(s)$. Then, the $l_k$ and $r_k$ of the parents of node $i$
must be adjusted. This is done by the following algorithm which, on
average, requires $\mathcal{O}(\log_2 \mathcal{N})$ iterations.
\begin{algorithm} Update the rate of event $i$
\begin{enumerate}
\item	Start at node $i$. Set $k \leftarrow i$, $w_i \leftarrow \rho_i(s)$.  
\item 	If $k = 1$, terminate. Else, go to the parent node by setting $k \leftarrow 
	\lfloor k/2 \rfloor$. 
\item 	Adjust $l_k$ and $r_k$. Set $l_k \leftarrow w_{2k} + l_{2k} + r_{2k}$ and 
	$r_k \leftarrow w_{2k+1} + l_{2k+1} + r_{2k+1}$. 
\item 	Go to step 2.
\end{enumerate}
\label{updatebaum}
\end{algorithm}
If the rates of $\mathcal{O}(1)$ events change in each iteration, the
time required for the complete update is $\mathcal{O}(
\log_2 \mathcal{N})$.

These algorithms work in the general case where there is a unique rate
for every event. However, in many lattice gas models the rate of an
event depends only on the local environment of the sites which are
modified by the event. Then, the number of different rates
$\mathcal{M}$ is independent of the system size and much smaller than
the number of events $ \mathcal{N}$. This feature can be exploited to
make the running time both of the random selection of events and the
update of data structures $\mathcal{O}(1)$.  The idea is to keep {\em
lists} of events the rates of which are equal.  Consider the case
where we have $\mathcal{M}$ lists $ \{ l_g \}_{g = 1}^{\mathcal{M}}$.
If the $n_g(s)$ events in list $l_g$ have the rate $\rho_g$, the total
rate of these events is $R_g(s) = n_g(s) \rho_g$. The probability for
one member of the list to be selected is
$$ \frac{\rho_g}{R(s)} = \frac{R_g(s)}{R(s)} \cdot \frac{1}{n_g(s)}
. $$ Consequently, we can perform the selection of events in a
two-step process.  First, we select a list $l_g$ with probability
$R_g(s)/R(s)$. This can be done with a binary tree where each node is
attributed to one {\em list} using algorithm \ref{geteventbaum}.
Then, we select one of the members of $l_g$ randomly with equal
probability for all events.  On update, events the rates of which have
changed are moved to their new group. Then, the $R_g(s)$ of lists the
length of which has changed are updated using algorithm
\ref{updatebaum}. Since the members of the lists can be accessed in
$\mathcal{O}(1)$ time, the running time both for the selection of
events and the updates is independent of $\mathcal{N}$.  However,
compared to the direct storage of the rates of events in a binary
tree, the manipulation of the lists requires some computational
overhead.  In our experience, in a two-dimensional lattice gas at
system sizes of a few hundred lattice constants, this overhead
outweighs the gain of speed whenever $\mathcal{M}$ is greater than a
few dozen.

\section{Application to lattice gases in thermal equilibrium}

The application to grand-canonical simulations of the lattice gases
discussed in chapter \ref{dreikapitel} is straightforward.  In a model
where each site is in one of $n$ possible states $\in \{0, ...,
n-1\}$, there are $n - 1$ events $m \in \{1, ..., n-1\}$ per lattice
site which change the state of the site: $x_{i,j} \leftarrow (x_{i,j}
+ m) \, \mbox{mod} \, n$. This can be done with
Metropolis-type rates $r_{s \rightarrow s'} = \mbox{min} \{ 1,
\exp[-(H(s') - H(s))/(kT)]\}$ or symmetrical rates $r_{s \rightarrow s'} =
\exp[-(H(s') - H(s))/(2kT)]$. Both possibilities fulfil the detailed
balance condition \ref{balance} and yield identical
results.

A canonical simulation requires an algorithm where the number of
particles is fixed. In this case, in every event one particle jumps to
a different site. We choose a nonlocal dynamics where the range of
particle jumps is unlimited. This yields considerably faster
equilibration compared to a Kawasaki dynamics with nearest neighbour
diffusion only. For simplicity, we permit only jumps to a site where
the binding energy of the particle is independent of the state of its
initial site, i.e.\ we forbid jumps to nearest and next nearest
neighbour sites.  If the particle jumps from site $i$ to site $j$, the
energy difference between the final and the initial state is $\Delta H
= \Delta H_j - \Delta H_i$, where $\Delta H_{x}$ is the energy
difference of the system with site $x$ occupied and empty. The rates
\begin{equation}
r_{i \rightarrow j} = \exp \left[ \left( \Delta H_i - \Delta H_j
\right) / \left( 2 k T \right) \right]
\end{equation}
fulfil the detailed balance condition. Then, the probability for a
jump from site $i$ to site $j$ factorizes, i.e.\
$$
p_{i \rightarrow j} = p_{i}^- \cdot p_{j}^+ \; \; \; \mbox{where} \;
\; \; p_{x}^\pm = \frac{r_x^\pm}{\sum_x r_x^\pm}.
$$ 
Here, we have introduced the rate for deposition of a particle at site
$x$ ($r_{x}^+$) and for removal of a particle at site $x$
($r_x^-$). $r_x^+ = \exp[-\Delta H_x/(2kT)]$ if site $x$ is empty
and zero otherwise. Conversely, $r_x^- = \exp[\Delta H_x/(2kT)]$ on
occupied sites and zero on empty sites. Due to this factorization
property we can proceed in two steps: In the first step, we select
the site $i$ from which the particle starts with probability
$p_{i}^-$. Then, we select the site $j$ where the particle is landing
with probability $p_j^+$. If the distance between site $i$ and site
$j$ is $\geq 2$ the particle is moved. Otherwise, the event is
rejected and the system remains unchanged. Since the number of
rejected events is small on large systems, the loss of speed can be
neglected.

In our model of flat CdTe(001) surfaces only the number of Cd atoms is
fixed, while the number of Te dimers may change. This requires a more
elaborate algorithm which uses both canonical and grand-canonical
techniques. A Cd atom may jump to any site which is not occupied by a
Cd atom. If the arrival site is occupied by a dimer, the dimer is
destroyed. The removal of the Cd atom at the starting site creates a
pair of Te atoms. We consider both the case where these Te atoms
dimerize immediately and the case where they remain unbound.
Additionally, Te dimers may break up and dimers may be created on
empty sites. In the following, these processes will be denoted as
``dimer flips''. In each timestep, a random number is drawn to decide
whether a Cd jump with immediate formation of a dimer at the starting
site, a Cd jump without dimerization or a dimer flip will occur. The
probabilities are proportional to the sums of the rates of the
processes in each group. A $L \times N$ system contains $\mathcal{O}(L
N)$ Cd atoms and Te dimers. Thus, there are $\mathcal{O}(L N)$ dimer
flip events. Since there are $\mathcal{O}(L N)$ possible arrival sites
for each Cd atom, there are $\mathcal{O}(L^2 N^2)$ Cd jump events. To
keep the ratio of dimer flips and Cd jumps independent of the system
size, the dimer flips have been weighted with a prefactor $L N$. Then,
one event in the selected group is perfomed. For dimer flips, this is
done with the grand-canonical algorithm while Cd jumps are performed
by the canonical two-step algorithm.

\section{Application to kinetical Monte Carlo simulations}

In kinetical Monte Carlo simulations of processes like crystal growth,
the set of events $i \in \{1, ..., \mathcal{N}\}$ which can occur in a
particular state of the system and their rates $\rho_i$ are determined
by the physics of the investigated model.  Typically, this set of
events includes the deposition of adatoms at the surface, diffusion of
particles at the surface and desorption. In all full diffusion Monte
Carlo simulations which are presented in this thesis, we assume that
the rates of diffusion and desorption are given by the Arrhenius law
\ref{arrheniusgesetz}. For a physically reasonable choice of the
activation energies, the rates for diffusion processes fulfil the
detailed balance condition \ref{balance} such that a system where
there is neither adsorption nor desorption will finally reach thermal
equilibrium.








\germanTeX
\selectlanguage{ngerman}
\chapter*{Danksagung}
Ich m"ochte allen, die zum Gelingen dieser Arbeit beigetragen haben, 
herzlich danken, insbesondere
\begin{itemize}
\item Dr.\ Michael Biehl, der diese Arbeit betreut hat und 
sich immer Zeit nahm, auf meine Fragen einzugehen und "uber 
meine Ideen zu diskutieren.  
\item Prof.\ Dr.\ Wolfgang Kinzel, an dessen Lehrstuhl ich diese 
Arbeit anfertigen durfte.
\item Dr.\ Michael Biehl, Martin Kinne, Prof.\ Dr.\ Wolfgang Kinzel, 
Florian Much, Dr.\ Stefan Schinzer, Prof.\ Dr.\ Moritz Sokolowski und 
Thorsten Volkmann f"ur die fruchtbare Zusammenarbeit,
deren Ergebnisse wir gemeinsam ver"offentlicht haben.
\item Prof.\ Dr.\ Joachim Krug und seiner Arbeitsgruppe f"ur die 
Gastfreundschaft an der Universit"at Essen. 
\item Florian Much und Thorsten Volkmann f"ur das gr"undliche Korrekturlesen
dieser Arbeit.
\item Ansgar Freking und allen anderen Systembetreuern f"ur den guten 
Zustand unserer Computer.
\item allen Mitgliedern unserer Arbeitsgruppe f"ur das freundliche und 
kreative Arbeitsklima.  
\item der Deutschen Forschungsgemeinschaft, die diese Arbeit finanziert hat.
\end{itemize}

\begin{thebibliography}{99}


\bibitem{ab00} M. Ahr, M. Biehl. {\em Singularity spectra of rough
growing surfaces from wavelet analysis.} Phys. Rev. E {\bf 62} (2),
1773 (2000)

\bibitem{ab01} M. Ahr, M. Biehl. {\em Modelling sublimation and atomic
layer epitaxy in the presence of competing surface reconstructions.}
Surf. Sci. {\bf 488} L553 (2001)

\bibitem{ab02} M. Ahr, M. Biehl. {\em Flat (001) surfaces of II-VI
semiconductors: A lattice gas model.} Surf. Sci. In print. 

\bibitem{abkk00} M. Ahr, M. Biehl, M. Kinne, W. Kinzel. {\em The
influence of the crystal lattice on coarsening in unstable epitaxial
growth.} Surf. Sci. {\bf 465}, 339 (2000)

\bibitem{abv02} M. Ahr, M. Biehl, T. Volkmann. {\em Modelling (001)
surfaces of II-VI semiconductors.} Comp. Phys. Comm. In print.


\bibitem{amar99} J. G. Amar. {\em Mechanisms of mound coarsening in
unstable epitaxial growth.} Phys. Rev. B {\bf 60} (16), R11317 (1999)


\bibitem{adr99} A. Arn\'{e}odo, N. Decoster, S. G. Roux. {\em
Intermittency, Log-Normal Statistics, and Multifractal Cascade Process
in High-Resolution Satellite Images of Cloud Structure.} 
Phys. Rev. Lett. {\bf 83} (6) 1255 (1999)

\bibitem{adr00} A. Arn\'{e}odo, N. Decoster, S. G. Roux. {\em A
Wavelet-based method for multifractal image analysis. I. Methodology
and test appilications on isotropic and anisotropic random rough
surfaces.}  Eur. Phys. J. B {\bf 15} (3), 567 (2000)


\bibitem{a1889} S. Arrhenius. {\em \"{U}ber die
Reaktionsgeschwindigkeit bei der Inversion von Rohrzucker durch
S\"{a}uren.} Zeitschrift f\"{u}r physikalische Chemie {\bf 4}, 226 (1889)


\bibitem{bma92} E. Bacry, J. F. Muzy, A. Arn\'{e}odo. {\em
Singularity Spectrum of Fractal Signals from Wavelet Analysis: Exact
Results.}  Journ. Stat. Phys. {\bf 70} (3/4) 635 (1993)


\bibitem{bbjkvz92} A.-L. Barab\'{a}si, R. Bourbonnais, M. Jensen,
J. Kert\'{e}sz, T. Vicsek, Y.-C. Zhang. {\em Multifractality of
growing surfaces.} Phys. Rev. A {\bf 45} (10)
R6951, (1992)

\bibitem{bs95} A.-L. Barab\'{a}si, H. E. Stanley. {\it Fractal
Concepts in Surface Growth.} Cambridge University Press, Cambridge
(1995)


\bibitem{br88} S. A. Barnett, A. Rockett. {\em Monte Carlo simulations
of Si(001) growth and reconstruction during molecular beam epitaxy.}
Surf. Sci. {\bf 198}, 133 (1988)


\bibitem{ber86} N. C. Bartelt, T.L. Einstein, L.D. Roelofs. {\em
Transfer-matrix approach to estimating coverage discontinuities and
multicritical-point positions in two-dimensional lattice-gas phase
diagrams.} Phys. Rev. B {\bf 34}, 1616 (1986)




\bibitem{bkks99} M. Biehl, M. Kinne, W. Kinzel, S. Schinzer. {\em A
simple model of epitaxial growth: the influence of step edge
diffusion.} in {\em Proceedings of the 1998 Conference on
Computational Physics.} Comp. Phys. Comm. {\bf 121} - {\bf 122}, 347
(1999)

\bibitem{baksv01} M. Biehl, M. Ahr, W. Kinzel, M. Sokolowski,
T. Volkmann.  {\em A lattice gas model of II-VI(001) semiconductor
surfaces.} Europhys. Lett. {\bf 53} (2), 169 (2001)




\bibitem{bbs95} J. L. Blue, I. Beichl, F. Sullivan. {\em Faster Monte
Carlo simulations.} Phys. Rev. E {\bf 51} (2), R867 (1995)




\bibitem{bkl75} A. B. Bortz, M. H. Kalos, J. L. Lebowitz. {\em New
Algorithm for Monte Carlo Simulation of Ising Spin Systems.}
J. Comput. Phys. {\bf 17}, 10 (1975)


\bibitem{cf72} W. J. Camp, M. E. Fisher. {\em Decay of Order in
Classical Many-Body Systems. I. Introduction and Formal Theory.}
Phys. Rev. B {\bf 6} (3), 946 (1972)


\bibitem{cn01} C.-S. Chin, M. den Nijs. {\em Reconstructed rough
growing interfaces: Ridge line trapping of domain walls.} Phys. Rev. E
{\bf 64}, 031606 (2001)


\bibitem{ct97} J. Cibert, S. Tatarenko. {\em The Surface Structure of
a II-VI Compound: CdTe.} Defect and Diffusion Forum Vols. {\bf 150-151}, 1
(1997) and references therein


\bibitem{ccd88} H. J. Cornelissen, D. A. Gammack, R. J. Dalby. {\em
Reflection high-energy electron diffraction observations during growth
of ZnS$_x$Se$_{1-x}$ $(0 \leq x \leq 1)$ by molecular-beam epitaxy.}
J. Vac. Sci. Technol. B {\bf 6} (2), 769 (1988)


\bibitem{dslkg96} S. Das Sarma, C. J. Lanczycki, R. Kotlyar, S. V.
Ghaisas. {\em Scale invariance and dynamical correlations in growth
models of molecular beam epitaxy.} Phys. Rev. E {\bf 53} (1), 359 (1996)

\bibitem{dsp97} S. Das Sarma, P. Punyindu. {\em Dynamic scaling in
a $(2 + 1)$-dimensional limited mobility model of epitaxial
growth.} Phys. Rev. E {\bf 55} (5), 5361 (1997).

\bibitem{sp00} S. Das Sarma, P. Punyindu. {\em A discrete model for
non-equilibrium growth under surface diffusion bias.} Surf. Sci. {\bf
424} (2--3), L339 (1999)


\bibitem{dkdds97} C. Dasgupta, J. M. Kim, M. Dutta, S. Das Sarma. {\em
Instability, intermittency, and multiscaling in discrete growth models
of kinetic roughening.}  Phys. Rev. E {\bf 55} (3), 2235 (1997)


\bibitem{dbt96} B. Daudin, D. Brun-Le Cunff, S. Tatarenko.  {\em
Stoichiometry determination of the Te-rich (100) CdTe and (100) ZnTe
surfaces.} Surf. Sci. {\bf 352--354}, 99 (1996)

\bibitem{dtb95} B. Daudin, S. Tatarenko, D. Brun-Le Cunff. {\em
Surface stoichiometry determination using reflection high-energy
electron diffraction and atomic-layer epitaxy: The case of ZnTe(100).}
Phys. Rev. B {\bf 52} (11), 7822 (1995)


\bibitem{dra00} N. Decoster, S. G. Roux, A. Arn\'{e}odo. {\em A
Wavelet-based method for multifractal image analysis. II. Applications
to synthetic multifractal rough surfaces.}  Eur. Phys. J. B {\bf 15}
(4), 739 (2000)


\bibitem{n90} M. den Nijs. {\em Preroughening of Crystal Surfaces and
Energy Differences between Inside and Outside Corners.}
Phys. Rev. Lett. {\bf 64} (4), 435 (1990)


\bibitem{ev94} I. Elkinani, J. Villain. {\em Growth roughness and
instabilities due to the Schwoebel effect: a one-dimensional model.}
J. Phys. I France {\bf 4}, 949 (1994)


\bibitem{e91} J. W. Evans. {\em Factors mediating smoothness in
epitaxial thin-film growth.} Phys. Rev. B {\bf 43} (5), 3897 (1991)

\bibitem{estp90} J. W. Evans, D. E. Sanders, P. A. Thiel,
A. E. DePristo. {\em Low-temperature epitaxial growth of thin metal
films.} Phys. Rev. B {\bf 41} (8), 5410 (1990)


\bibitem{fv85} F. Family, T. Vicsek. {\em Scaling of the active zone
in the Eden process on percolation networks and the ballistic
deposition model.} J. Phys. A {\bf 18}, L75 (1985)


\bibitem{fs90} W. Faschinger, H. Sitter. {\em Atomic-layer epitaxy of
(100) CdTe on GaAs substrates.} Journ. Cryst. Growth {\bf 99}, 566
(1990)


\bibitem{f98} P. J. Feibelman. {\em Interlayer Self-Diffusion on
Stepped Pt(111).} Phys. Rev. Lett. {\bf 81} (1), 168 (1998)


\bibitem{gn94} A. Garcia, J. Northrup. {\em First-principles study of
Zn- and Se-stabilized ZnSe(001) surface reconstructions.}
J. Vac. Sci. Technol. B {\bf 12}, 2678 (1994)


\bibitem{gs91} C. M. Gilmore, J. A. Sprague. {\em Molecular-dynamics
simulation of the energetic deposition of Ag thin films.} Phys. Rev. B
{\bf 44} (16), 8950 (1991)


\bibitem{g97} S. Gundel. {\em {\em ab initio}-Simulationen von
II-VI-Halbleitern und ZnSe-Oberfl\"{a}chen.} Di\-plom\-ar\-beit,
Universit\"{a}t W\"{u}rzburg (1997)

\bibitem{gffh99} S. Gundel, A. Fleszar, W. Faschinger, W. Hanke.  {\em
Atomic and electronic structure of the CdTe(001) surface: LDA and GW
calculations.} Phys. Rev. B {\bf 59} (23), 15261 (1999)


\bibitem{h82} C. J. Hamer. {\em Magnetisations from finite-size
scaling.} J. Phys. A {\bf 15}, L675 (1982)

\bibitem{h83} C. J. Hamer. {\em Latent heats from finite-size scaling.}
J. Phys. A {\bf 16}, 3085 (1983)


\bibitem{h79} W. A. Harrison. {\em Theory of polar semiconductor
surfaces.} J. Vac. Sci. Technol. {\bf 16} (5), 1492 (1979)


\bibitem{hs96} M. A. Herman, M. Sitter. {\em Molecular Beam
Epitaxy. Fundamentals and Current Status.} 2d ed.
Springer-Verlag Berlin Heidelberg New York (1996)


\bibitem{ibajjv98} M. Itoh, G. R. Bell, A. R. Avery, T. S. Jones,
B. A. Joyce, D. D. Vvedensky. {\em Island Nucleation and Growth on
Reconstructed GaAs(001) Surfaces.} Phys. Rev. Lett. {\bf 81} (3), 633
(1998)

\bibitem{itoh99} M. Itoh. {\em Atomic-scale homoepitaxial growth
simulations of reconstructed III-V surfaces.} PhD thesis, Imperial
College of Science, London (1999) 


\bibitem{jjs94} J. Jacobsen, K. W. Jacobsen, P. Stoltze. {\em
Nucleation of the Pt(111) reconstruction: a simulation study.}
Surf. Sci. {\bf 317}, 8 (1994)


\bibitem{kscm99} M. Kalff, P. \v{S}milauer, G. Comsa, T. Michely. {\em
No coarsening in Pt(111) homoepitaxy.} Surf. Sci. {\bf 426}, L447
(1999)


\bibitem{kw88} J. Kert\'esz, D. E. Wolf. {\em Noise reduction in Eden
models: II. Surface structure and intrinsic width.} J. Phys. A:
Math. Gen. {\bf 21}, 747 (1988)


\bibitem{k99} M. Kinne. {\em Computersimulationen von epitaktischen
Wachstumsprozessen: Ein einfaches Modell.} Diplomarbeit,
Universit\"{a}t W\"{u}rzburg (1999)


\bibitem{ks81} W. Kinzel, M. Schick. {\em Extent of exponent variation
in a hard-square lattice gas with second-neighbor repulsion.}
Phys. Rev. B {\bf 24}, 324 (1981)


\bibitem{ksb82} W. Kinzel, W. Selke, K. Binder. {\em Phase transitions
on centred rectangular lattice gases: A model for the adsorption of H
on Fe(110)} Surf. Sci. {\bf 121}, 13 (1982)


\bibitem{knuth} D. E. Knuth. {\em The Art of Computer Programming.} 2d
ed. Addison-Wesley, Reading, Massachusetts (1981)


\bibitem{kms99} P. Kratzer, C. G. Morgan, M. Scheffler. {\em Model for
nucleation in GaAs homoepitaxy derived from first principles.}
Phys. Rev. B {\bf 59} (23), 15246 (1999)

\bibitem{kps01} P. Kratzer, E. Penev, M. Scheffler. {\em
First-principles studies of kinetics in epitaxial growth of III-V
semiconductors.} Preprint, cond-mat/0105430

\bibitem{ks02} P. Kratzer, M. Scheffler. {\em Reaction-Limited Island
Nucleation in Molecular Beam Epitaxy of Compound Semiconductors.}
Phys. Rev. Lett. {\bf 88} (3), 036102 (2002)


\bibitem{k94} J. Krug. {\em Turbulent interfaces.}
Phys. Rev. Lett. {\bf 72} (18) 2907 (1994)

\bibitem{ks95} J. Krug, M. Schimschak. {\em Metastability of Step Flow
Growth in $1 + 1$ Dimensions.} J. Phys. I (France) {\bf 5}, 1065
(1995)


\bibitem{lbadebt00} V. P. LaBella, D. W. Bullock, M. Anser, Z. Ding,
C. Emery, L. Bellaiche, P. M. Thibado. {\em Microscopic View of a
Two-Dimensional Lattice-Gas Ising System within the Grand Canonical
Ensemble.} Phys. Rev. Lett. {\bf 84} (18), 4152 (2000)


\bibitem{l99} J. M. L\'{o}pez. {\em Scaling Approach to Calculate
Critical Exponents in Anomalous Surface Roughening.}
Phys. Rev. Lett. {\bf 83} (22) 4594 (1999)

\bibitem{lopezmail} J. M. L\'{o}pez, private communication. 

\bibitem{lr96} J. M. L\'{o}pez, M. A. Rodr\'{\i}guez. {\em Lack of
self-affinity and anomalous roughening in growth processes.}
Phys. Rev. E {\bf 54} (3) R2189 (1996)

\bibitem{lrc97} J. M. L\'{o}pez, M. A. Rodr\'{\i}guez, R. Cuerno. {\em
Super-roughening versus intrinsic anomalous scaling of surfaces.}
Phys. Rev. E {\bf 56} (4) 3993 (1997)


\bibitem{m88} P. A. Maksym. {\em Fast Monte Carlo simulation of MBE
growth.} Semicond. Sci. Technol. {\bf 3}, 594 (1988)


\bibitem{mh92} S. Mallat, W. L. Hwang. {\em Singularity Detection and
Processing with Wavelets.} IEEE Transactions on Information Theory
{\bf 38} (2), 617 (1992)

\bibitem{mz92} S. Mallat, S. Zhong. {\em Characterization of Signals
from Multiscale Edges.} IEEE Transactions on Pattern Analysis and
Machine Intelligence, {\bf 14} (7), 710 (1992)


\bibitem{megm98} D. Martrou, J. Eymery, P. Gentile, N. Magnea.  {\em
Epitaxial growth of CdTe(001) studied by scanning tunneling
microscopy.} Journ. Cryst.  Growth {\bf 184/185}, 203 (1998)

\bibitem{mem99} D. Martrou, J. Eymery, N. Magnea. {\em Equilibrium
Shape of Steps and Islands on Polar II-VI Semiconductors Surfaces.}
Phys. Rev. Lett.  {\bf 83} (12), 2366 (1999)


\bibitem{mg00} D. Moldovan, L. Golubovic. {\em Interfacial coarsening
dynamics in epitaxial growth with slope selection.} Phys. Rev. E {\bf
61} (6), 6190 (2000)


\bibitem{m00} F. Much. {\em Strukturbildung in der Heteroepitaxie.}
Diplomarbeit, Universit\"{a}t W\"{u}rzburg (2000)

\bibitem{mabk01} F. Much, M. Ahr, M. Biehl, W. Kinzel. {\em Kinetic
Monte Carlo simulations of dislocations in heteroepitaxial growth.}
Europhys. Lett. {\bf 56} (6), 791 (2001)


\bibitem{mba93} J. F. Muzy, E. Bacry, A. Arn\'{e}odo. {\em
Multifractal formalism for fractal signals: The structure-function
approach versus the wavelet-transform modulus-maxima method.}
Phys. Rev. E {\bf 47} (2) 875, (1993)


\bibitem{nskts00} H. Neureiter, S. Schinzer, W. Kinzel, S. Tatarenko,
M. Sokolowski.  {\em Simultaneous layer-by-layer and step-flow
sublimation on the CdTe(001) surface derived from a diffraction
analysis.} Phys. Rev. B {\bf 61} (8), 5408 (2000)

\bibitem{n98} H. Neureiter. {\em Struktur und Kinetik polarer
II-VI-Halbleiteroberfl\"{a}chen: Experimentelle Untersuchungen an
CdTe(001).} PhD thesis, Universit\"{a}t W\"{u}rzburg (1998)

\bibitem{ntss00} H. Neureiter, S. Tatarenko, S. Spranger,
M. Sokolowski. {\em Domain wall formation at the $c(2 \times
2)$-$(2\times 1)$ phase transition of the CdTe(001) surface. }
Phys. Rev. B. {\bf 62} (4), 2542 (2000)


\bibitem{nb99} M. E. J. Newman, G. T. Barkema. {\em Monte Carlo
Methods in Statistical Physics.} Clarendon Press, Oxford (1999)


\bibitem{n82} P. Nightingale. {\em Finite-size scaling and
phenomenological renormalization.} J. Appl. Phys. {\bf 53} (11), 7927
(1982)

\bibitem{na78} M. P. Nightingale. {\em Phenomenological Renormalization
Group Theory}. PhD Thesis, University of Amsterdam (1978)


\bibitem{pc94} C. H. Park, D. J. Chadi. {\em First-principles study of
the atomic reconstructions of ZnSe(100) surfaces.} Phys. Rev. B {\bf
49}, 16467 (1994)


\bibitem{p89} M. D. Pashley. {\em Electron counting model and its
application to island structures on molecular-beam epitaxy grown
GaAs(001) and ZnSe(001).} Phys. Rev. B {\bf 40}, 10481 (1989)


\bibitem{ppct98} P. Peyla, A. Pimpinelli, J. Cibert,
S. Tatarenko. {\em Deposition and growth with desorption for CdTe
molecular beam epitaxy.} Journ. Cryst. Growth {\bf 184/185}, 75 (1998)


\bibitem{pv97} A. Pimpinelli, J. Villain. {\em Physics of Crystal
Growth.} Cambridge University Press, Cambridge (1998)

\bibitem{pv94} A. Pimpinelli, J. Villain. {\em What does an
evaporating surface look like?} Physica A {\bf 204}, 521 (1994)


\bibitem{psk93} M. Plischke, M. Siegert, J. Krug. {\em Surface
diffusion currents and the universality classes of growth.}
Phys. Rev. Lett. {\bf 70} (21), 3271 (1993)


\bibitem{p97} P. Politi. {\em Different Regimes in the
Ehrlich-Schwoebel Instability.} J. Phys. I France {\bf 7}, 797 (1997)

\bibitem{pgmpv00} P. Politi, G. Grenet, A. Marty, A. Ponchet,
J. Villain. {\em Instabilities in crystal growth by atomic or
molecular beams.} Phys. Rep. {\bf 324}, 271 (2000)

\bibitem{pv96} P. Politi, J. Villain. {\em Ehrlich-Schwoebel
instability in molecular-beam epitaxy: A minimal model.} Phys. Rev. B
{\bf 54} (7), 5114 (1996) 


\bibitem{pon90} V. Pontikis. {\em Thermally Activated Processes in
Solids.} in A. L. Laskar (ed.) {\em Diffusion in Materials}, Kluwer
(1990)


\bibitem{pf83} V. Privman, M. E. Fisher. {\em Finite-Size Effects at
First-Order Transitions.} Journ. Stat. Phys. {\bf 33} (2), 385 (1983)

\bibitem{ps82} V. Privman, L. S. Schulman. {\em Analytic Continuation
at First-Order Phase Transitions.} Journ. Stat. Phys. {\bf 29} (2),
205 (1982)


\bibitem{rlr00} J. J. Ramasco, J. M. L\'{o}pez, M. A. Rodriguez. {\em
Generic Dynamic Scaling in Kinetic Roughening.}  Phys. Rev. Lett. {\bf
84} (10), 2199 (2000)


\bibitem{reif} F. Reif. {\em Statistische Physik und Theorie der
W\"{a}rme.} de Gruyter, Berlin, New York (1987)


\bibitem{rkgk83} P. A. Rikvold, W. Kinzel, J. D. Gunton,
K. Kaski. {\em Finite-size-scaling study of a two-dimensional
lattice-gas model with a tricritical point.} Phys. Rev. B {\bf 28}
(5), 2686 (1983)


\bibitem{r90} A. Rockett. {\em The influence of surface structure on
growth of Si(001) $2\times 1$ from the vapor phase.} Surf. Sci. {\bf
227}, 208 (1990)


\bibitem{rk96} M. Rost, J. Krug. {\em Coarsening of Surface Structures
in Unstable Epitaxial Growth.} Phys. Rev. E {\bf 55} (4), 3952
(1997)


\bibitem{rad99} S. G. Roux, A. Arn\'{e}odo, N. Decoster. {\em A
Wavelet based method for multifractal image analysis. III. Application
to high-resolution sattelite images of cloud structure.}
Eur. Phys. J. B {\bf 15} (4), 765 (2000)


\bibitem{saan99} M. A. Salmi, M. Alatalo, T. Ala-Nissila,
R. M. Nieminen. {\em Energetics and diffusion paths of gallium and
arsenic adatoms on flat and stepped GaAs(001) surfaces.}
Surf. Sci. {\bf 425}, 31 (1999)


\bibitem{s80} M. Schick. in {\em Phase Transitions in Surface Films.}
by J. G. Dash and J. Ruvalds (eds.) Plenum Press, New York (1980)


\bibitem{s99} A. Schindler. {\em Theoretical aspects of growth on one
and two dimensional strained crystal surfaces.} PhD thesis,
Gerhard-Mercator-Universit\"{a}t Duisburg (1999)


\bibitem{schi99} S. Schinzer. {\em Kinetic Monte-Carlo Simulations of
Crystal Surfaces: Applications to II-VI Semiconductors.} PhD thesis,
Universit\"{a}t W\"{u}rzburg (1999)

\bibitem{skbk99} S. Schinzer, M. Kinne, M. Biehl, W. Kinzel. {\em The
role of step edge diffusion in epitaxial crystal growth.} Surf.
Sci. {\bf 439}, 191 (1999)

\bibitem{sk98} S. Schinzer, W. Kinzel. {\em Modeling sublimation by
computer simulation: morphology-dependent effective energies.}
Surf. Sci. {\bf 401} (1), 96 (1998)

\bibitem{skr00} S. Schinzer, S. K\"{o}hler, G. Reents. {\em
Ehrlich-Schwoebel barrier controlled slope selection in epitaxial
growth.} Eur. Phys. J. B {\bf 15}, 161 (2000)

\bibitem{ssbk99} S. Schinzer, M. Sokolowski, M. Biehl, W.
Kinzel. {\em Unconventional MBE strategies from computer simulations
for optimized growth conditions.} Phys. Rev. B {\bf 60}, 2893 (1999)


\bibitem{s90} M. Schroeder. {\it Fractals, Chaos, Power Laws: Minutes
from an Infinite Paradise.} Freeman, New York (1990)


\bibitem{s69} R. L. Schwoebel. {\em Step Motion on Crystal
Surfaces. II.} Journ. Appl. Phys. {\bf 40} (2), 614 (1969)

\bibitem{ss66} R. L. Schwoebel, E. J. Shipsey. {\em Step Motion on
Crystal Surfaces.} Journ. Appl. Phys. {\bf 37} (10), 3682 (1966)


\bibitem{sfjetbd95} L. Seehover, G. Falkenberg, R. L. Johnson,
V. H. Etgens, S. Tatarenko, D. Brun, B. Daudin.  {\em Scanning
tunneling microscopy study of CdTe(001).} Appl. Phys. Lett. {\bf 67}
(12), 1680 (1995)


\bibitem{sbk83} W. Selke, K. Binder, W. Kinzel. {\em Lattice gas
models with competing interactions.} Surf. Sci. {\bf 125}, 74 (1983)


\bibitem{s98} M. Siegert. {\em Coarsening Dynamics of Crystalline Thin
Films.} Phys. Rev. Lett. {\bf 81} (25), 5481 (1998)

\bibitem{spz97} M. Siegert, M. Plischke, R. K. P. Zia. {\em Contrasts
between coarsening and relaxational dynamics of surfaces.}
Phys. Rev. Lett.  {\bf 78} (19), 3705 (1997)


\bibitem{sb85} J. Singh, K. K. Bajaj. {\em Theoretical studies of the
intrinsic quality of GaAs/AlGaAs interfaces grown by MBE: Role of
kinetic processes.} J. Vac. Sci. Technol. B {\bf 3} (2), 520 (1985)

\bibitem{sm83} J. Singh, A. Madhukar. {\em Prediction of Kinetically
Controlled Surface Roughening: A Monte Carlo Computer-Simulation
Study.} Phys. Rev. Lett. {\bf 51} (9), 794 (1983)


\bibitem{sb93} J. Stoer, R. Bulirsch. {\em Introduction to
numerical analysis.} Springer-Verlag, New York (1993)


\bibitem{ss94} R. Stumpf, M. Scheffler. {\em Theory of Self-Diffusion
at and Growth of Al(111).} Phys. Rev. Lett. {\bf 72} (2), 254 (1994)


\bibitem{si00} I. Suemune, A. Ishibashi (eds.). {\em Proceedings of
the Ninth International Conference on II-VI Compounds, Kyoto 1999.}
J. Cryst. Growth {\bf 214-215} (2000)


\bibitem{t98} L.-H. Tang. {\em Unstable Growth and Coarsening in
Molecular-Beam Epitaxy.} Physica A {\bf 254}, 135 (1998)

\bibitem{tsv97} L.-H. Tang, P. \v{S}milauer, D. D. Vvedensky. {\em
Noise-assisted mound coarsening in epitaxial growth.}  Eur. J. Phys. B
{\bf 2}, 409 (1998)


\bibitem{tdbev94} S. Tatarenko, B. Daudin, D. Brun, V. H. Etgens, M.
B. Veron. {\em Cd and Te desorption from (001), (111)B, and (110)
CdTe surfaces.} Phys. Rev. B {\bf 50} (24), 18479 (1994)


\bibitem{t99} J. M. Thijssen. {\em Computational Physics.} Cambridge
University Press, Cambridge (1999)


\bibitem{wave} J. C. Van Den Berg. {\it Wavelets in Physics.} Cambridge
University Press, Cambridge (1999)


\bibitem{vsh84} J. A. Venables, G. D. T. Spiller,
M. Hanb\"{u}cken. {\em Nucleation and growth of thin films.}
Rep. Prog. Phys. {\bf 47}, 399 (1984)


\bibitem{vadt96} M. B. Veron, A. Arnoult, B. Daudin,
S. Tatarenko. {\em Reversibility of the elementary mechanisms of
atomic-layer epitaxy and sublimation of (001) CdTe.} Phys. Rev. B {\bf
54} (8), R5267 (1996)


\bibitem{vptw92} J. Villain, A. Pimpinelli, L. Tang, D. Wolf. {\em
Terrace sizes in molecular beam epitaxy.}
J. Phys. I (France) {\bf 2}, 2107 (1992)


\bibitem{v92} T. Vicsek. {\it Fractal Growth Phenomena.} World
Scientific, Singapore (1992)


\bibitem{v00} T. Volkmann. {\em Anisotropie-Effekte an
Halbleiteroberfl\"{a}chen.} Diplomarbeit, Universit\"{a}t W\"{u}rzburg
(2000)


\bibitem{wr91} J. Wang, A. Rockett. {\em Simulating diffusion on
Si(001) $2 \times 1$ surfaces using a modified interatomic potential.}
Phys. Rev. B {\bf 43} (15), 12571 (1991)


\bibitem{w95} D. E. Wolf. {\em Computer simulation of molecular beam
epitaxy.} In M. Droz, A. J. McKane, J. Vannimenus, D. E. Wolf (eds.)
{\em Scale Invariance, Interfaces, and Non-Equilibrium Dynamics.}
NATO-ASI Series, Plenum Press, New York (1994) 


\bibitem{wewr00} D. Wolfframm, D. A. Evans, D.I. Westwood, J. Riley.
{\em A detailed surface phase diagram for ZnSe MBE growth and
ZnSe/GaAs(001) interface studies.} Journ.  Cryst. Growth {\bf 216}, 119
(2000)


\bibitem{yhd98} Y. Yue, Y. K. Ho, Z. Y. Pan. {\em Molecular-dynamics
study of transient-diffusion mechanisms in low-temperature epitaxial
growth.} Phys. Rev. B {\bf 57} (11), 6685 (1998)

\end{thebibliography}
\end{document}